 \documentclass[final,3p,times]{elsarticle}
\usepackage{amsmath,hyperref}
\usepackage{lineno}

\usepackage{epstopdf}
\journal{*****}

\usepackage{lineno,hyperref}
\usepackage{graphicx}
\usepackage{dcolumn}
\usepackage{bm}
\usepackage{epsfig}
\usepackage{booktabs}
\usepackage{subfigure}
\usepackage{graphics}
\usepackage{amssymb}
\usepackage{amsmath}
\usepackage{array}
\usepackage{color}
\usepackage{booktabs}
\usepackage{multirow}
\usepackage{caption}
\usepackage{chngpage}
\usepackage{subfigure}
\usepackage{mathrsfs,gensymb,float}

\biboptions{numbers,sort&compress}
\modulolinenumbers[1]

\begin{document}

\captionsetup[figure]{labelfont={bf},name={Fig.},labelsep=period}        

\begin{frontmatter}
	
	\title{Three-dimensional study of double droplets impact on a wettability-patterned surface}
	\author[1,2]{Jiangxu Huang}
	\author[1,2]{Kun He}
	\author[1,2]{Lei Wang\corref{mycorrespondingauthor}}
	\cortext[mycorrespondingauthor]{Corresponding author}
	\ead{leiwang@cug.edu.cn}

	\address[1]{School of Mathematics and Physics, China University of Geosciences, Wuhan 430074, China}
	\address[2]{Center for Mathematical Sciences, China University of Geosciences, Wuhan 430074, China}
	

	
\begin{abstract}
The directional movement and rebound behaviors of two droplets simultaneously impacting a designed flat surface with wettability difference is investigated based on the three-dimensional multi-relaxation-time pseudopotential lattice Boltzmann model. The effects of several factors, such as wettability difference, Weber number and droplet spacing on the directional movement and rebound behaviors are investigated in detail. The numerical results show that the unbalanced Young 's force caused by the wetting difference will cause the droplets to rebound or migrate laterally toward to the side with lower hydrophobicity on the surface, and the contact time of the droplets is found to decrease with the increase of the wetting difference. In addition, it is noted that there exists a secondary spreading behavior in the case of a lower Weber number, which in turn leads to an increase in contact time.  Further, as far as the influence of the  droplet spacing is concerned, we found that the coalescence intensity of the droplets decreases with the increase of droplet spacing, and in particular, the coalescing droplets are found to divide into two sub-droplets during asymmetric contraction, and three detachment patterns are then defined to reveal the effects of the  droplet spacing. 

\end{abstract}

	
\end{frontmatter}

\section{Introduction}
Spontaneous directional movement of impinging droplet is widely existed in the nature and industrial applications, such as water collection by cactus spines and spider silk \cite{JuNC2012, ZhengNature2010}, inkjet printing \cite{GAnAM2004}, spray cooling \cite{HoracekIJHMT2005}, microfluidics chips \cite{BhatiaNB2014} and self-cleaning surfaces \cite{TutejaScience2007,MishchenkoAN2010, DengScience2012}. Due to its fundamental effect in many practical applications, the investigation of impinging droplet has attracted increasing interests from the scientific community. Among various droplet self-propulsion strategies, such as gradients of chemical \cite{YangLangmuir2008}, employing external force fields\cite{ZhouLangmuir2021, TakedaSS2012, DouSR2021, BenAS2019}, structural topography \cite{LinkePRL2006, LiuAN2014}, temperature \cite{WangACS2016}, mechanical vibration \cite{DanielLangmuir2005}, PH-induced \cite{BannoLangmuir2012,LagziACS2010} and gradient wetting surfaces and so on, the surfaces with wettability gradient have attracted considerable attention due to their ability to control the directional spreading or movement of droplets without external mechanical or auxiliary force fields. Therefore, it is great significance for promoting the industrial manufacturing of wetting gradient to obtain the wetting mechanism of droplets on the surface with wettability difference. 

In recent years, some experimental and numerical studies \cite{MockJOPCM2005,VaikuntanathanCSAPEA2010,KimJFM2013,RamanIJHMT2016,ZhangLangmuir2016,JiLangmuir2021,YuanPRF2021} have been conducted to investigate the droplet impacting a gradient wetting surface. Mock et al. \cite{MockJOPCM2005} experimentally studied the droplets impacting on a superhydrophobic region with a circular hydrophilic zone and found that the droplets eventually cover the hydrophilic region completely if the droplet touches the hydrophilic region during the spreading phase. Vaikuntanathan et al. \cite{VaikuntanathanCSAPEA2010} presented an experimental analysis on the impact of droplet onto the junction line between the hydrophobic (textured) and hydrophilic (smooth) portions of a dual-textured surface. The experimental results suggest that the droplet will migrate toward the hydrophilic smooth portions. Kim et al. \cite{KimJFM2013} conducted experiments of impacting on superhydrophobic background with a superhydrophilic annulus and observed that the remaining part of the droplet forms a water ring in the hydrophilic region after splashing. Ashoke Raman et al. \cite{RamanIJHMT2016} used the phase-field lattice Boltzmann (LB) method to study a droplet impingement on a surface with varying wettabilities, and found that wettability difference resulted in droplet exhibits the inchworm type motion due to transfer of the vertical momentum of the droplet in the horizontal direction. Zhang et al. \cite{ZhangLangmuir2016} numerical investigate droplet impacting dynamics on a nonuniform superhydrophobic textured surface with a wettability gradient. They concluded that the trajectory of the droplet rebounce is the result of the competition between the lateral recoil of the liquid and the penetration and capillary emptying of the liquid in the vertical direction. More recently, Ji et al. \cite{JiLangmuir2021} numerically studied the rebound mechanism of a drop impacting a hydrophobic and chemically heterogeneous surface based on the multiphase LB method. Results indicated that the main reason for the lateral rebounding is the asymmetry between the left and right unbalanced Young’s forces on the heterogeneous surface and this asymmetry force exists in the whole process of the drop contacting the heterogeneous surface. Yuan et al. \cite{YuanPRF2021} explored the effects of the surface wettability difference, Weber number (We), and offset impinging on droplet asymmetric spreading and direction migration using the coupled level-set and volume of fluid (CLSVOF) method. It also found that the asymmetric spreading behavior of the droplet was caused by the unbalanced net force in the lateral direction.

Although the impact dynamics of a single droplet have been thoroughly studied in the preceding researchs, multi-droplet impact is more representative due to its more cloase to the practical conditions in industrial applications. In this setting, several investigations \cite{FujimotoIJMF2001,RayIJNMHFF2013,GrahamPOF2012,RamanCES2016,GaoLangmuir2021,WangAIPA2021} have been conducted to understand the interaction dynamics of multi droplets impacting the homogeneous surface. Fujimoto et al. \cite{FujimotoIJMF2001} experimentally investigated the influence of spacing between drops impacting successively on solid substrates. Ray et al. \cite{RayIJNMHFF2013} performed axisymmetric computations of consecutive droplets impact on a liquid pool and investigated the effect of time separation, ratio of drop size and drop impact velocity. They found that the time separation determines whether the drops coalesce before impact or second drop impacts after the first drop. Graham et al. \cite{GrahamPOF2012} experimentally and numerically investigated the coalescence of two droplets on a solid surface with various wettabilities and a correlation equation is proposed to predict the maximum diffusion length of merged droplets on different surfaces under different collision conditions. Raman et al. \cite{RamanCES2016} conducted three dimensional numerical simulations to investigate the impact dynamics of double impacting droplets on a solid surface and the result show that increase in advancing contact angle leads to greater recoiling of the droplet system for in-phase interactions. Gao et al. \cite{GaoLangmuir2021} investigated the impact and rebound dynamics of multiple droplets simultaneously impacting a superhydrophobic surface, they found that three different rebound morphologies emerged due to the strong interactions generated by the central droplets and the central ridges. Wang et al. \cite{WangAIPA2021} studied the energy conversion of double droplets impacting on superhydrophobic surface based on numerical simulation. They found that the coalescence between the two impacting droplets leads to the release of surface energy and the viscous dissipation strongly depends on the coalescence strength.

From the published literature, it can be found that when multiple droplets impact a solid surface, they may contact with each other during diffusion, thus causing coalescence. Then the internal velocity distribution and exterior appearance of the droplet will invariably change as a result of agglomeration. Therefore, the multiple-droplet impact is not a simple superposition of the individual single droplet impact \cite{BrianEF2019}. However, previous research on droplets impacting non-uniformly wetted surfaces usually focus on an indicidual droplet \cite{XiongCF2018,CuiCF2021}, and the dynamics of multiple droplets impacting on a  gradient-wetted surface still poorly understand. Framed in this general background, in the present work, the impacting dynamics for two same droplets simultaneously impacting flat surface with wettability difference is numerically studied by using an three-dimensional multiple-relaxation time (MRT) pseudopotential LB method, which has been widely used in previous studies of droplet impact dynamics problems due to its remarkable computational efficiency, clear representation of the underlying microscopic physics, simplicity and versatility \cite{ChenIJHMT2014}.

The remainder of the paper is organized as follows. In the following section, the three-dimensional MRT pseudopotential LB model is introduced. Section 3 describes the problem of double droplet impact on the flat surface with wettability difference. In Section 4, the three-dimensional MRT pseudopotential LB model is validated with three tests. The numerical results and discussion are presented in Section 5. Finally, we provide a brief conclusion in Section 6.

\section{Three-dimensional MRT pseudopotential lattice Boltzmann model}
\label{section2}
	In order to improve the stability of the model and obtain higher density ratios and low viscosities, a three-dimensional fifteen-velocity (D3Q15)  pseudopotential LB model with a multiple-relaxation-time (MRT) collision operator
	with a tunable surface tension source term is adopted in this work, which is defined as \cite{McCrackenPRE2005, YuPRE2010, LallemandPRE2000}
	\begin{equation}
	\begin{aligned}
	f_{\alpha}\left(x+\boldsymbol{e_{\alpha}} \Delta t, t+\Delta t\right)=f_{\alpha}(x, t)-\left.\sum_{\alpha} \bar{\Lambda}_{\alpha \beta}\left(f_{\beta}-f_{\beta}^{\mathrm{eq}}\right)\right|_{(x, t)} -\left.\sum_{j}\left(I_{\alpha \beta}-\bar{\Lambda}_{\alpha \beta}\right) \bar{S}_{\beta} \Delta t\right|_{(x, t)}+C_{\alpha}
	\end{aligned} ,
	\label{Eq1}
	\end{equation}	
	where $ f_{\alpha}(\mathbf{x},t) $ is the density distribution function with velocity at spatial position $ 	\mathbf{x} $ and time $ t $ , $ f_{\beta}^{\mathrm{eq}} $ is the equilibrium distribution function,  $ \boldsymbol{e_{\alpha}} $ is the discrete velocity along the $ \alpha $th direction given by
	\begin{equation}
		\mathbf{e}_{\alpha}=\left[\begin{array}{ccccccccccccccc}
			0 & 1 & -1 & 0 & 0 & 0 & 0 & 1 & -1 & 1 & -1 & 1 & -1 & 1 & -1 \\
			0 & 0 & 0 & 1 & -1 & 0 & 0 & 1 & 1 & -1 & -1 & 1 & 1 & -1 & -1 \\
			0 & 0 & 0 & 0 & 0 & 1 & -1 & 1 & 1 & 1 & 1 & -1 & -1 & -1 & -1
		\end{array}\right] ,
	\end{equation}
	$ \Delta t $ is the time increment, $ \mathbf{I} $ is the unit matrix, and $\overline{\boldsymbol{\Lambda}}=\mathbf{M}^{-1} \boldsymbol{\Lambda} \mathbf{M}$ is the collision matrix, in which $ \mathbf{M} $ is an orthogonal transformation matrix given by \cite{dHumieresSA2002}
	\begin{equation}
	\mathbf{M}=\left[\begin{array}{ccccccccccccccc}
	1 & 1 & 1 & 1 & 1 & 1 & 1 & 1 & 1 & 1 & 1 & 1 & 1 & 1 & 1 \\
	-2 & -1 & -1 & -1 & -1 & -1 & -1 & 1 & 1 & 1 & 1 & 1 & 1 & 1 & 1 \\
	16 & -4 & -4 & -4 & -4 & -4 & -4 & 1 & 1 & 1 & 1 & 1 & 1 & 1 & 1 \\
	0 & 1 & -1 & 0 & 0 & 0 & 0 & 1 & -1 & 1 & -1 & 1 & -1 & 1 & -1 \\
	0 & -4 & 4 & 0 & 0 & 0 & 0 & 1 & -1 & 1 & -1 & 1 & -1 & 1 & -1 \\
	0 & 0 & 0 & 1 & -1 & 0 & 0 & 1 & 1 & -1 & -1 & 1 & 1 & -1 & -1 \\
	0 & 0 & 0 & -4 & 4 & 0 & 0 & 1 & 1 & -1 & -1 & 1 & 1 & -1 & -1 \\
	0 & 0 & 0 & 0 & 0 & 1 & -1 & 1 & 1 & 1 & 1 & -1 & -1 & -1 & -1 \\
	0 & 0 & 0 & 0 & 0 & -4 & 4 & 1 & 1 & 1 & 1 & -1 & -1 & -1 & -1 \\
	0 & 2 & 2 & -1 & -1 & -1 & -1 & 0 & 0 & 0 & 0 & 0 & 0 & 0 & 0 \\
	0 & 0 & 0 & 1 & 1 & -1 & -1 & 0 & 0 & 0 & 0 & 0 & 0 & 0 & 0 \\
	0 & 0 & 0 & 0 & 0 & 0 & 0 & 1 & -1 & -1 & 1 & 1 & -1 & -1 & 1 \\
	0 & 0 & 0 & 0 & 0 & 0 & 0 & 1 & 1 & -1 & -1 & -1 & -1 & 1 & 1 \\
	0 & 0 & 0 & 0 & 0 & 0 & 0 & 1 & -1 & 1 & -1 & -1 & 1 & -1 & 1 \\
	0 & 0 & 0 & 0 & 0 & 0 & 0 & 1 & -1 & -1 & 1 & -1 & 1 & 1 & -1 
	\end{array}\right].
	\end{equation}
	$ \boldsymbol{\Lambda} $ is a diagonal matrix given by
	\begin{equation}
	\begin{aligned}
	&\boldsymbol{\Lambda}=\operatorname{diag}\left(s_{0},s_{1}, s_{2}, s_{3}, s_{4}, s_{5}, S_{6}, s_{7}, s_{8}, s_{9}, s_{10}, s_{11}, s_{12}, s_{13}, s_{14}\right)\\
	&=\operatorname{diag}\left(s_{\rho}, s_{e}, s_{\epsilon}, s_{j}, s_{q}, s_{j}, s_{q}, s_{j}, s_{q}, s_{\upsilon}, s_{\upsilon}, s_{\upsilon}, s_{\upsilon}, s_{\upsilon}, s_{x y z}\right).
	\end{aligned}
	\end{equation}	
	The non-dimensional relaxation time is defined as
	\begin{equation}
	\tau_{\upsilon}=\frac{1}{\mathrm{~s}_{\upsilon}}=\frac{v}{c_{\mathrm{s}}^{2}}+0.5,
	\end{equation}
	where $ \nu $ is the kinematic viscosity of the liquid, $ c_{s}=c/\sqrt{3} $ is the lattice sound speed.

	By using the transformation matrix $ \mathbf{M} $, the density distribution function $ f $ and the equilibrium distribution function $ f^{eq} $ can be transferred into the space of moments, i.e.,  $\mathbf{m}=\mathbf{M} \mathbf{f}$ and $\mathbf{m}^{eq}=\mathbf{M} \mathbf{f}^{eq}$. For the D3Q15 lattice model, the equilibrium distribution functions $ \mathbf{m}^{eq} $ in the moment space are given by \cite{XuIJMFF2015}
	\begin{equation}
		\begin{aligned}
			\mathbf{m}^{(\mathrm{eq})}= \rho\left(1,-1+|\mathbf{u}|^{2}, 1-5|\mathbf{u}|^{2}, u_{x},-\frac{7}{3} u_{x}, u_{y},-\frac{7}{3} u_{y}, u_{z}\right.
			\left.-\frac{7}{3} u_{z}, 2 u_{x}^{2}-u_{y}^{2}-u_{z}^{2}, u_{y}^{2}-u_{z}^{2}, u_{x} u_{y}, u_{y} u_{z}, u_{x} u_{z}, 0\right)^{T}
		\end{aligned} ,
	\end{equation}
	where $ \mathbf{u} $ is the macroscopic velocity of the fluid. Thus, Eq. (\ref{Eq1}) can be rewritten as 
	\begin{equation}
		\mathbf{m}^{*}=\mathbf{m}-\boldsymbol{\Lambda} \left(\mathbf{m}-\mathbf{m}^{e 	q}\right)+\delta_{t}\left(\mathbf{I}-\frac{\boldsymbol{\Lambda}}{2}\right) \mathbf{S}+\boldsymbol{C} ,
	\end{equation}
	in which $ \mathbf{I} $ is the unit tensor and $ \mathbf{S} $ is the forcing term in the moment. The streaming process is then given by
	\begin{equation}
		f_{\alpha}\left(\mathbf{x}+\mathbf{e}_{\alpha} \Delta t, t+\Delta t\right)=f ^{*}(\mathbf{x}, t)=\mathbf{M}^{-1} \mathbf{m}^{*}.
	\end{equation}
	The corresponding macroscopic density and velocity are calculated by
	\begin{equation}
		\rho=\sum_{\alpha} f_{\alpha}, \quad \rho \mathbf{u}=\sum_{\alpha} \mathbf{e}_{\alpha} f_{\alpha}+\frac{\Delta t }{2} \mathbf{F},
	\end{equation}
	where $ F = (F_{x}, F_{y}) $ is the total force acting on the system. The modified external forcing scheme proposed by Li et al. is adopted in the present study, which is expressed as \cite{XuIJMFF2015}
	\begin{equation}
		\overline{\mathbf{S}}=\left[\begin{array}{c}
		0 \\
		2 \mathbf{u} \cdot \mathbf{F}+\frac{6 \eta \mid \mathbf{F_{m}}\mid^{2}}{\psi^{2} \delta_{t}\left(\left(s_{e}^{-1}-0.5\right)\right.} \\
		-10 \mathbf{u} \cdot \mathbf{F} \\
		F_{x} \\
		-\frac{7}{3} F_{x} \\
		F_{y} \\
		-\frac{7}{3} F_{y} \\
		F_{z} \\
		-\frac{7}{3} F_{z} \\
		4 u_{x} F_{x}-2 u_{y} F_{y}-2 u_{z} F_{z} \\
		2 u_{y} F_{y}-2 u_{z} F_{z} \\
		u_{x} F_{y}+u_{y} F_{x} \\
		u_{y} F_{z}+u_{z} F_{y} \\
		u_{x} F_{z}+u_{z} F_{x} \\
		0
		\end{array}\right],
		\label{eq9}
	\end{equation}	
	where $ \eta  $ is utilized to adjust the numerical stability in the proposed pseudopotential MRT LB model, and $\left|\boldsymbol{F}_{\mathrm{m}}\right|^{2}=\left(F_{\mathrm{m} x}^{2}+F_{\mathrm{my}}^{2}\right.)$. In addition, the fluid–fluid interaction force $ F_{m} $  in the Shan–Chen model for multiphase flow can be given by \cite{LiPRE2014} 
	\begin{equation}
		\mathbf{F}_{\mathrm{m}}=-G \psi(\mathbf{x}, t)\left[\sum_{\alpha=1}^{14} w\left(\left|\boldsymbol{e}_{\alpha}\right|^{2}\right) \psi\left(\mathbf{x}+\boldsymbol{e}_{\alpha}, t\right) \boldsymbol{e}_{\alpha}\right],
	\end{equation}
	where $ G $ is the interaction strength, $ w(|e_{\alpha}|^{2}) $ is the
	weight factor. For the case of nearest-neighbor interactions on the D3Q15 model, the weights are $ w(1) = 1/3 $ and $ w(2) = 1/24 $. $ \psi $ represents the effective density, which can be obtained by introducing a non-ideal equation of state (EOS) \cite{YuanPOF2006}
	\begin{equation}
		\psi=\sqrt{\frac{2\left(P_{EOS}-\rho c_{s}^{2}\right)}{G c^{2}}},
	\end{equation}
	where $ P_{EOS} $ is the pressure from the Carbahan-Starling (C-S) in the present work, which is defined as 
	\begin{equation}
	P_{\mathrm{EOS}}=\rho R T \frac{1+(b \rho / 4)+(b \rho / 4)^{2}-(b \rho / 4)^{3}}{(1-b \rho / 4)^{3}}-a \rho^{2},
	\end{equation}
	where $ a=0.4963R^{2}T_{c}^{2}/p_{c} $ and $ b=0.18727RT_{c}/p_{c} $, $ T_{c} $ is the critical temperature,  , and $ p_{c} $ is the critical pressure. In our simulations we set  $ a = 0.25 $, $ b = 4 $ and $ R = 1 $, then $T_{c}$ would be given by 0.02358.

	Further, the influence of the gravitational force is also consideded here, which is given by 
	\begin{equation}
		\boldsymbol{F}_{b}(\boldsymbol{x}, t)=\rho(\boldsymbol{x}, t) g, 
	\end{equation}	
	and the gravity acceleration g is fixed at $1.0 \times 10^{-5}$ according to the work of Li et al. \cite{LiSM2016}. In such a cse, the total force $\boldsymbol{F}$ can be expressed as 
	\begin{equation}
		\boldsymbol{F}(\boldsymbol{x}, t)=\boldsymbol{F}_{m}+\boldsymbol{F}_{b}. 
	\end{equation}	
	Moreover, the additional source term $ \mathbf{C} $ in the MRT-LB is used to modified the surface tension, and it is defined as \cite{XuIJMFF2015}
	\begin{equation}
		\boldsymbol{C}=\left[\begin{array}{c}
		0 \\
		\frac{4}{5} s_{e}\left(Q_{x x}+Q_{y y}+Q_{z z}\right) \\
		0 \\
		0 \\
		0 \\
		0 \\
		0 \\
		0 \\
		0 \\
		-s_{\upsilon}\left(2 Q_{x x}-Q_{y y}-Q_{z z}\right) \\
		-s_{\upsilon}\left(Q_{y y}-Q_{z z}\right) \\
		-s_{\upsilon} Q_{x y} \\
		-s_{\upsilon} Q_{y z} \\
		-s_{\upsilon} Q_{x z} \\
	0
	\end{array}\right],
	\end{equation}
	where $\boldsymbol{Q}$ is the discrete pressure tensor, and it can be obtined via the following equation
	\begin{equation}
		\boldsymbol{Q}=\kappa \frac{G}{2} \psi(\boldsymbol{x}, t)\left\{\sum_{\alpha=1}^{14} w\left(\left|\boldsymbol{e}_{\alpha}\right|^{2}\right)\left[\psi\left(\boldsymbol{x}+\boldsymbol{e}_{\alpha}, t\right)-\psi(\boldsymbol{x}, t)\right] \boldsymbol{e}_{\alpha} \boldsymbol{e}_{\alpha}\right\},
		\label{eq14}
	\end{equation}	
	in which $0 \leq \kappa \leq 1$ is the parameter to adjuest the durface tension.

	Finally, the geometric formula proposed by Ding et al. \cite{DingPRE2007}  is adopted in this work to model the wetting condition, and the expression of this scheme in three-dimensional space is given by \cite{WangPRE2013}
	\begin{equation}
	\rho_{i, j, 0}=\rho_{i, j, 2}+\tan \left(\frac{\pi}{2}-\theta \right) \sqrt{\left(\rho_{i+1, j, 1}-\rho_{i-1, j, 1}\right)^{2}+\left(\rho_{i, j+1,1}-\rho_{i, j-1,1}\right)^{2}},
	\end{equation}
	where $ \theta $ is an an analytically prescribed static contact angle, $ \rho_{i, j, 0} $ represents the
	density at the ghost layer $ (i, j, 0) $ beneath the solid wall. The first and the second indexes denote the coordinates along the x- and y-direction, respectively, while the third index denotes the coordinate normal to the solid wall.

\section{Problem Description}
\begin{figure}[ht]
		\centering
		\includegraphics[width=0.6\textwidth]{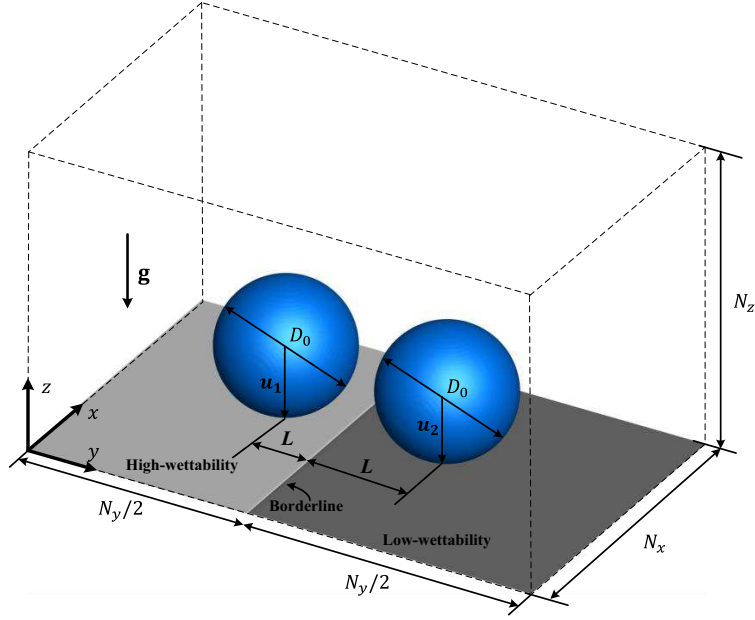}
		\caption{ Schematic illustration of double droplets impact on a wettability-patterned surface.}
		\label{fig1}
	\end{figure}

	Fig. \ref{fig1} shows the schematics of two droplets simultaneously impacting flat surfaces with wettability difference. Two droplets with the same diameter $ D_{0} $, initial velocity $ u_{0} $ and initial height are located in the high-wettability region (light gray) and low-wettability region (dark gray) impact the surface, respectively. In addition, the two droplets have the same lateral offset $ L^{*}=L/D_{0} $ with respect to the borderline, in which $ L $ is the length of the droplet from the borderline. The periodic boundary conditions are imposed at the horizontal direction, while the no-slip boundary condition is adopted for the top and bottom boundaries. The dimensionless parameters that control the impact process include Weber number (We) and dimensionless time ($ t^{*} $), which are defined as
		\begin{equation}
			W e=\frac{\rho_{l} u_{0}^{2} D_{0}}{\sigma},\quad t^{*}=\frac{t}{\sqrt{\rho R_{0}^{3} / \sigma}},
		\end{equation}
	where $ R_{0} $, $ \rho_{l} $ and $ \upsilon_{l} $ are the radius, density and kinematic viscosity of droplet, respectively. 
	
	Further, in our simulations, the kinematic viscosity ratio is fixed at 15 ($ \upsilon^{*}=\upsilon_{g} / \upsilon_{l}=15.0 $), and  the reduced temperature is set to $ T/T_{c} = 0.5 $, at which the density ratio between liquid and vapor phases is nearly 700, i.e., $ \rho^{*}=\rho_{g} / \rho_{l}=700.0 $, and we note that these values are very close to the water to air kinematic viscosity ratio and density ratio in reality. 
    
\section{Model validation}	

\begin{figure}[H]
		\begin{minipage}[t]{0.5\textwidth}
			\centering
			\includegraphics[width=\textwidth]{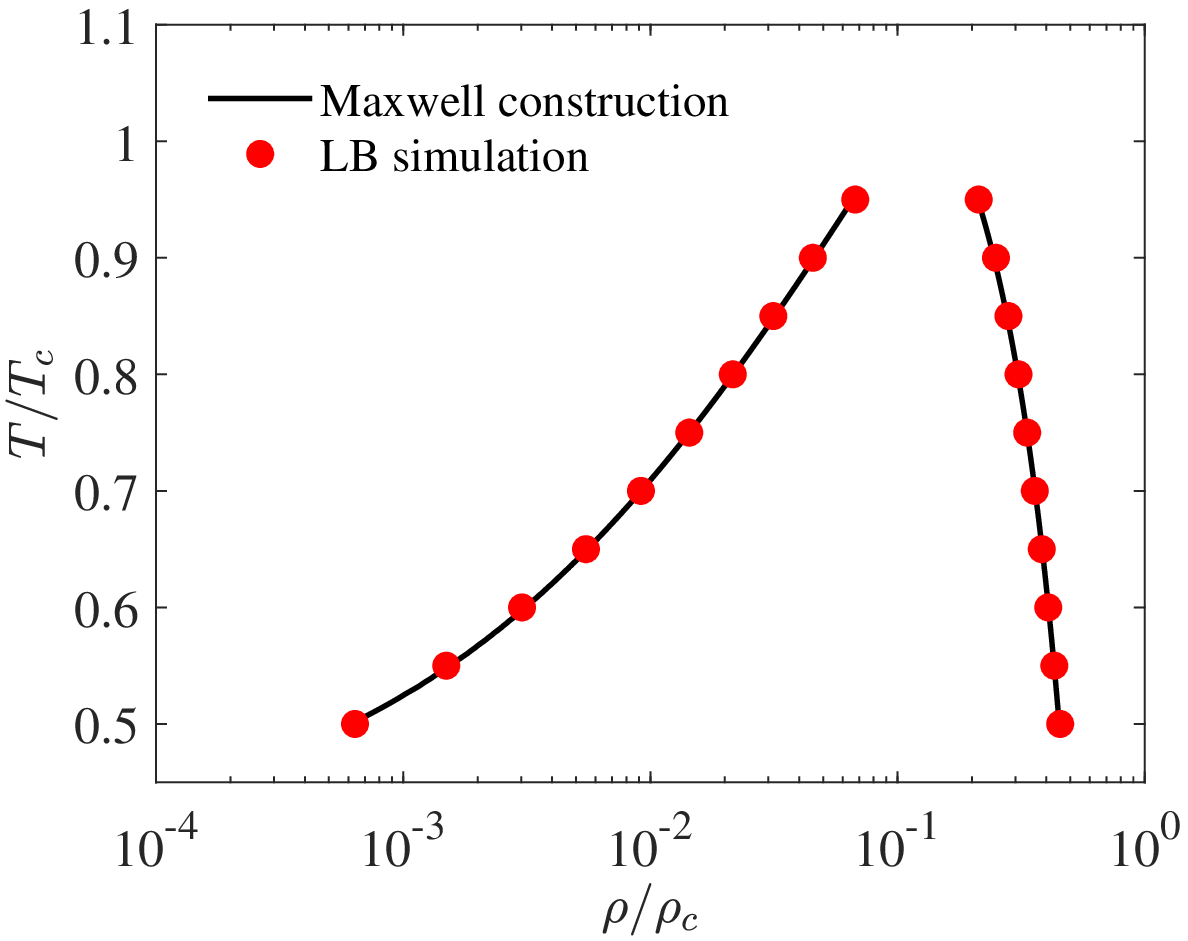}
			\caption{Comparison of coexistence values from simulations with those of the Maxwell construction.}
			\label{fig2}
		\end{minipage}%
		\hfill
		\begin{minipage}[t]{0.5\textwidth}
			\centering
			\includegraphics[width=\textwidth]{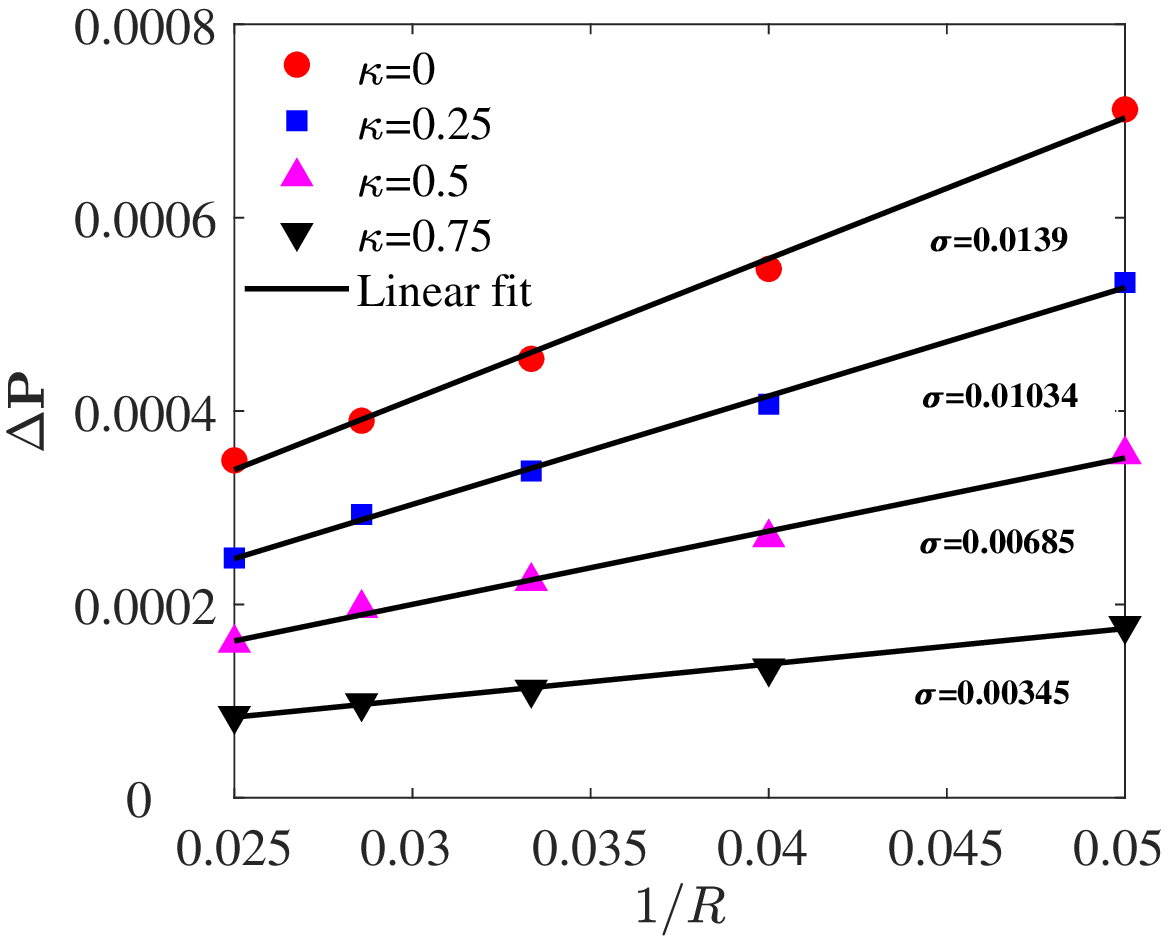}
			\caption{Evaluation of Laplace law at $ T/T_{c} = 0.5 $.}
			\label{fig3}
		\end{minipage}
	\end{figure}

	In this section, three tests are considered to validate the three-dimensional MRT pseudopotential LB model. First, we consider a static liquid droplet problem to compare the analytical solution by the Maxwell construction with the LB results results. A circular droplet with radius $ R $ is placed at the center of the $ 100lu \times 100lu \times 100lu $ computational domain, and the period boundary is applied to all surrounding boundaries of the computational domain. The initial density field is defined as follows
	\begin{equation}
		\rho(x, y, z)=\frac{\rho_{l}+\rho_{g}}{2}-\frac{\rho_{l}-\rho_{g}}{2} \tanh \left[\frac{2 \sqrt{\left(x-x_{0}\right)^{2}+\left(y-y_{0}\right)^{2}+\left(z-z_{0}\right)^{2}}-R}{W}\right],
	\end{equation}
	where the gas–liquid initial interface width $ W $ is set as $ 5lu $. The parameter $ \kappa $ in Eq. \ref{eq14} is set to be $ \kappa = 0 $. As show in Fig. \ref{fig2}, it can be clearly seen that the present numerical results are in good agreement with the analytical solution, indicating that the present LB model capable of achieving thermodynamic consistency.

	Secondly, the Laplace law is also considered to validate the present LB method, and it is known that the Laplace law states that the pressure difference $ \Delta P$ inside and outside the droplet is linear with the reciprocal of the droplet radius $ R $, and the surface tension is proportional coefficient. Fig. \ref{fig3} shows the variation of the pressure difference between the inside and outside of the droplet with the inverse of the droplet radius for the four cases at $ T/T_{c} = 0.5 $ corresponding to a density ratio of nearly 700. As it can be seen from this figure, it can be clearly seen that there is a linear relationship between the pressure difference and the reciprocal of droplet radius for different values of $ \kappa $, indicating that the present LB model is consistent with those of the Laplace law and the surface tension independent of density ratio can be adjusted in the numerical model.

\begin{figure}[H]
		\centering 
		\subfigure[]{ 
			\includegraphics[width=0.45\textwidth]{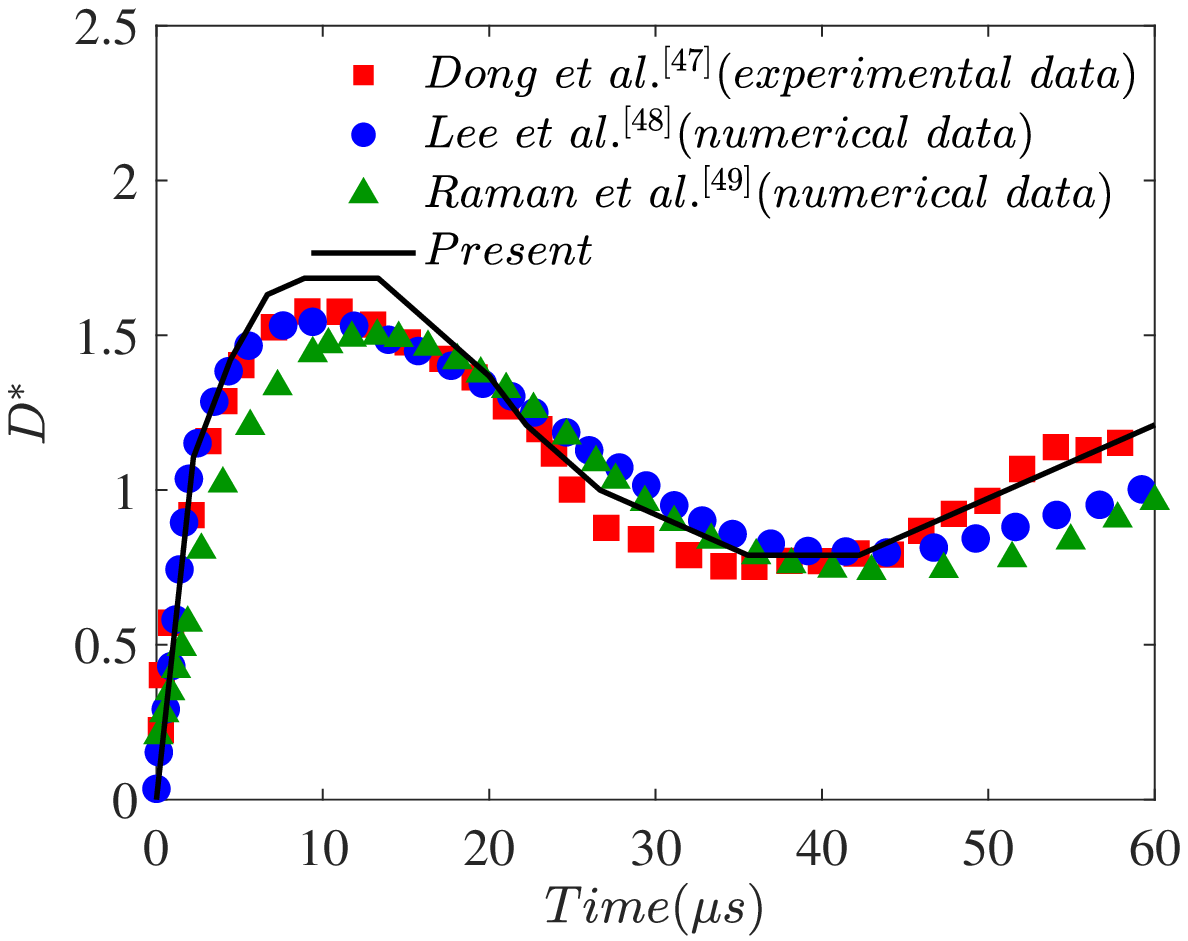}} 
		\subfigure[]{ 
			\includegraphics[width=0.45\textwidth]{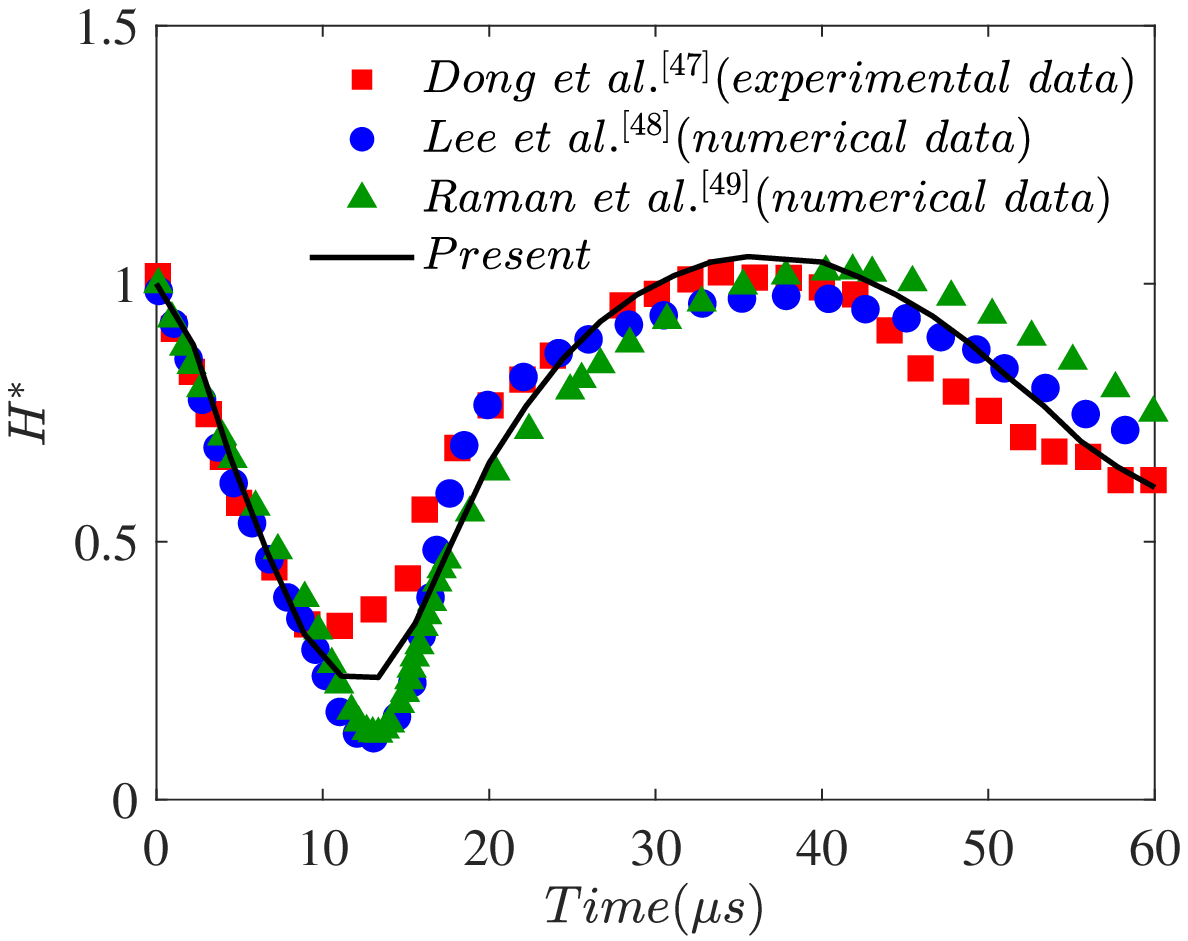}}
		\caption{Comparison between present simulation results and previously published data \cite{DongAJ2007, LeeJCP2010, RamanPRE2016}. (a) Dimensionless spreading factor and (b) dimensionless height.}
		\label{fig4}
	\end{figure}

	Finally, we conduct the simulations of a water droplet impact on flat surfaces with wettabilities $ \theta =107 \degree $ and comparing the results to similar experimental \cite{DongAJ2007} and numerical data \cite{LeeJCP2010,RamanPRE2016} from the previous literature. Fig. \ref{fig4} illustrates the variations of the spreading factor and the dimensionless height of the droplet with time based on the present LB method in comparison with the experimental data and numerical results. The dimensionless diameter or spreading factor $ D^{*}=D/D_{0} $ and dimensionless height $ H^{*}=H/D_{0} $ are defined as the ratio of spread length $ D $ and height $ H $ to initial droplet diameter $ D_{0} $, respectively. As show in Fig. \ref{fig4}, it can be clearly seen that the numerical model can better simulate the shape change and spreading of droplets during the droplet impact process.

\section{Results and discussion}
In what follows, the numerical results obtained for two droplets impact on a wettability-patterned surface are presented and discussed, and the influences of the surface wettability difference, Weber number as well as the droplet spacing during the droplet impact are all considered. It will be shown later that the droplet dynamics reported here are largely different from a single-droplet impact observed in previous work \cite{XiongCF2018,CuiCF2021}.

\subsection{The effect of wettability difference}

	\begin{figure}[ht]
		\centering
		
		\begin{minipage}[c]{0.1\textwidth}
			\centering
			\caption*{(a) $ t^{*}=0.2 $ }
		\end{minipage}
		\begin{minipage}[c]{0.2\textwidth}
			\includegraphics[width=\textwidth]{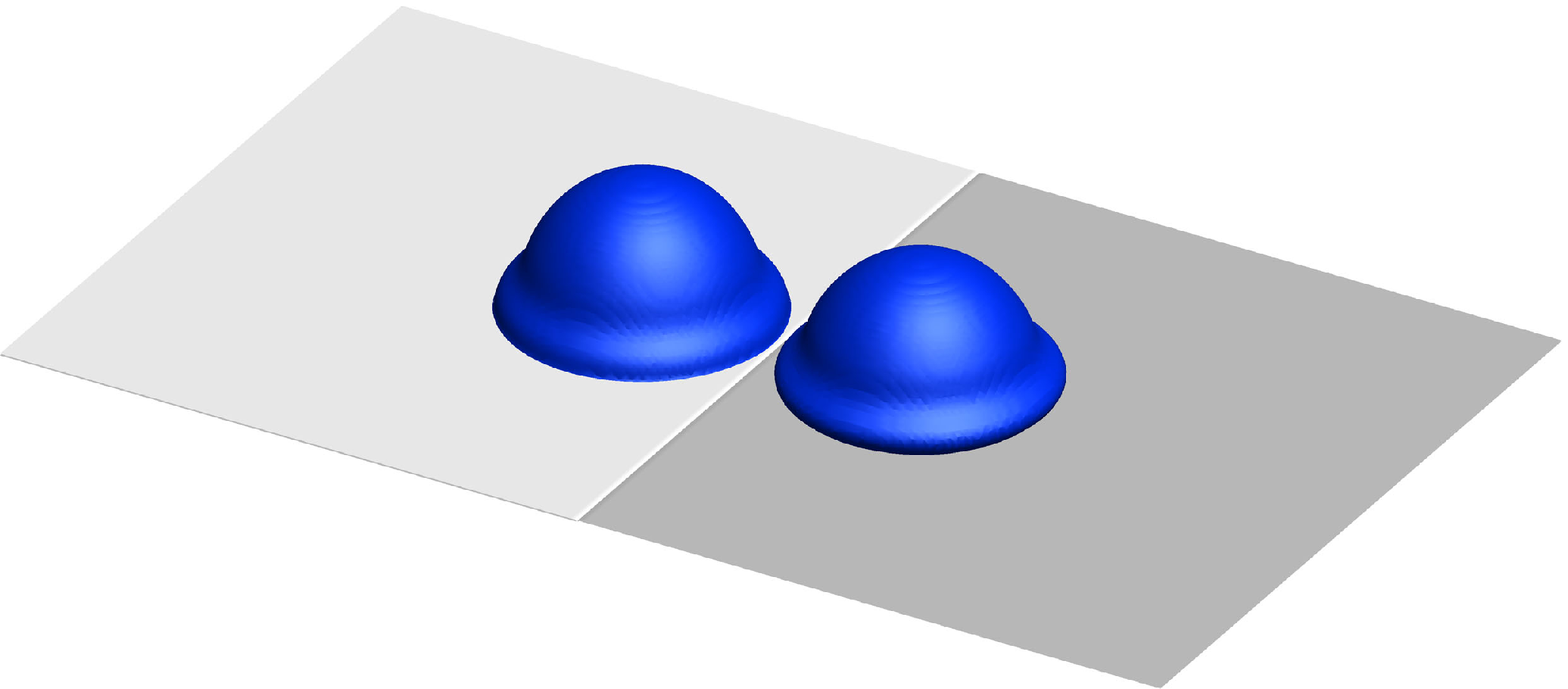}
		\end{minipage}
		\begin{minipage}[c]{0.2\textwidth}
			\includegraphics[width=\textwidth]{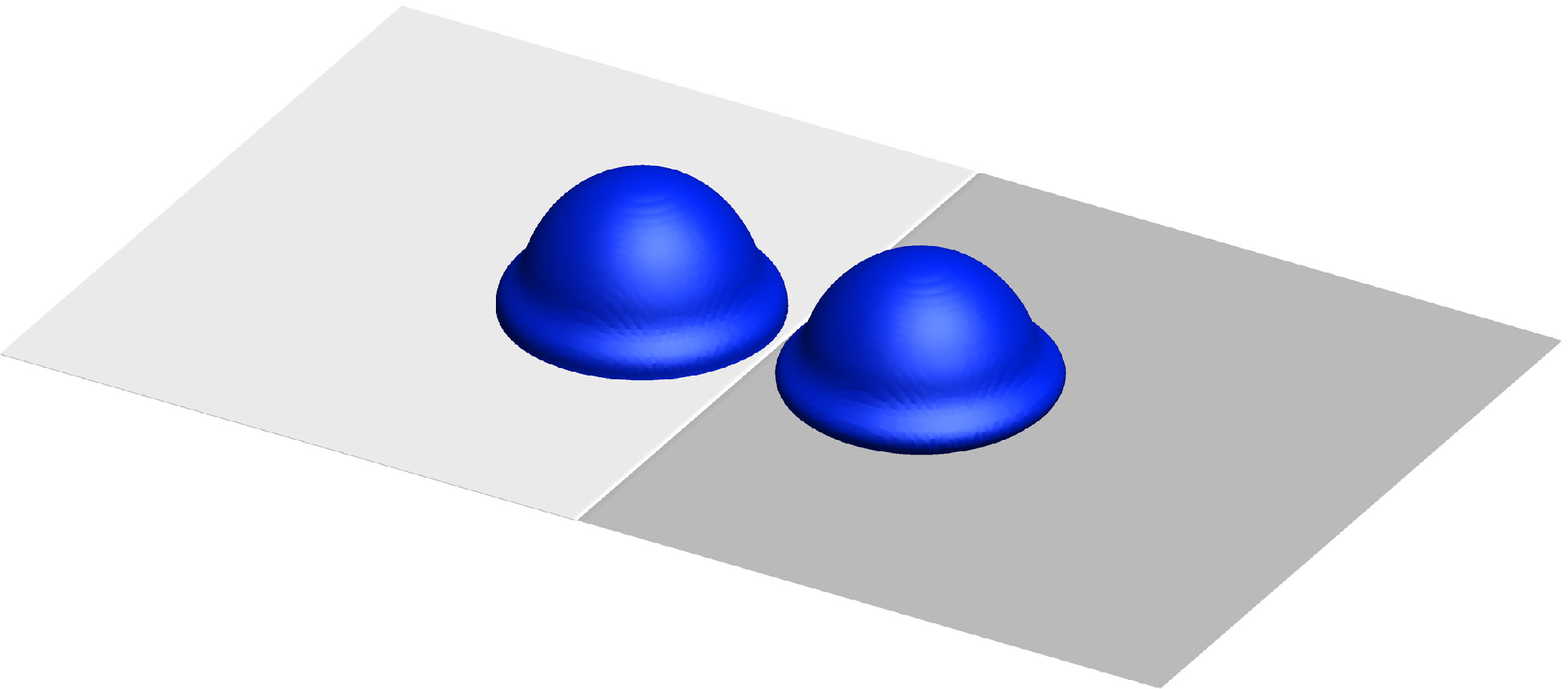}
		\end{minipage}
		\begin{minipage}[c]{0.2\textwidth}
			\includegraphics[width=\textwidth]{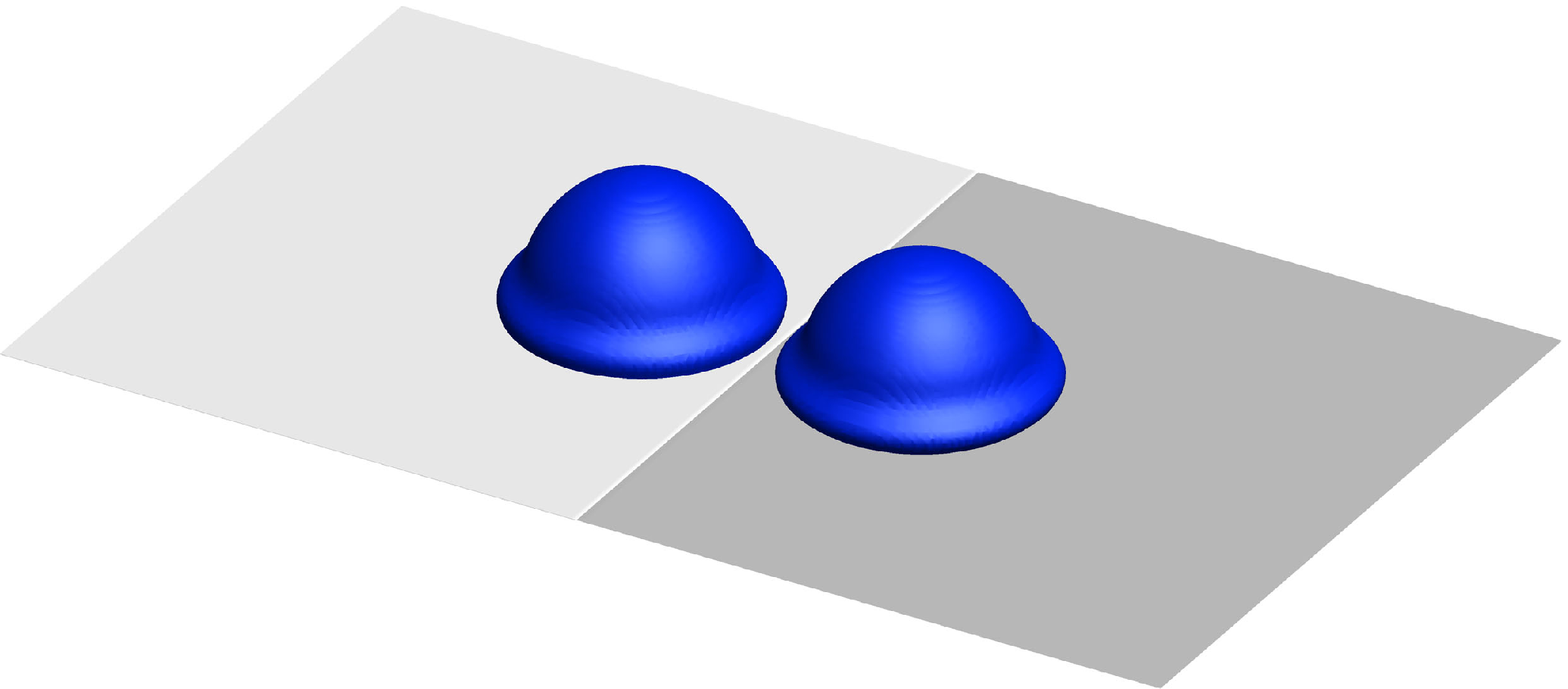}
		\end{minipage}
		\begin{minipage}[c]{0.2\textwidth}
			\includegraphics[width=\textwidth]{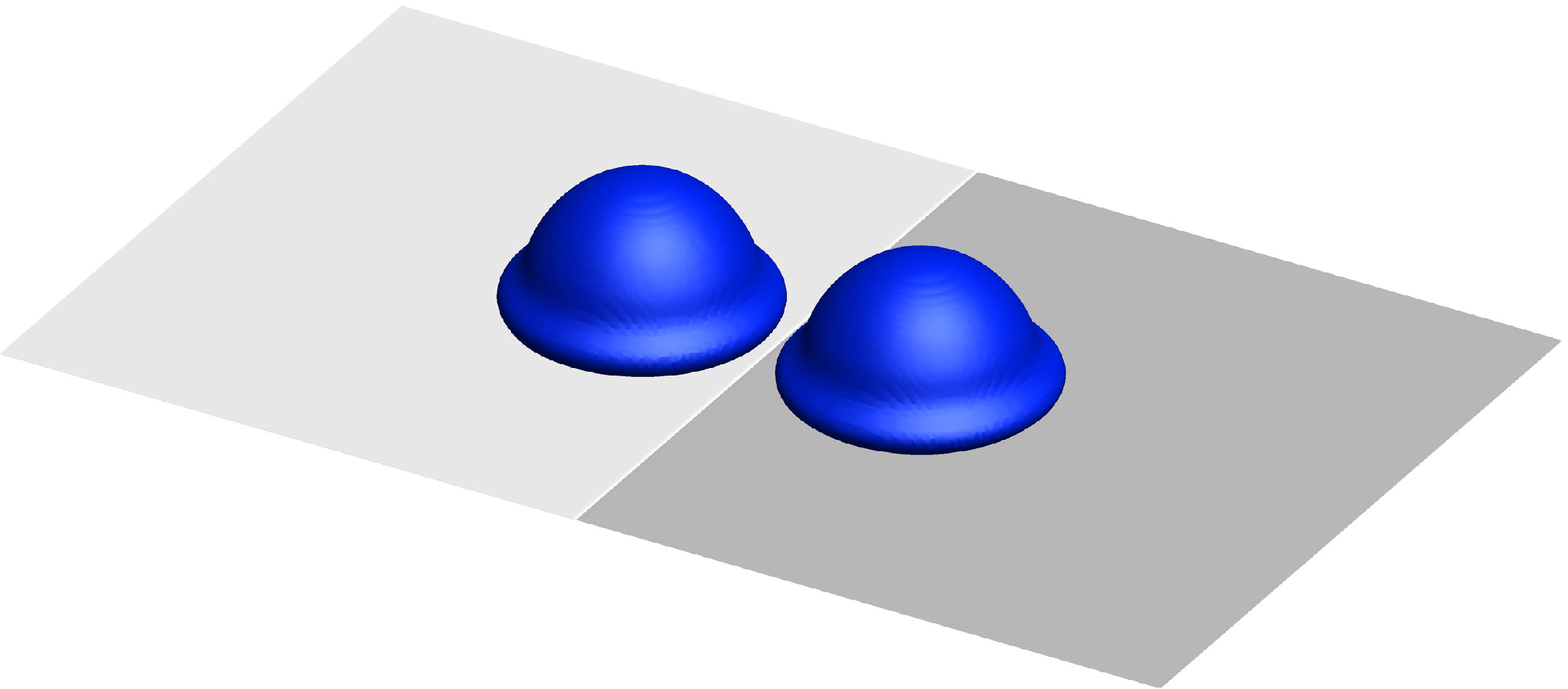}
		\end{minipage}

		\begin{minipage}[c]{0.1\textwidth}
			\centering
			\caption*{(b) $ t^{*}=0.6 $  }
		\end{minipage}
		\begin{minipage}[c]{0.2\textwidth}
			\includegraphics[width=\textwidth]{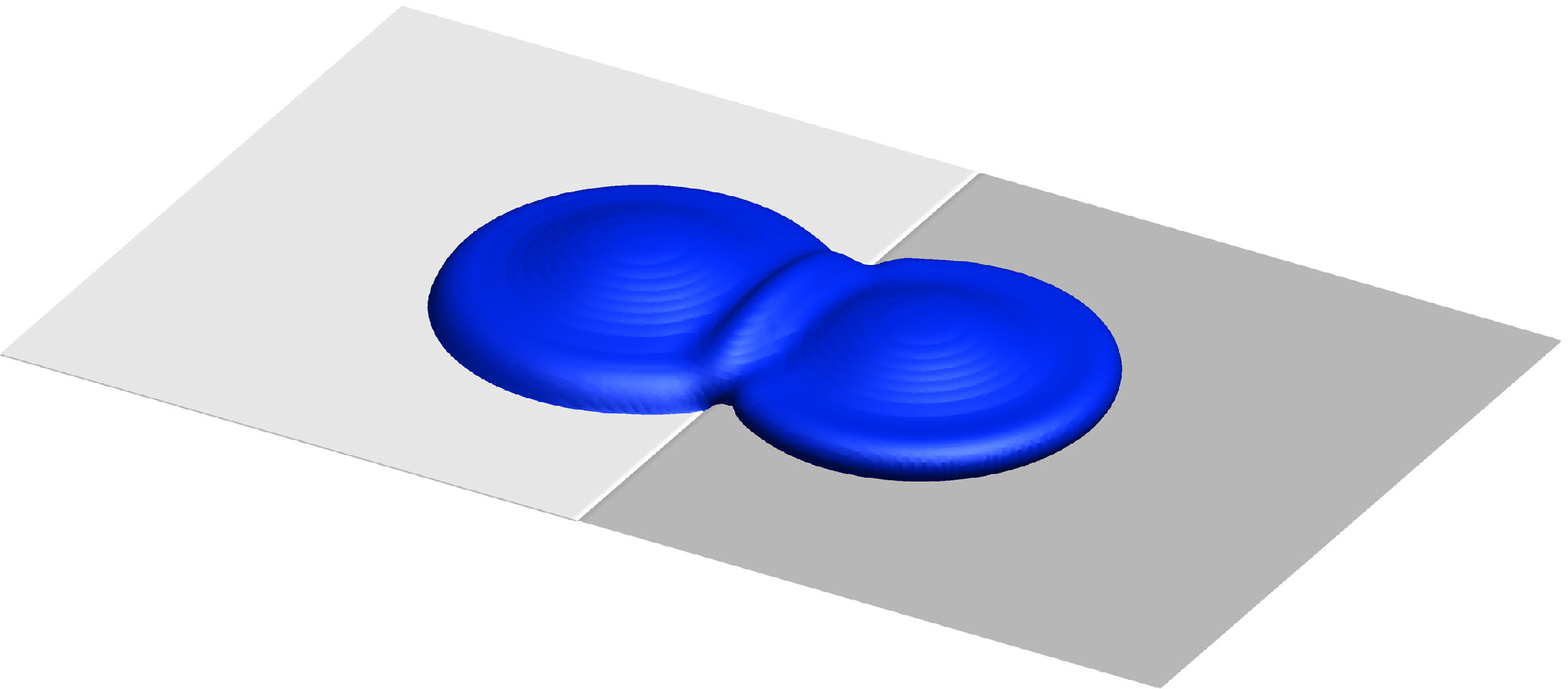}
		\end{minipage}
		\begin{minipage}[c]{0.2\textwidth}
			\includegraphics[width=\textwidth]{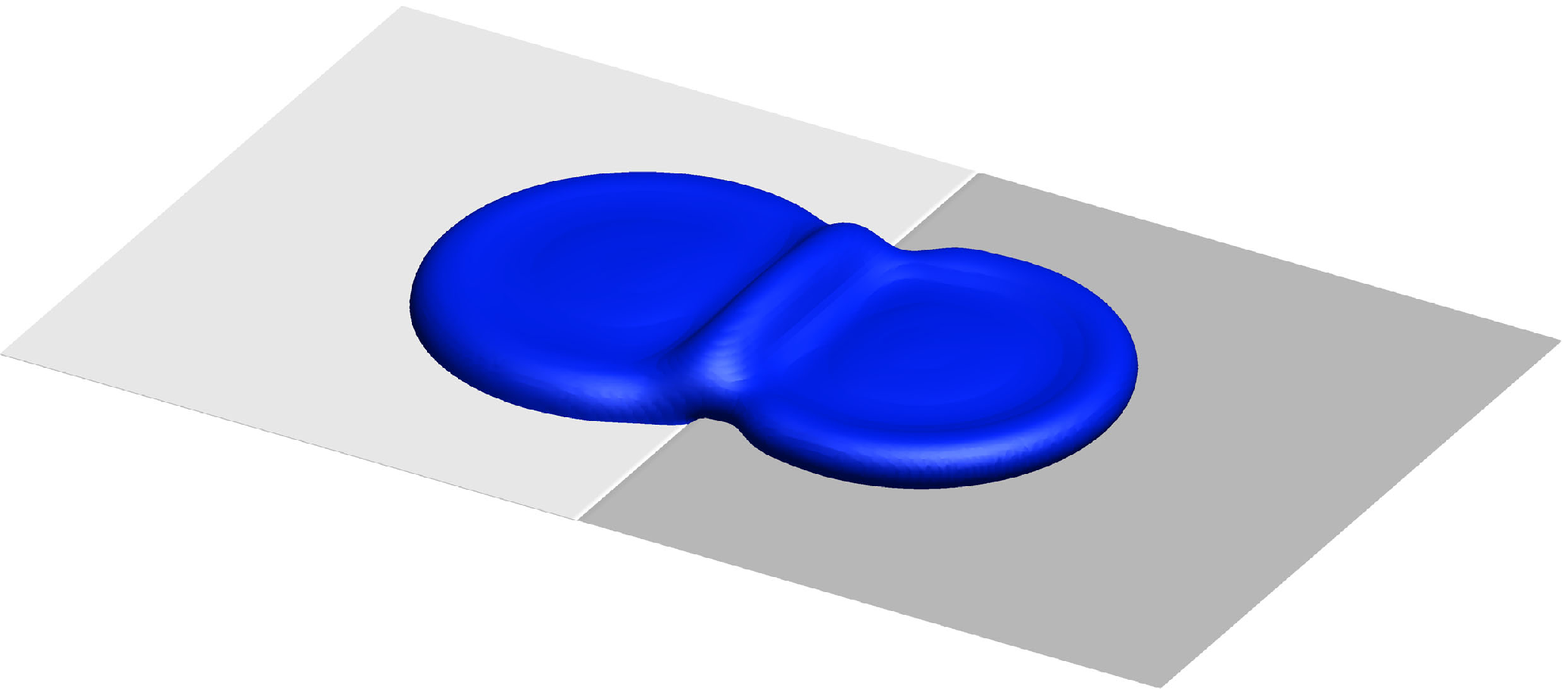}
		\end{minipage}
		\begin{minipage}[c]{0.2\textwidth}
			\includegraphics[width=\textwidth]{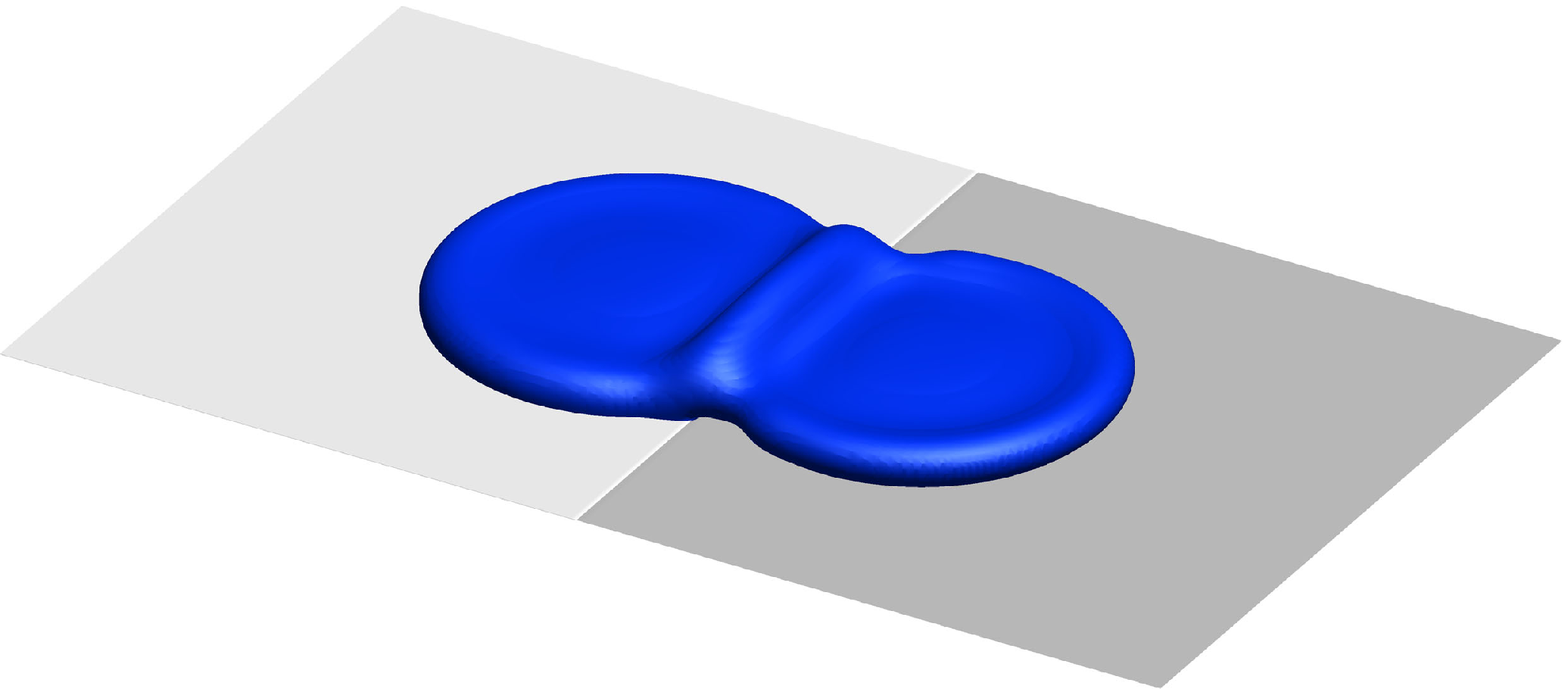}
		\end{minipage}
		\begin{minipage}[c]{0.2\textwidth}
			\includegraphics[width=\textwidth]{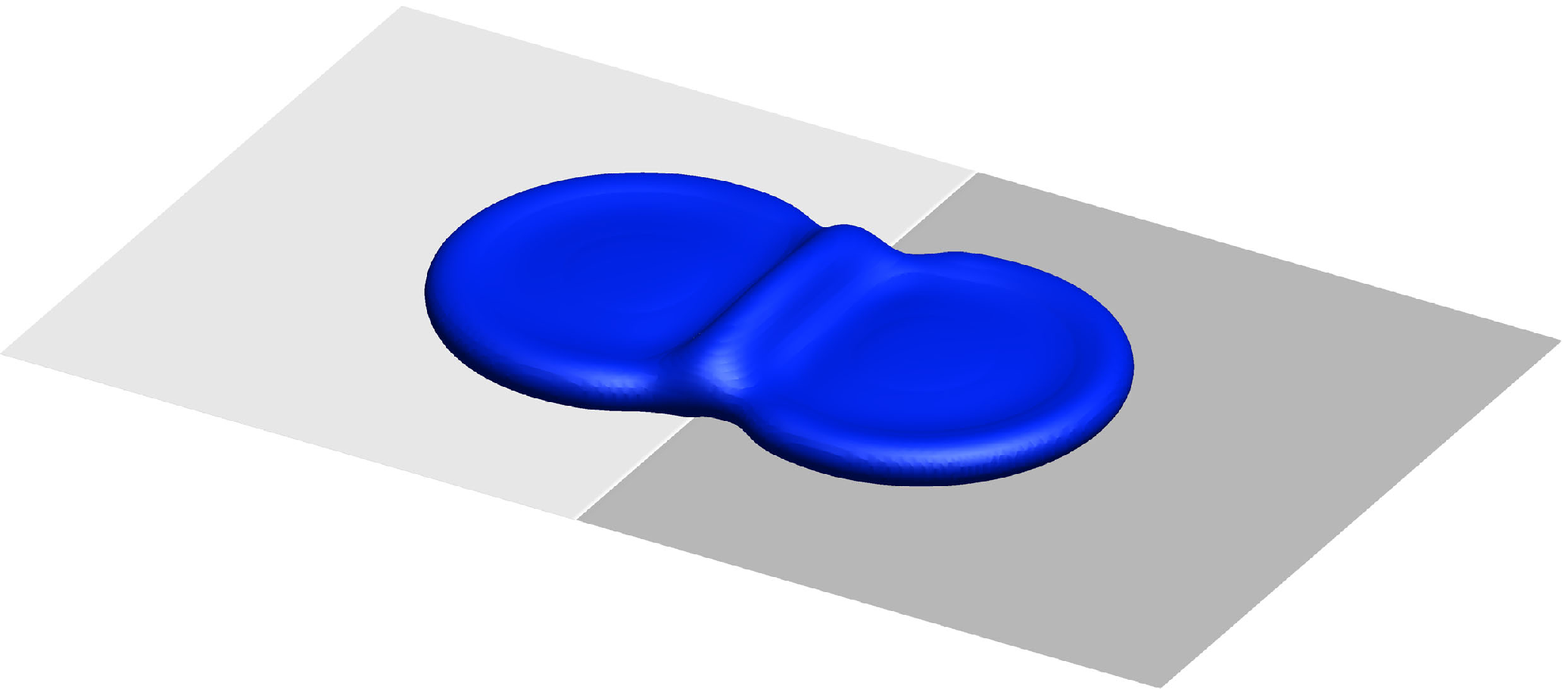}
		\end{minipage}

		\begin{minipage}[c]{0.1\textwidth}
			\centering
			\caption*{(c) $ t^{*}=1.0 $  }
		\end{minipage}
		\begin{minipage}[c]{0.2\textwidth}
			\includegraphics[width=\textwidth]{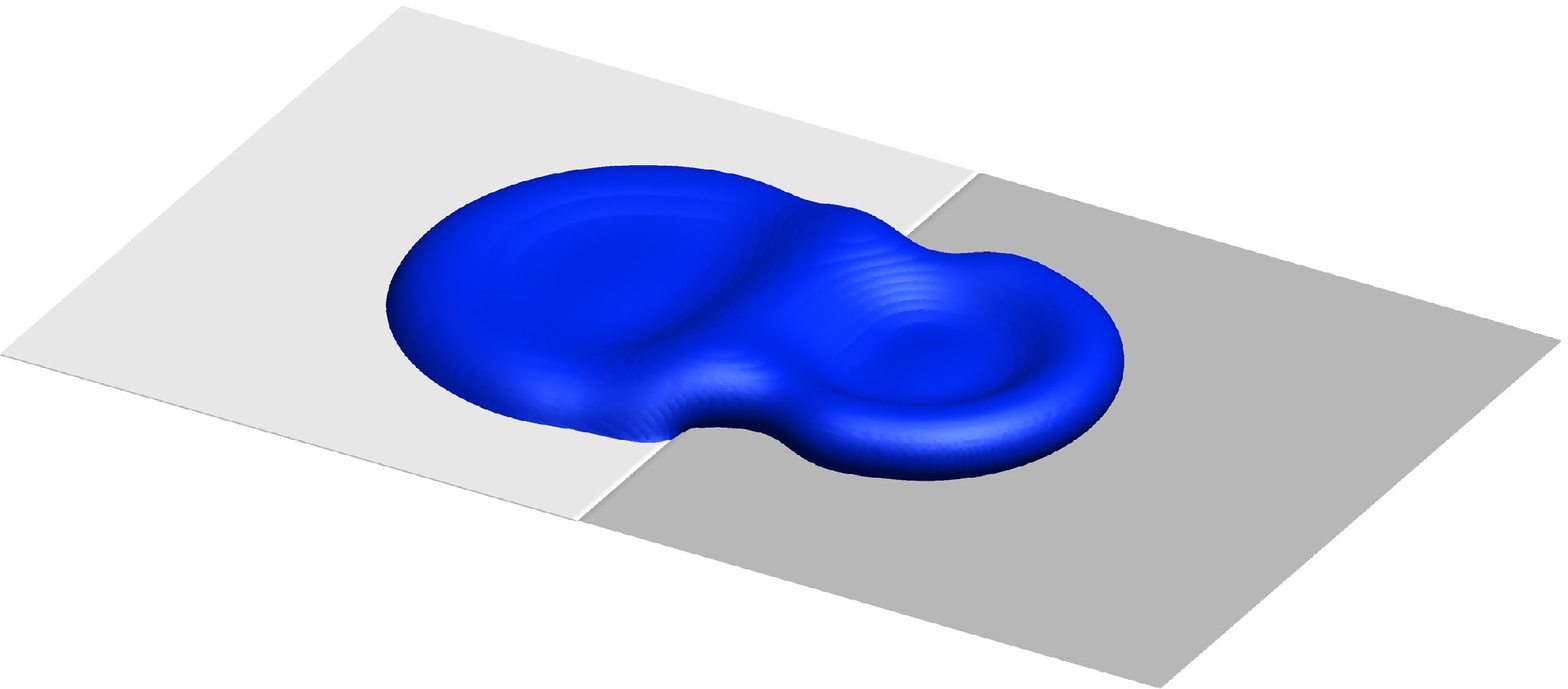}
		\end{minipage}
		\begin{minipage}[c]{0.2\textwidth}
			\includegraphics[width=\textwidth]{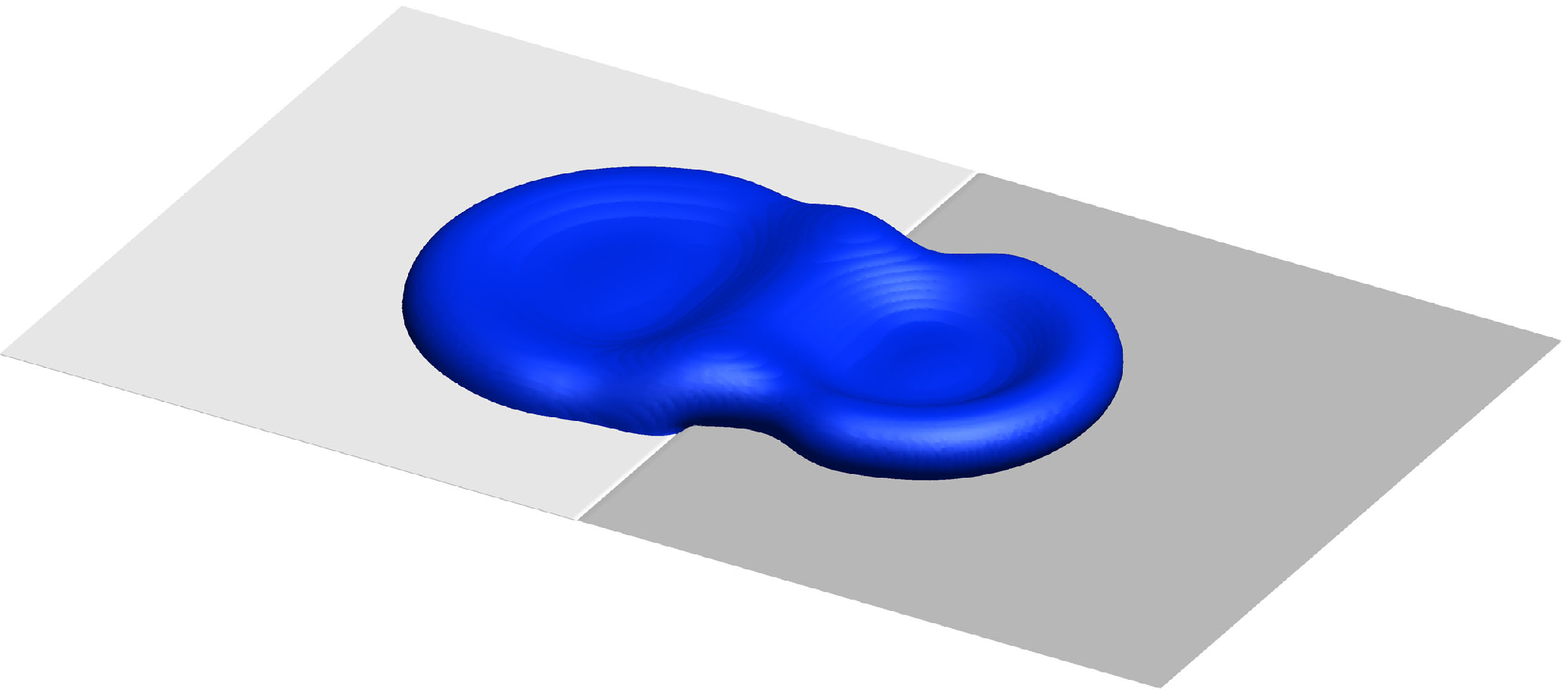}
		\end{minipage}
		\begin{minipage}[c]{0.2\textwidth}
			\includegraphics[width=\textwidth]{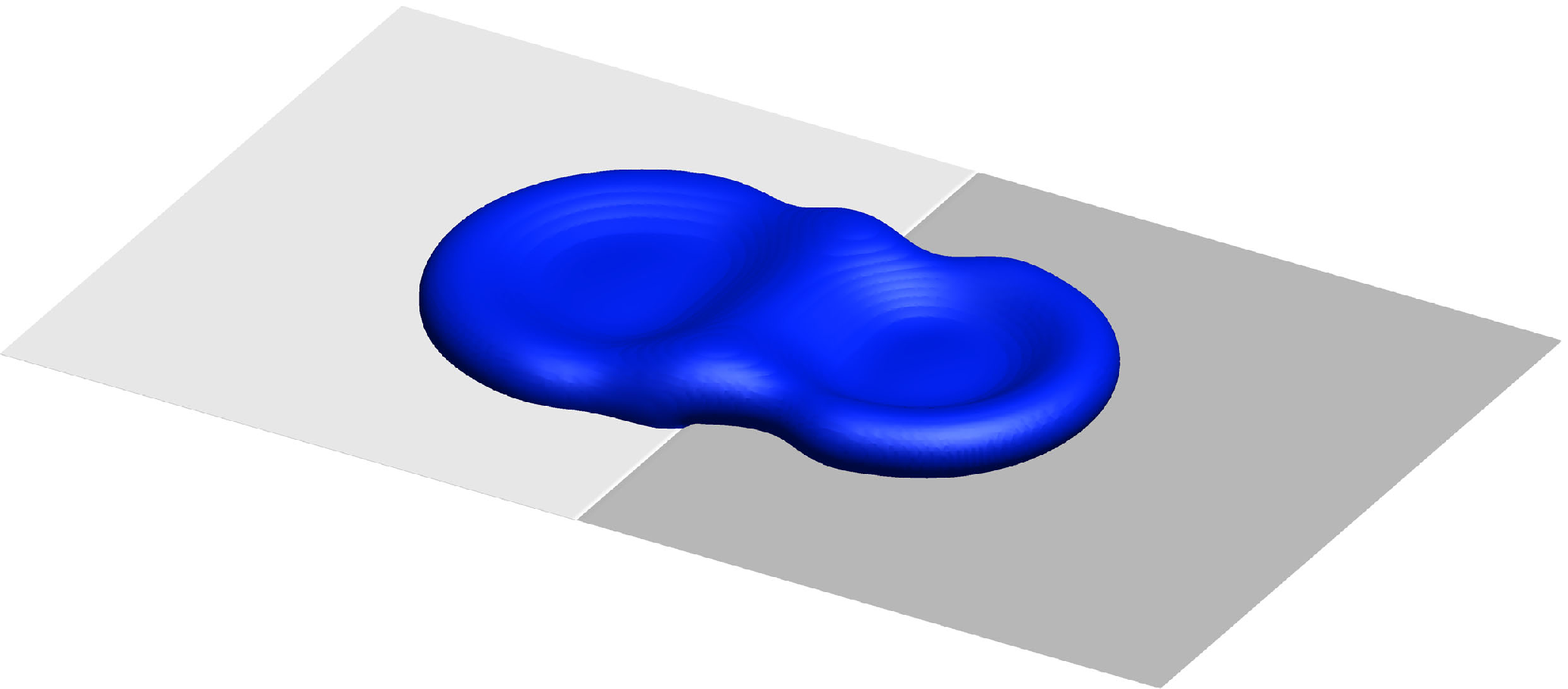}
		\end{minipage}
		\begin{minipage}[c]{0.2\textwidth}
			\includegraphics[width=\textwidth]{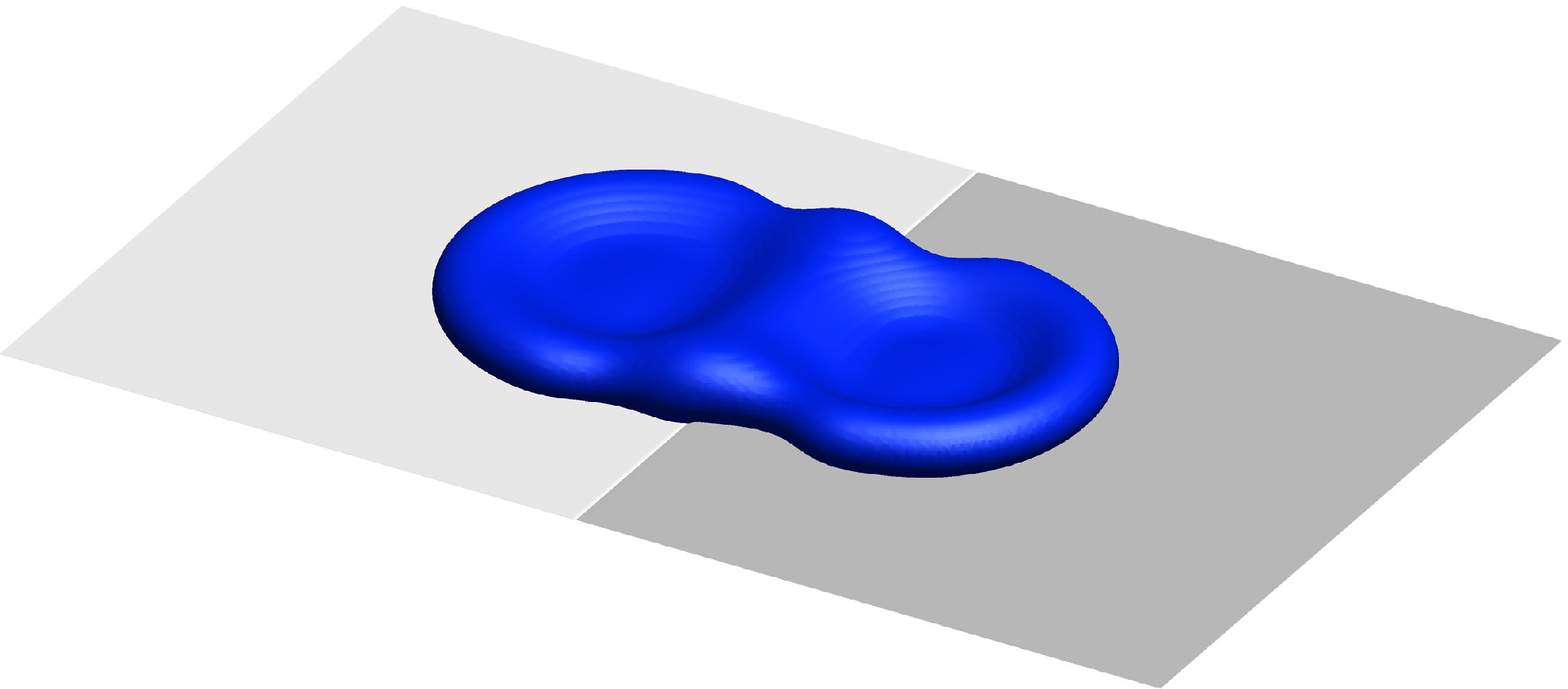}
		\end{minipage}

		\begin{minipage}[c]{0.1\textwidth}
			\centering
			\caption*{(d) $ t^{*}=1.4 $  }
		\end{minipage}
		\begin{minipage}[c]{0.2\textwidth}
			\includegraphics[width=\textwidth]{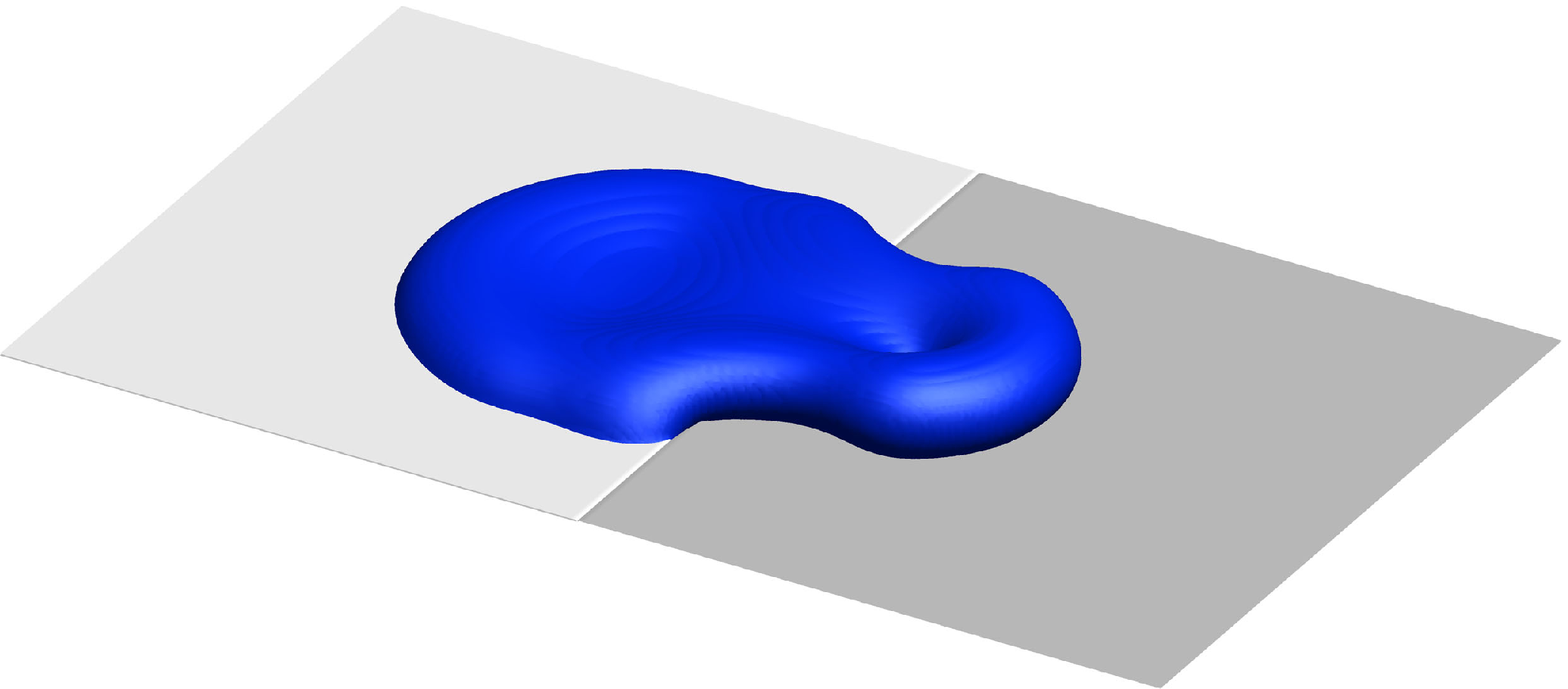}
		\end{minipage}
		\begin{minipage}[c]{0.2\textwidth}
			\includegraphics[width=\textwidth]{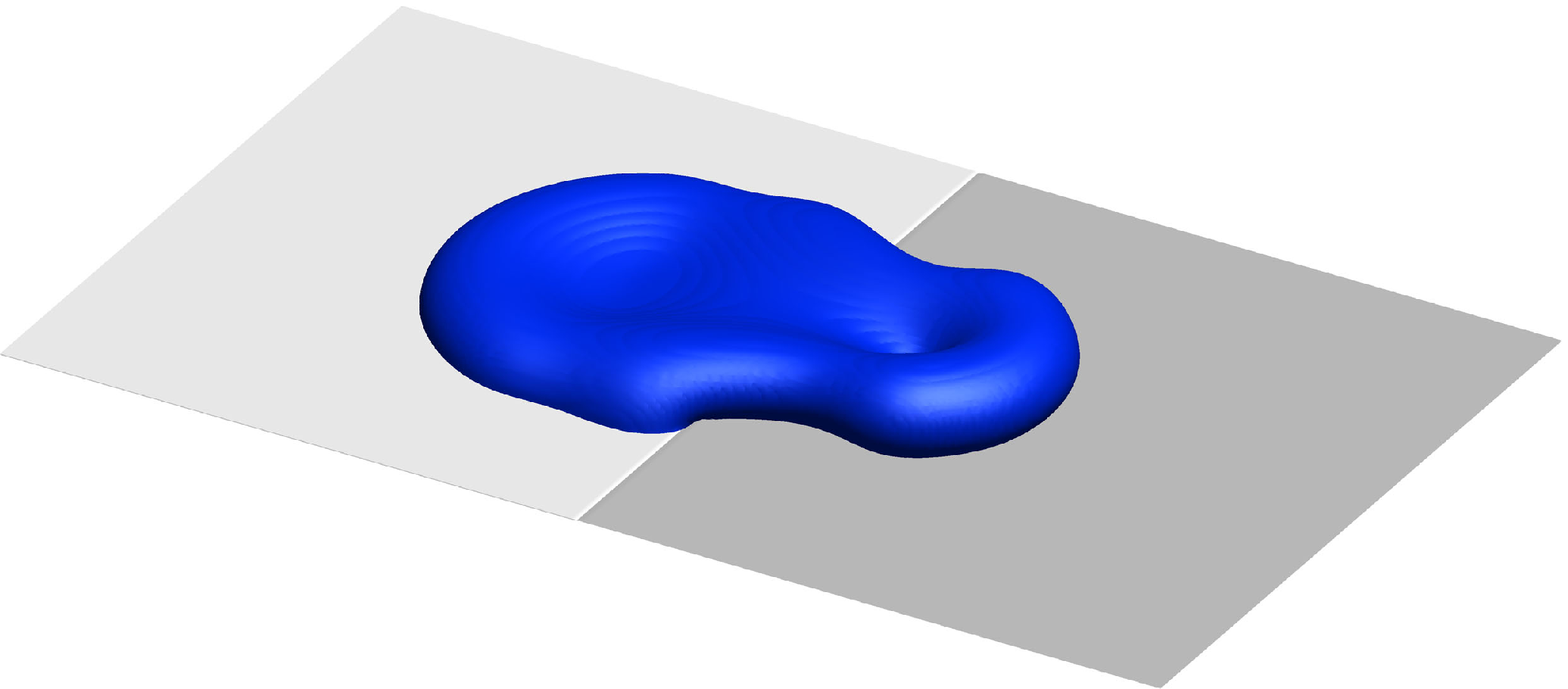}
		\end{minipage}
		\begin{minipage}[c]{0.2\textwidth}
			\includegraphics[width=\textwidth]{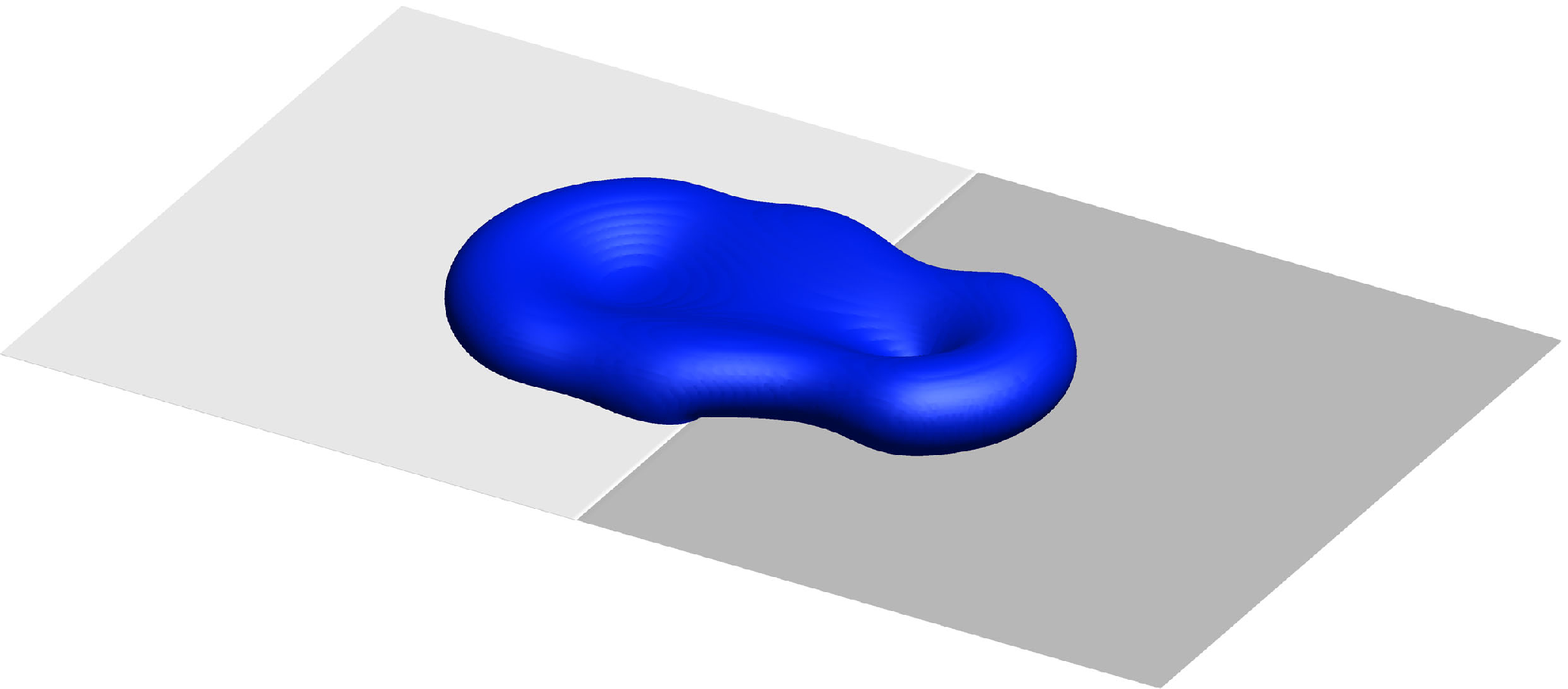}
		\end{minipage}
		\begin{minipage}[c]{0.2\textwidth}
			\includegraphics[width=\textwidth]{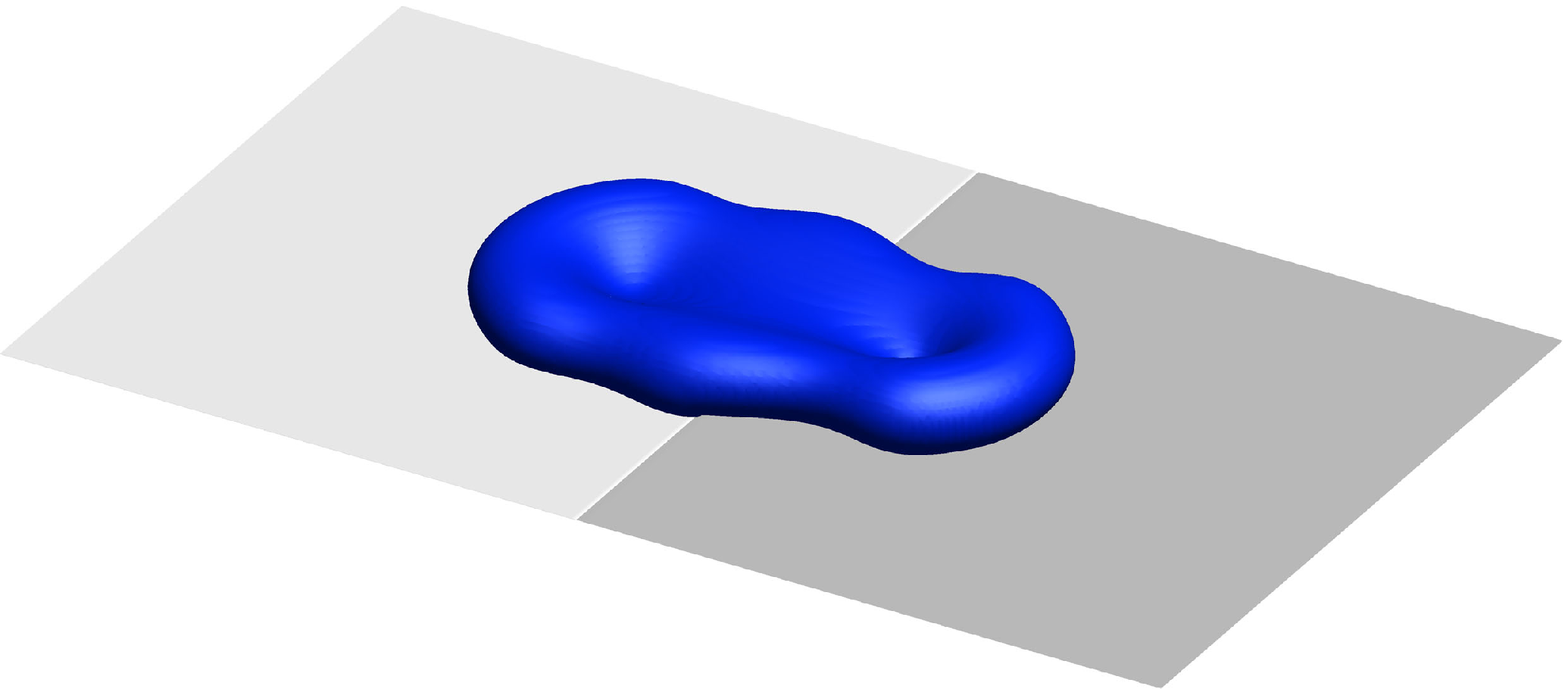}
		\end{minipage}

		\begin{minipage}[c]{0.1\textwidth}
			\centering
			\caption*{(e) $ t^{*}=2.0 $  }
		\end{minipage}
		\begin{minipage}[c]{0.2\textwidth}
			\includegraphics[width=\textwidth]{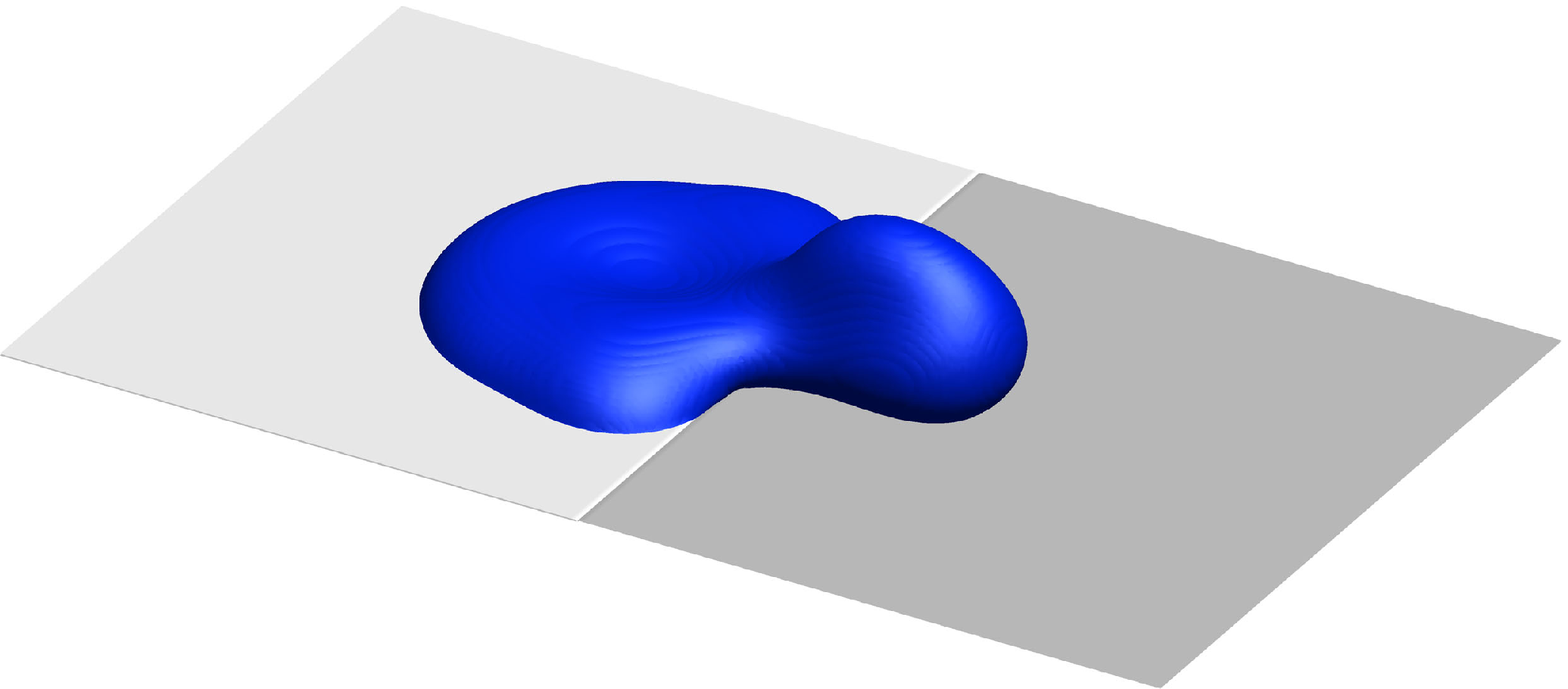}
		\end{minipage}
		\begin{minipage}[c]{0.2\textwidth}
			\includegraphics[width=\textwidth]{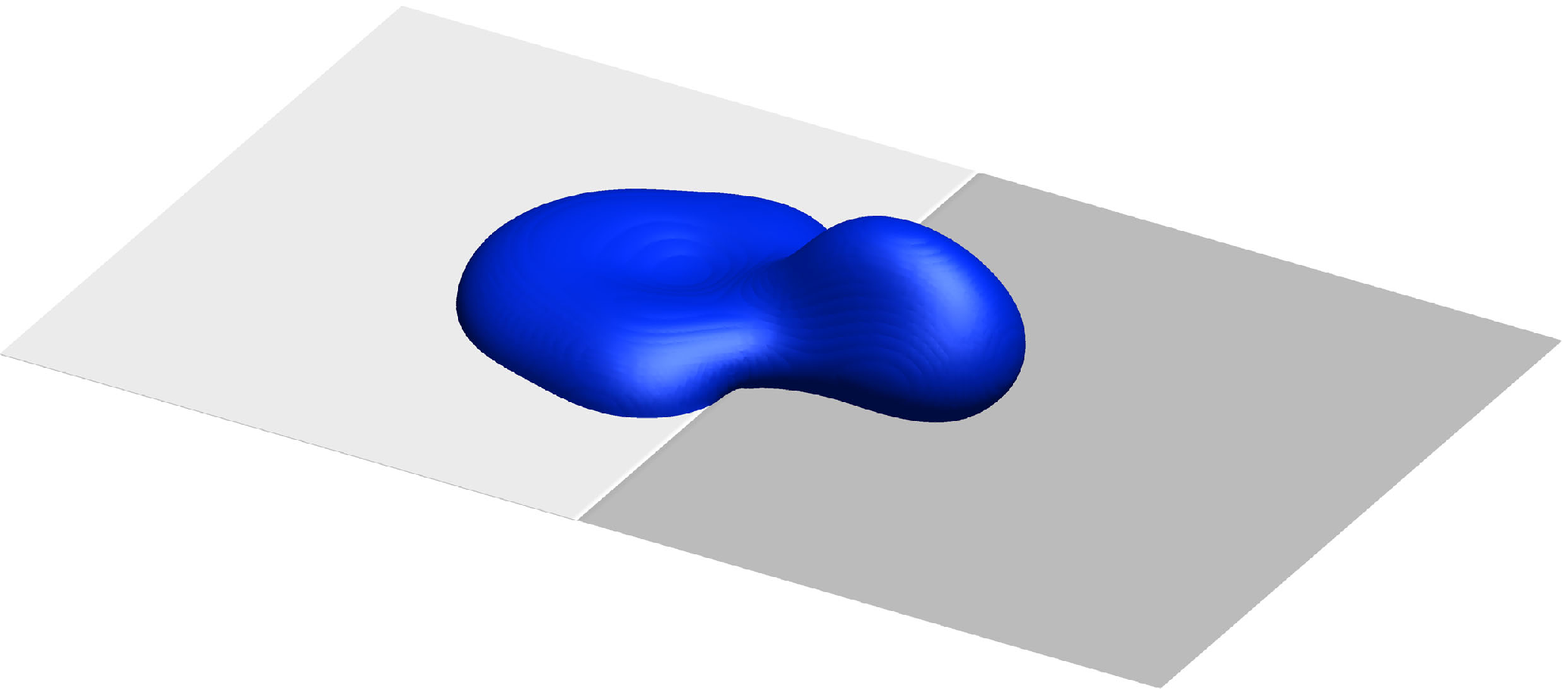}
		\end{minipage}
		\begin{minipage}[c]{0.2\textwidth}
			\includegraphics[width=\textwidth]{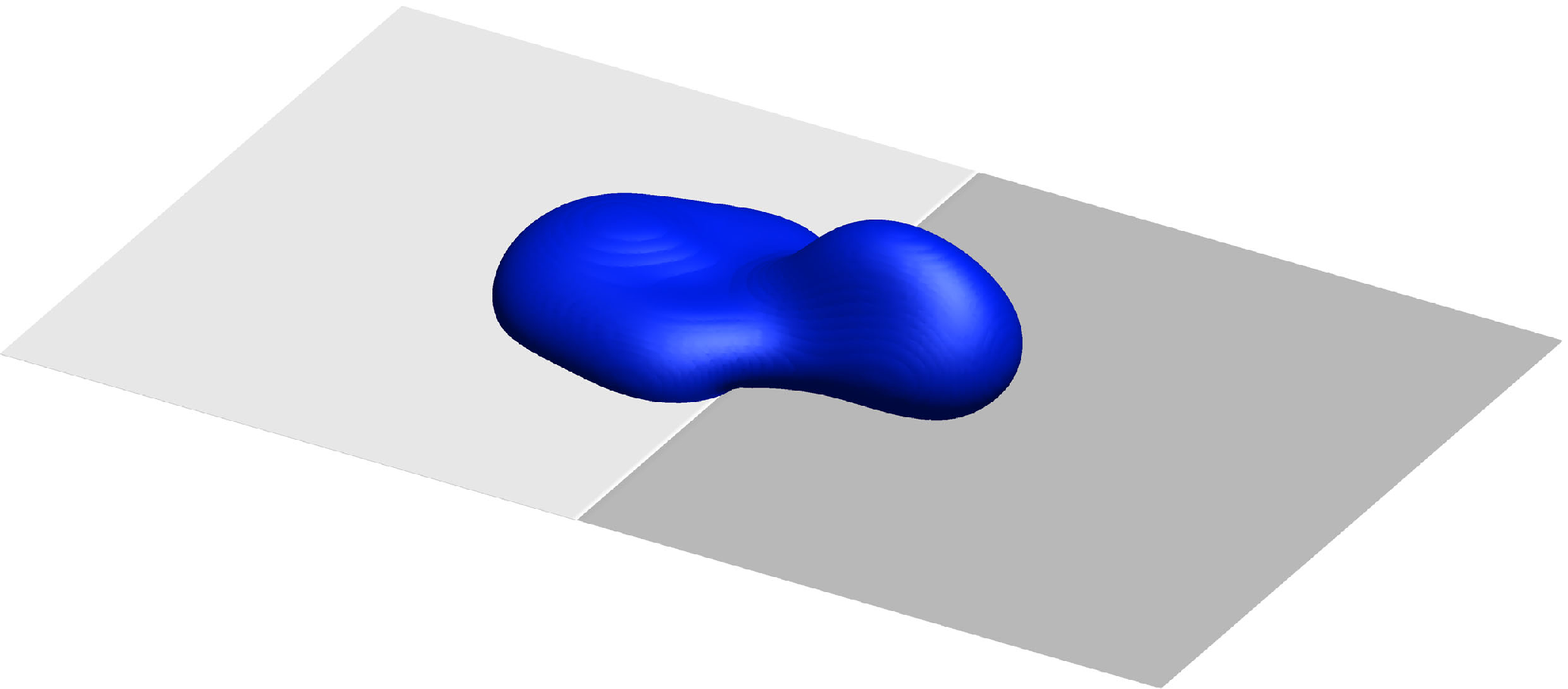}
		\end{minipage}
		\begin{minipage}[c]{0.2\textwidth}
			\includegraphics[width=\textwidth]{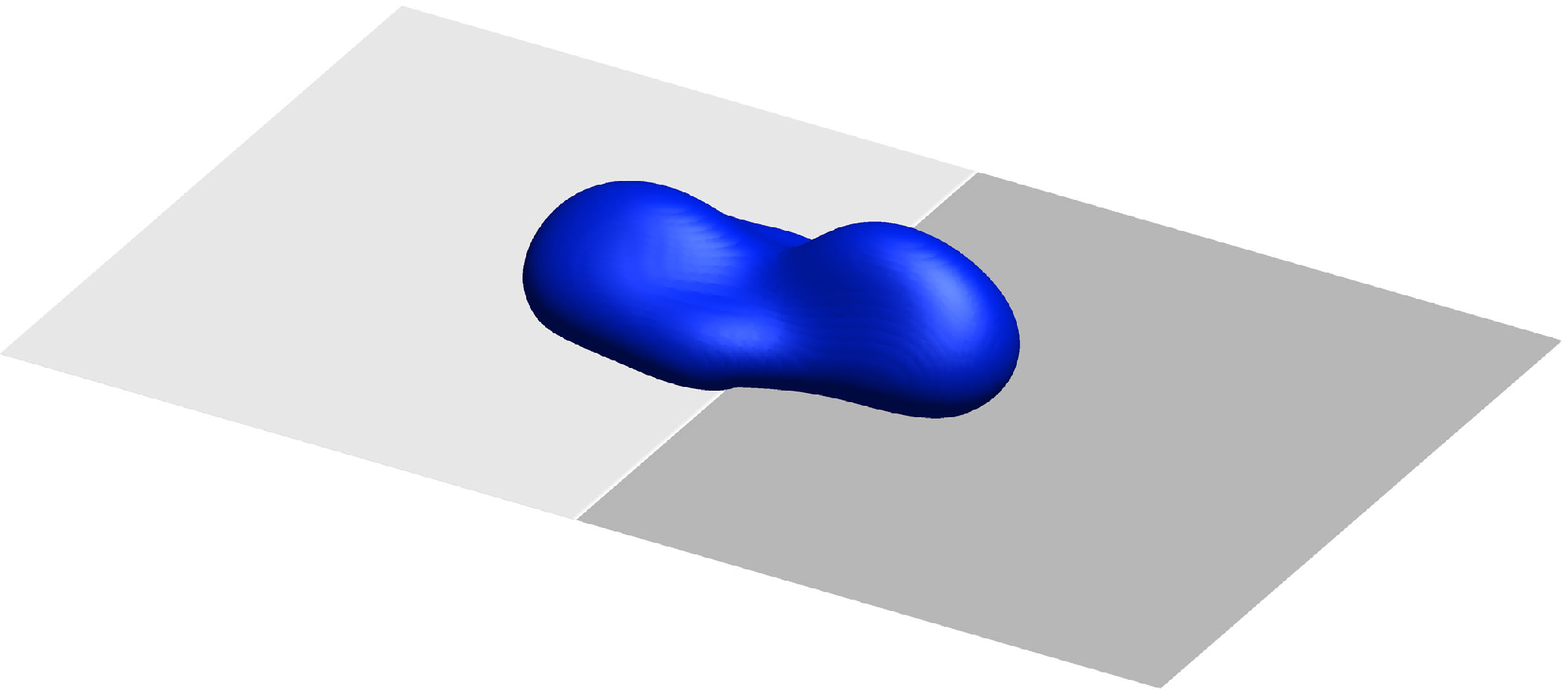}
		\end{minipage}

		\begin{minipage}[c]{0.1\textwidth}
			\centering
			\caption*{(f) $ t^{*}=3.0 $  }
		\end{minipage}
		\begin{minipage}[c]{0.2\textwidth}
			\includegraphics[width=\textwidth]{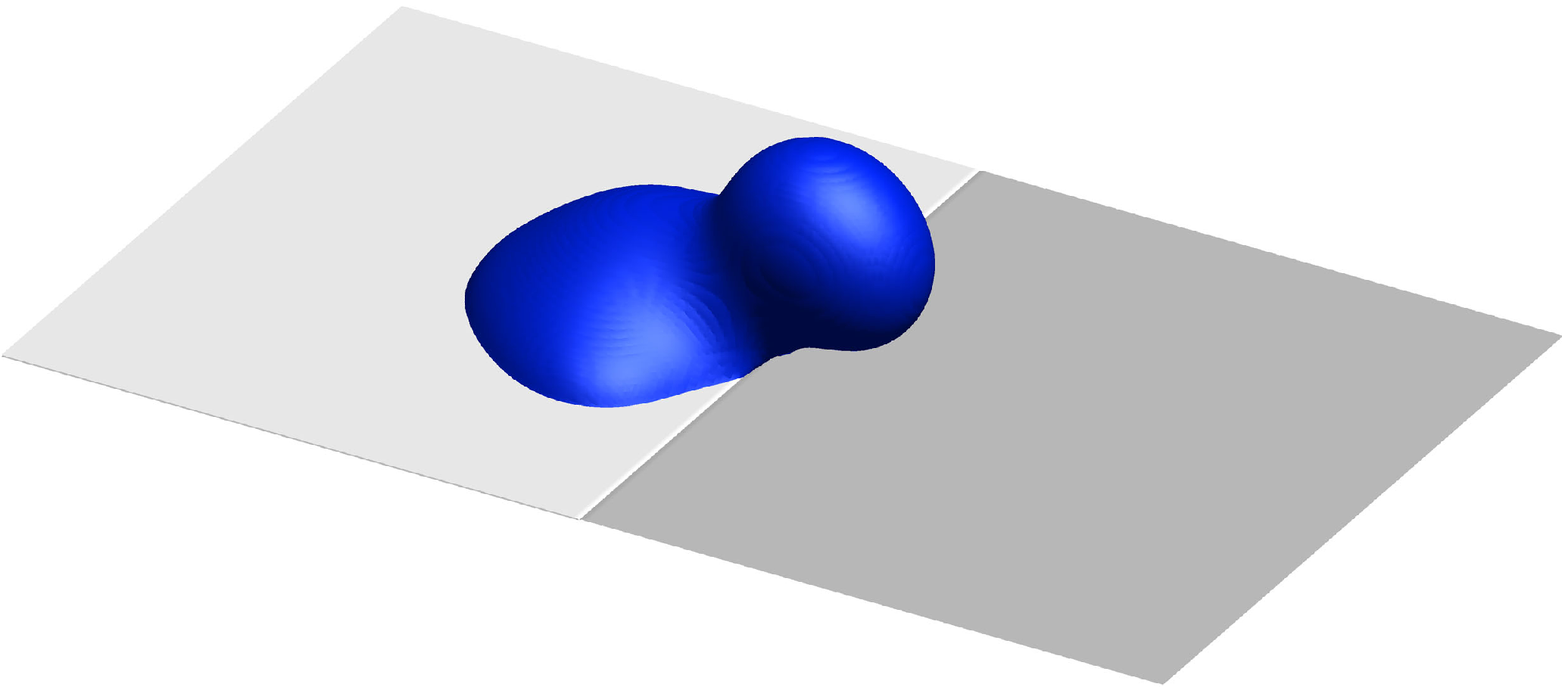}
		\end{minipage}
		\begin{minipage}[c]{0.2\textwidth}
			\includegraphics[width=\textwidth]{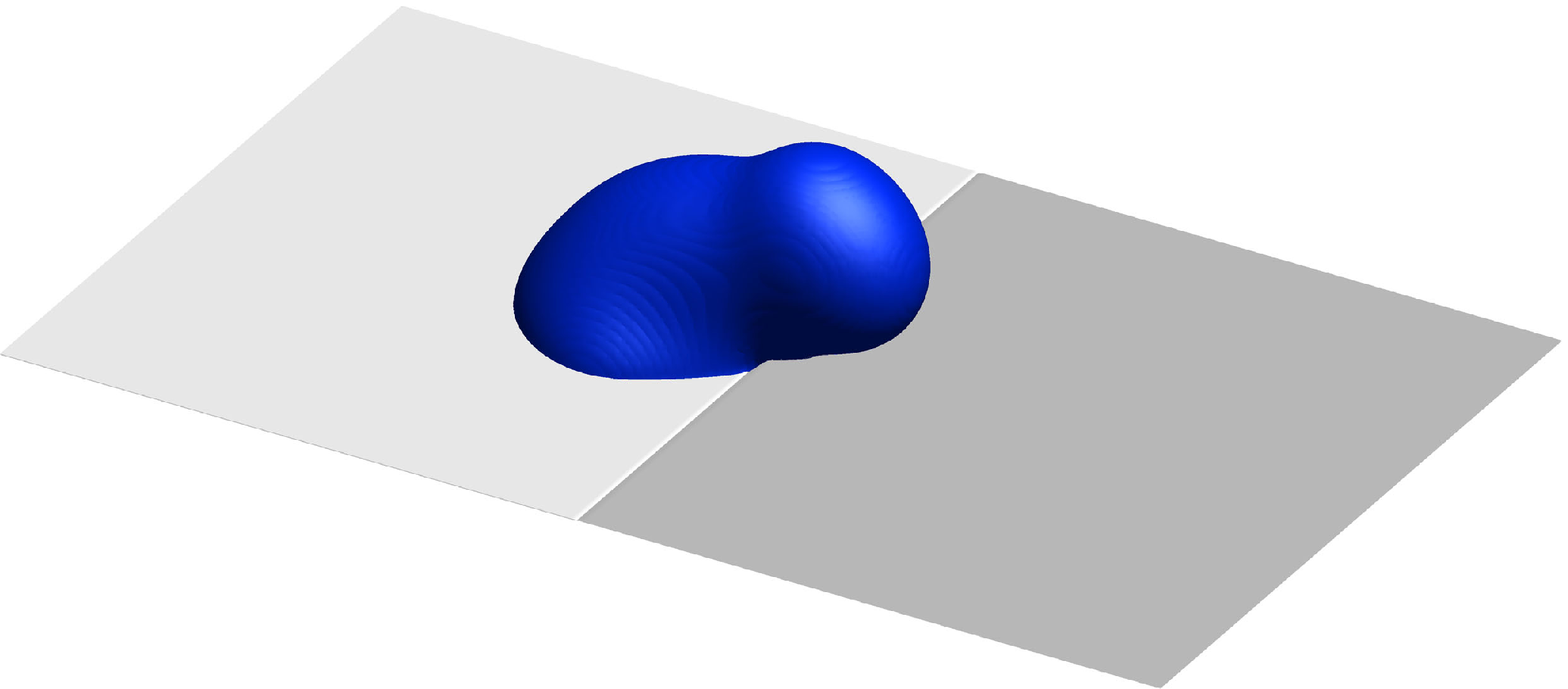}
		\end{minipage}
		\begin{minipage}[c]{0.2\textwidth}
			\includegraphics[width=\textwidth]{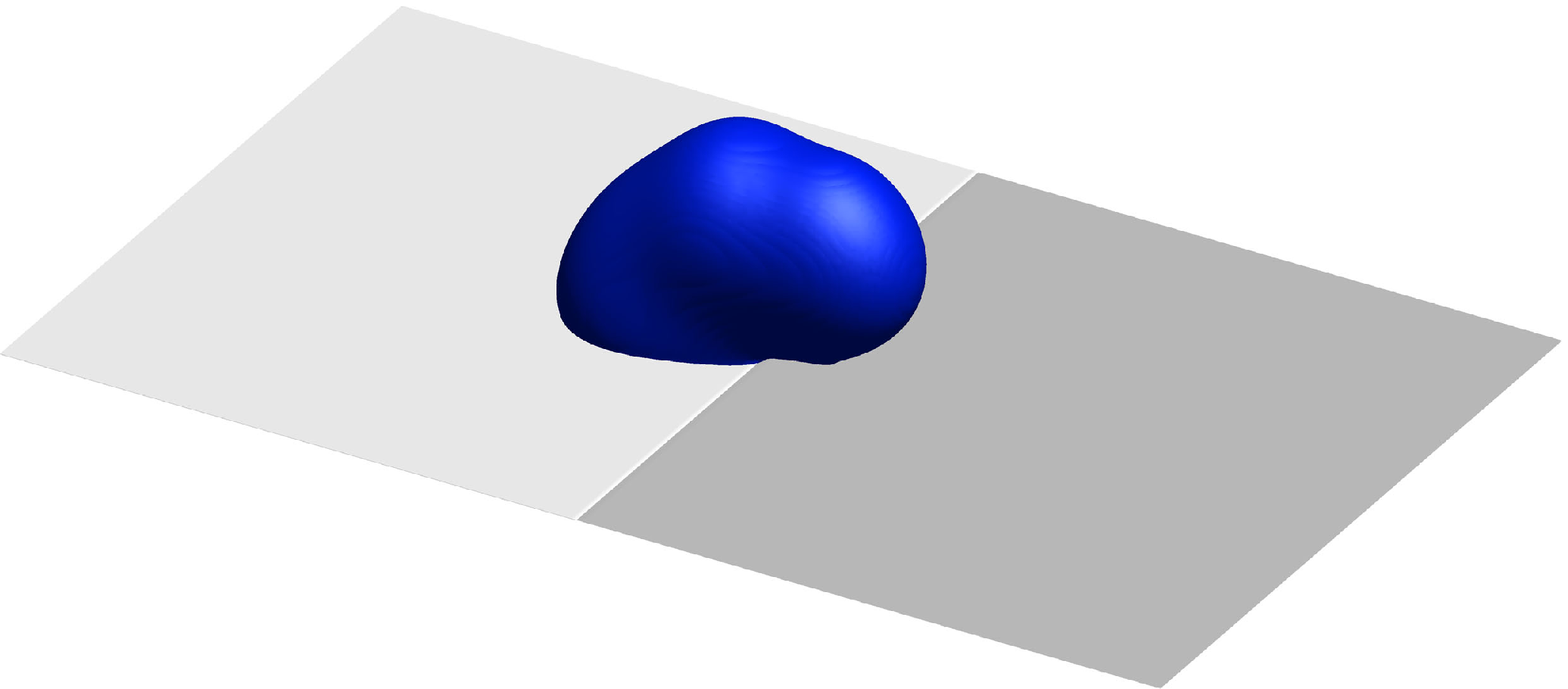}
		\end{minipage}	
		\begin{minipage}[c]{0.2\textwidth}
			\includegraphics[width=\textwidth]{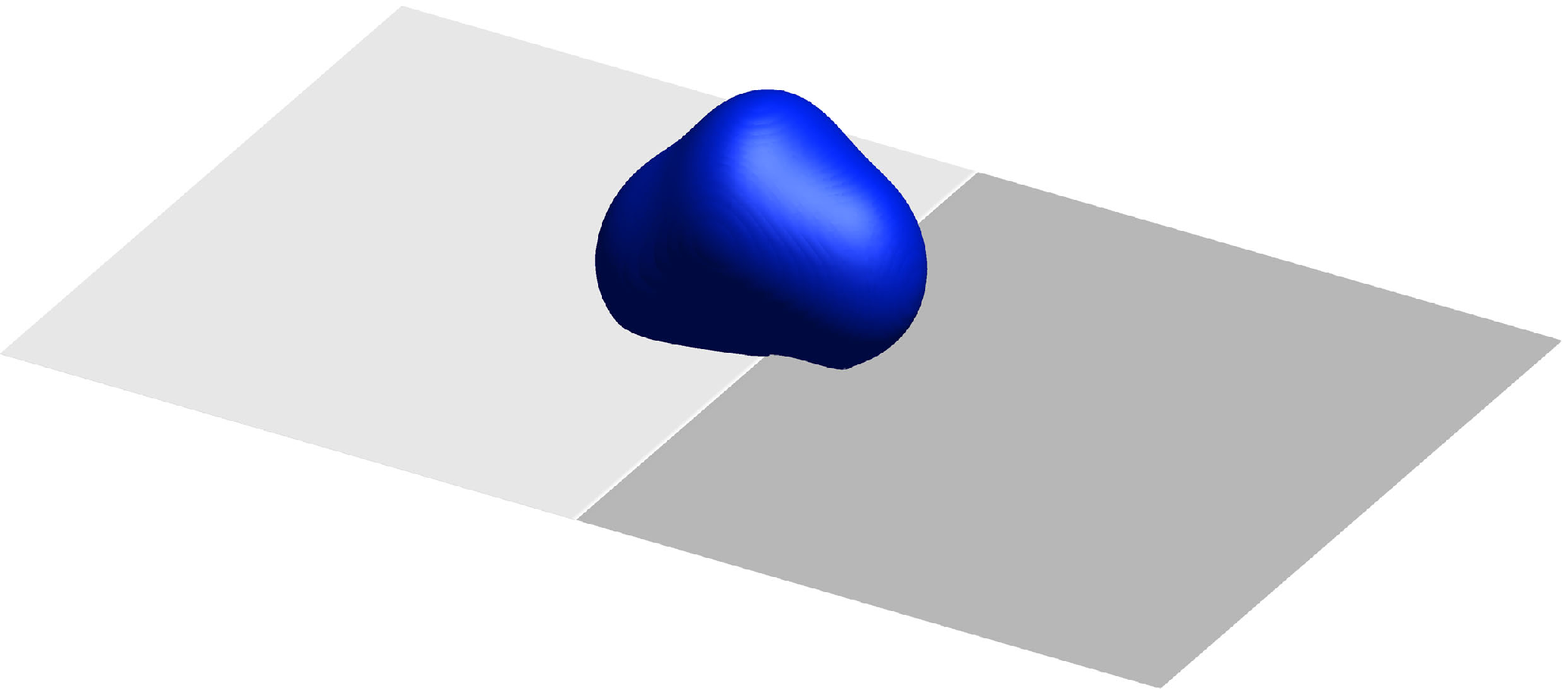}
		\end{minipage}

		\begin{minipage}[c]{0.1\textwidth}
			\centering
			\caption*{(g) $ t^{*}=4.0 $  }
		\end{minipage}
		\begin{minipage}[c]{0.2\textwidth}
			\includegraphics[width=\textwidth]{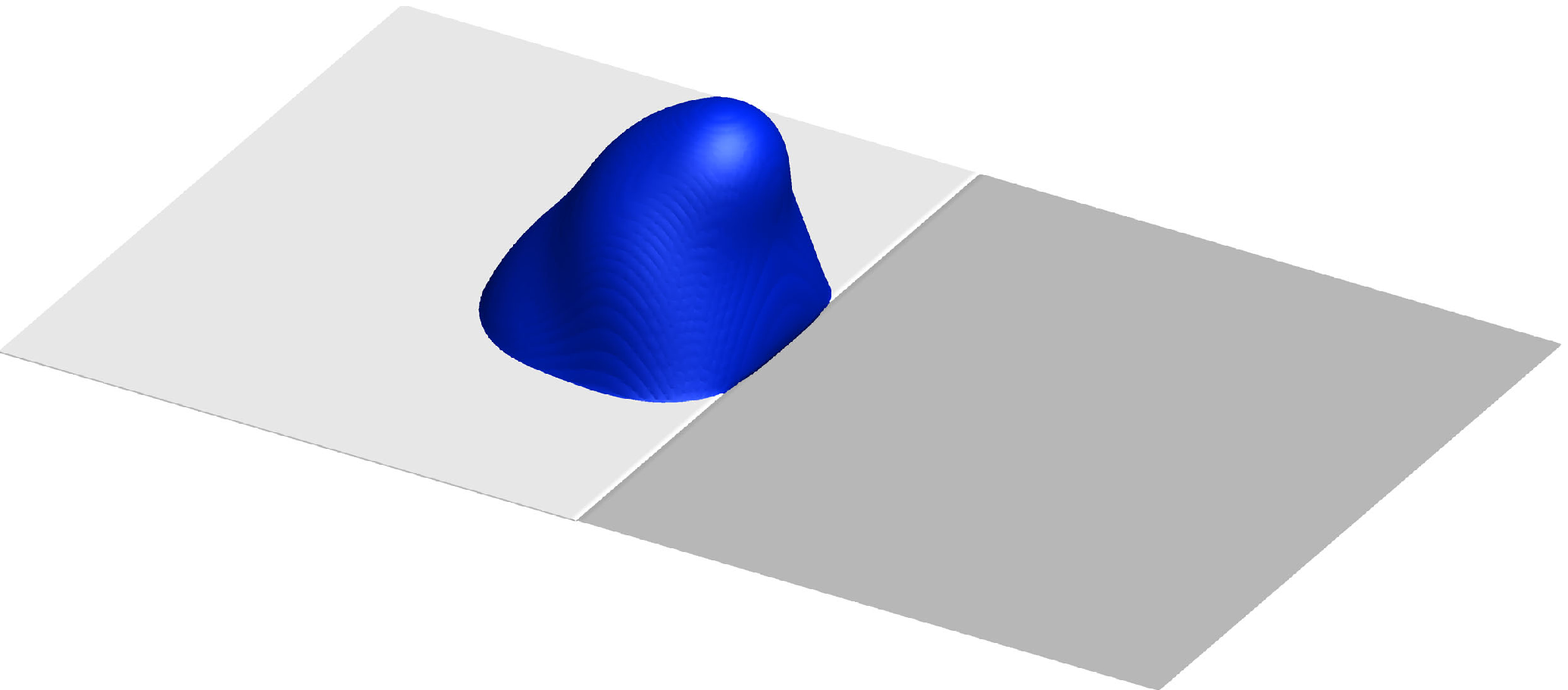}
		\end{minipage}
		\begin{minipage}[c]{0.2\textwidth}
			\includegraphics[width=\textwidth]{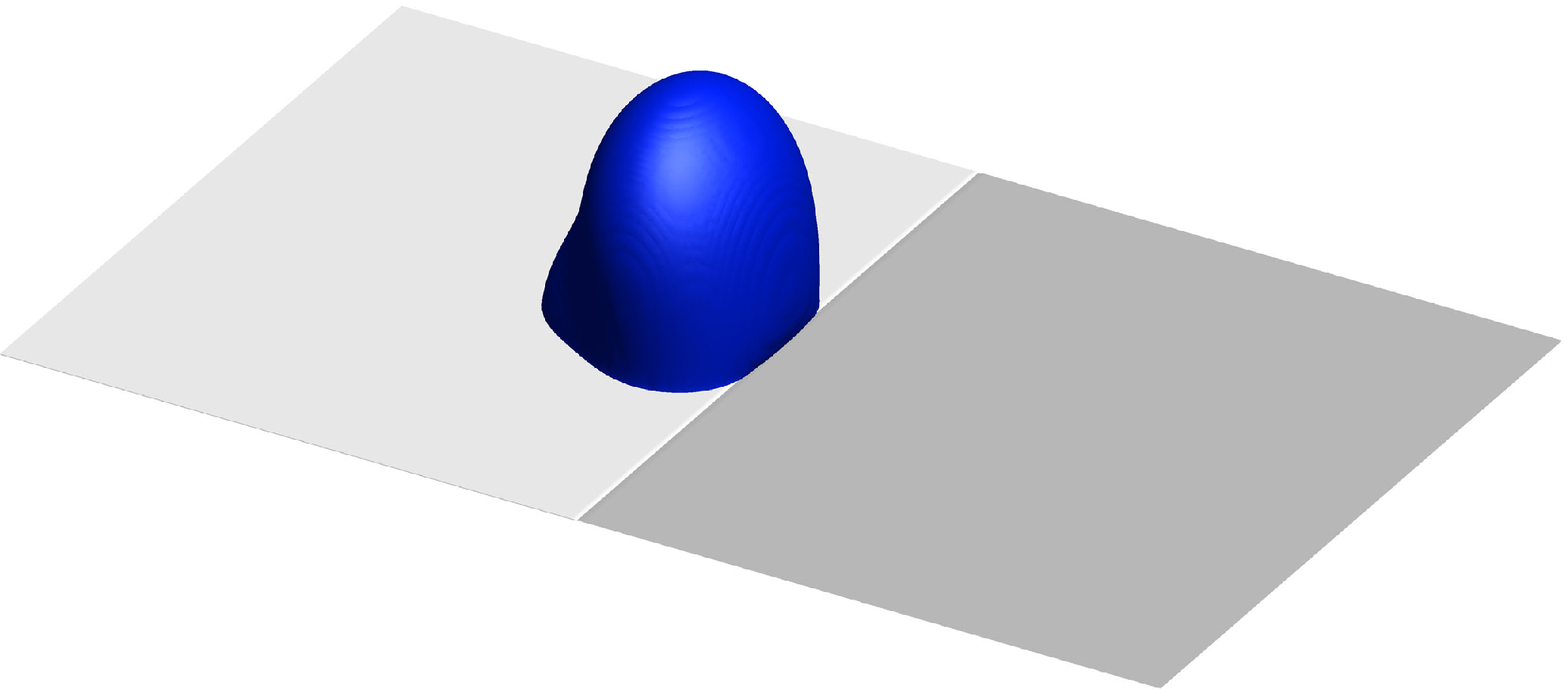}
		\end{minipage}
		\begin{minipage}[c]{0.2\textwidth}
			\includegraphics[width=\textwidth]{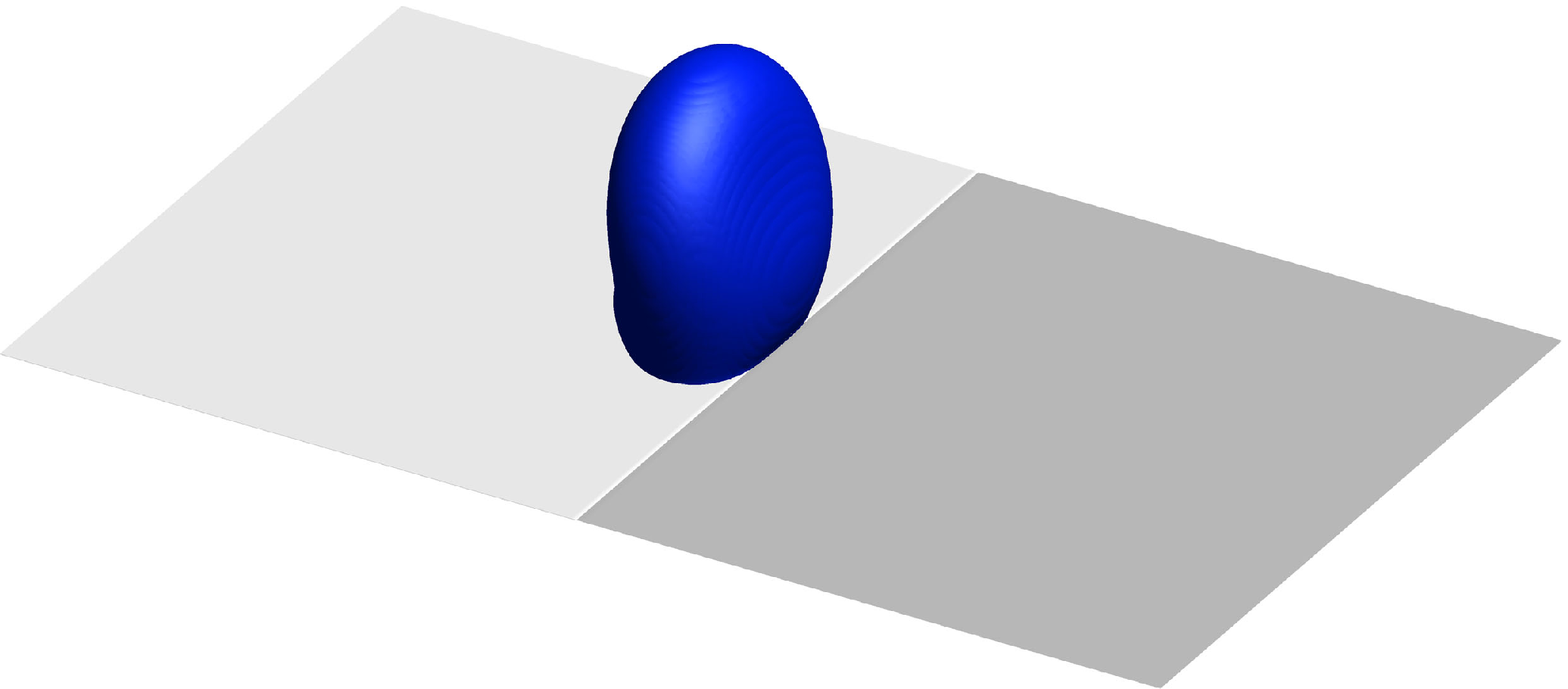}
		\end{minipage}			
		\begin{minipage}[c]{0.2\textwidth}
			\includegraphics[width=\textwidth]{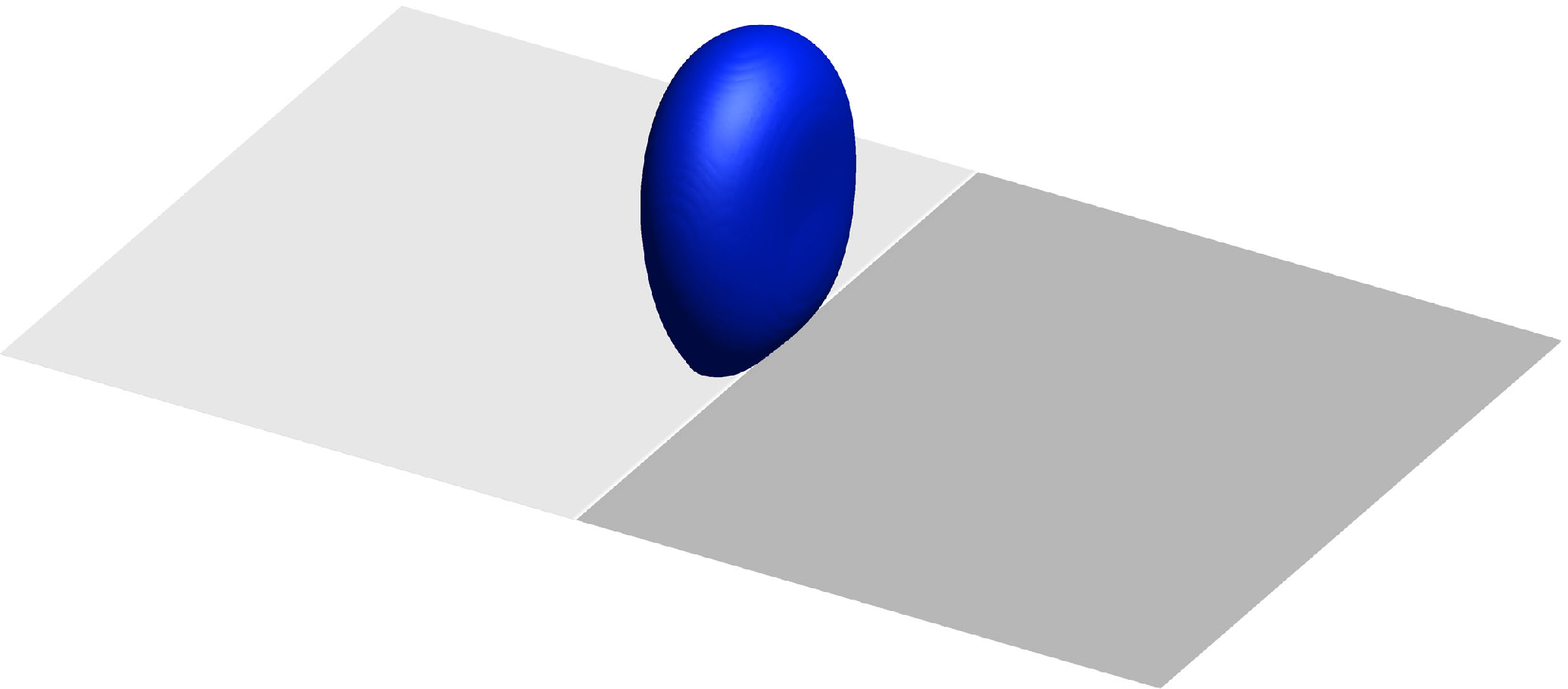}
		\end{minipage}

		\begin{minipage}[c]{0.1\textwidth}
			\centering
			\caption*{(h) $ t^{*}=5.0 $  }
		\end{minipage}
		\begin{minipage}[c]{0.2\textwidth}
			\includegraphics[width=\textwidth]{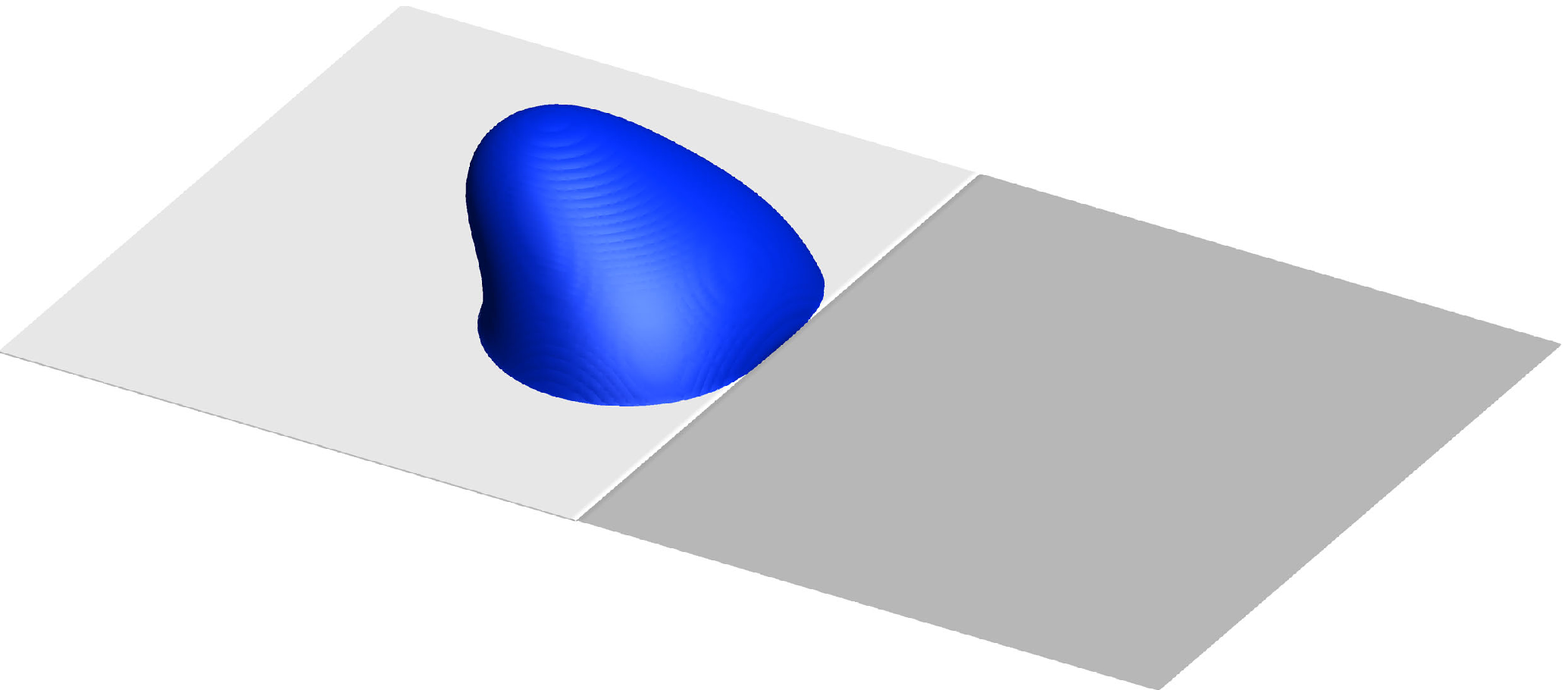}
			\caption*{$ \Delta \theta=80 \degree $}
		\end{minipage}
		\begin{minipage}[c]{0.2\textwidth}
			\includegraphics[width=\textwidth]{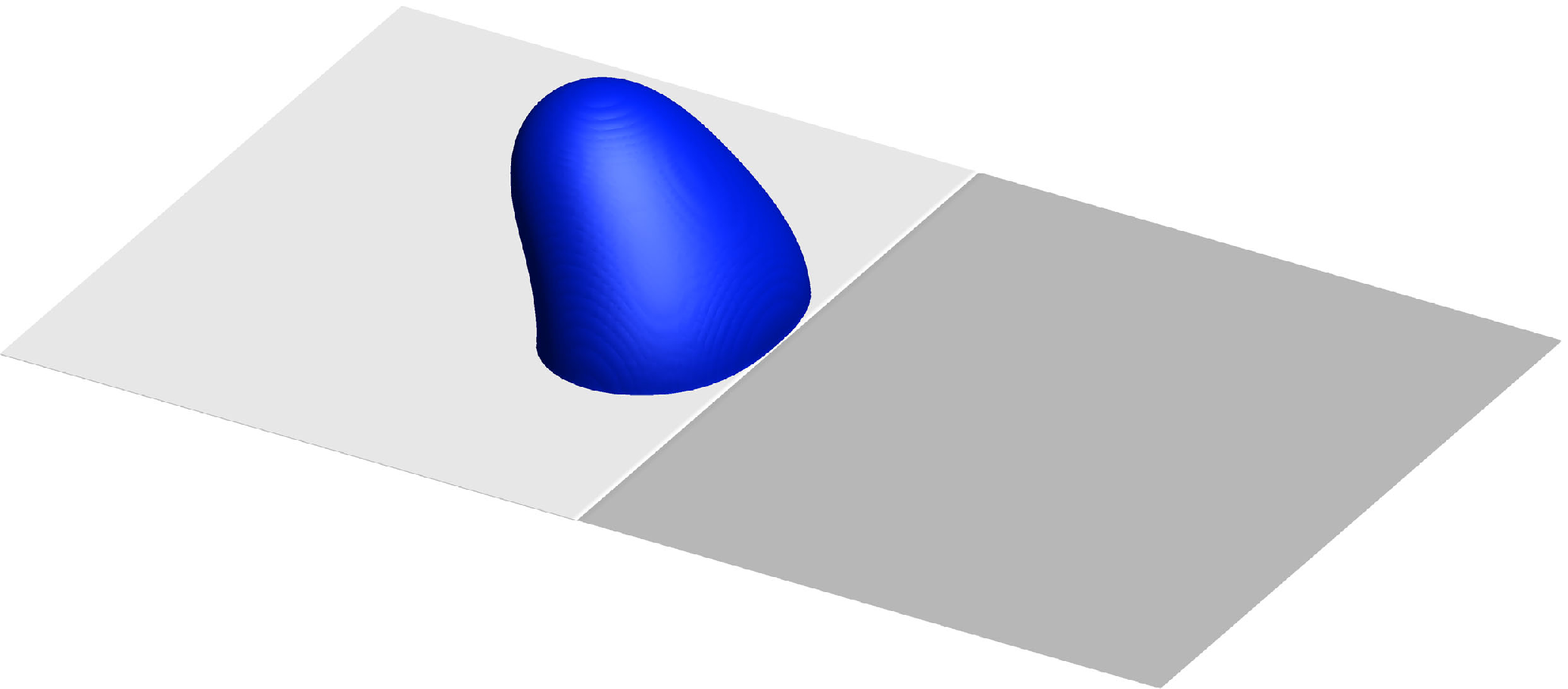}
			\caption*{$ \Delta \theta=60 \degree $}
		\end{minipage}
		\begin{minipage}[c]{0.2\textwidth}
			\includegraphics[width=\textwidth]{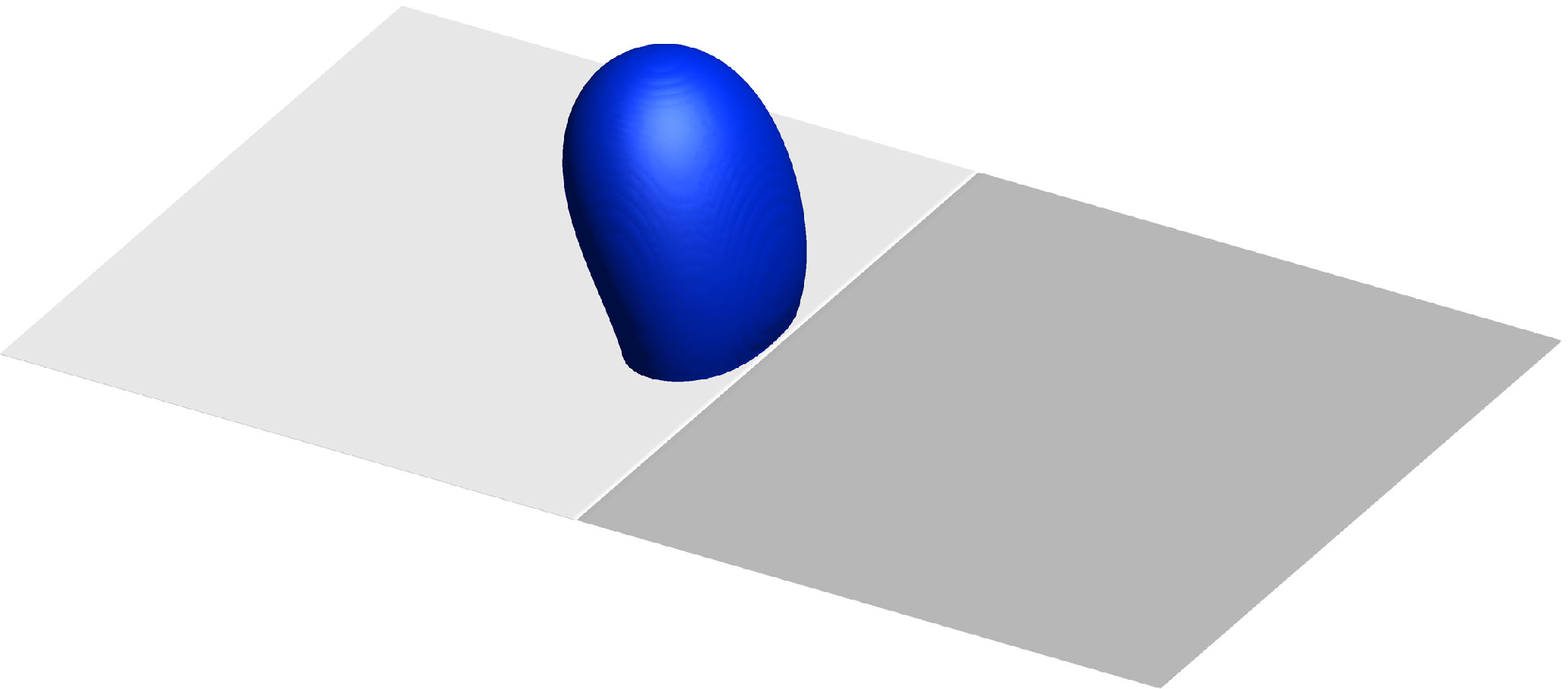}
			\caption*{$ \Delta \theta=40 \degree $}
		\end{minipage}		
		\begin{minipage}[c]{0.2\textwidth}
			\includegraphics[width=\textwidth]{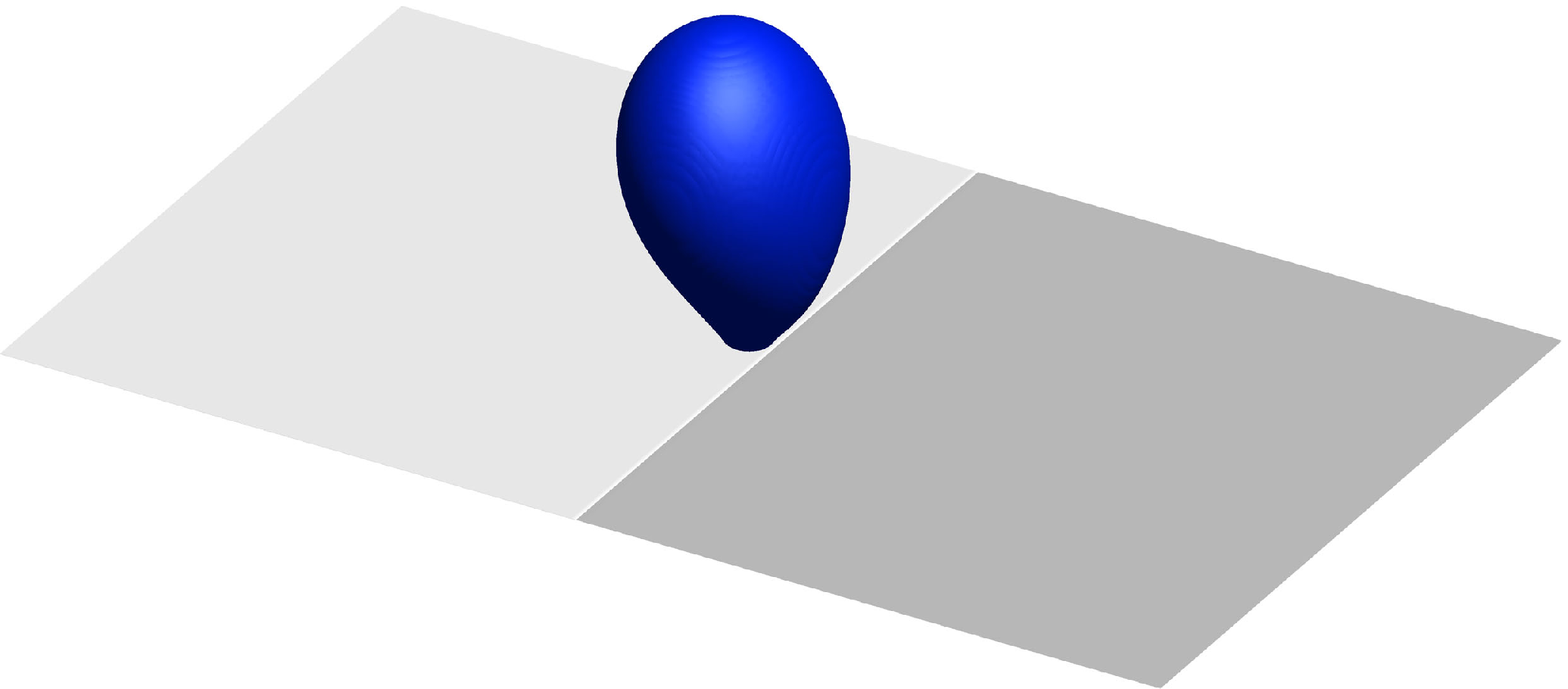}
			\caption*{$ \Delta \theta=20 \degree $}
		\end{minipage}

		\caption{The morphologies of double droplets impinging upon the surface with different wettability difference  $ \Delta \theta $ at $We=52.48$ and $ L^{*}=0.75 $.} 
		\label{fig5}
	\end{figure}

The wettability difference of the heterogeneous surface has a significant influence on the dynamics and behavior of the impinging droplets. In this subsection, the influence of wettability difference on droplet impact and interaction dynamics are discussed. As reported in some previous works \cite{RiobooEF2001}, the evolutionary process of a single droplet impinging on a homogeneous surface can be divided into four phases: the motion phase, the diffusion phase, the relaxation phase, and the wetting/equilibrium phase. According to our numerical results, however, three different evolutionary phases are observed for the  double droplets impact on a wettability-patterned surface, i.e., the asymmetric spreading phase, asymmetric retracting phase and wetting equilibrium phase.

\begin{figure}[H]
		\centering 
		\subfigure[]{ \label{fig6a}
			\includegraphics[width=0.3\textwidth]{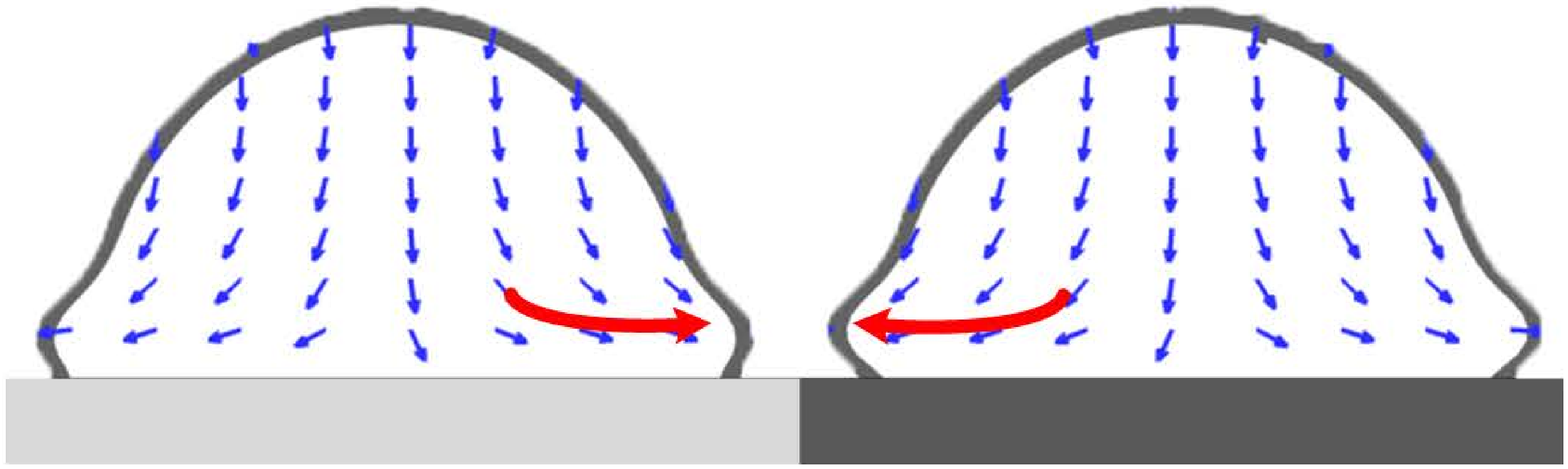}} 
		\subfigure[]{ \label{fig6b}
			\includegraphics[width=0.3\textwidth]{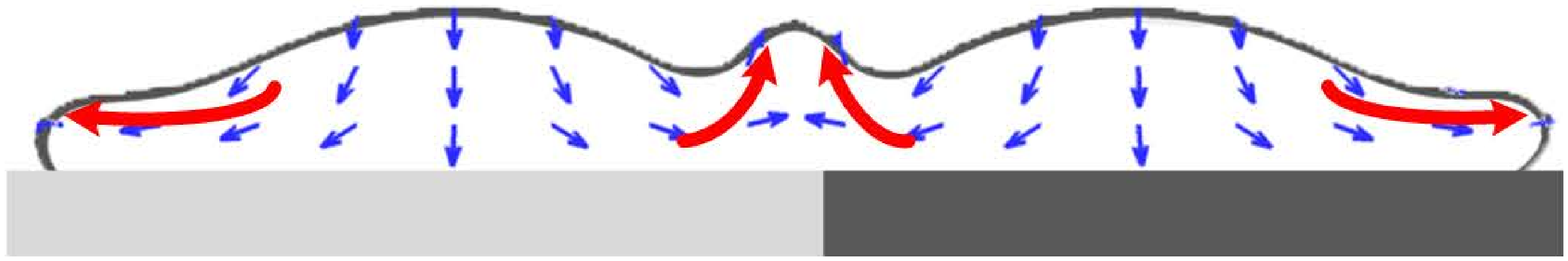}}
		\subfigure[]{ \label{fig6c}
			\includegraphics[width=0.3\textwidth]{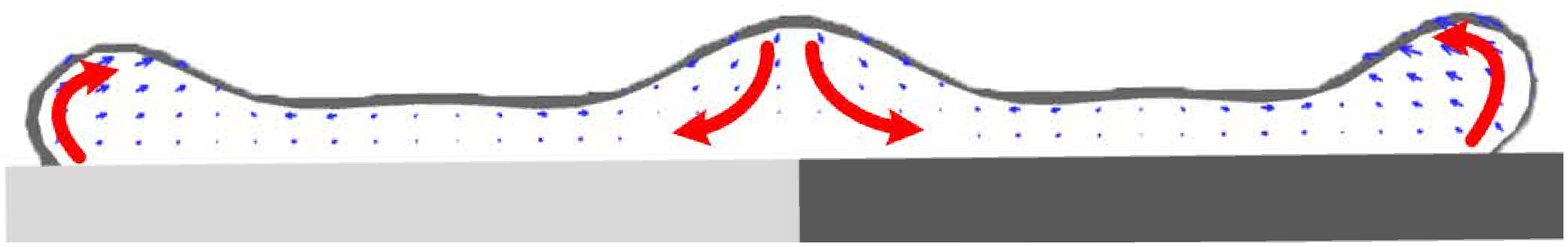}}			
		\caption{ Velocity field inside the droplet at different instants on a surface with wettability difference $ \Delta \theta $ .}
		\label{fig6}
	\end{figure}

Fig. \ref{fig5} illustrates the evolution of the two impinging droplets on surfaces with different wettability differences, in which the contact angle of the high-wettability region (light gray) for all cases is fixed at $140 \degree$. As shown in this figure, the above mentioned three evolutionary phases can be clearly observed. Specifically, at the initial stage, the droplets form a crown-like film after impacting the surface and spread freely in the $ x $ and $ y $ directions (see Fig. \ref{fig5}(a)). However, the low wettability side diffuses faster and has a larger spreading area compared to the high wettability side (see Fig. \ref{fig5}(c)). The two droplets coalesce rapidly when the edges of the droplets get in touch with each other, this mode of interaction between the droplets is called out-of-phase \cite{TongNHT2007}. Fig. \ref{fig6} shows the velocity field inside the droplets in this interaction mode, the two droplets prevent each other from spreading further and combine to form a liquid ridge due to the fact that the velocity vectors inside the two droplets in the spreading phase are moving in opposite directions. The liquid ridge gradually increases in the spreading phase due to the high momentum fluid inside the droplet pumps upwards (see Fig. \ref{fig6}(b)). When the droplets keep expanding outward to reach the maximum spreading diameters, two puddles and peripheral bulges would form on the two sides of the liquid ridges. Subsequently, the droplet starts to retract toward the center driven by surface tension to reduce its interfacial energy and enter the asymmetric shrinkage phase. During the recoiling process, the wettability difference of the surface causes the retracting process to be asymmetric, and the right side of the droplet retracts faster due to high-hydrophobic on the right side (see Fig. \ref{fig5}(e)). According to the Young-Laplace equation, the fluid in both the peripheral bulge and the liquid ridge flows into the puddle since the pressure in the puddle is lower compared to the outer bulge. As show in Fig. \ref{fig6c}, this flow model caused the central liquid ridge to disappear first and the peripheral bulge evolves into an elliptical puddle by retracting along the ridge. With the progress of contraction process, the droplet has completely left the low-wettability surface and migrate completely to the high-wettability side driven by the unbalanced force created by the wettability differences (see Fig. \ref{fig5}(f)). For a small wettability difference $ \Delta \theta=20 \degree $, two bulges are formed on both sides of the borderline line, and the liquid around the retraction point is continuously lifted up and merged into a new retraction point. Finally, the coalescing droplets rebound towards the hydrophilic region. Meanwhile, the coalesced droplets are stays in an elongated shape along the z-direction. And for the case of large wettability difference, the droplet forms a bump in the hydrophobic region, which gradually rises and moves towards the high wetting region, and the droplet migrates to the high-wettability surface without leaving the surface eventually and tend to be stable gradually (see Fig. \ref{fig5}(h)). We also note that in the process of contraction, because of the existence of droplets and the out-of-phase motion in y-direction, the contraction of droplets in y-direction is greatly suppressed, and the droplet retraction in the x-direction is much faster than that in the y-direction, leading to the formation of an elongated droplet and thus further enhancing the asymmetry.

	\begin{figure}[ht]
		\centering
		\includegraphics[width=0.5\textwidth]{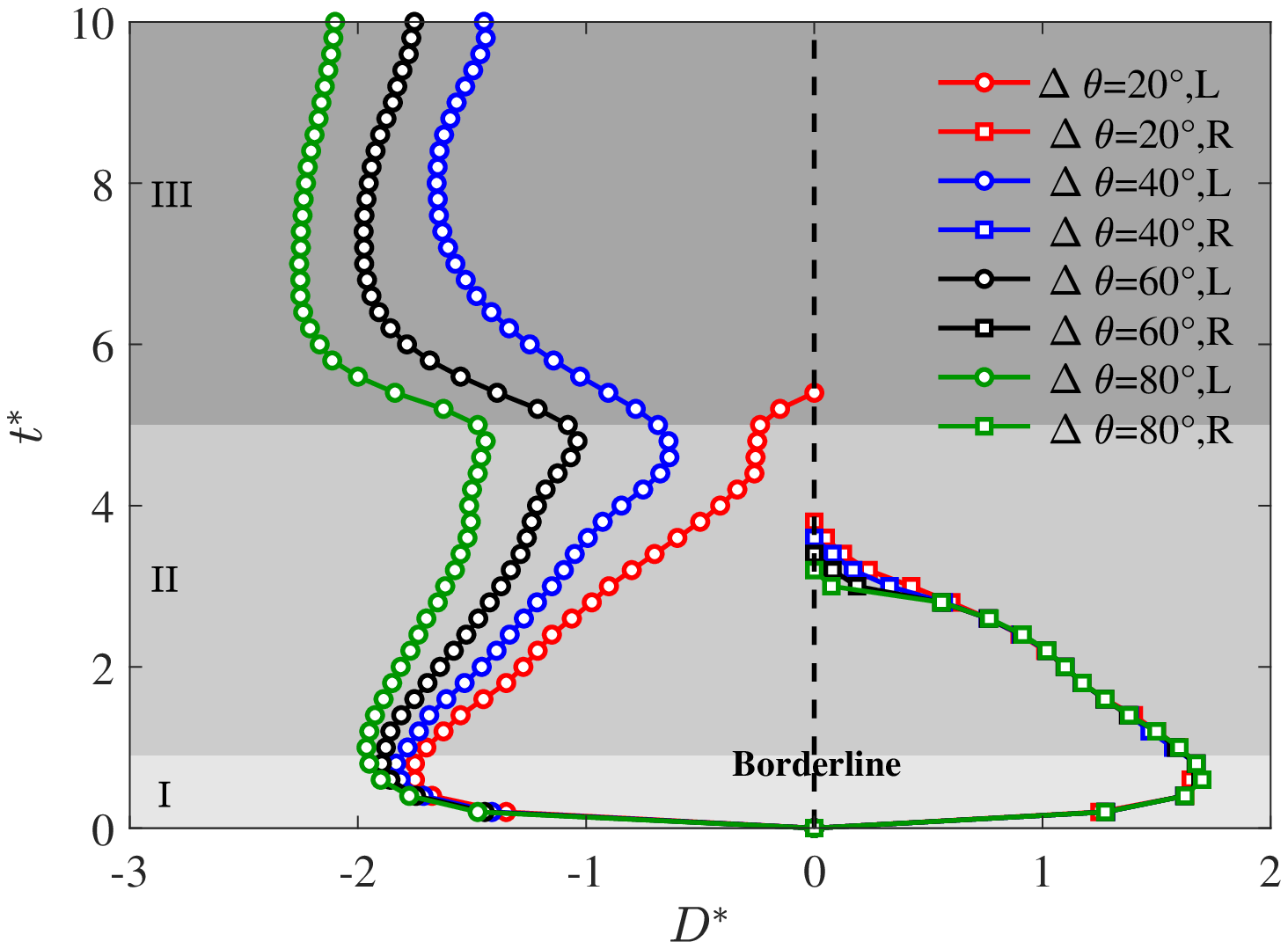}
		\caption{ Temporal evolution of the spread factor for different wettability differences $ \Delta \theta $ at $We=20$.}
		\label{fig7}
	\end{figure}
	
	\begin{figure}[ht]
		\centering
		\includegraphics[width=0.5\textwidth]{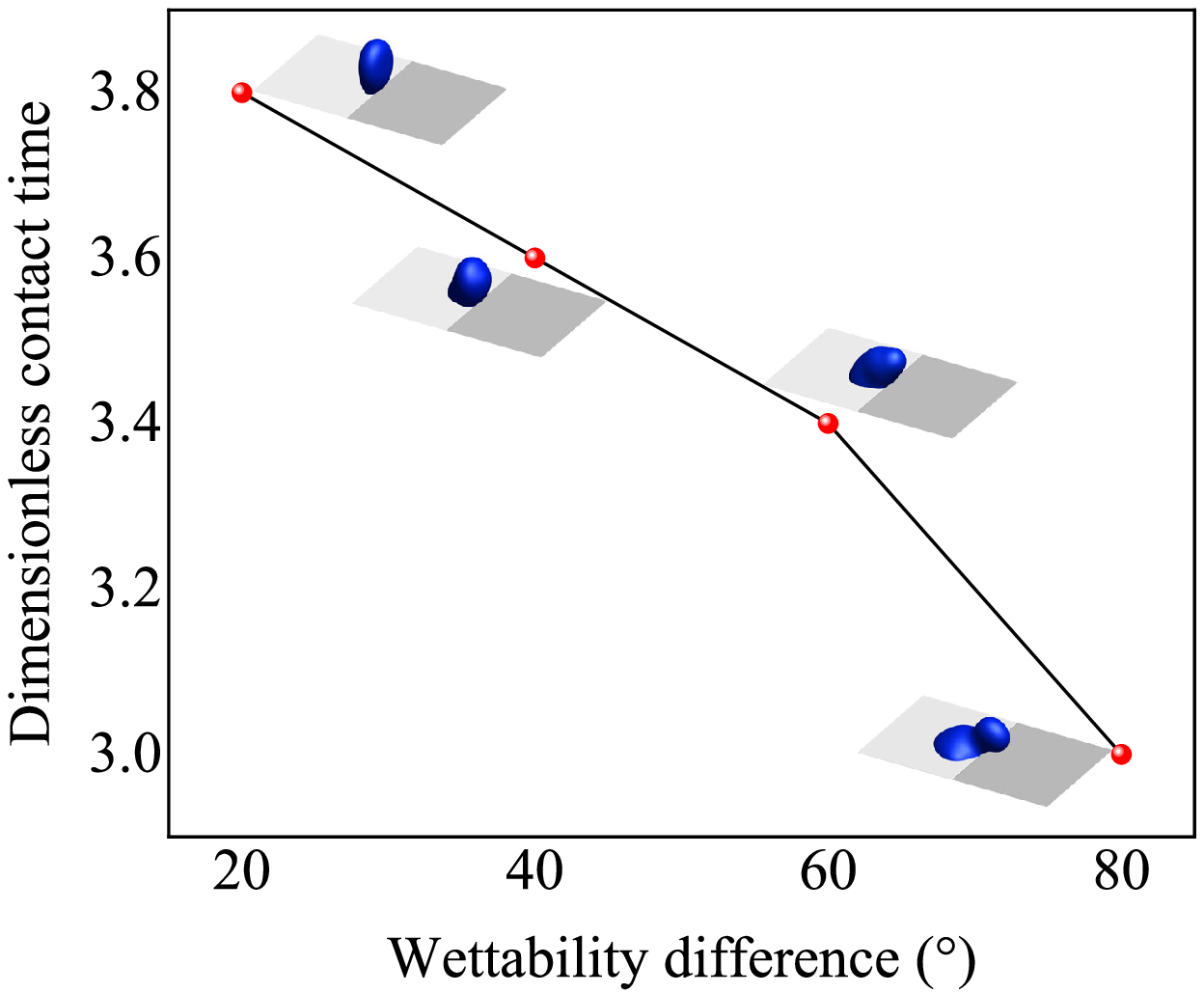}
		\caption{Variation of dimensionless contact time with different wettability differences $ \Delta \theta $.}
		\label{fig8}
	\end{figure}

To further understand the effects of wettability difference on the behaviors of double impinging droplets, Fig. \ref{fig7} illustrates the variation of diffusion factor $ D^{*} $ with dimensionless time $ t^{*} $ at different wettability differences. Here the spreading diameter is defined as the maximum distance of the three-phase contact line in the y-axis direction, because the wettability difference exists in the y-direction. It is apparent that the initial evolution of $D^{*}$ is nearly identical for all the three cases considered. This is attributed to the inertia dominated initial phase of impact which depends on the initial velocity. In the asymmetric spreading phase $ \textrm{I} $, the right contact angle decreases with the increase of wetting difference, and according to \cite{PasandidehFardPOF1996} 

\begin{equation}
D^{*}_{\max }=\frac{D_{\max }}{D}=\sqrt{\frac{\mathrm{We}+12}{3\left(1-\cos \theta \right)+4(\mathrm{We} / \sqrt{\mathrm{Re}})}}, 
\end{equation}
we can conclude that the right maximum spreading factor $ D^{*}_{\max } $ increases with the increase of wettability difference. The right spreading factor was the same under various wettability differences because the contact angles on the right side were consistent. In this process, part of the initial kinetic energy is converted into surface energy and the other part is dissipated by the viscous force. In the asymmetric retracting phase $ \textrm{II} $, the droplet detaches itself from the right hydrophobic region and gradually migrates to the left high-wettability side, the right spreading factor progressively decreases to zero, and the migration is completed first for $ \Delta \theta = 80 \degree $. For smaller wettability differences $ \Delta \theta = 20 \degree $, the right spread factor converges to zero values as the droplet rebounds off the substrate. In the equilibrium phase $ \textrm{III} $, the droplets do not rebound due to greater wettability for the remaining three cases. The spreading factor is gradually constant under the surface wetting effect.

\begin{figure}[H]
		\centering 
		\subfigure[$ \Delta \theta=80 \degree $]{ 
			\includegraphics[width=0.2\textwidth]{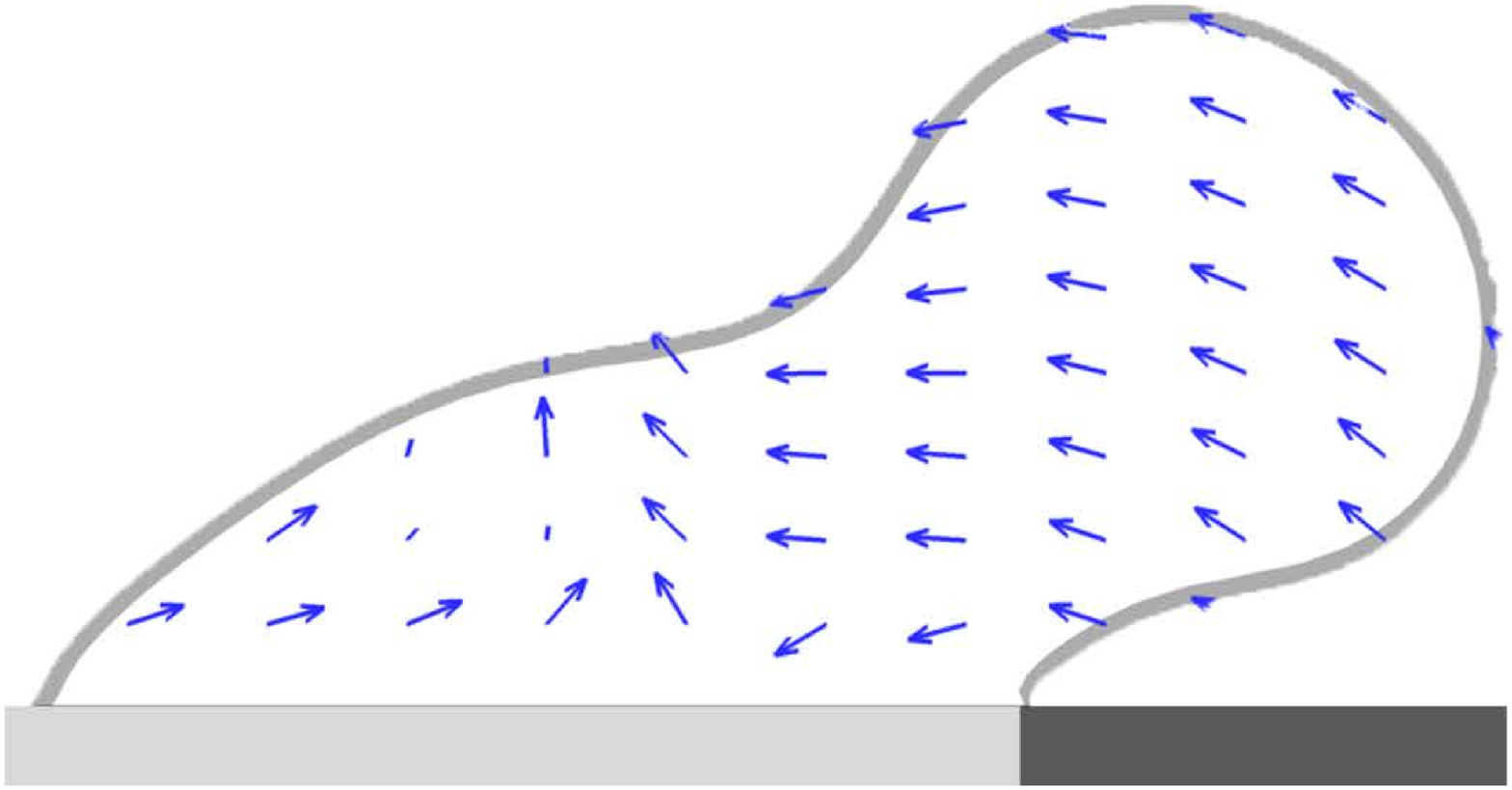}} 
		\subfigure[$ \Delta \theta=60 \degree $]{ 
			\includegraphics[width=0.2\textwidth]{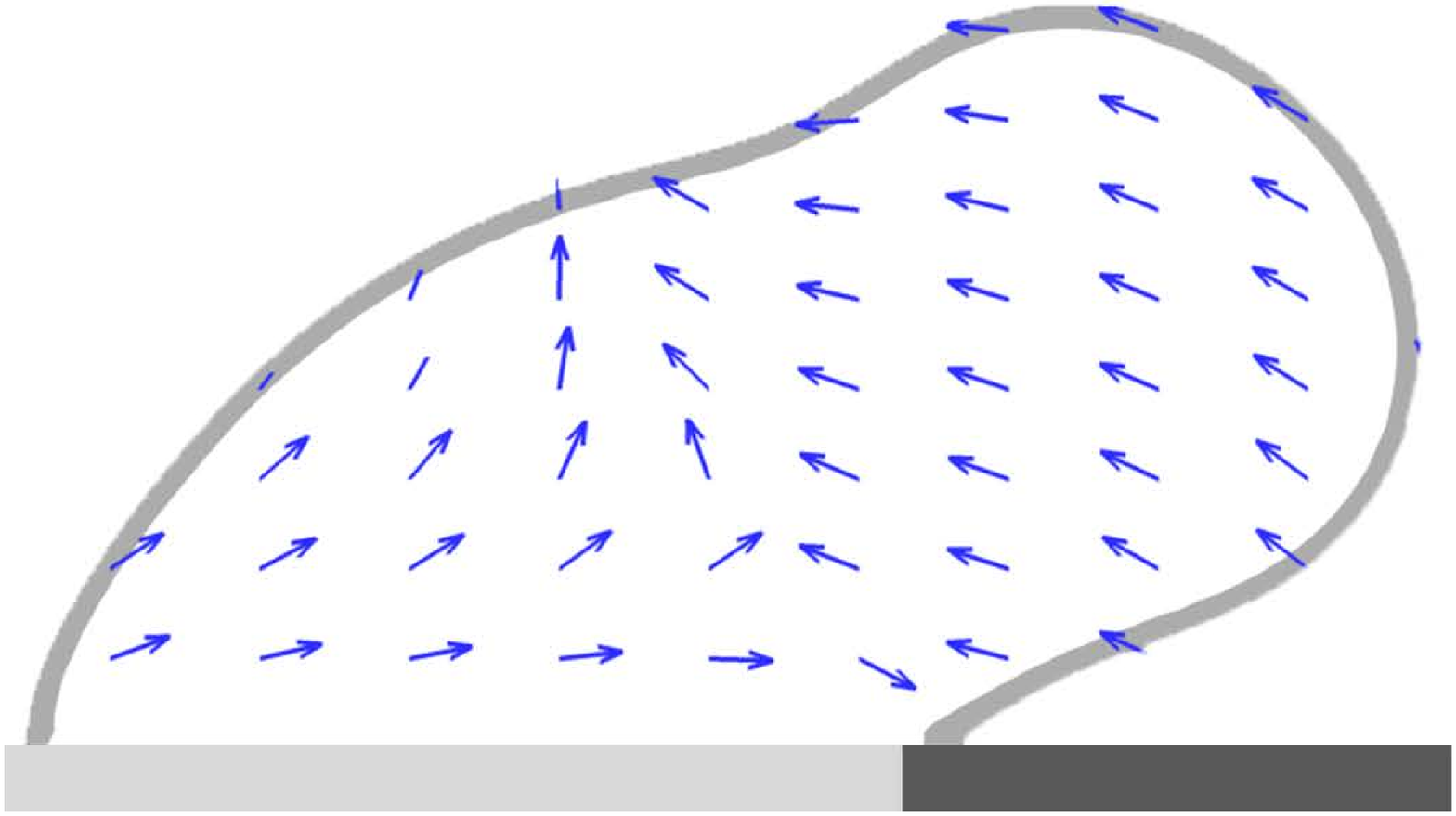}}
		\subfigure[$ \Delta \theta=40 \degree $]{ 
			\includegraphics[width=0.2\textwidth]{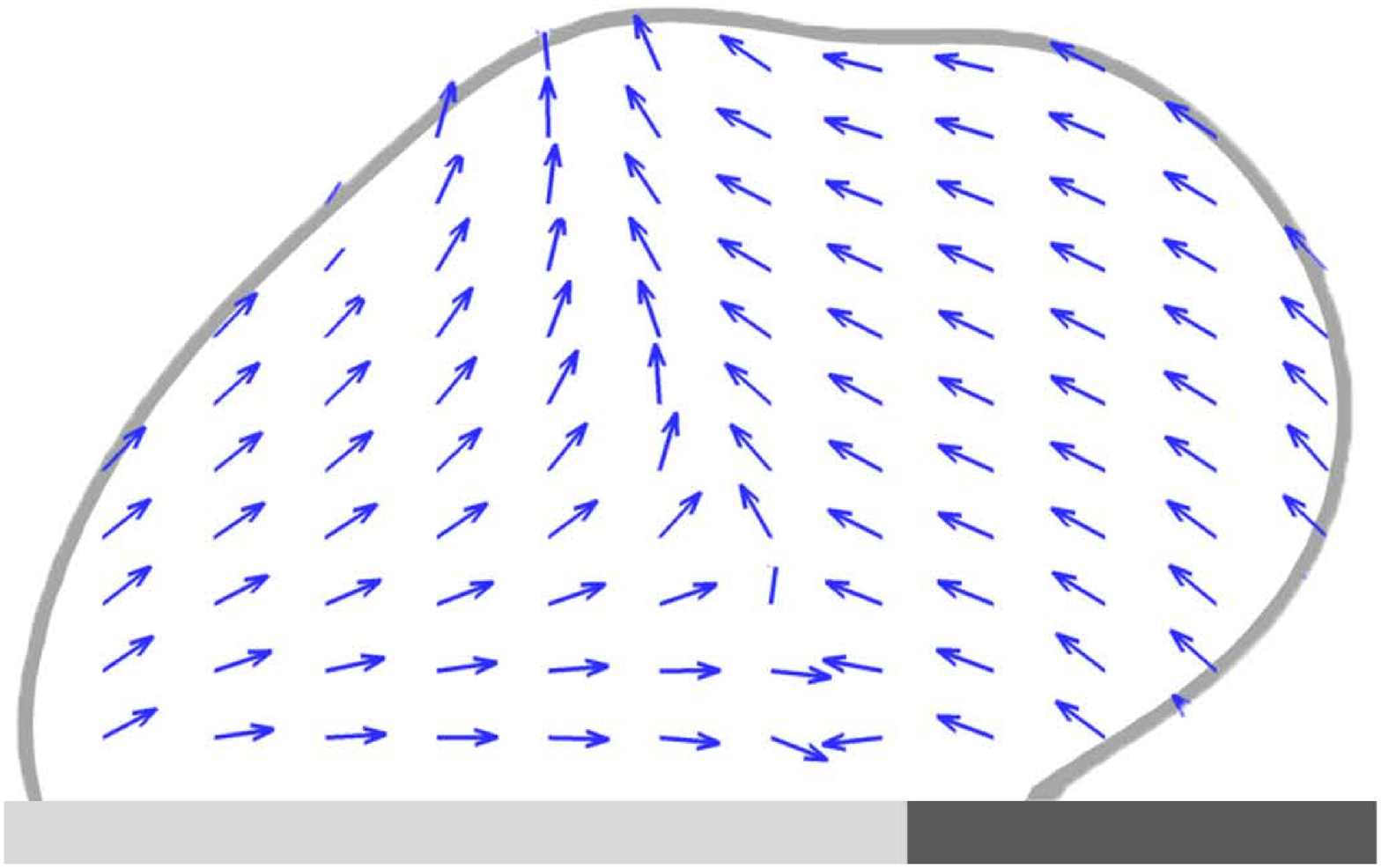}}	
		\subfigure[$ \Delta \theta=20 \degree $]{ 
			\includegraphics[width=0.2\textwidth]{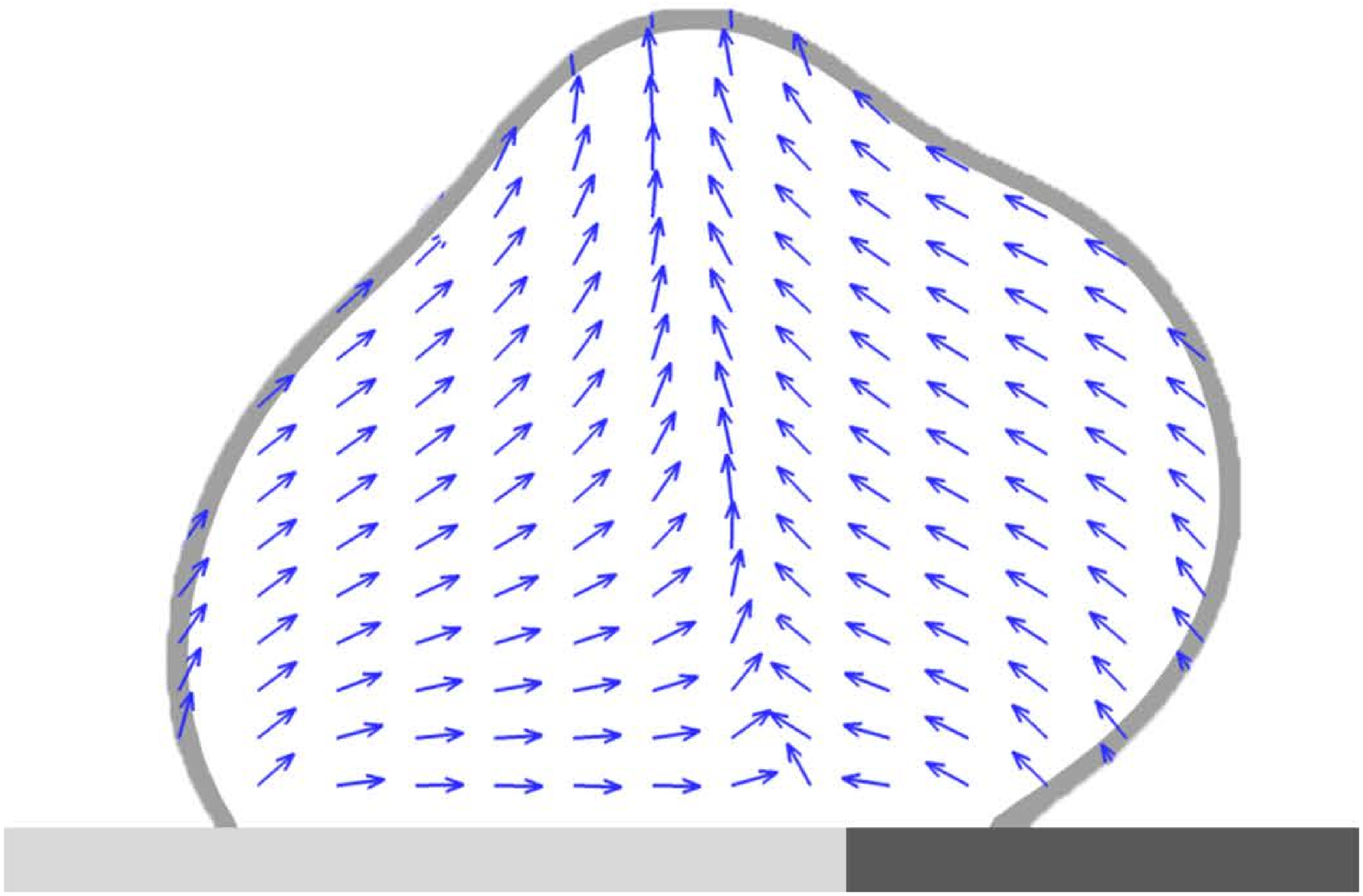}}				
		\caption{ Velocity field inside the droplet at asymmetric retracting phase $ \textrm{II} $ on a surface with different wettability difference $ \Delta \theta $ }
		\label{fig9}
	\end{figure}

Fig. \ref{fig8} shows the variation of contact time with wettability difference. The contact time represents the residence time of droplets in the right low-wettability region. It can be seen from Fig. \ref{fig8} that the contact time increases with the increase of the wettability differences. This is mainly due to the high wettability, which leads to the low recoil velocity and small resistance in the left part of the droplet, so that the right part of the droplet leaves the low-wettability surface faster. In order to intuitively understand the flow behavior inside the condensation droplet, Fig. \ref{fig9} shows the velocity field inside the droplet along the $ x $-$ z $ plane during the contraction process. As is shown in the Fig. \ref{fig9}, it can be seen that the motion direction of the velocity vector in the droplets on both sides of the borderline is opposite. With the increase of the wettability difference, the recoil velocity on the left side gradually decreases and the resistance becomes smaller, which makes the overall movement trend of the right droplet to the left side. This motion mode makes its contact time shorter. In addition, the effect of the higher tangential velocity component leads to greater droplet elongation. For lower wettability differences, the high hydrophobicity on the left side makes the droplet recoil faster and prevents the right droplet from recoiling. The combination of two velocity fields forms an upward trend. As a consequence, the coalesced droplet bounces off the surface eventually.

\subsection{The effect of Weber number} 

	\begin{figure}[htbp]
		\centering
		
		\begin{minipage}[c]{0.1\textwidth}
			\centering
			\caption*{(a) $ t^{*}=0.2 $ }
		\end{minipage}
		\begin{minipage}[c]{0.2\textwidth}
			\includegraphics[width=\textwidth]{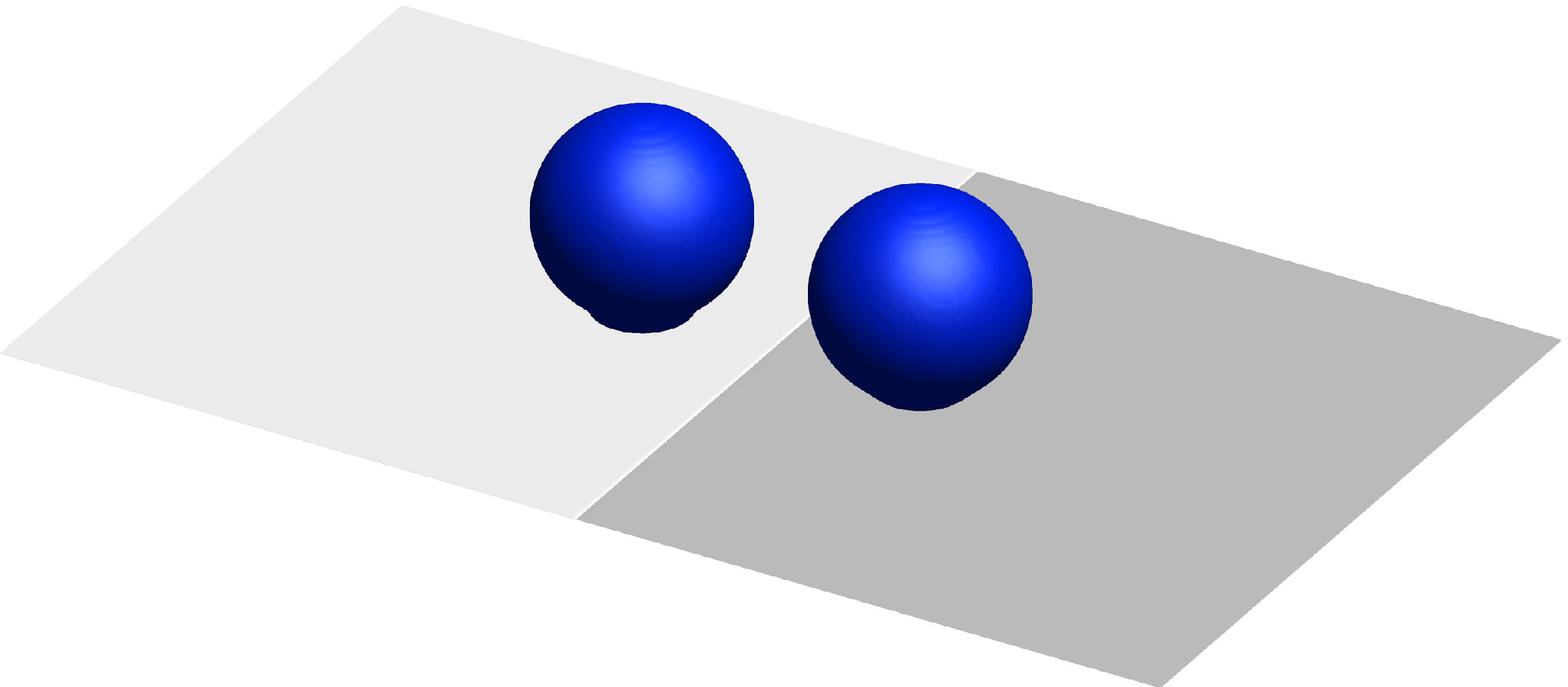}
		\end{minipage}
		\begin{minipage}[c]{0.2\textwidth}
			\includegraphics[width=\textwidth]{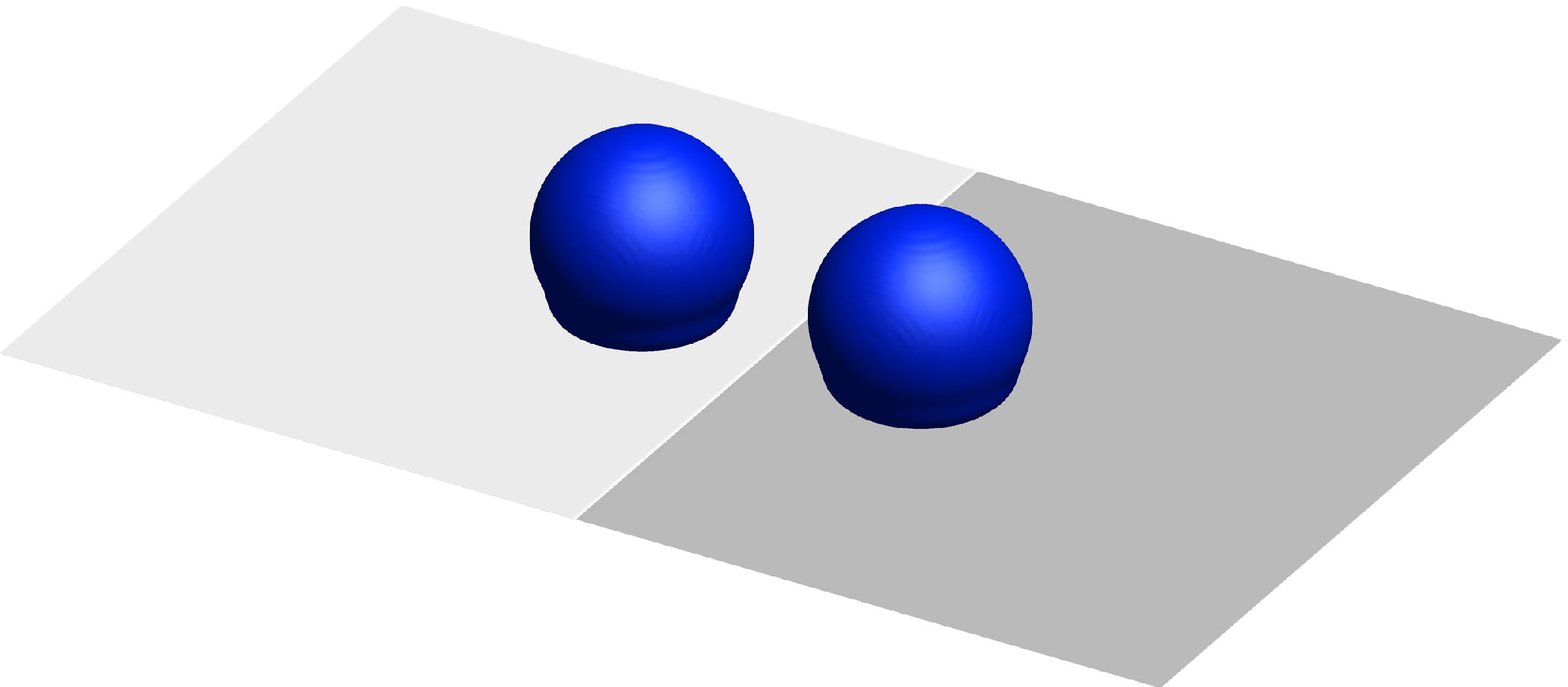}
		\end{minipage}
		\begin{minipage}[c]{0.2\textwidth}
			\includegraphics[width=\textwidth]{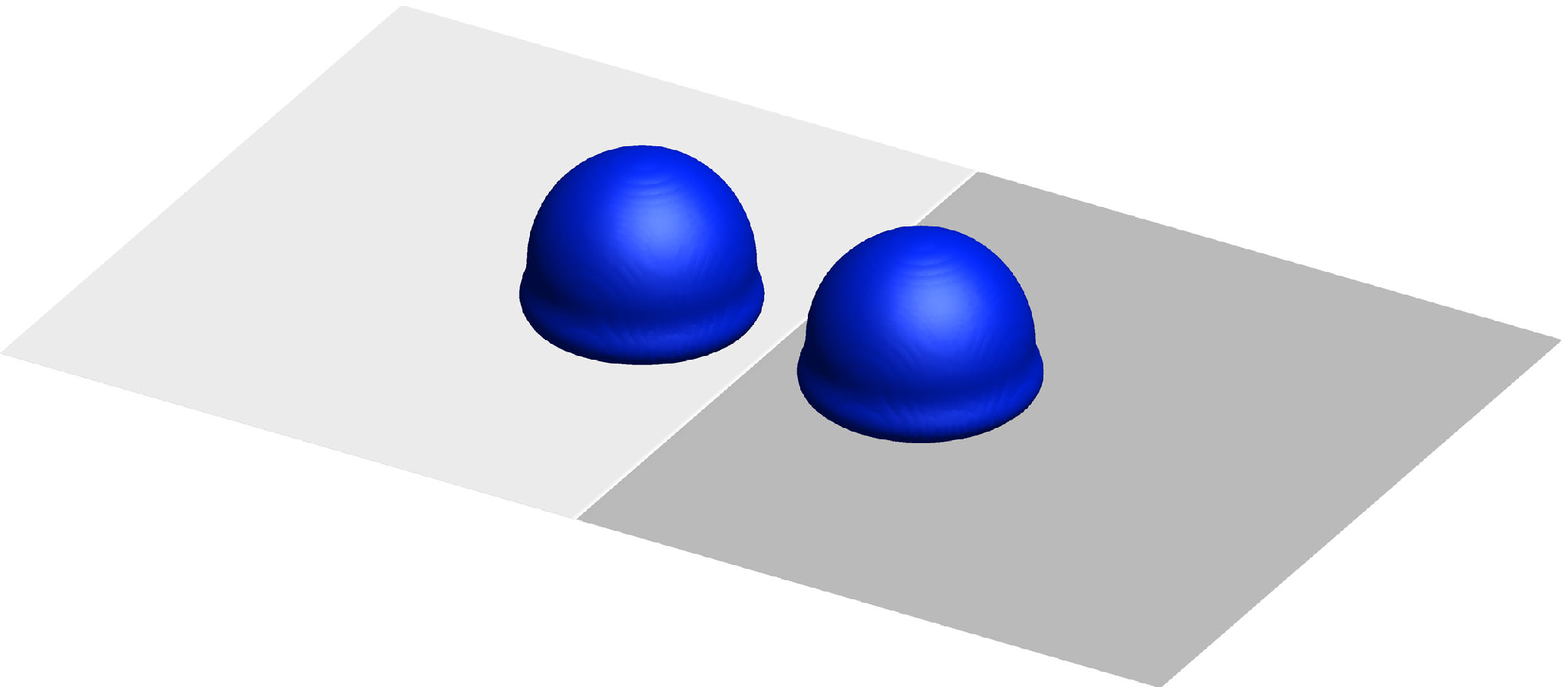}
		\end{minipage}
		\begin{minipage}[c]{0.2\textwidth}
			\includegraphics[width=\textwidth]{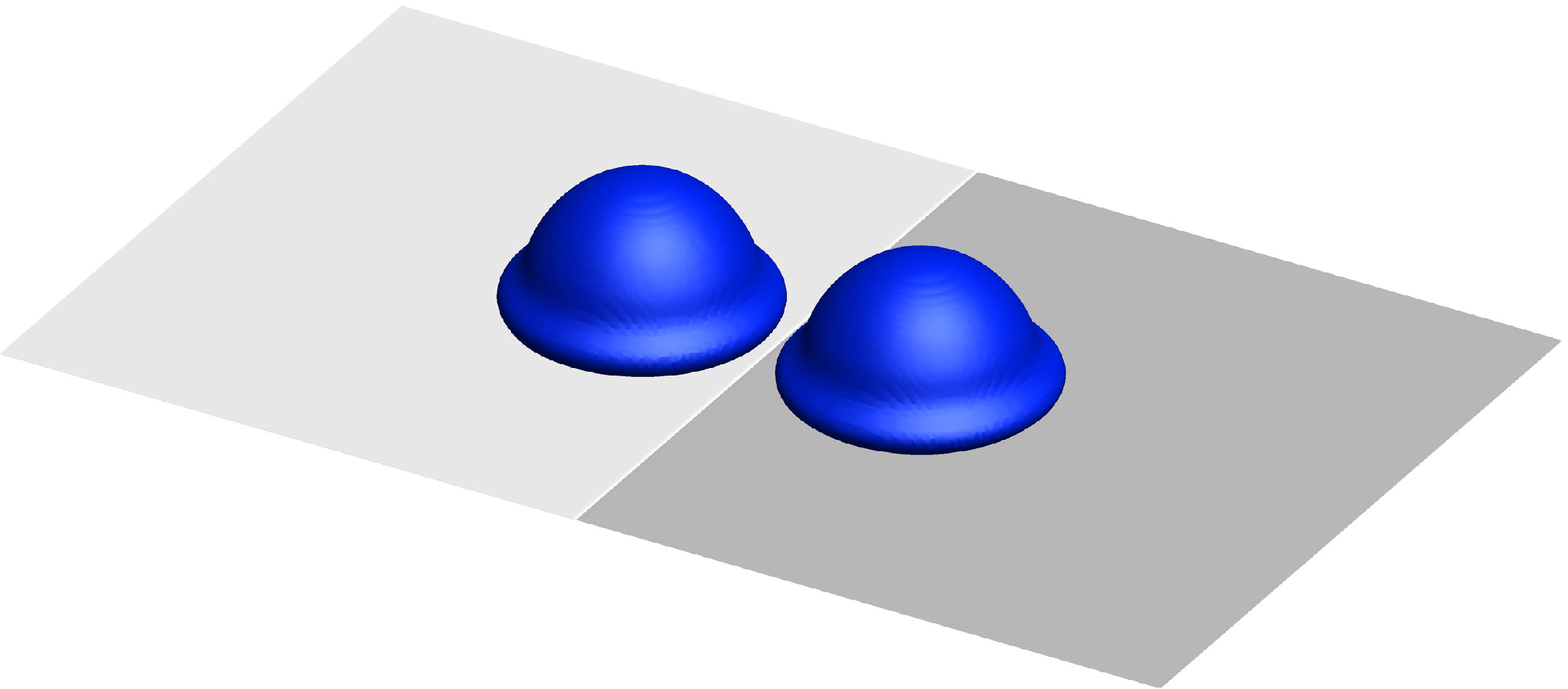}
		\end{minipage}

		\begin{minipage}[c]{0.1\textwidth}
			\centering
			\caption*{(b) $ t^{*}=0.6 $  }
		\end{minipage}
		\begin{minipage}[c]{0.2\textwidth}
			\includegraphics[width=\textwidth]{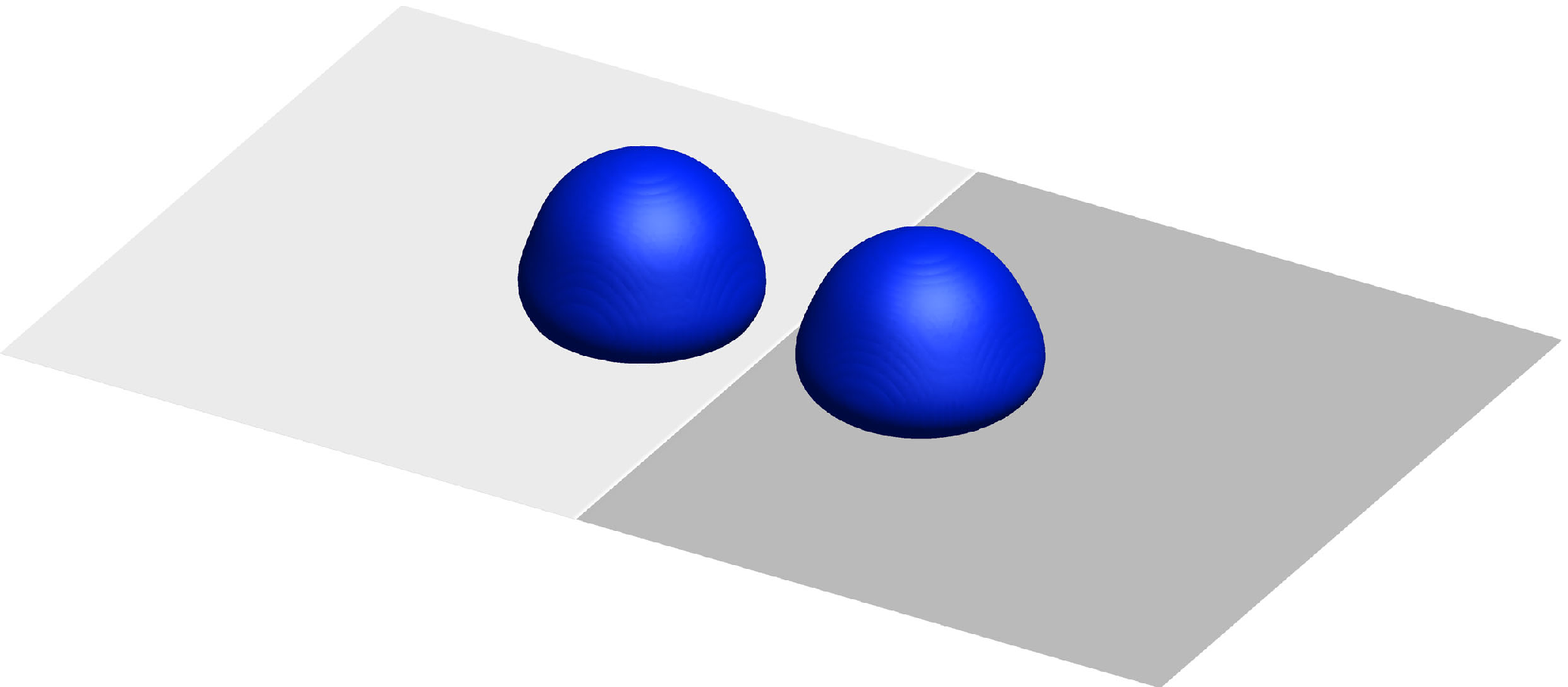}
		\end{minipage}
		\begin{minipage}[c]{0.2\textwidth}
			\includegraphics[width=\textwidth]{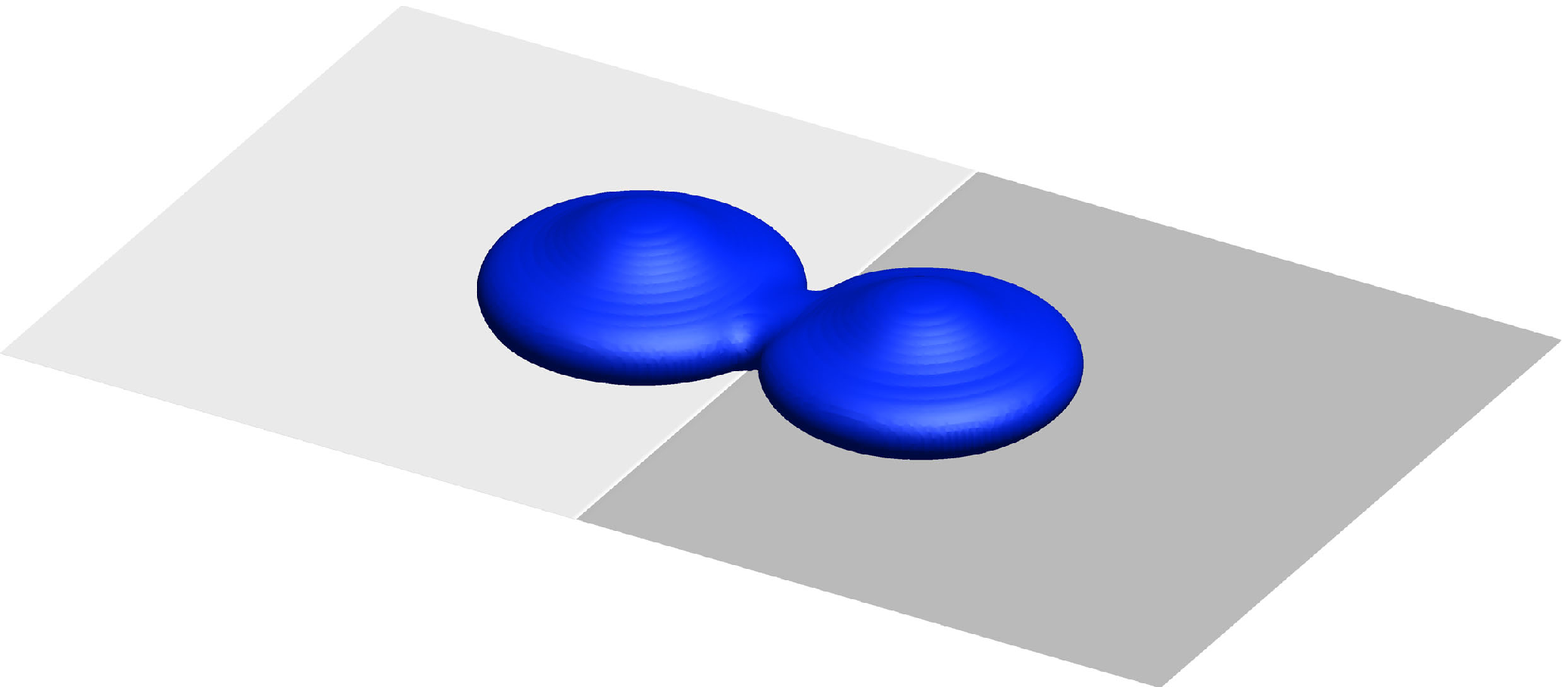}
		\end{minipage}
		\begin{minipage}[c]{0.2\textwidth}
			\includegraphics[width=\textwidth]{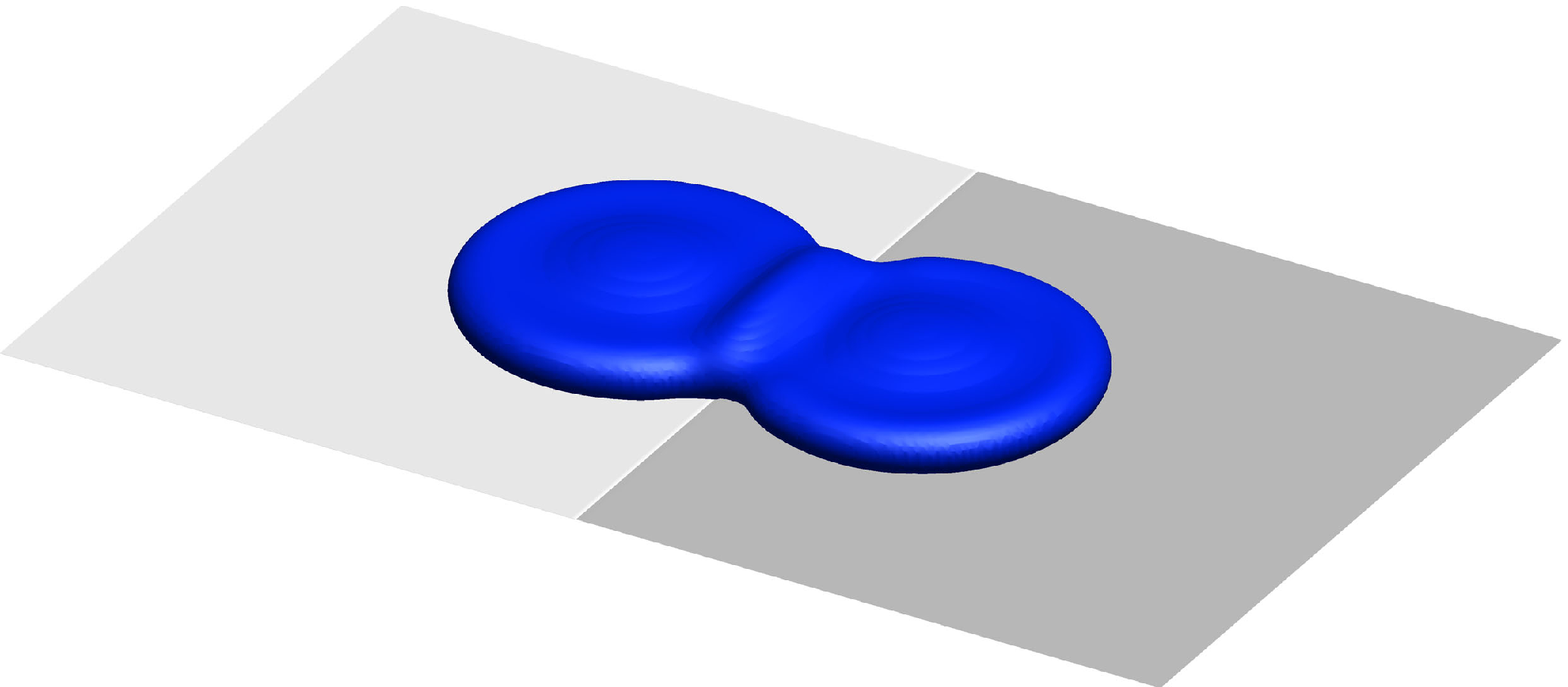}
		\end{minipage}
		\begin{minipage}[c]{0.2\textwidth}
			\includegraphics[width=\textwidth]{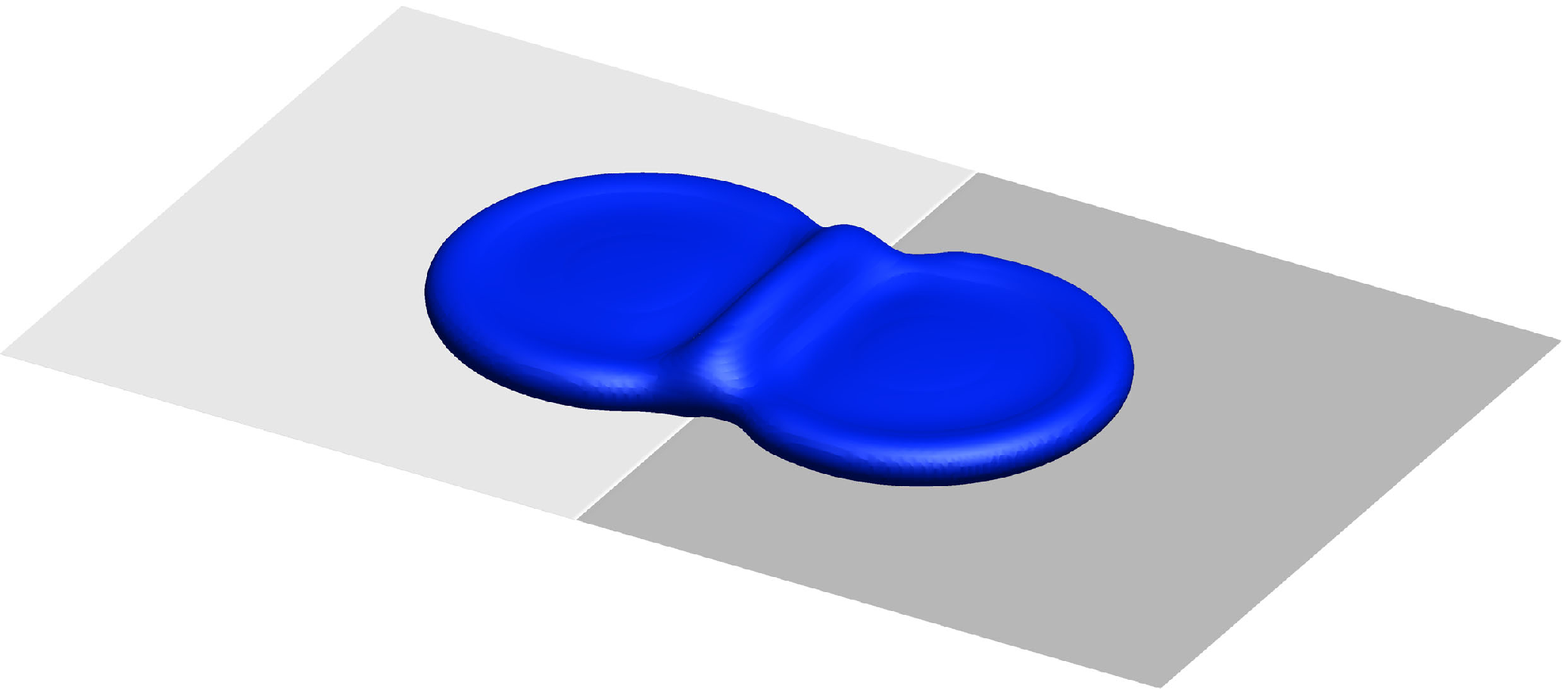}
		\end{minipage}

		\begin{minipage}[c]{0.1\textwidth}
			\centering
			\caption*{(c) $ t^{*}=1.0 $  }
		\end{minipage}
		\begin{minipage}[c]{0.2\textwidth}
			\includegraphics[width=\textwidth]{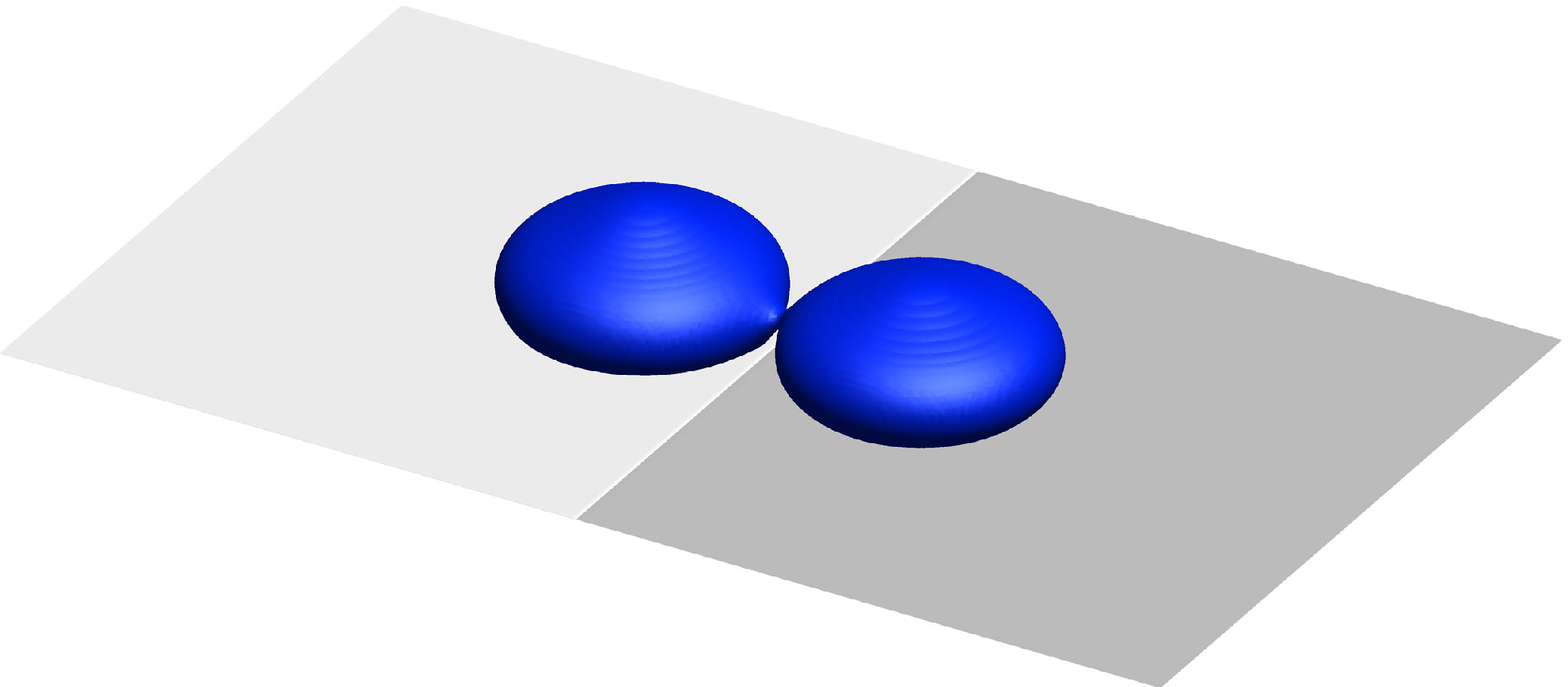}
		\end{minipage}
		\begin{minipage}[c]{0.2\textwidth}
			\includegraphics[width=\textwidth]{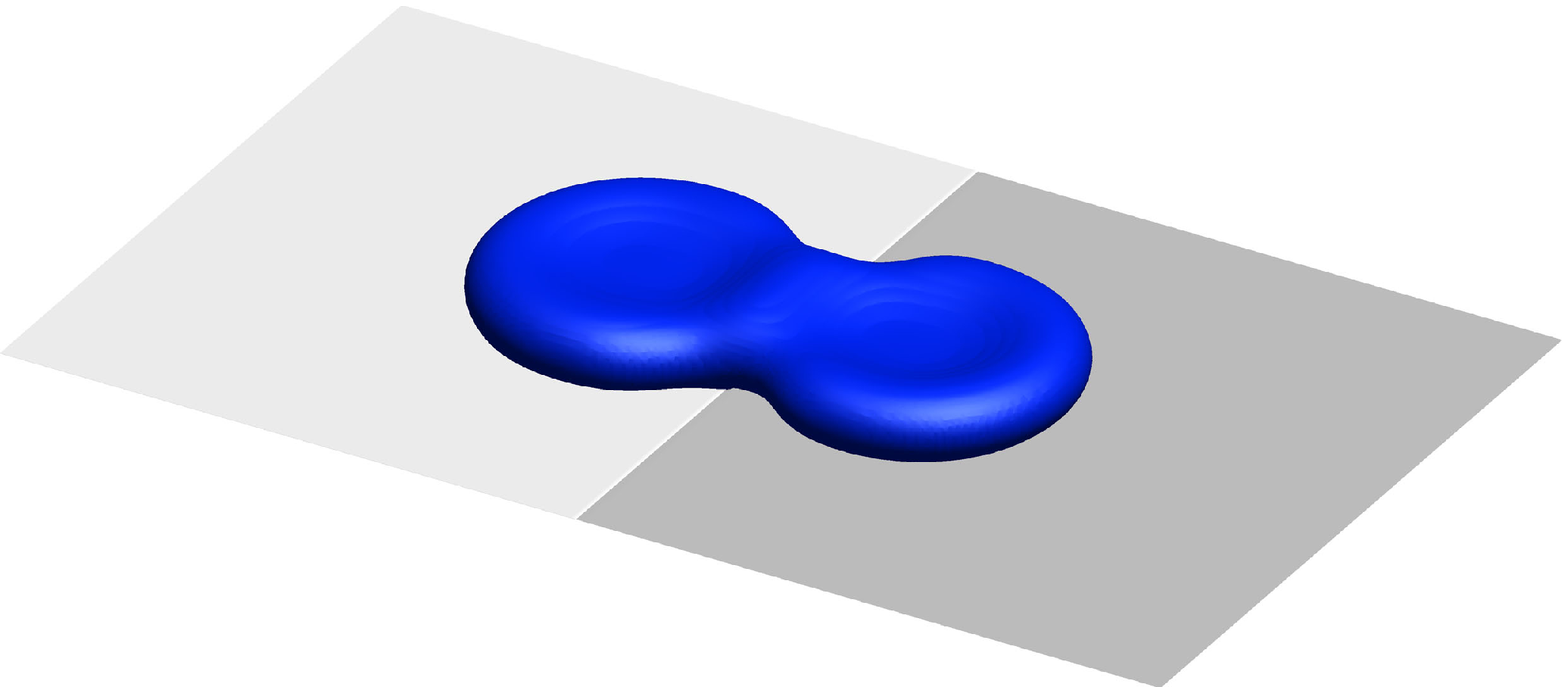}
		\end{minipage}
		\begin{minipage}[c]{0.2\textwidth}
			\includegraphics[width=\textwidth]{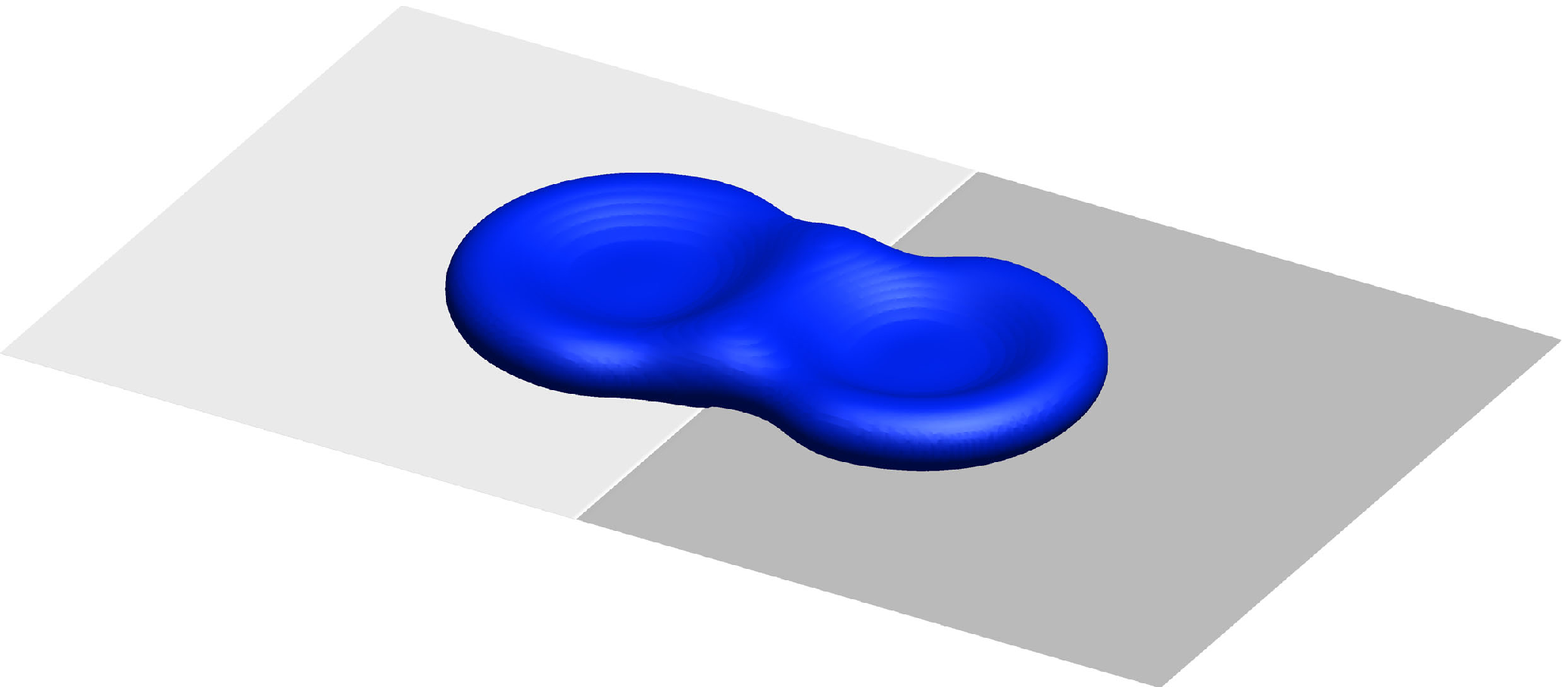}
		\end{minipage}
		\begin{minipage}[c]{0.2\textwidth}
			\includegraphics[width=\textwidth]{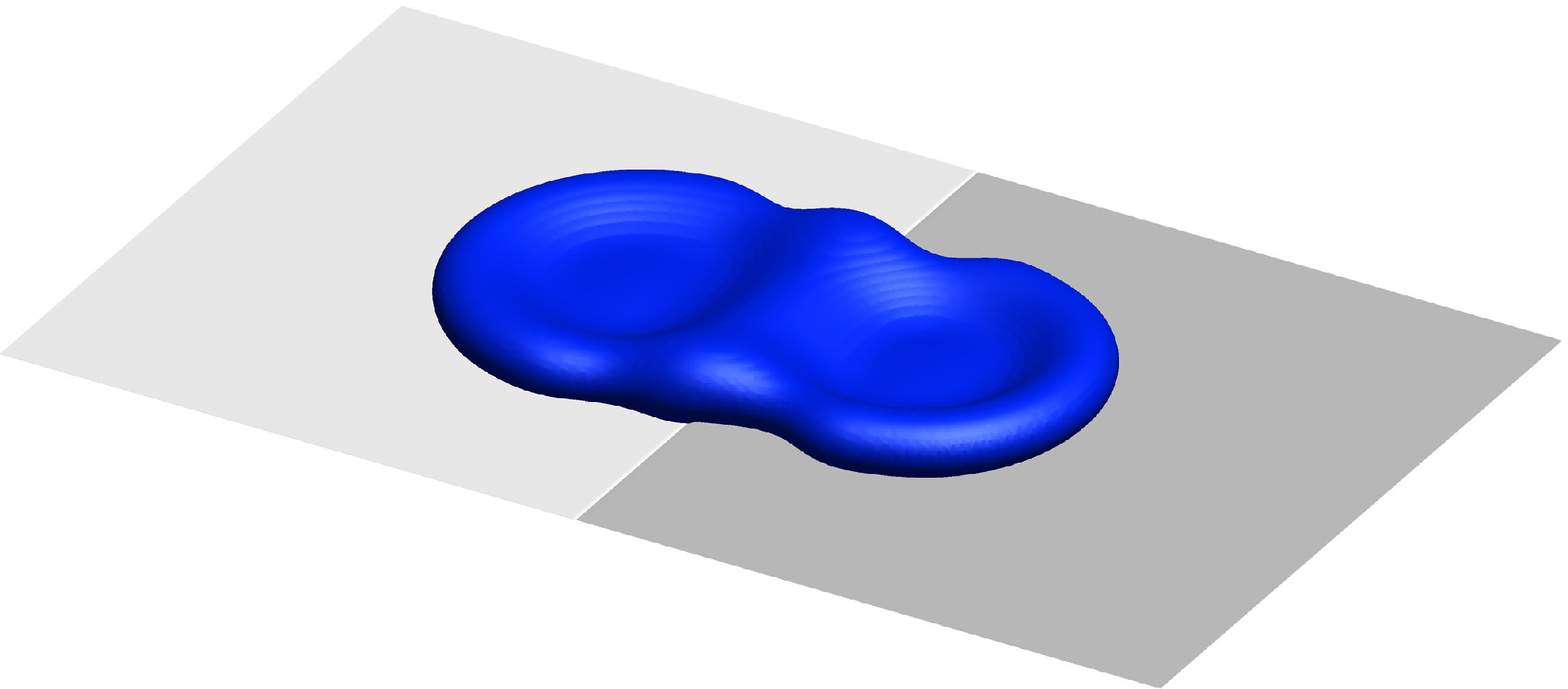}
		\end{minipage}

		\begin{minipage}[c]{0.1\textwidth}
			\centering
			\caption*{(d) $ t^{*}=1.4 $  }
		\end{minipage}
		\begin{minipage}[c]{0.2\textwidth}
			\includegraphics[width=\textwidth]{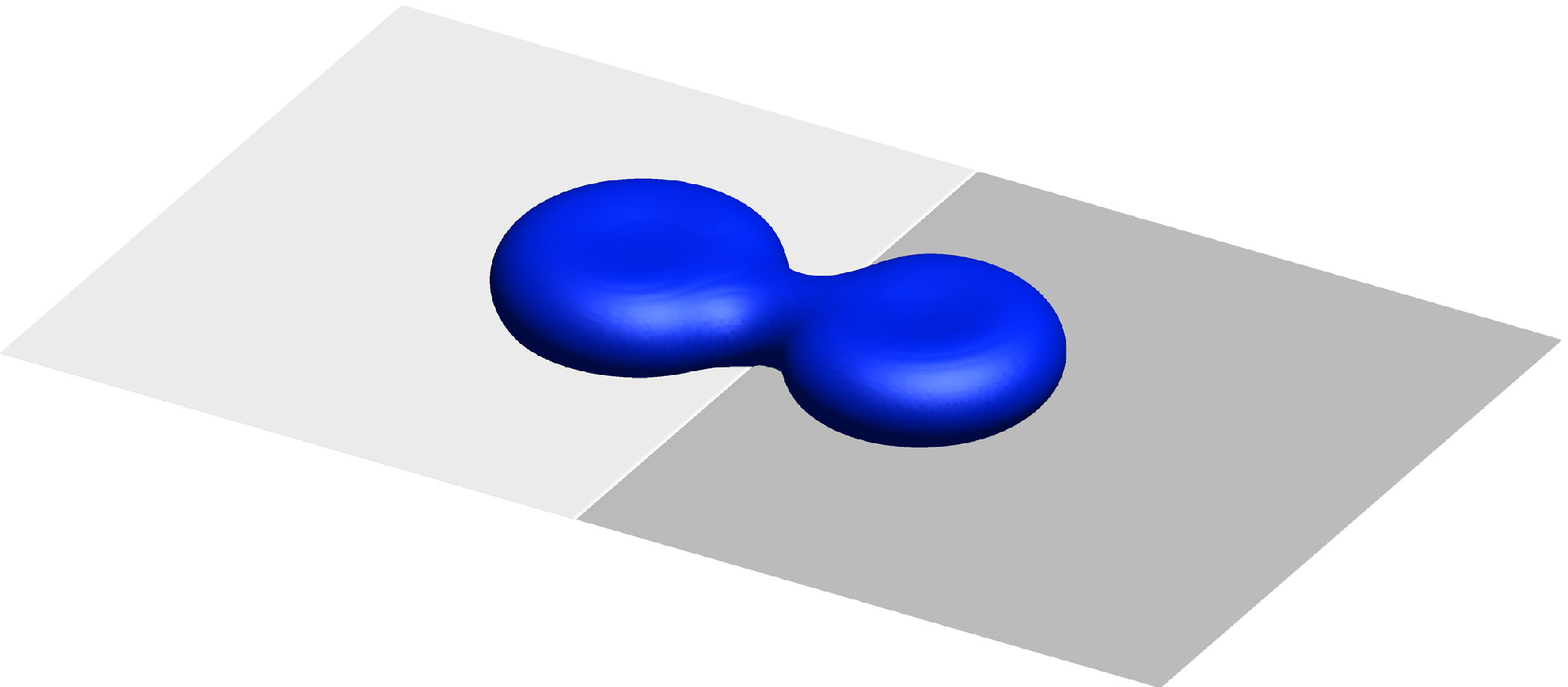}
		\end{minipage}
		\begin{minipage}[c]{0.2\textwidth}
			\includegraphics[width=\textwidth]{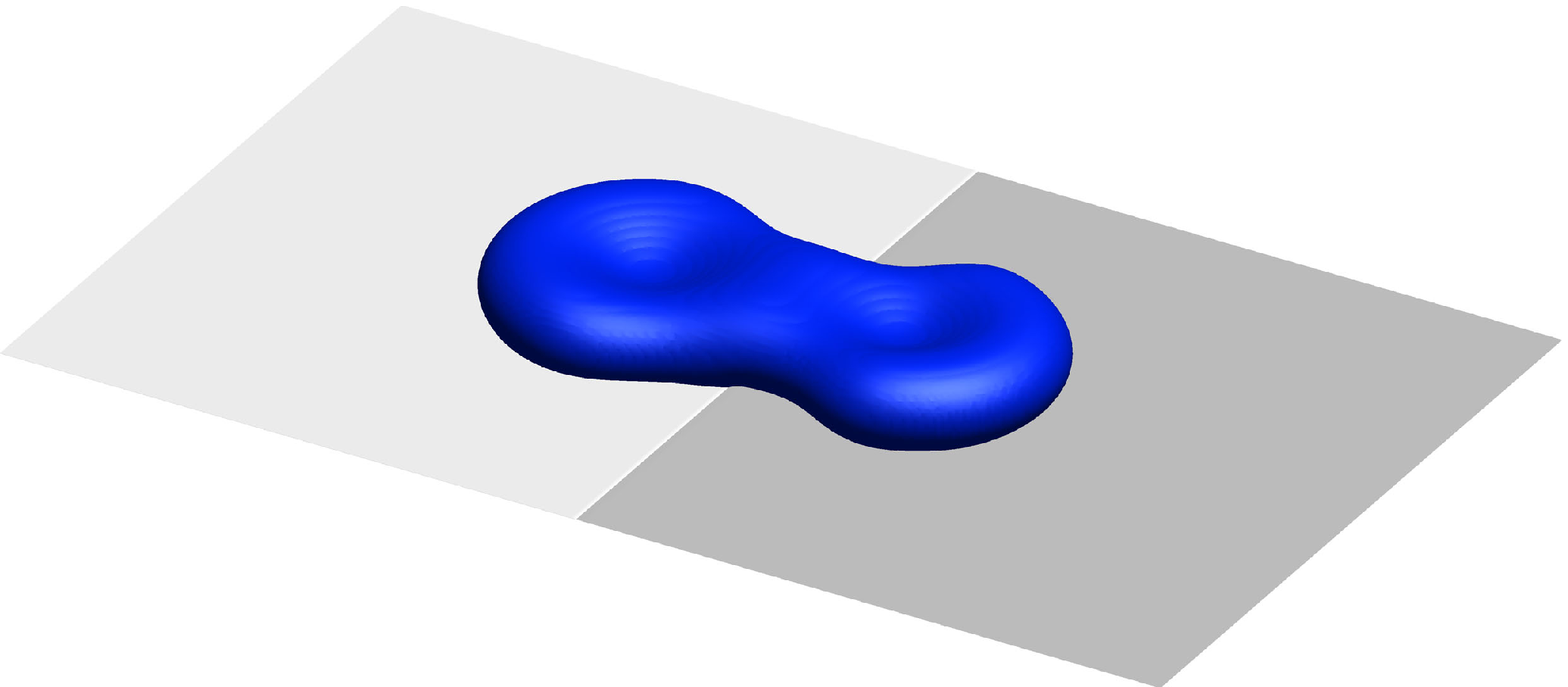}
		\end{minipage}
		\begin{minipage}[c]{0.2\textwidth}
			\includegraphics[width=\textwidth]{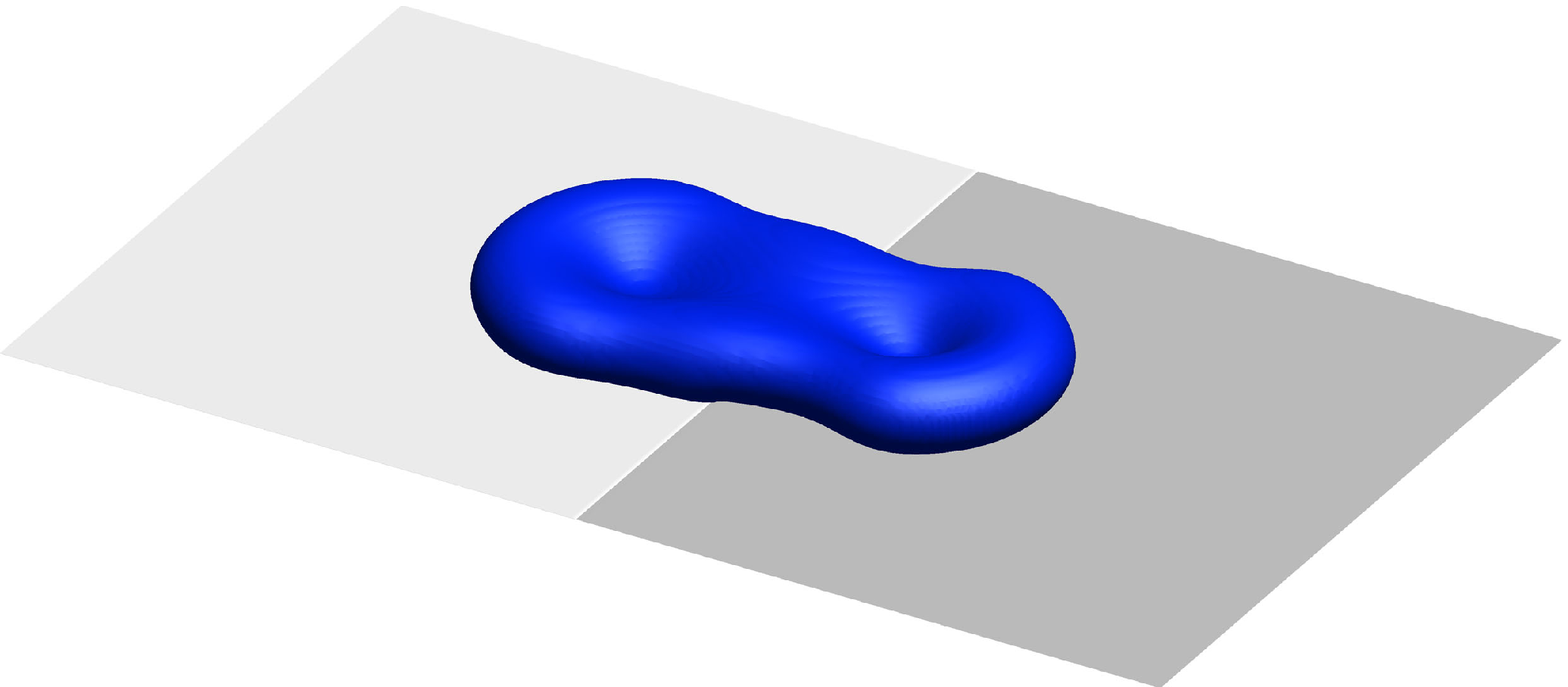}
		\end{minipage}
		\begin{minipage}[c]{0.2\textwidth}
			\includegraphics[width=\textwidth]{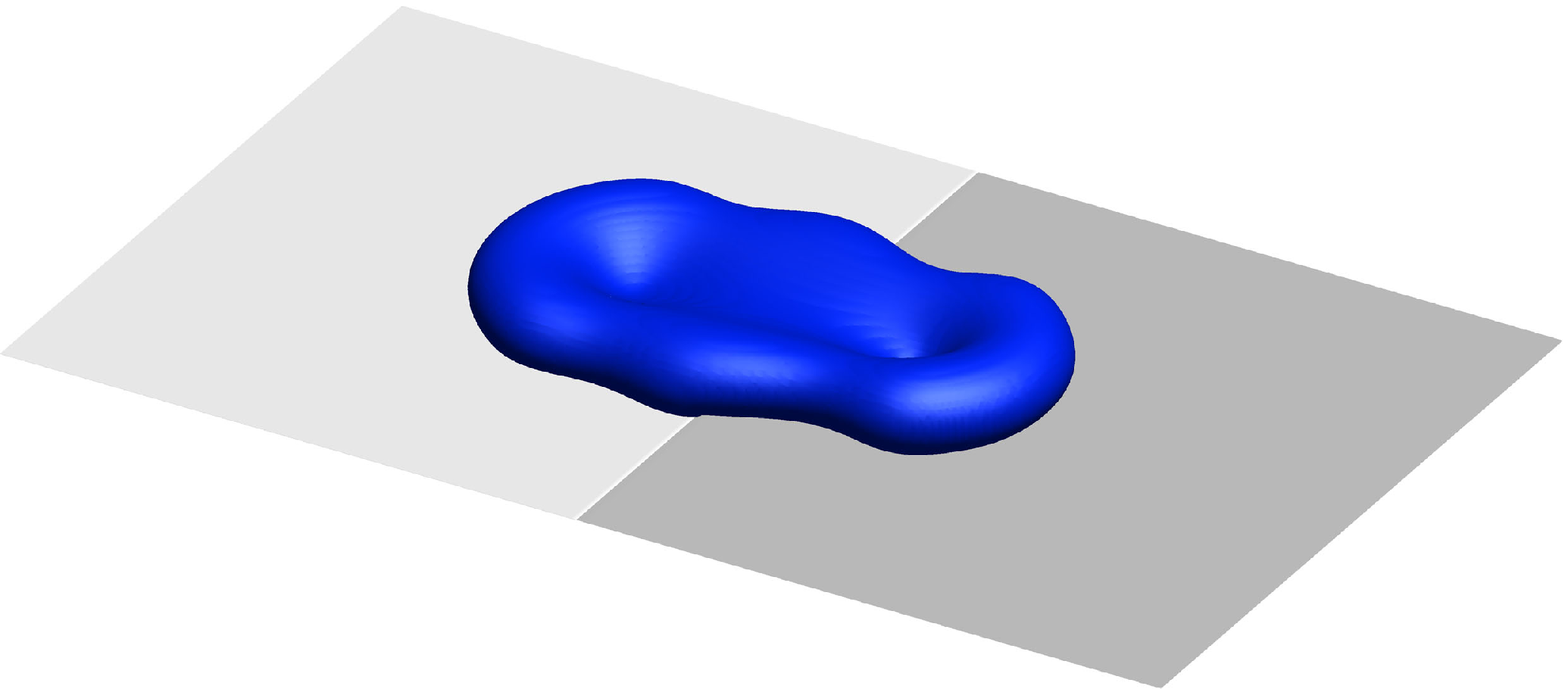}
		\end{minipage}

		\begin{minipage}[c]{0.1\textwidth}
			\centering
			\caption*{(e) $ t^{*}=2.0 $  }
		\end{minipage}
		\begin{minipage}[c]{0.2\textwidth}
			\includegraphics[width=\textwidth]{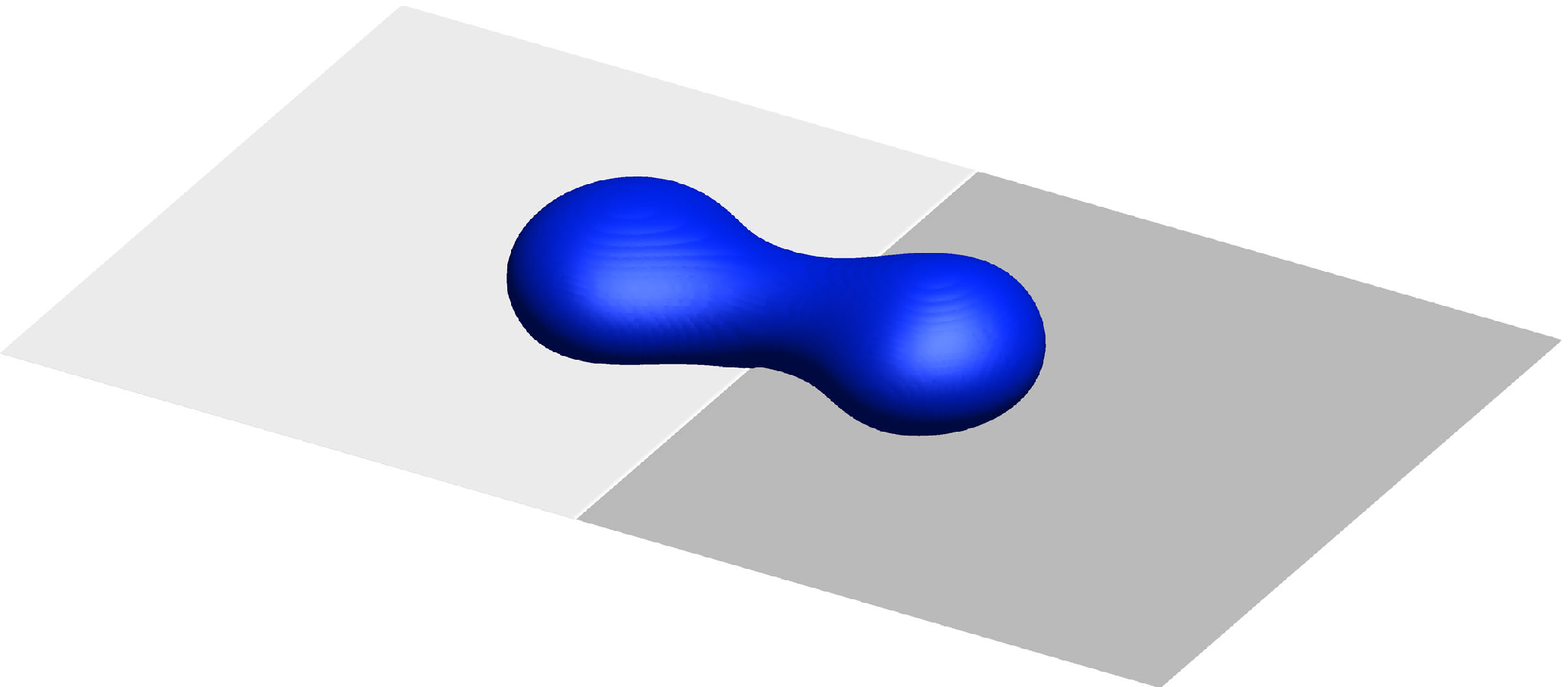}
		\end{minipage}
		\begin{minipage}[c]{0.2\textwidth}
			\includegraphics[width=\textwidth]{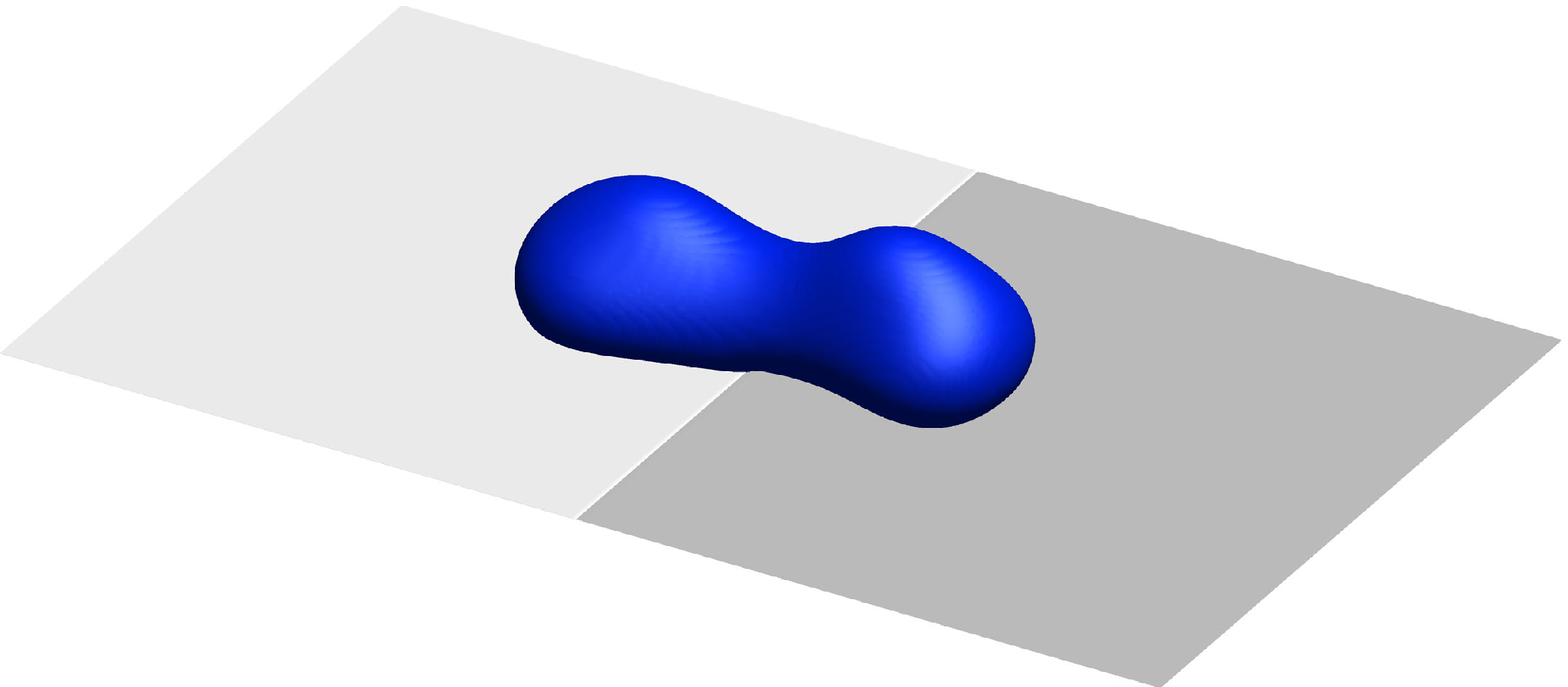}
		\end{minipage}
		\begin{minipage}[c]{0.2\textwidth}
			\includegraphics[width=\textwidth]{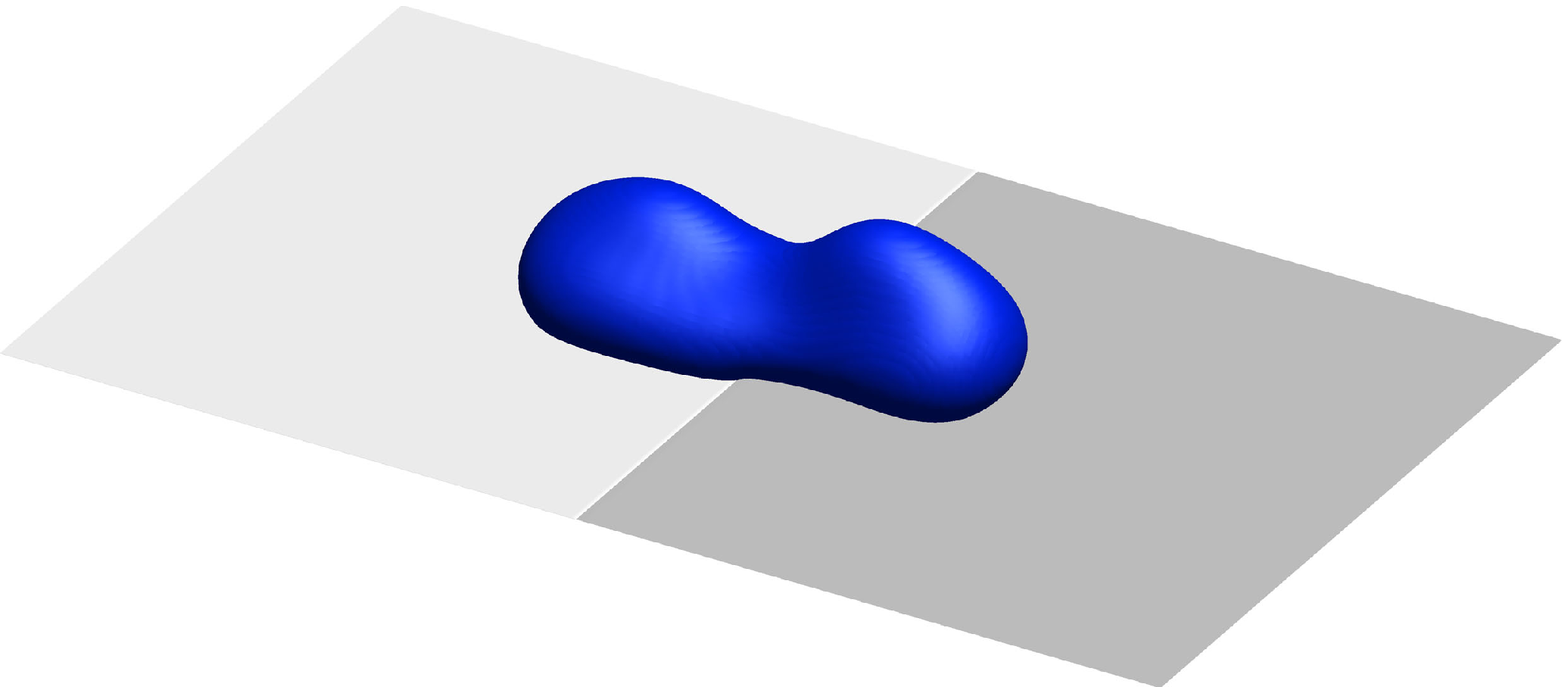}
		\end{minipage}
		\begin{minipage}[c]{0.2\textwidth}
			\includegraphics[width=\textwidth]{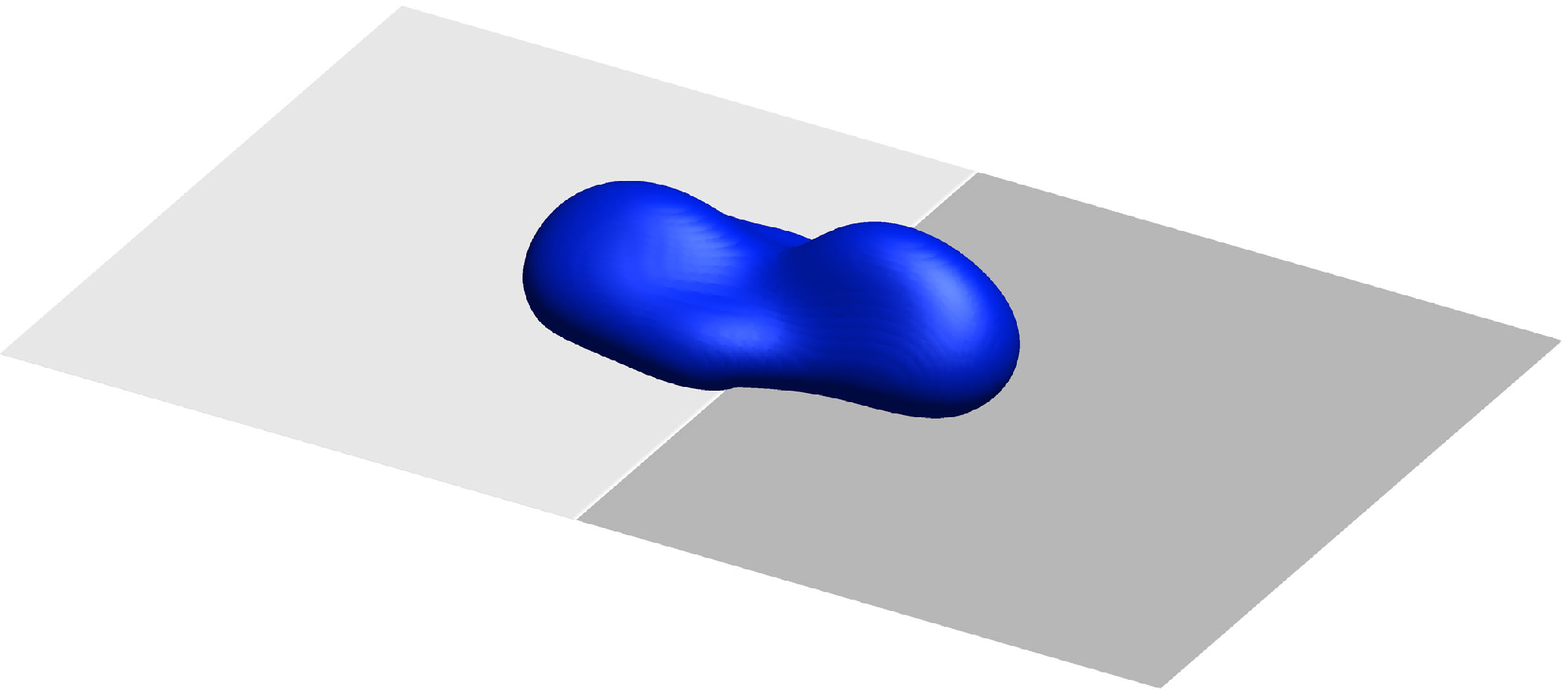}
		\end{minipage}

		\begin{minipage}[c]{0.1\textwidth}
			\centering
			\caption*{(f) $ t^{*}=3.0 $  }
		\end{minipage}
		\begin{minipage}[c]{0.2\textwidth}
			\includegraphics[width=\textwidth]{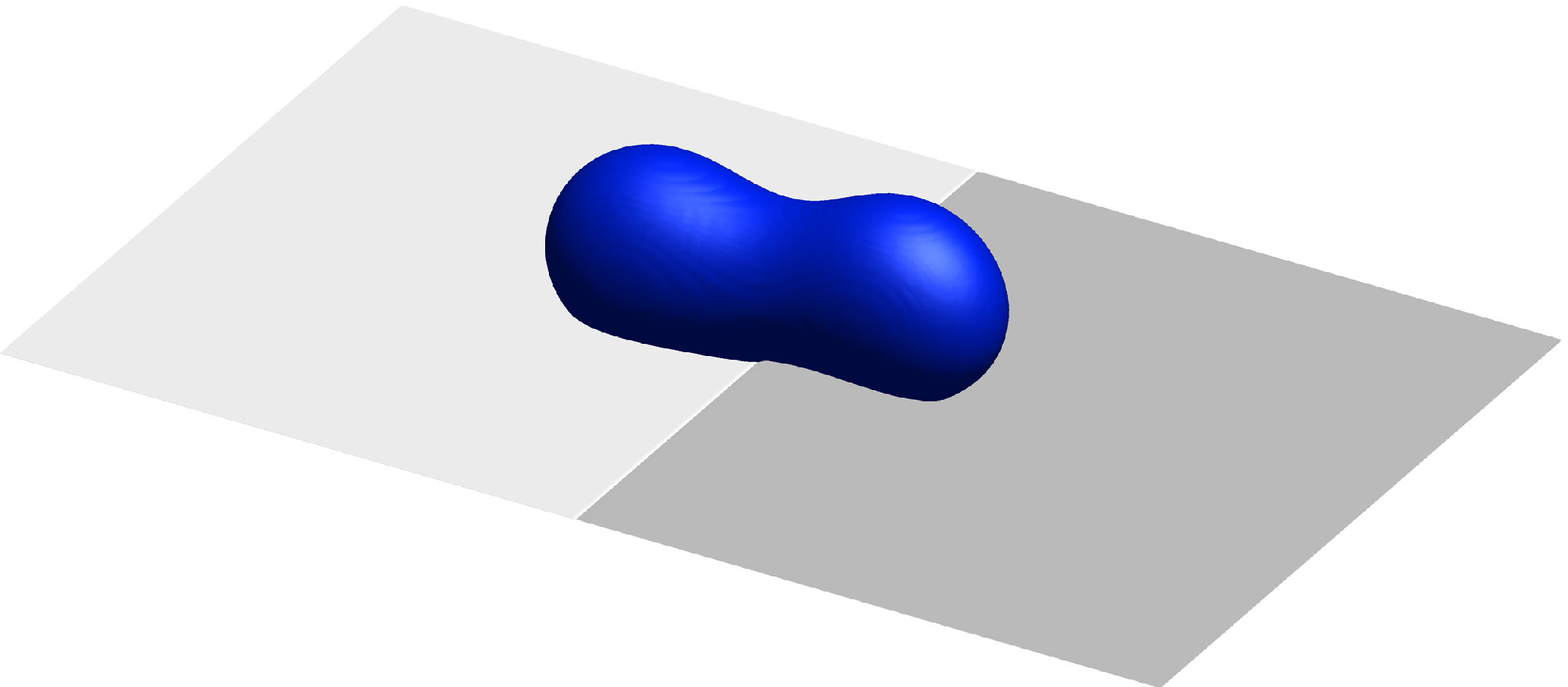}
		\end{minipage}
		\begin{minipage}[c]{0.2\textwidth}
			\includegraphics[width=\textwidth]{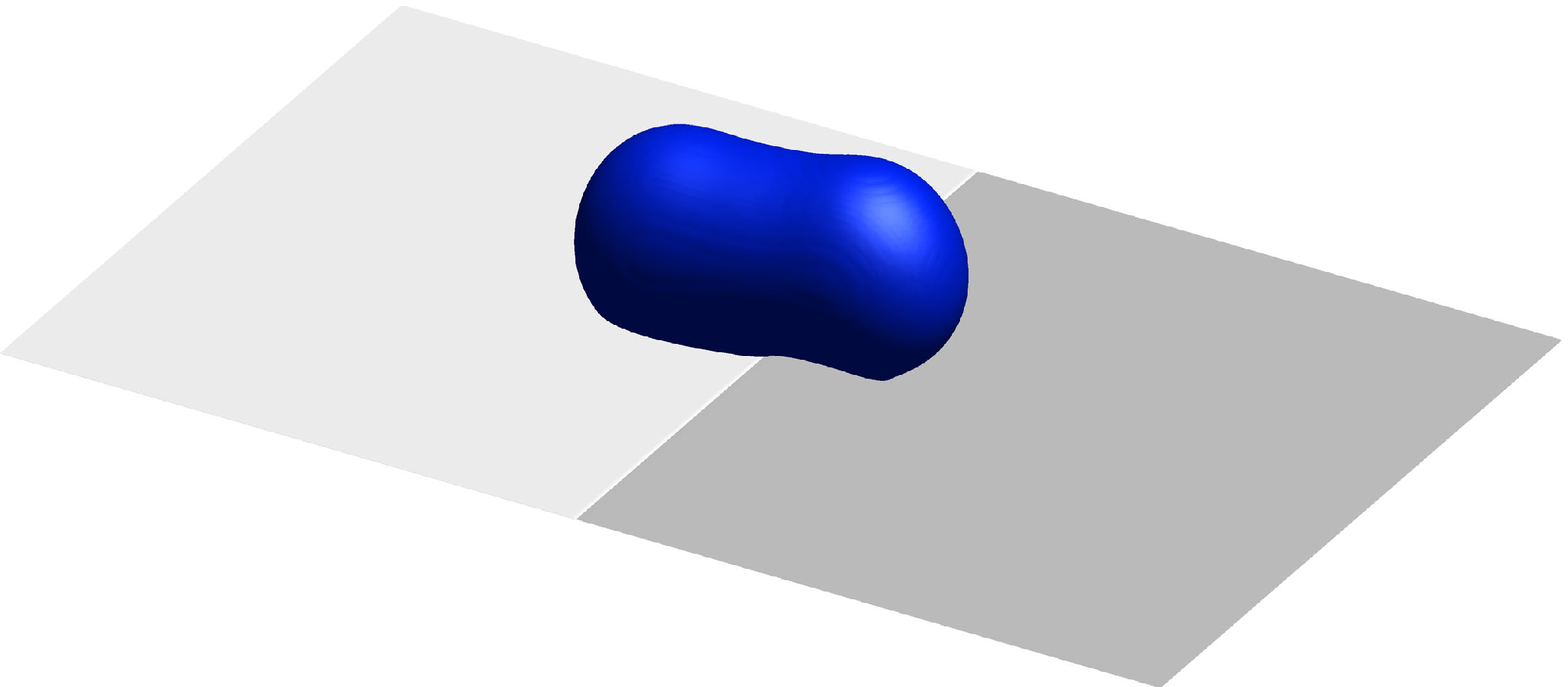}
		\end{minipage}
		\begin{minipage}[c]{0.2\textwidth}
			\includegraphics[width=\textwidth]{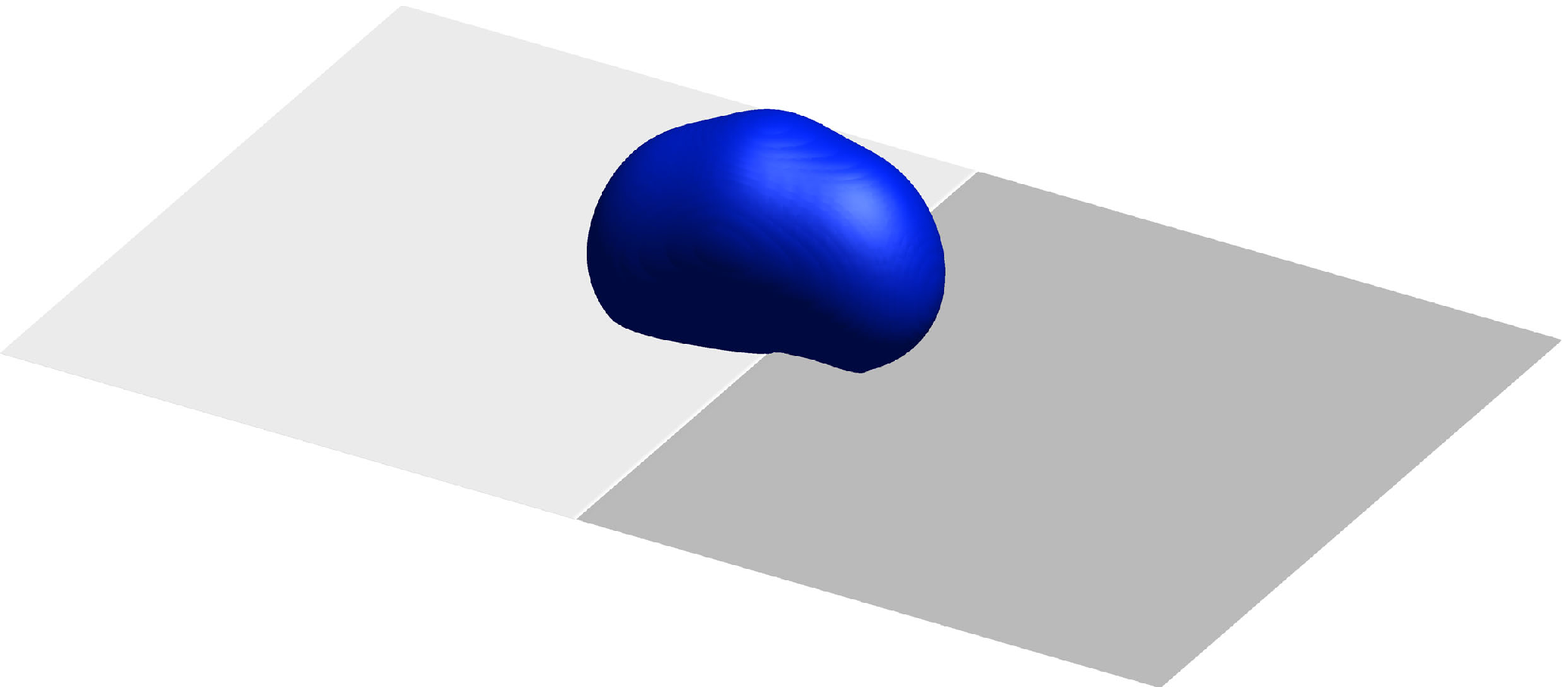}
		\end{minipage}	
		\begin{minipage}[c]{0.2\textwidth}
			\includegraphics[width=\textwidth]{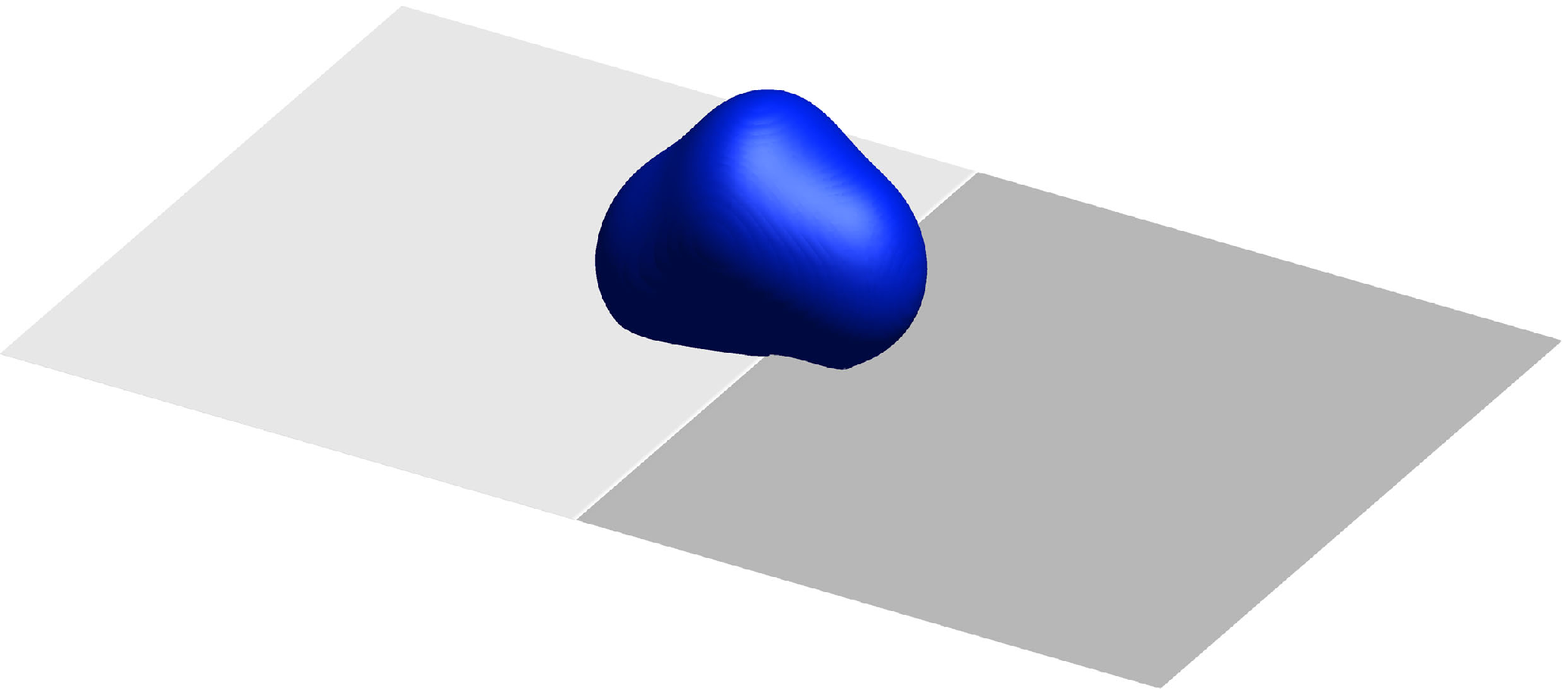}
		\end{minipage}

		\begin{minipage}[c]{0.1\textwidth}
			\centering
			\caption*{(g) $ t^{*}=4.0 $  }
		\end{minipage}
		\begin{minipage}[c]{0.2\textwidth}
			\includegraphics[width=\textwidth]{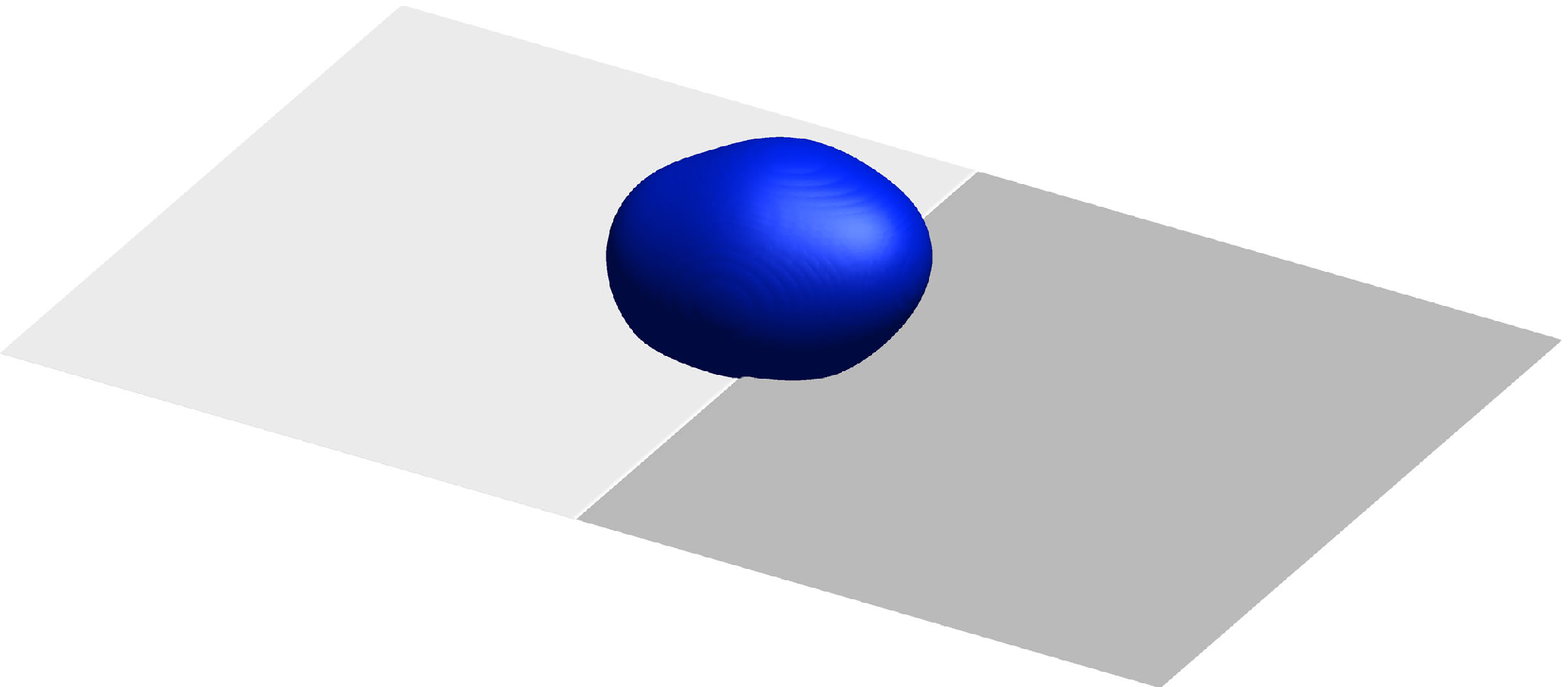}
		\end{minipage}
		\begin{minipage}[c]{0.2\textwidth}
			\includegraphics[width=\textwidth]{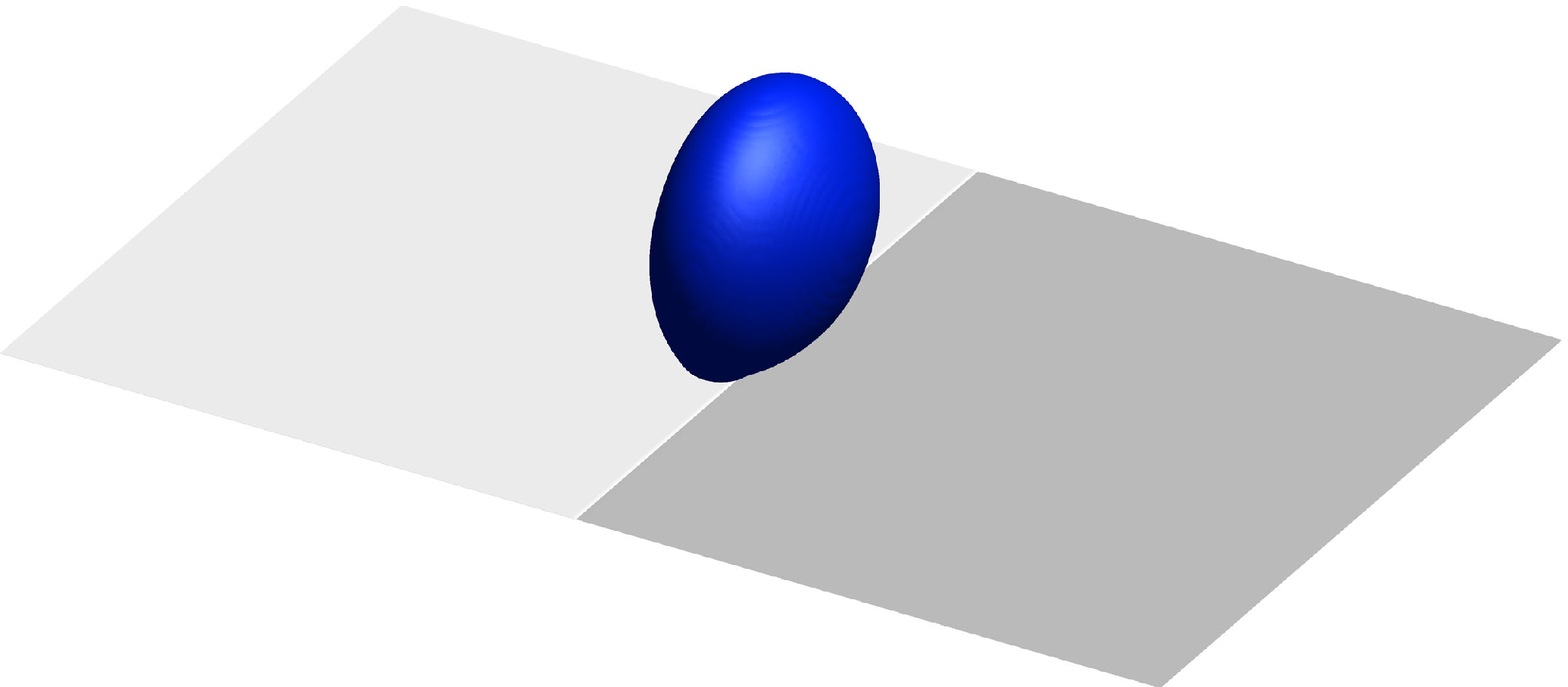}
		\end{minipage}
		\begin{minipage}[c]{0.2\textwidth}
			\includegraphics[width=\textwidth]{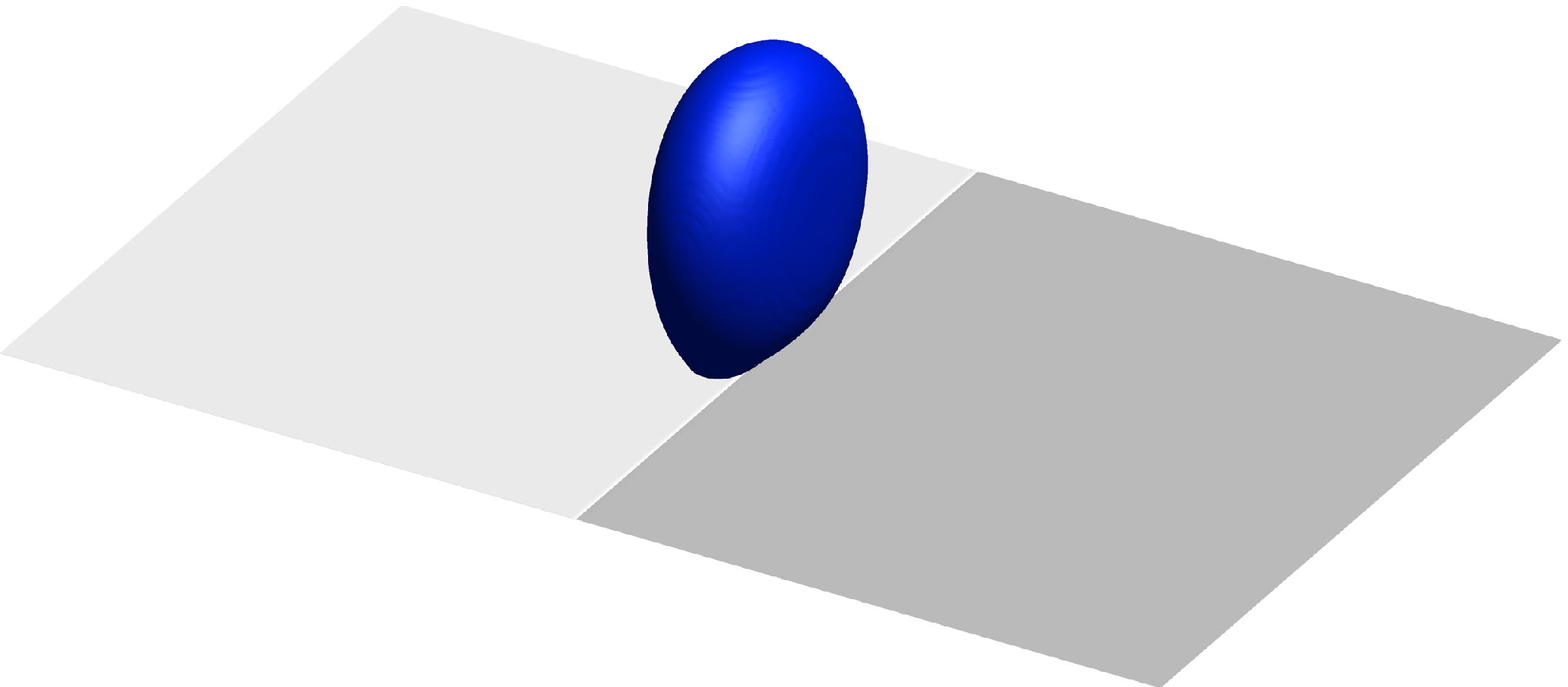}
		\end{minipage}			
		\begin{minipage}[c]{0.2\textwidth}
			\includegraphics[width=\textwidth]{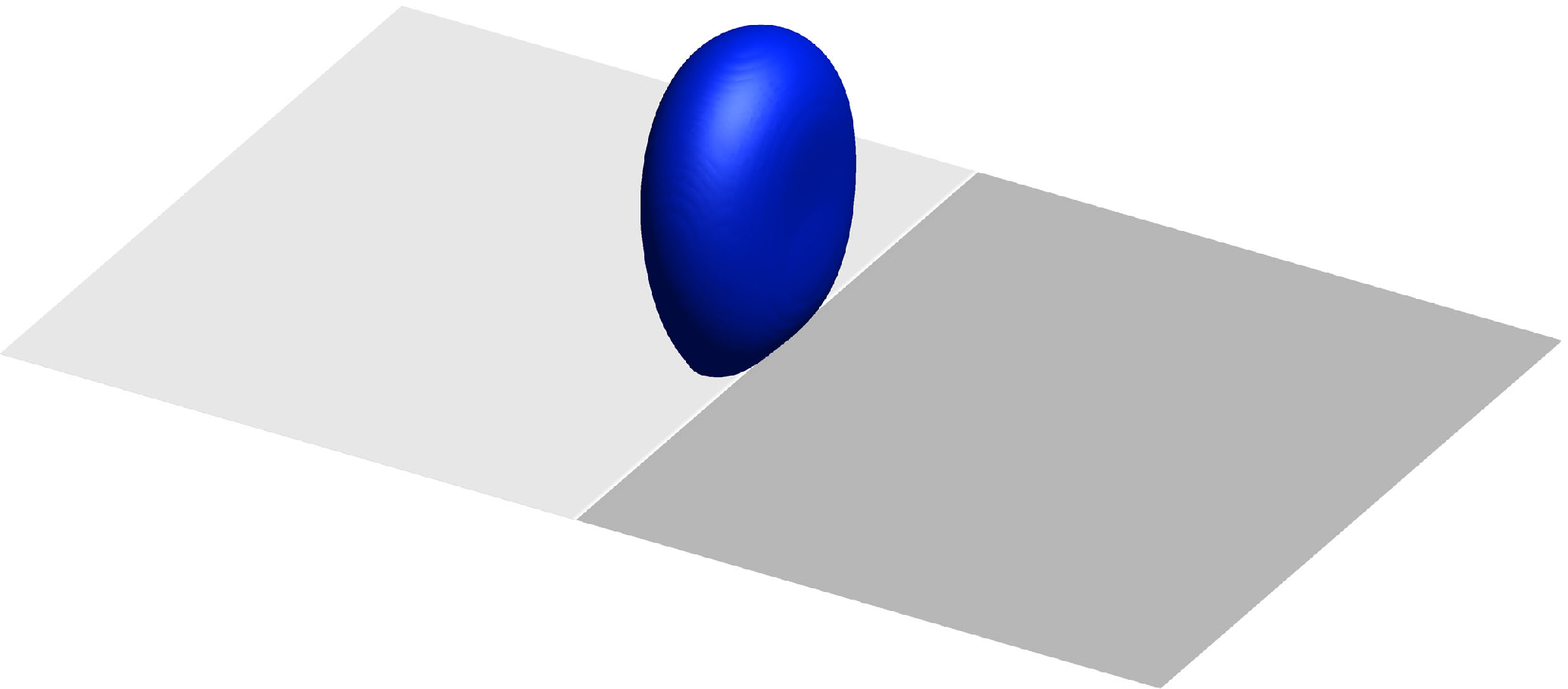}
		\end{minipage}

		\begin{minipage}[c]{0.1\textwidth}
			\centering
			\caption*{(h) $ t^{*}=5.0 $  }
		\end{minipage}
		\begin{minipage}[c]{0.2\textwidth}
			\includegraphics[width=\textwidth]{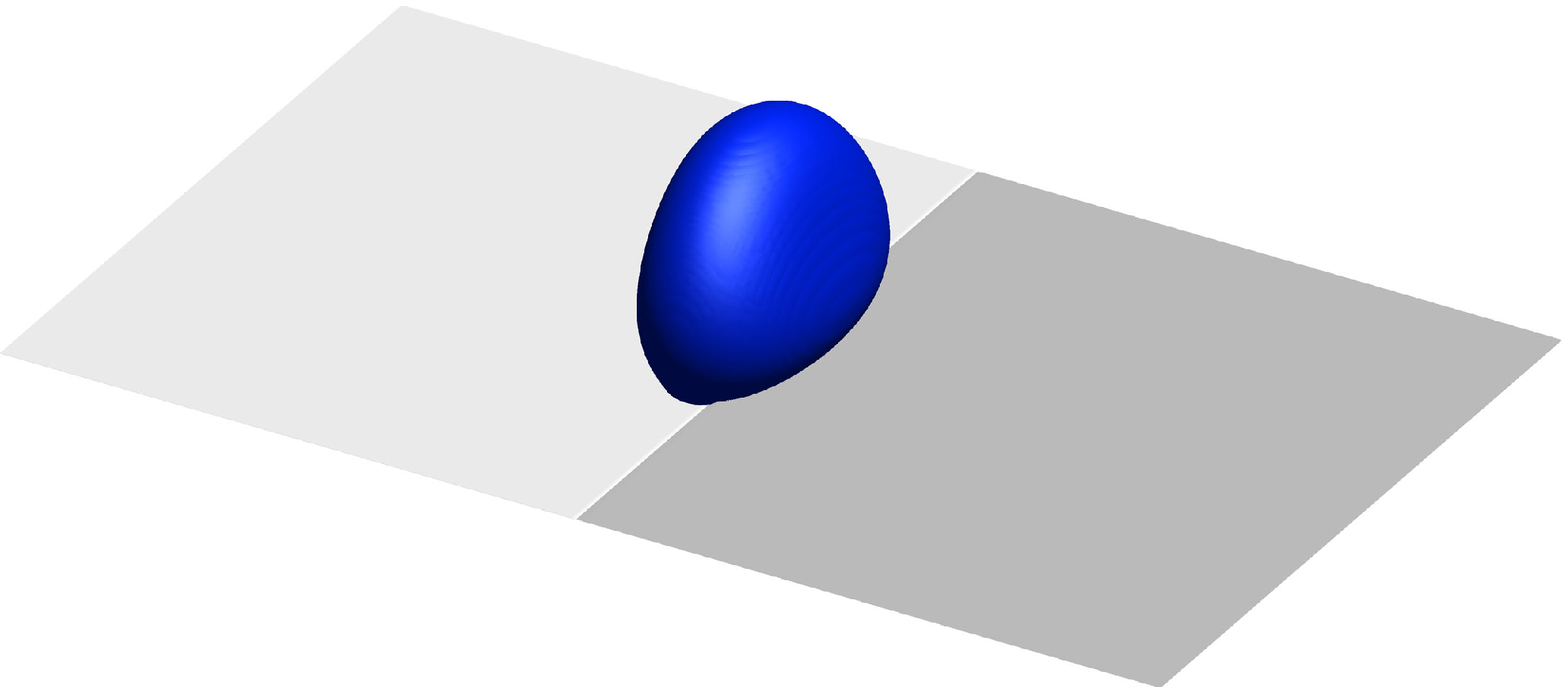}
			\caption*{$ We=3.3 $}
		\end{minipage}
		\begin{minipage}[c]{0.2\textwidth}
			\includegraphics[width=\textwidth]{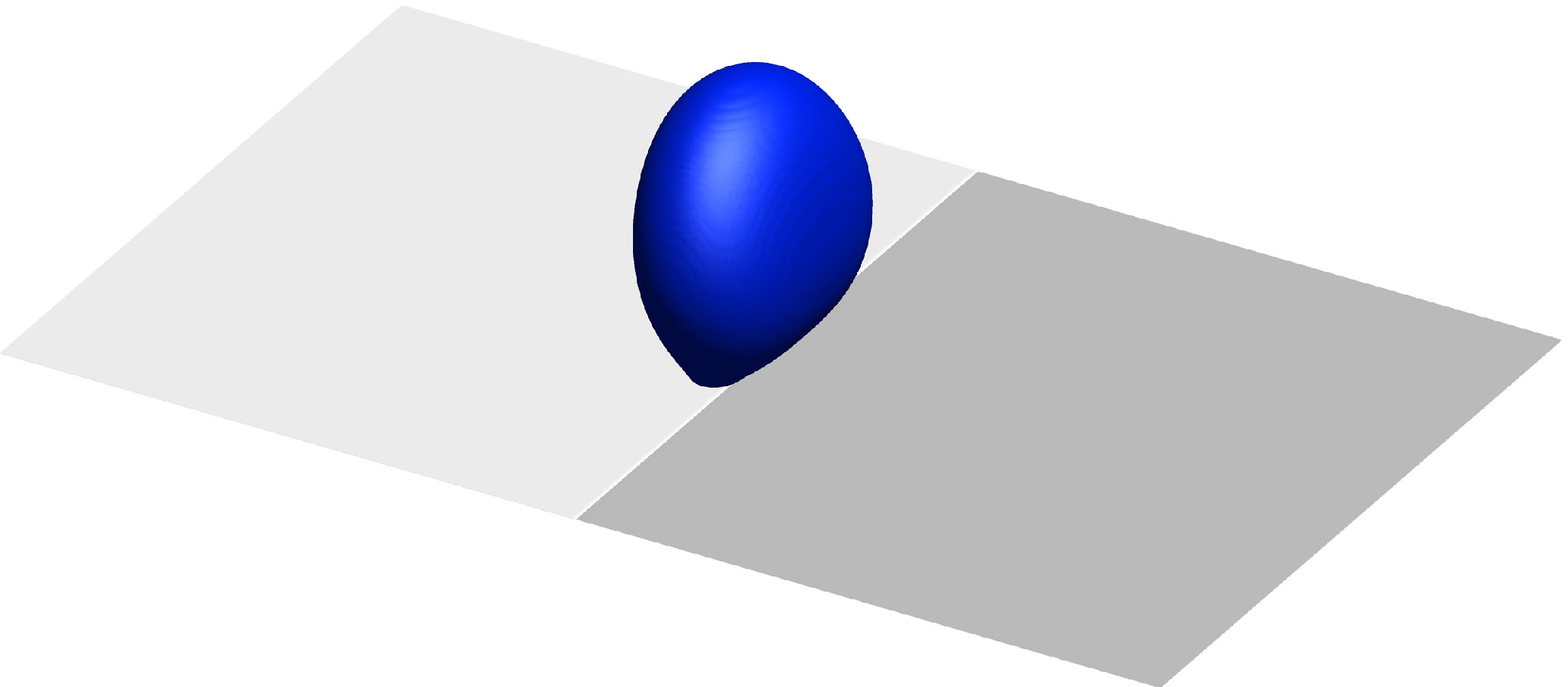}
			\caption*{$ We=13.1 $}
		\end{minipage}
		\begin{minipage}[c]{0.2\textwidth}
			\includegraphics[width=\textwidth]{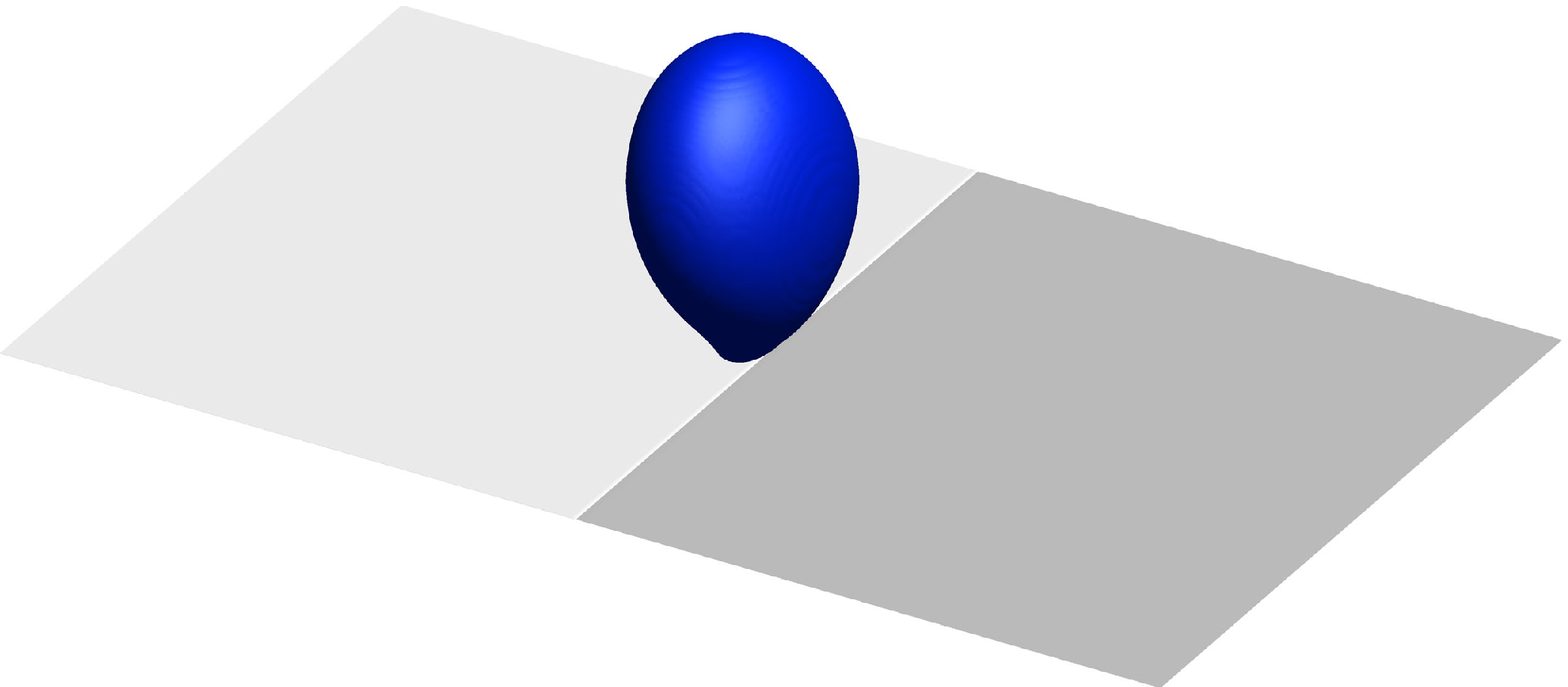}
			\caption*{$ We=29.5 $}
		\end{minipage}		
		\begin{minipage}[c]{0.2\textwidth}
			\includegraphics[width=\textwidth]{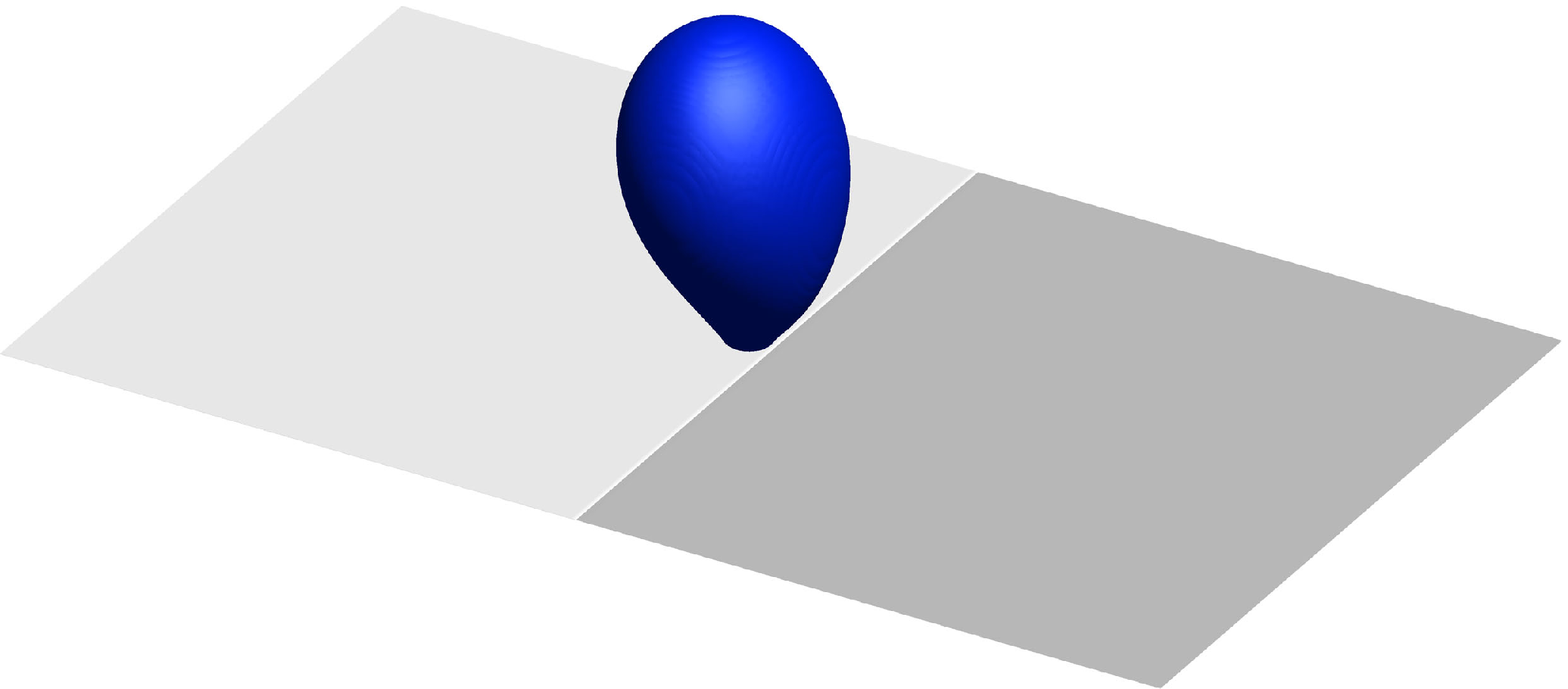}
			\caption*{$ We=52.5 $}
		\end{minipage}

		\caption{The morphologies of double droplets impinging upon the surface for various $We$ at wettability difference  $ \Delta \theta=20 \degree $ and $ L^{*}=0.75 $.} 
		\label{fig10}
	\end{figure}

	\begin{figure}[ht]
		\centering
		\includegraphics[width=0.6\textwidth]{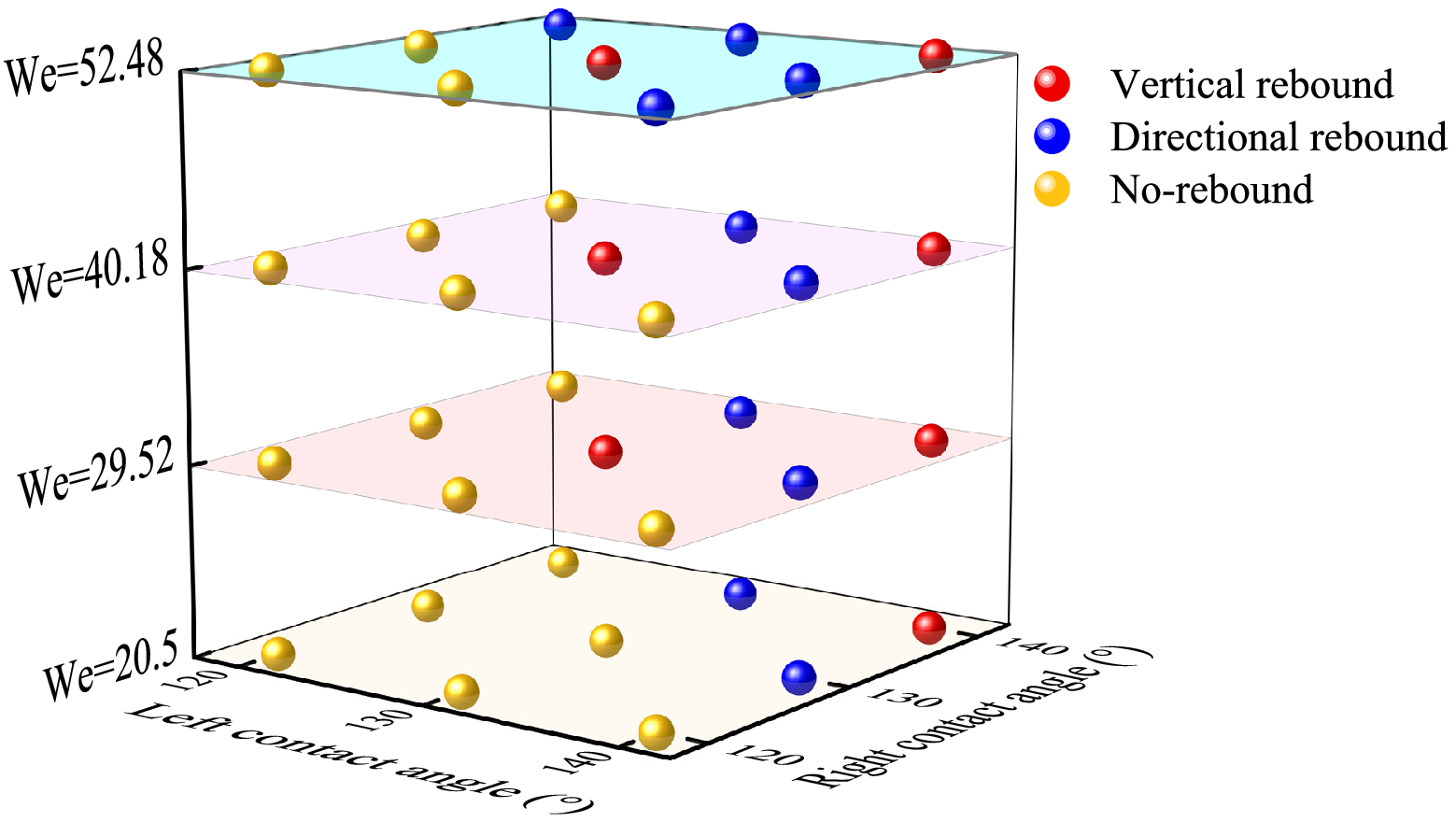}
		\caption{The rebound phase diagram of the double droplets impacting the surface.}
		\label{fig11}
	\end{figure}

We next consider the influence of Weber number on the impact of double droplets on heterogeneous surfaces. Fig. \ref{fig10} shows the evolution of two impinging droplets on the surface with the wettability difference $ \Delta \theta=20 \degree $ for various Weber number conditions. It can be seen from the Fig. \ref{fig10} that with the increase of Weber number, the droplet has greater kinetic energy, which increases the spreading velocity of the droplet and the contraction velocity after impacting the wall, and increases the deformation degree (see Fig. \ref{fig10}(c)). For the higher Weber number $ (We = 52.5) $, the droplet eventually completely migrated to the low-wettability surface and rebounded toward the left side, while the droplet did not rebound for the other three cases. Based on the above, we speculate that the rebound of droplets is the result of the combined action of Weber number and surface wettability.  Fig. \ref{fig11} shows the rebound phase diagram of the double droplets impinging on the surfaces, there are three possible outcomes. It can be seen that droplet rebound occurs at large Weber numbers and large hydrophobicity. For the surface with wettability difference, since the unbalanced force is produced by the wetting difference, the rebounds on heterogeneous surfaces tend to be directional bounces toward the high-wettability surface, while the rebounds are vertical on the uniform surfaces \cite{XiongCF2018,CuiCF2021}.

	\begin{figure}[H]
		\begin{minipage}[t]{0.45\textwidth}
			\centering
			\includegraphics[width=\textwidth]{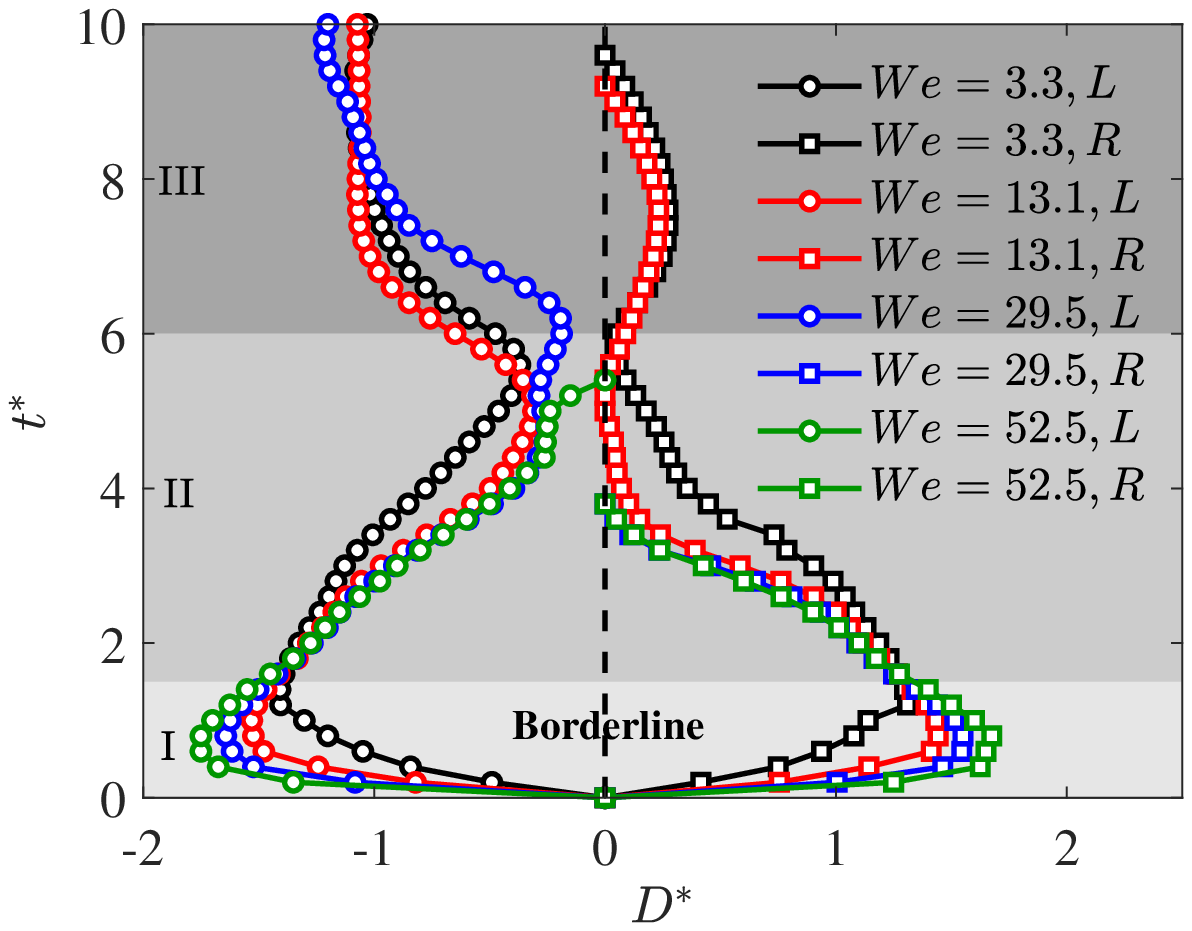}
			\caption{Temporal evolution of the spread factor for different $We$ at wettability differences $ \Delta \theta=20 \degree $.}
			\label{fig12}
		\end{minipage}%
		\hfill
		\begin{minipage}[t]{0.45\textwidth}
			\centering
			\includegraphics[width=\textwidth]{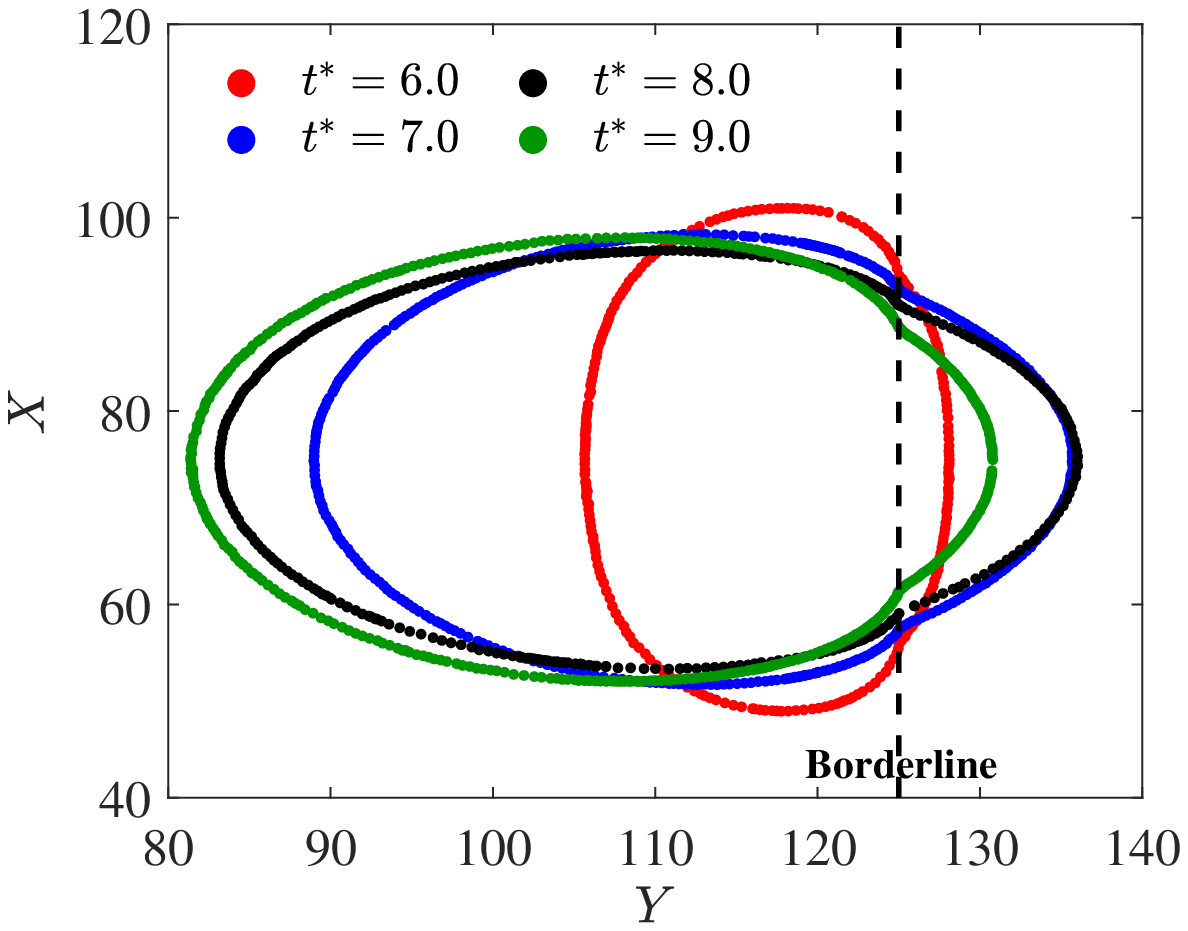}
			\caption{The contour of three-phase contact line of droplet on x-y plane at different moments for $We=3.3$.}
			\label{fig13}
		\end{minipage}
	\end{figure}

Fig. \ref{fig11} shows the evolution of the spreading factor with dimensionless time under different Weber numbers. It can be seen that in the asymmetric spreading stage $ \textrm{I} $, the spreading factor increases with the increase of Weber number due to the fact that initial kinetic energy increases with increasing Weber number, which is consistent with the morphological performance in Fig. \ref{fig10}. In the asymmetric contraction stage $ \textrm{II} $, for the large Weber number, the droplet bounces after migrating to the high wetting area due to the wettability difference, and thus the spreading factor goes to zero. For $ We = 29.5 $, the droplet completely migrates from the hydrophobic region to the left and enters the wetting stage $ \textrm{III} $ to reach equilibrium. However, for the low Weber number, part of the droplets did not completely migrate to the left high wetting area and remained in the right hydrophobic area at the beginning of wetting stage $ \textrm{III} $. In the wetting stage, the right diffusion factor of the droplet first increases and then decreases, and finally completely migrates to the left low hydrophobic region.

\begin{figure}[ht]
		\centering
		\includegraphics[width=0.5\textwidth]{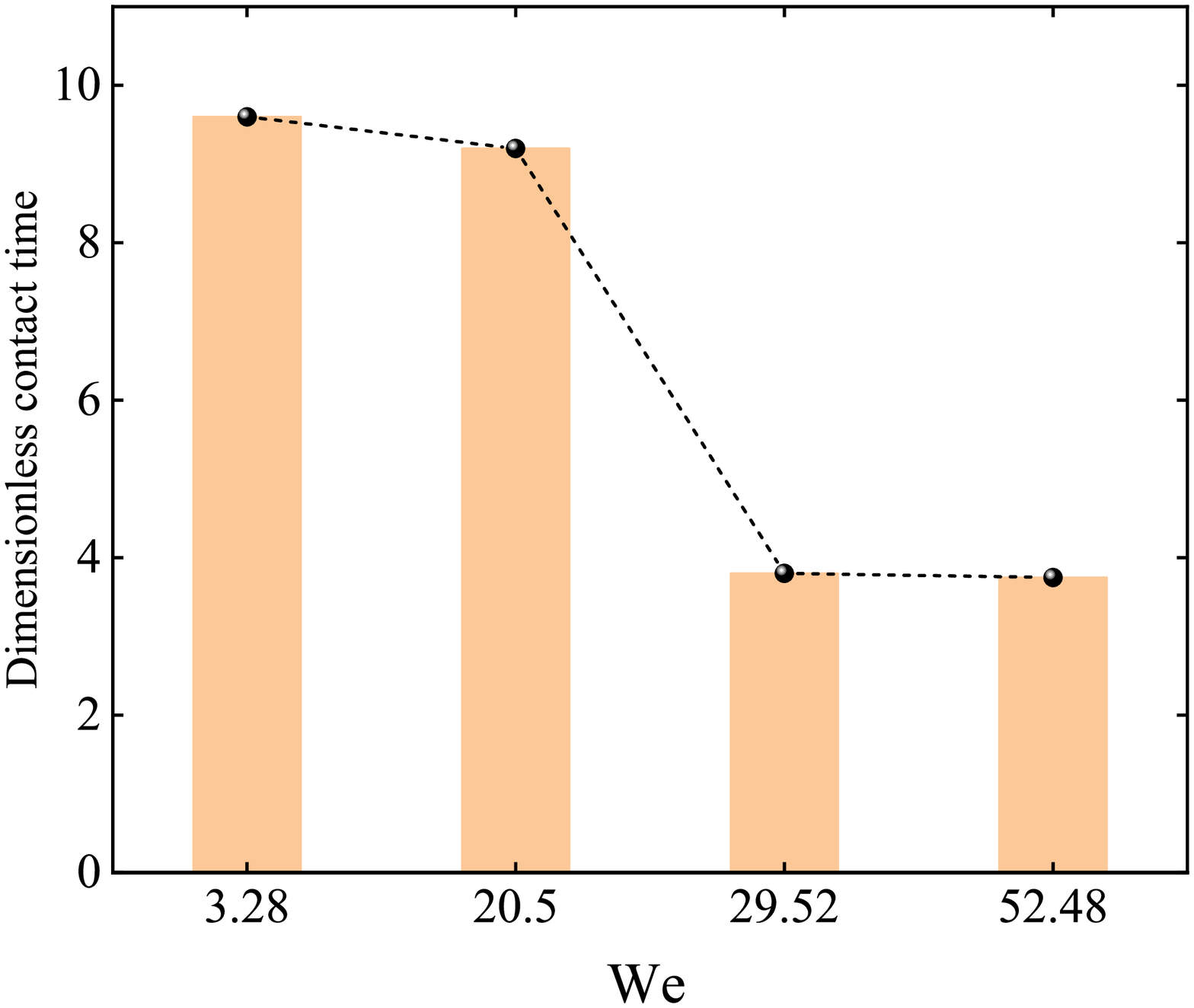}
		\caption{Dimensionless contact time for different $We$ at wettability difference $ \Delta \theta=20 \degree $ and droplet spacing $L^{*}=0.75$.}
		\label{fig14}
	\end{figure}

To intuitively visualize the variation of droplet spreading length for low Weber number conditions, Fig. \ref{fig13} shows the variation of droplet profile in the contact surface on $ x-y $ plane with time at the secondary spreading for $ We=3.3 $. When $ t^{*} = 6.0 $, the droplet is in the final stage of unbalanced contraction. Since the wettability difference exists in the $ y $ direction, the spreading factor in the y direction of the droplet is smaller than that in the $ x $ direction. Under the action of recoil, the coalesced droplets elongated along the $ x $ direction, the secondary spreading causes spreading factor to increase. Subsequently, the droplets slowly migrated to the low hydrophobic region under the action of the unbalanced force generated by the wetting difference, and the spreading factor gradually decreased to zero. It is the appearance of secondary spreading at low Weber number conditions that leads to a significant increase in contact time, as show in Fig. \ref{fig14}.

\subsection{The effect of droplet spacing} 

It is evident that the coalescence of the two droplets is influenced by the droplet spacing, and thus the morphologies of double droplets impinging upon the heterogeneous surface are expected to be differenent when varying the droplet spacing. In this section, the effect of the spacing between the two droplets is investigated. The different dimensionless horizontal distances are set as $ L^{*}=0.75 $, $ L^{*}=0.875 $, $ L^{*}=1.0 $ and $L^{*}=1.0625 $.
	\begin{figure}[ht]
		\centering
		
		\begin{minipage}[c]{0.1\textwidth}
			\centering
			\caption*{(a) $ t^{*}=1.0 $ }
		\end{minipage}
		\begin{minipage}[c]{0.2\textwidth}
			\includegraphics[width=\textwidth]{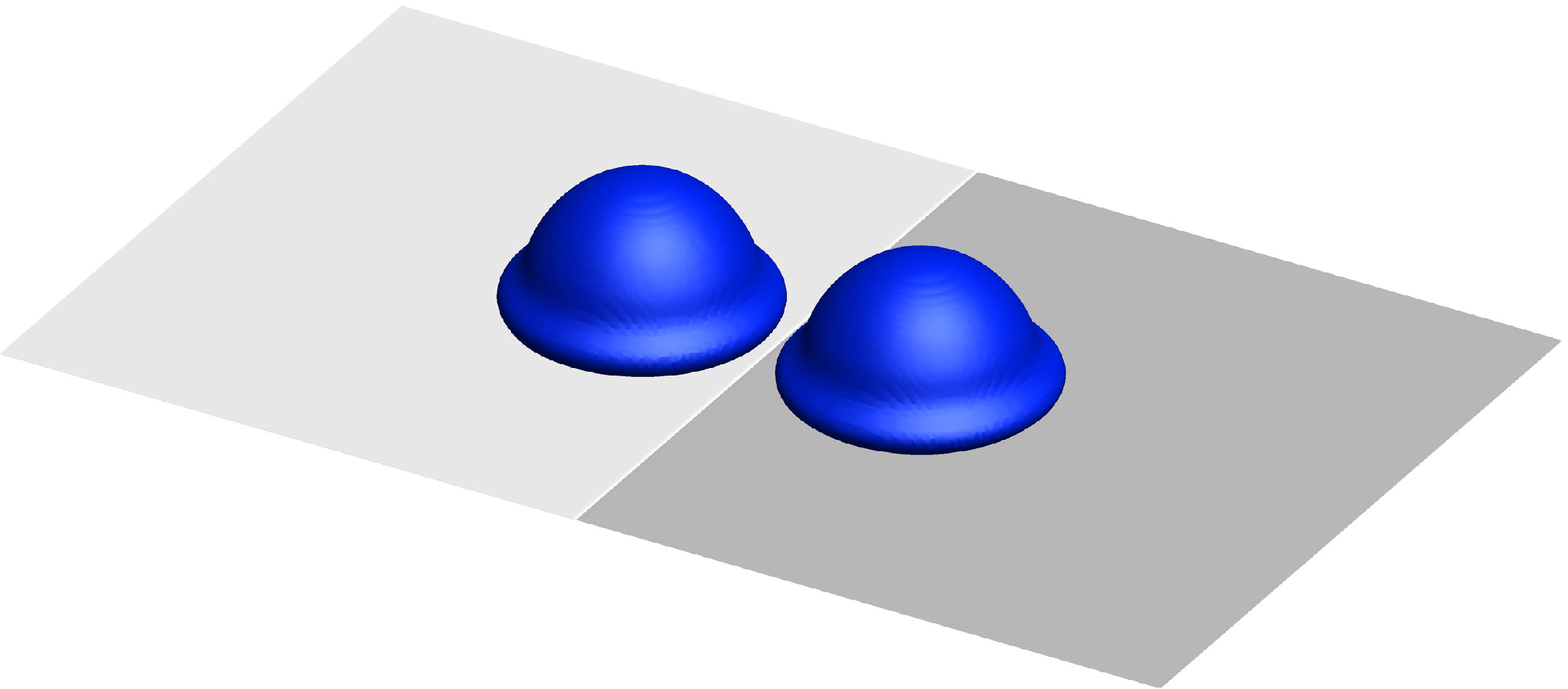}
		\end{minipage}
		\begin{minipage}[c]{0.2\textwidth}
			\includegraphics[width=\textwidth]{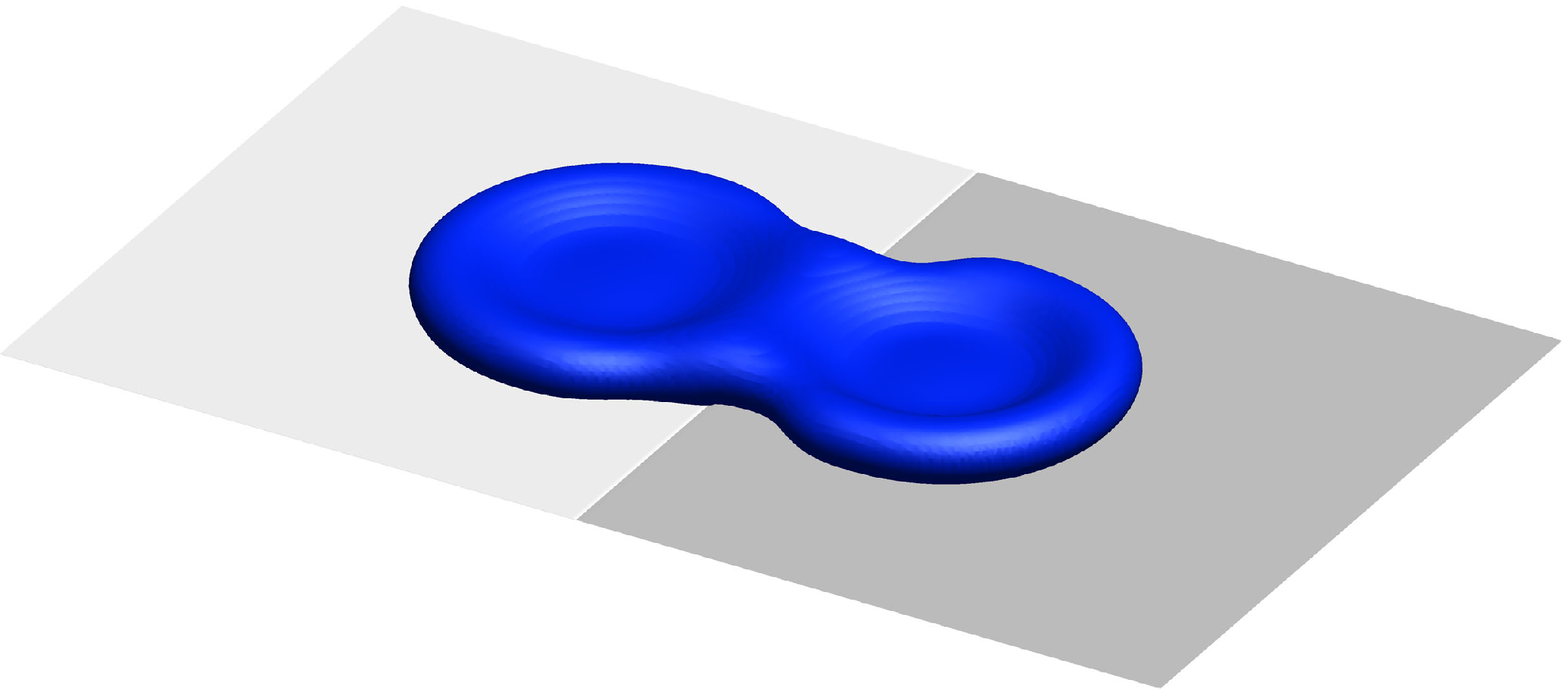}
		\end{minipage}
		\begin{minipage}[c]{0.2\textwidth}
			\includegraphics[width=\textwidth]{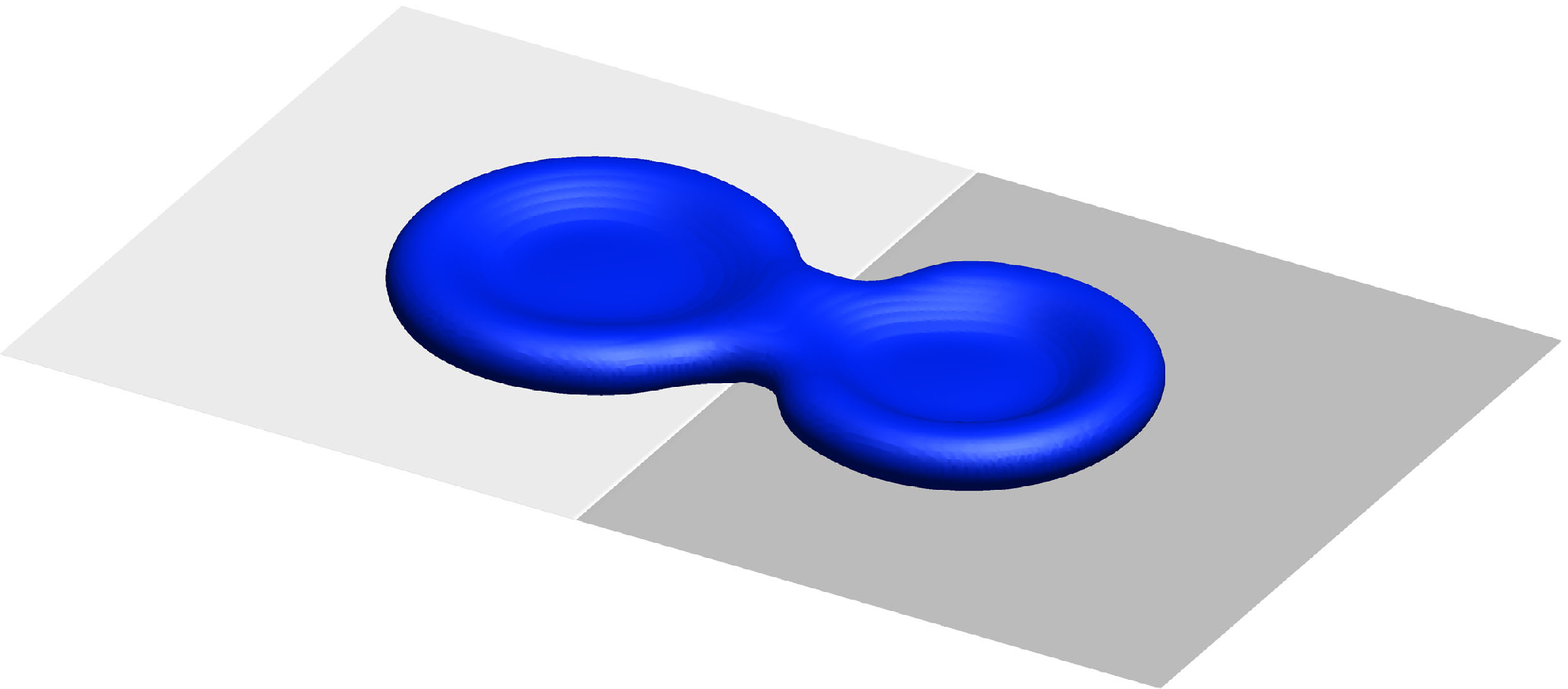}
		\end{minipage}
		\begin{minipage}[c]{0.2\textwidth}
			\includegraphics[width=\textwidth]{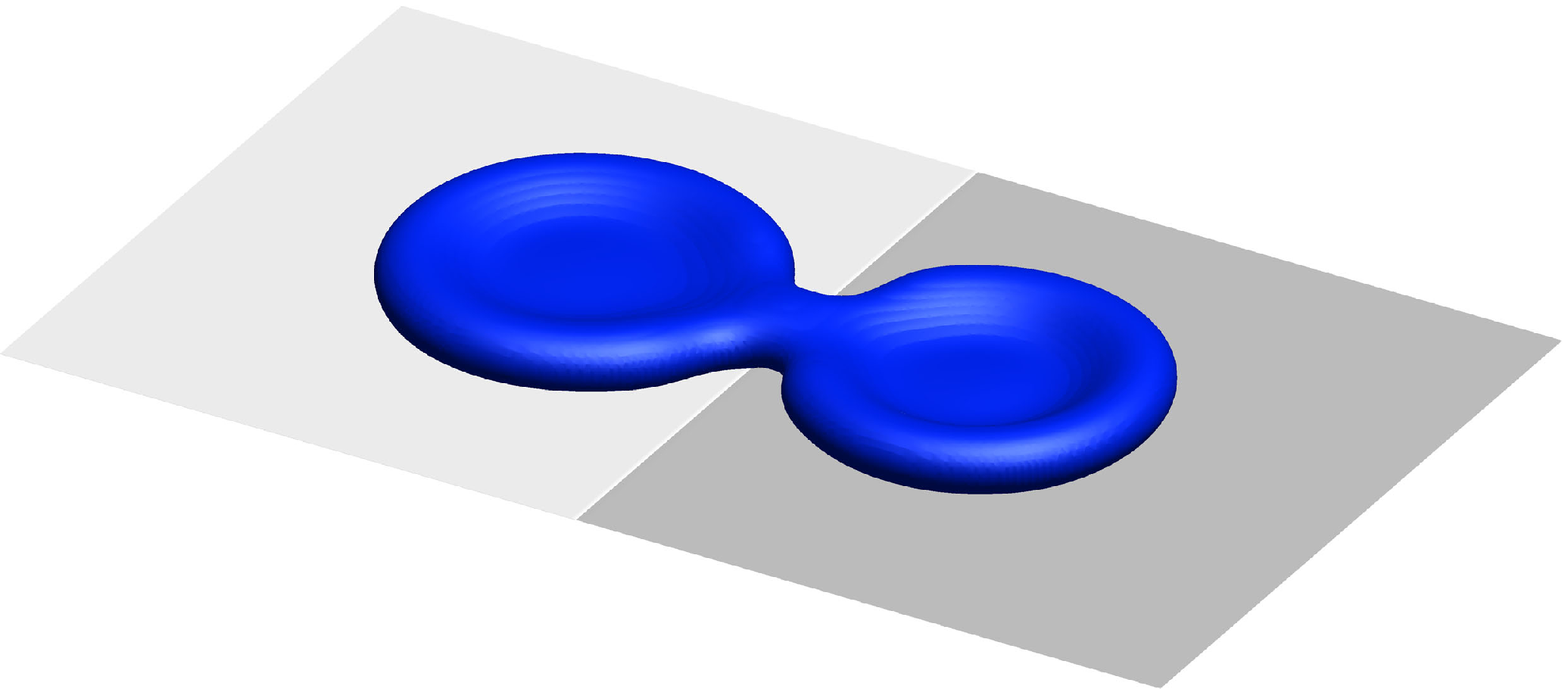}
		\end{minipage}

		\begin{minipage}[c]{0.1\textwidth}
			\centering
			\caption*{(b) $ t^{*}=1.4 $  }
		\end{minipage}
		\begin{minipage}[c]{0.2\textwidth}
			\includegraphics[width=\textwidth]{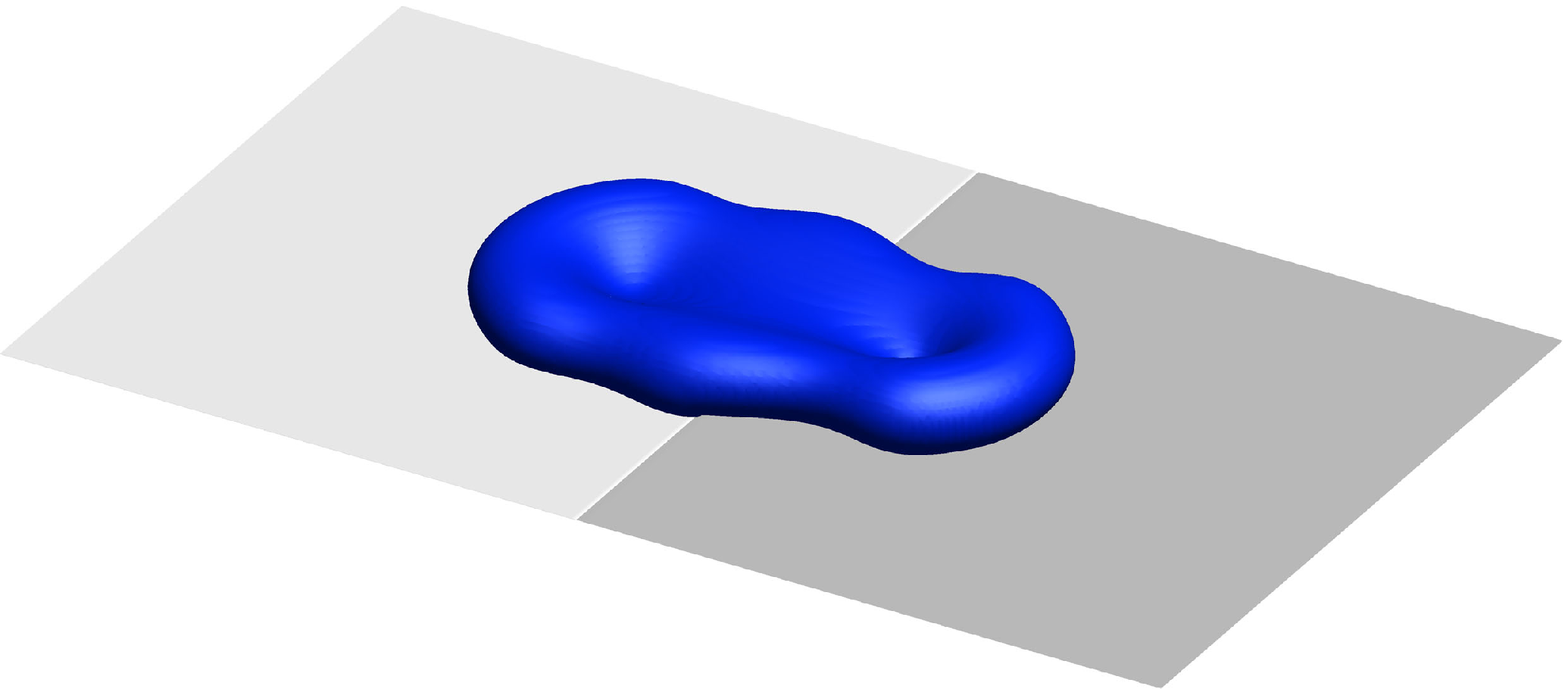}
		\end{minipage}
		\begin{minipage}[c]{0.2\textwidth}
			\includegraphics[width=\textwidth]{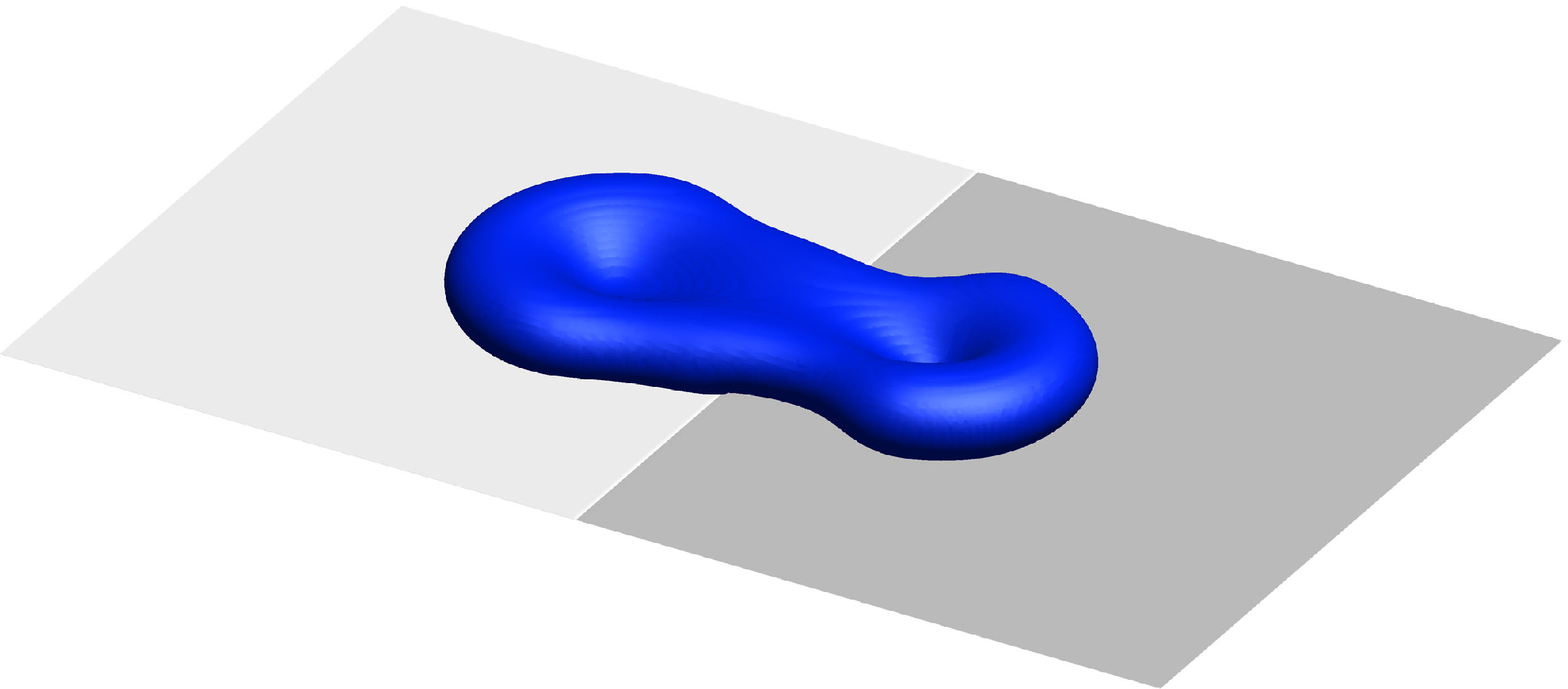}
		\end{minipage}
		\begin{minipage}[c]{0.2\textwidth}
			\includegraphics[width=\textwidth]{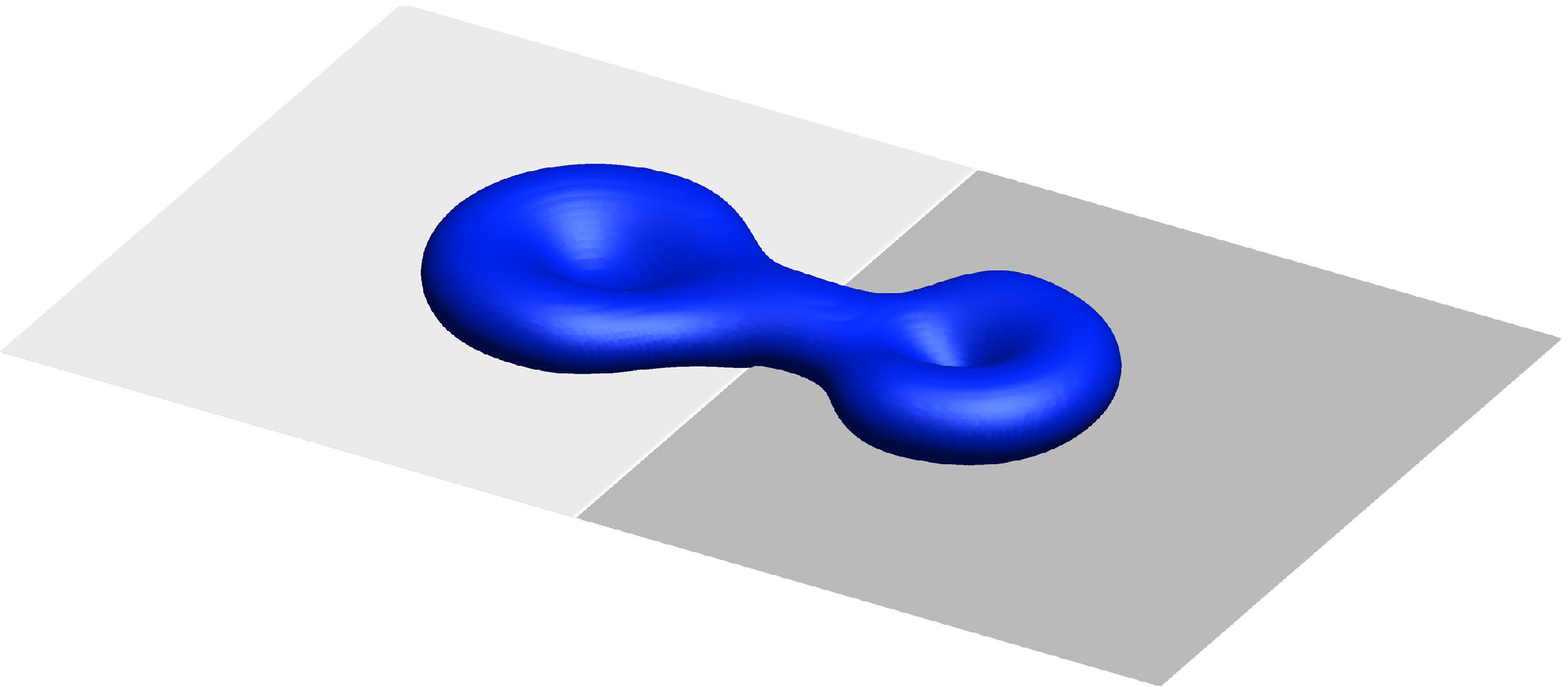}
		\end{minipage}
		\begin{minipage}[c]{0.2\textwidth}
			\includegraphics[width=\textwidth]{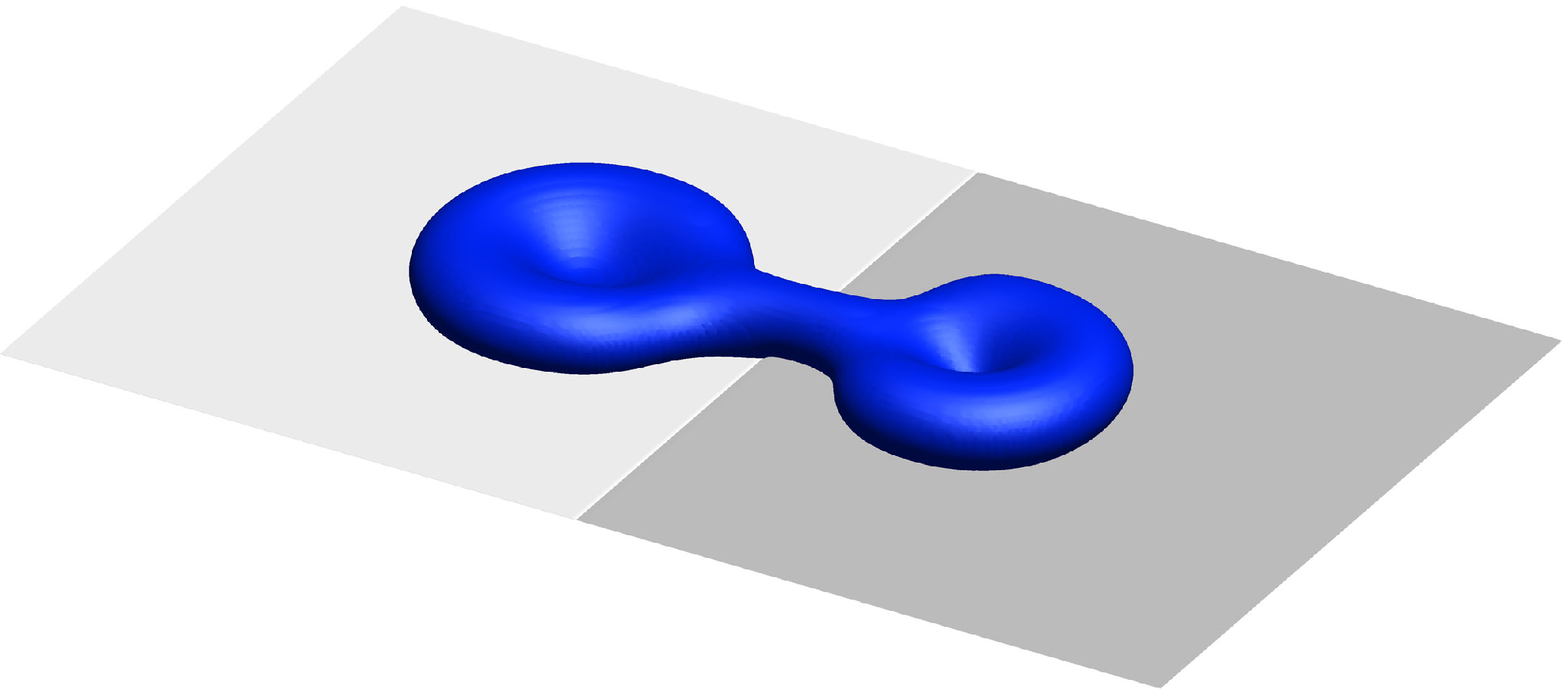}
		\end{minipage}

		\begin{minipage}[c]{0.1\textwidth}
			\centering
			\caption*{(c) $ t^{*}=2.0 $  }
		\end{minipage}
		\begin{minipage}[c]{0.2\textwidth}
			\includegraphics[width=\textwidth]{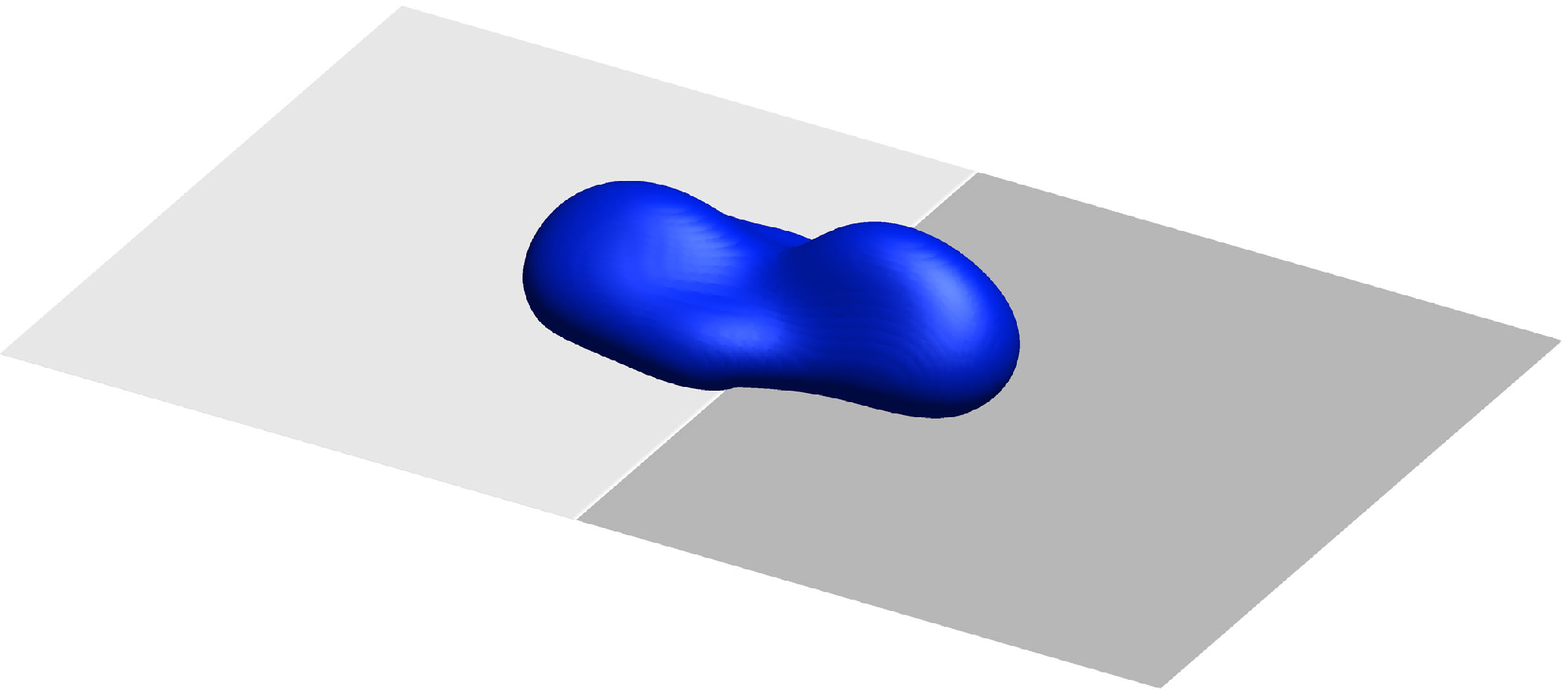}
		\end{minipage}
		\begin{minipage}[c]{0.2\textwidth}
			\includegraphics[width=\textwidth]{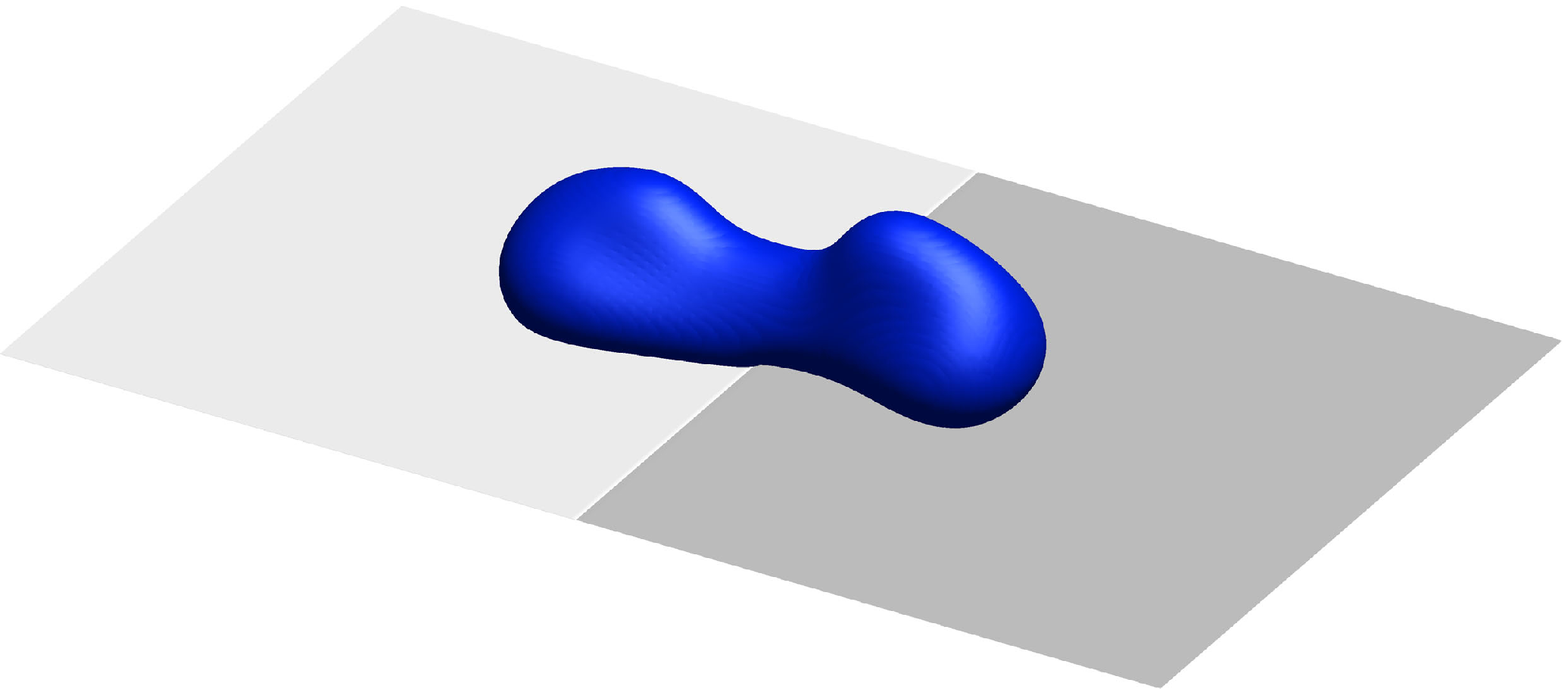}
		\end{minipage}
		\begin{minipage}[c]{0.2\textwidth}
			\includegraphics[width=\textwidth]{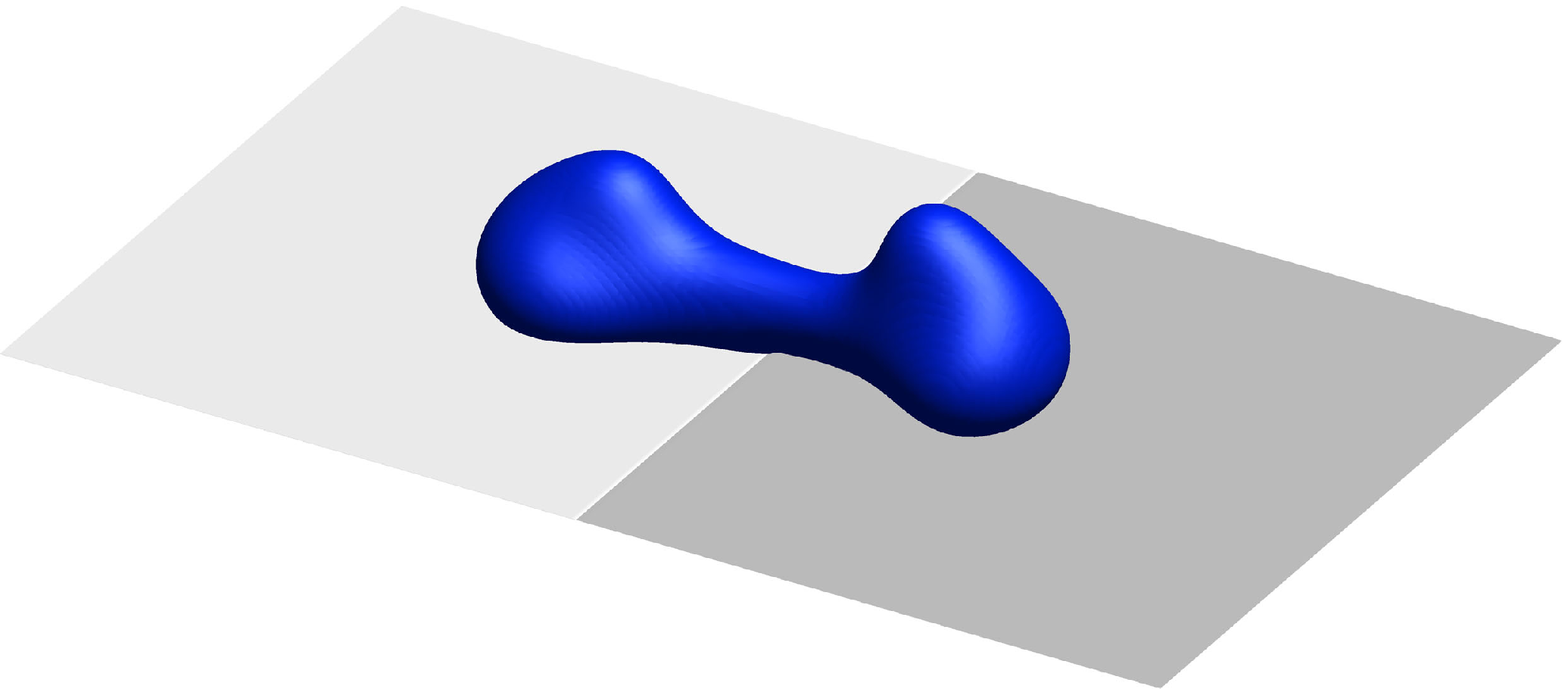}
		\end{minipage}
		\begin{minipage}[c]{0.2\textwidth}
			\includegraphics[width=\textwidth]{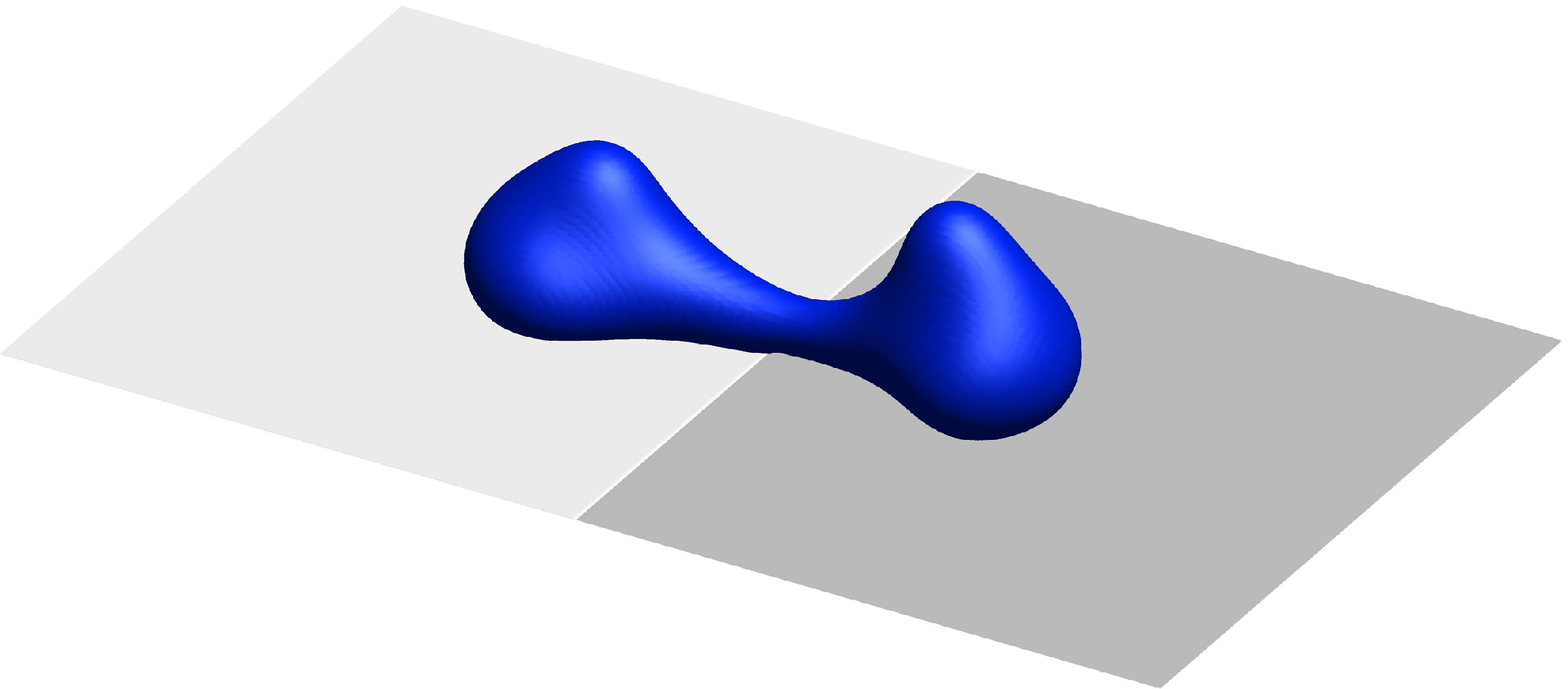}
		\end{minipage}

		\begin{minipage}[c]{0.1\textwidth}
			\centering
			\caption*{(d) $ t^{*}=2.6 $  }
		\end{minipage}
		\begin{minipage}[c]{0.2\textwidth}
			\includegraphics[width=\textwidth]{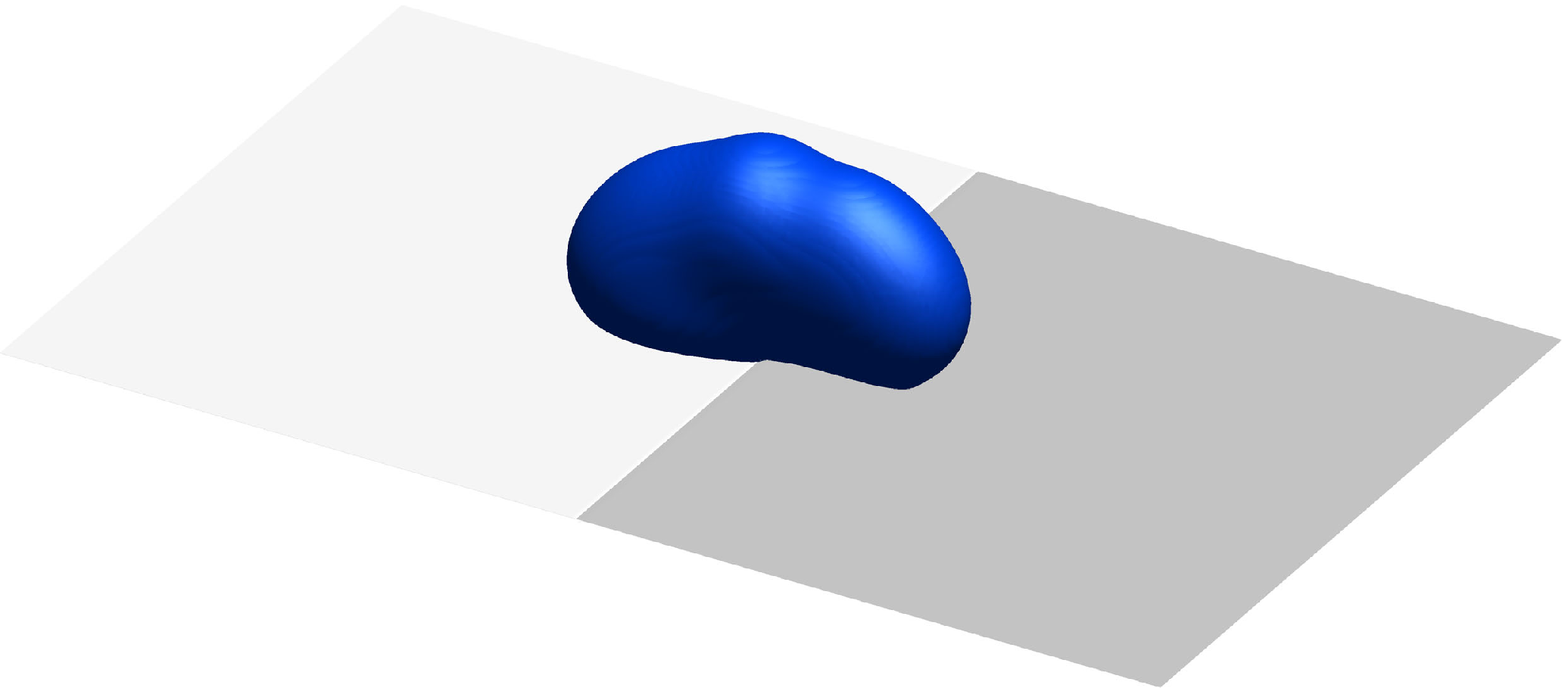}
		\end{minipage}
		\begin{minipage}[c]{0.2\textwidth}
			\includegraphics[width=\textwidth]{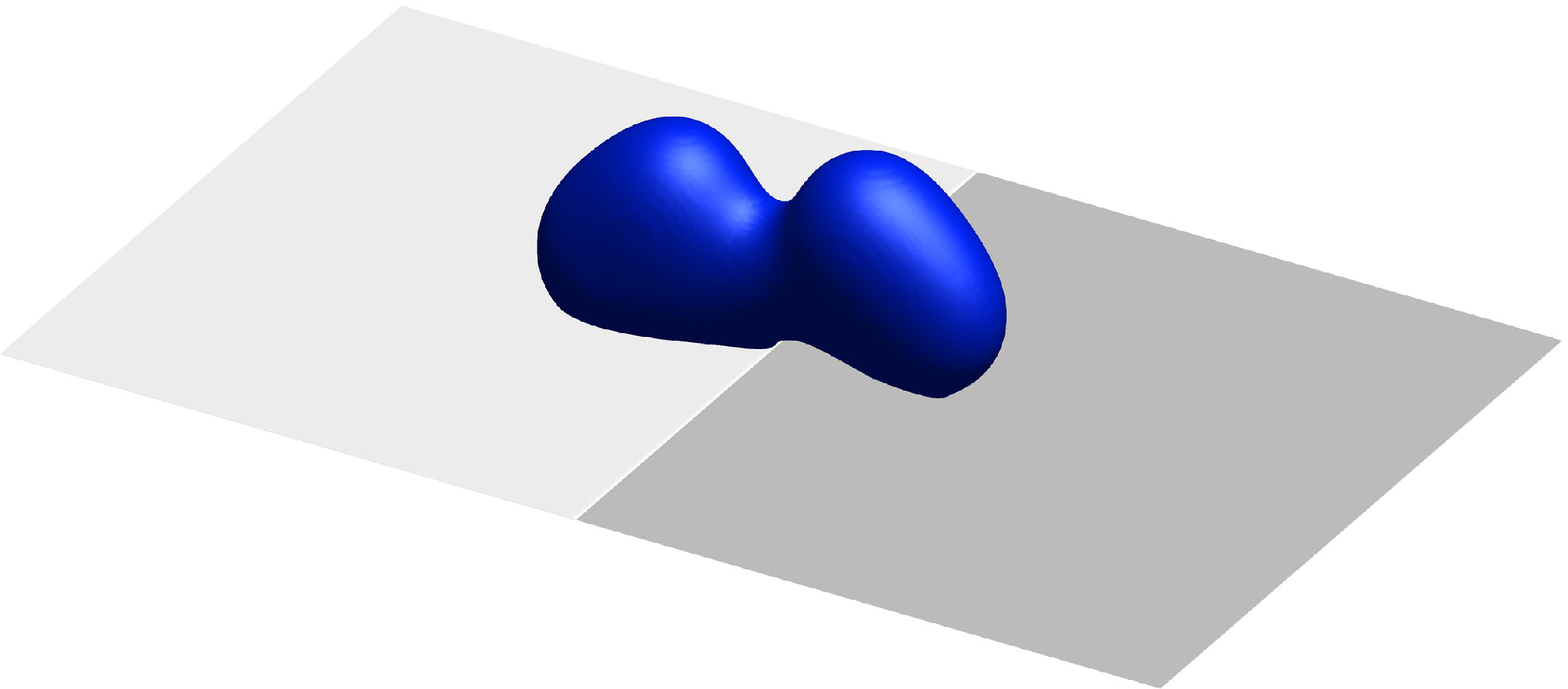}
		\end{minipage}
		\begin{minipage}[c]{0.2\textwidth}
			\includegraphics[width=\textwidth]{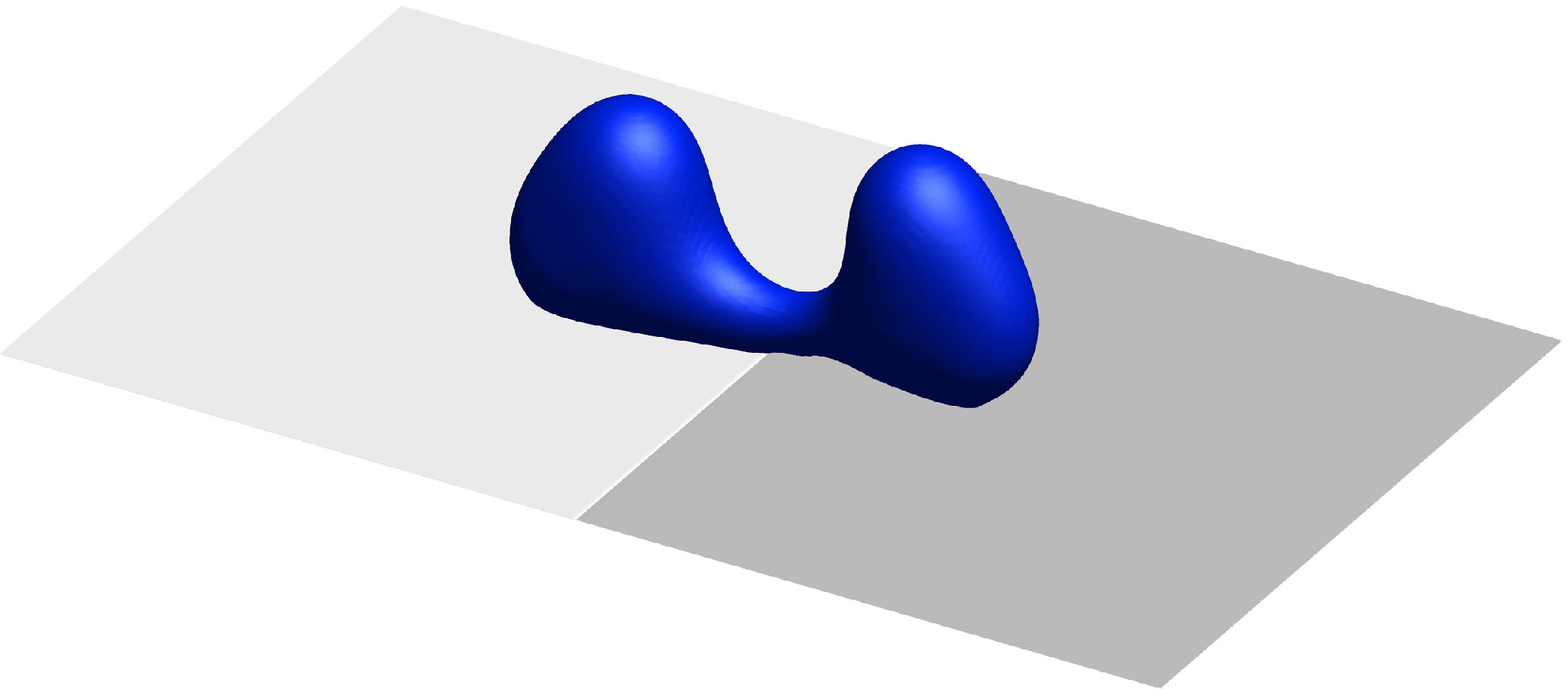}
		\end{minipage}
		\begin{minipage}[c]{0.2\textwidth}
			\includegraphics[width=\textwidth]{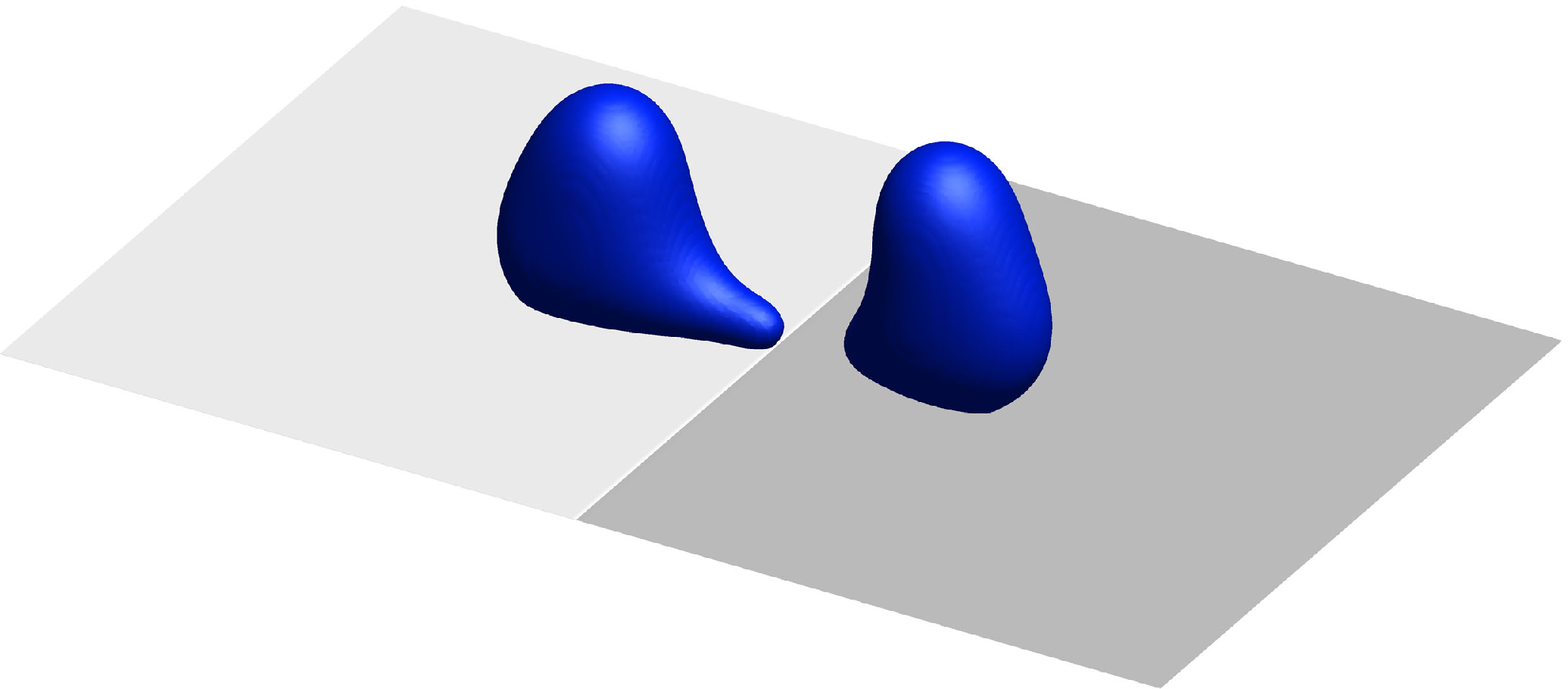}
		\end{minipage}

		\begin{minipage}[c]{0.1\textwidth}
			\centering
			\caption*{(e) $ t^{*}=3.0 $  }
		\end{minipage}
		\begin{minipage}[c]{0.2\textwidth}
			\includegraphics[width=\textwidth]{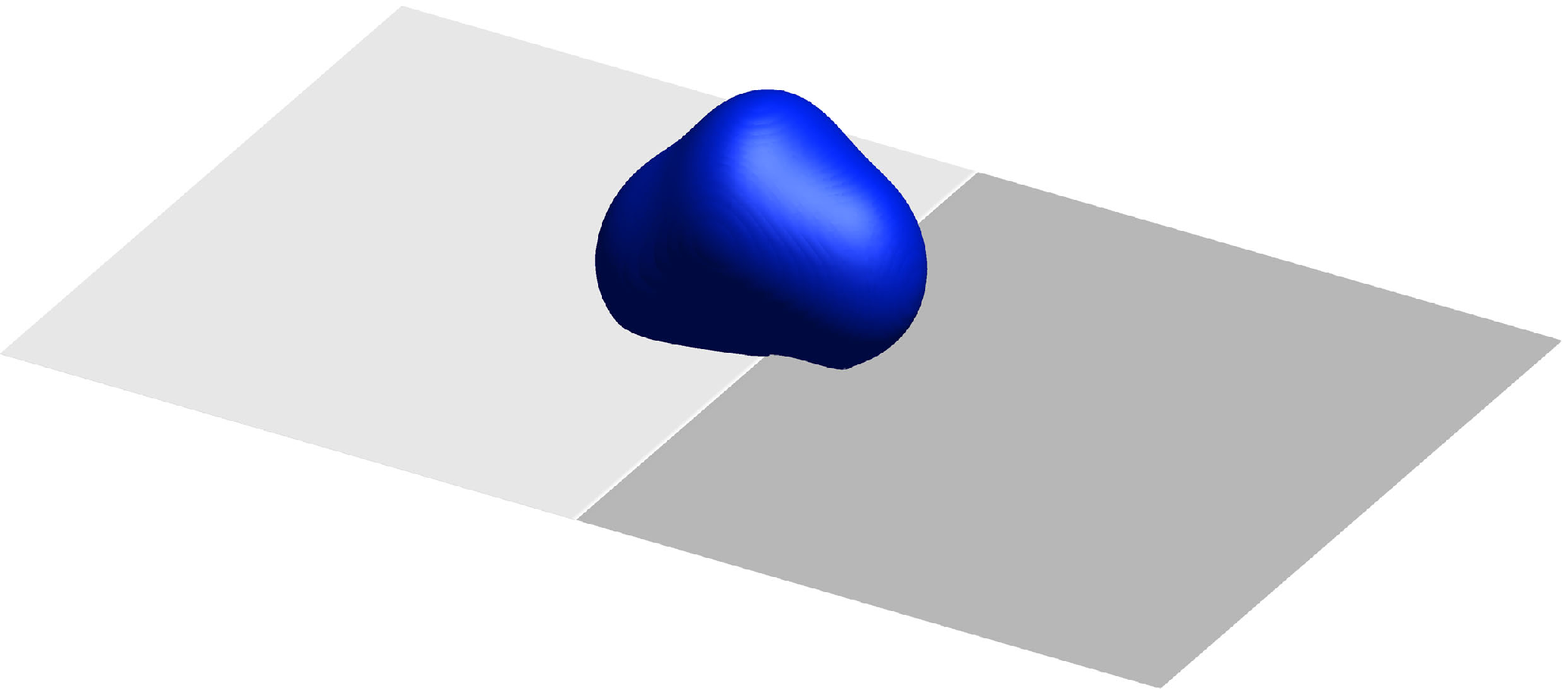}
		\end{minipage}
		\begin{minipage}[c]{0.2\textwidth}
			\includegraphics[width=\textwidth]{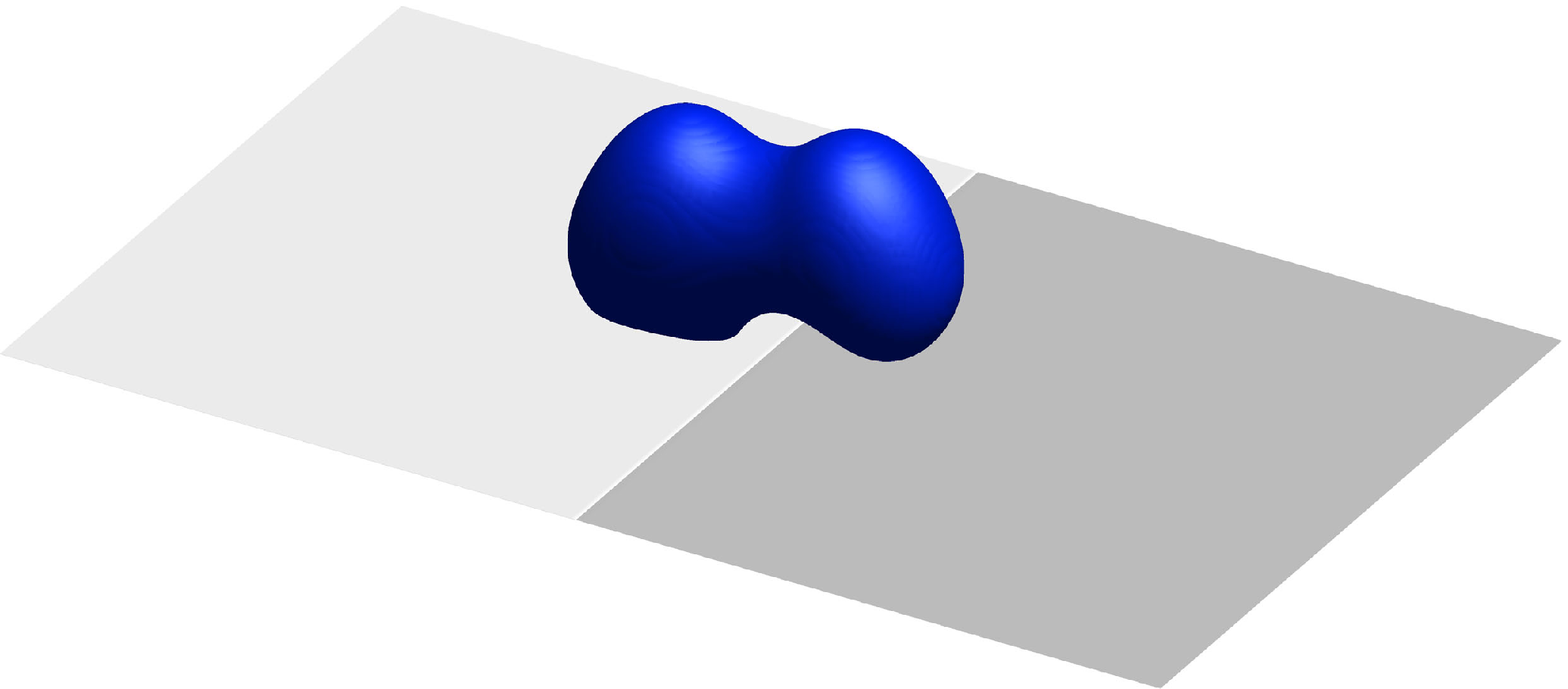}
		\end{minipage}
		\begin{minipage}[c]{0.2\textwidth}
			\includegraphics[width=\textwidth]{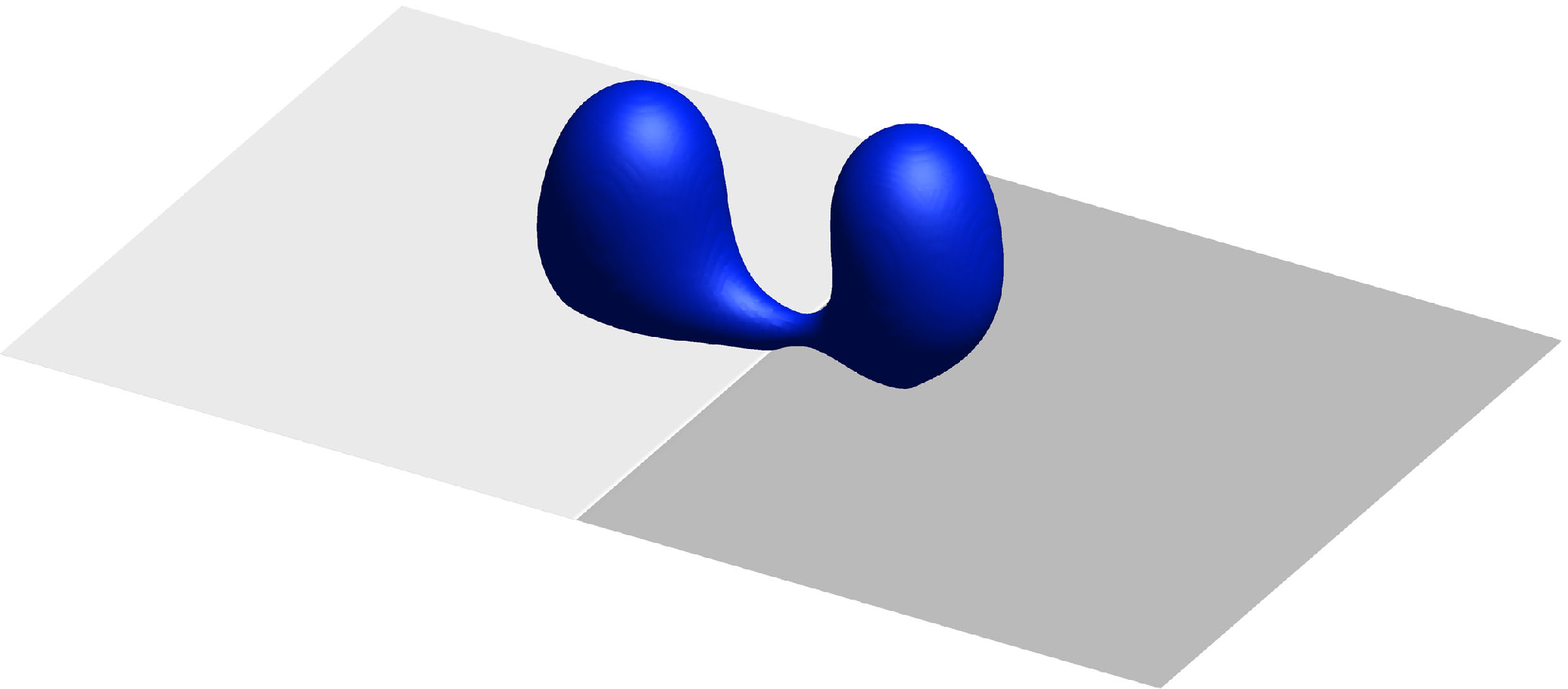}
		\end{minipage}
		\begin{minipage}[c]{0.2\textwidth}
			\includegraphics[width=\textwidth]{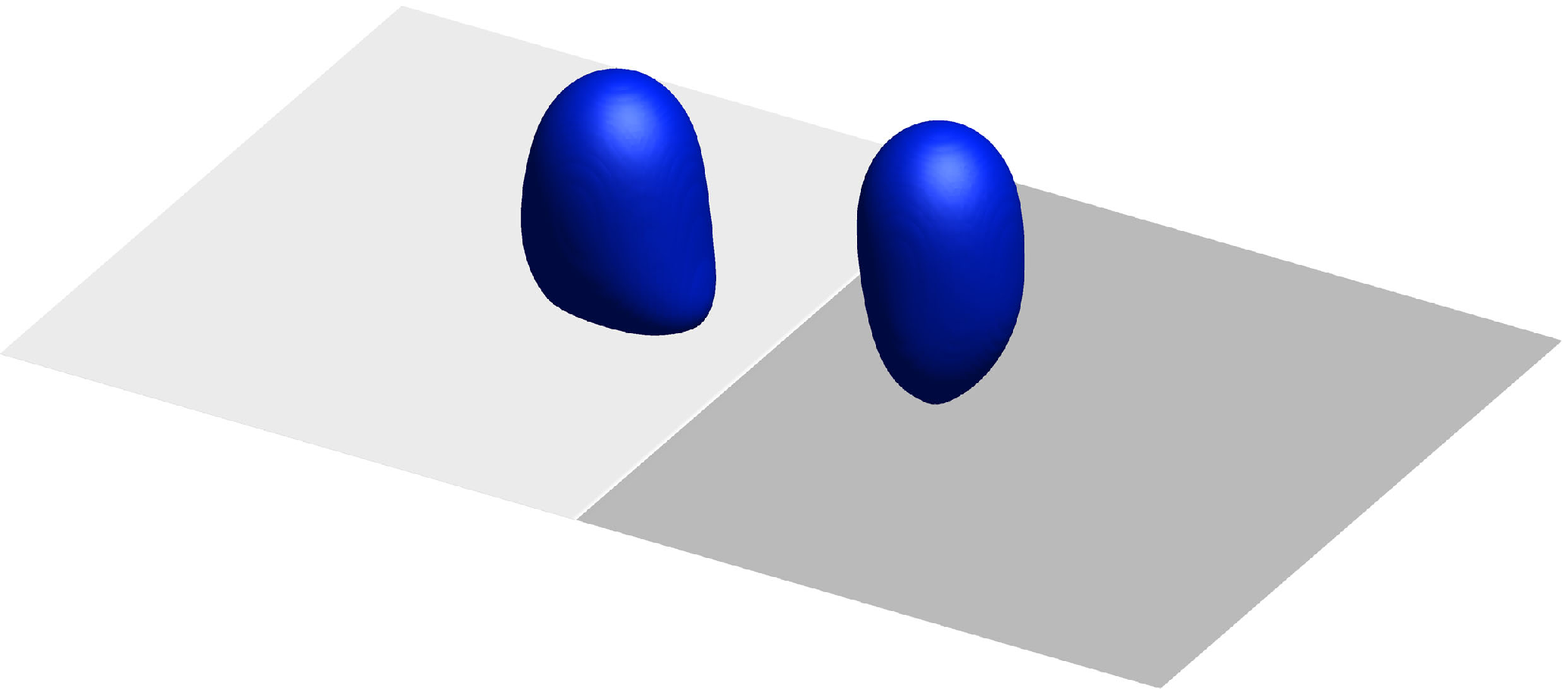}
		\end{minipage}

		\begin{minipage}[c]{0.1\textwidth}
			\centering
			\caption*{(f) $ t^{*}=4.0 $  }
		\end{minipage}
		\begin{minipage}[c]{0.2\textwidth}
			\includegraphics[width=\textwidth]{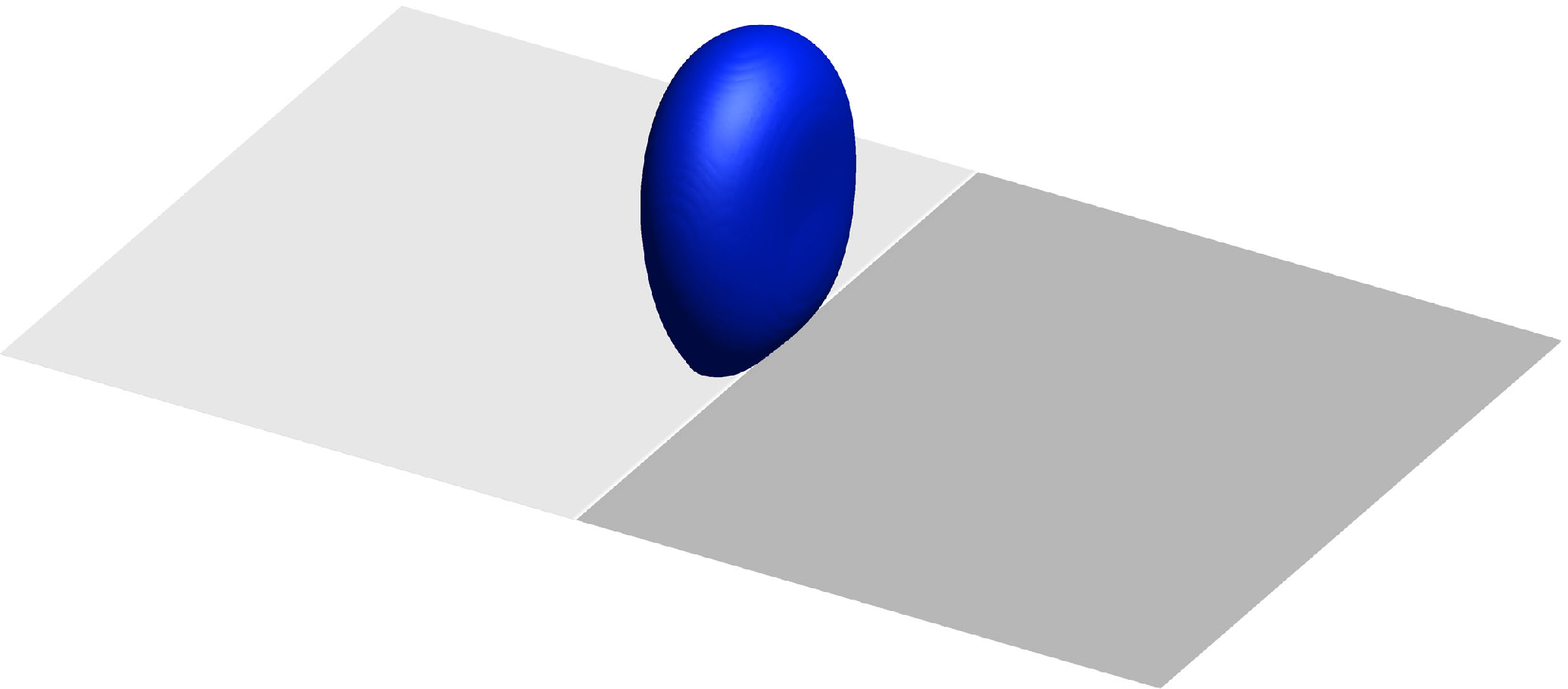}
			\caption*{$ L^{*}=0.75 $}
		\end{minipage}
		\begin{minipage}[c]{0.2\textwidth}
			\includegraphics[width=\textwidth]{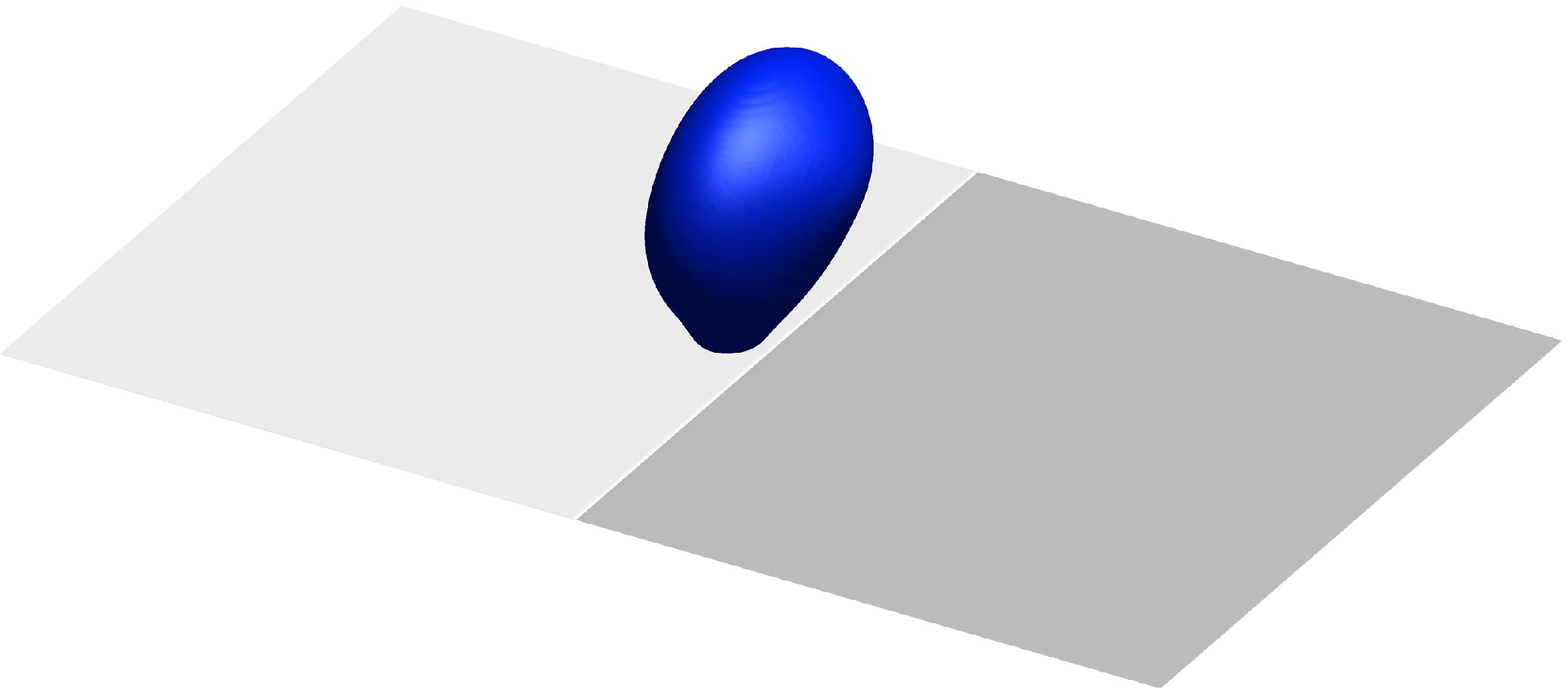}
			\caption*{$ L^{*}=0.875 $}
		\end{minipage}
		\begin{minipage}[c]{0.2\textwidth}
			\includegraphics[width=\textwidth]{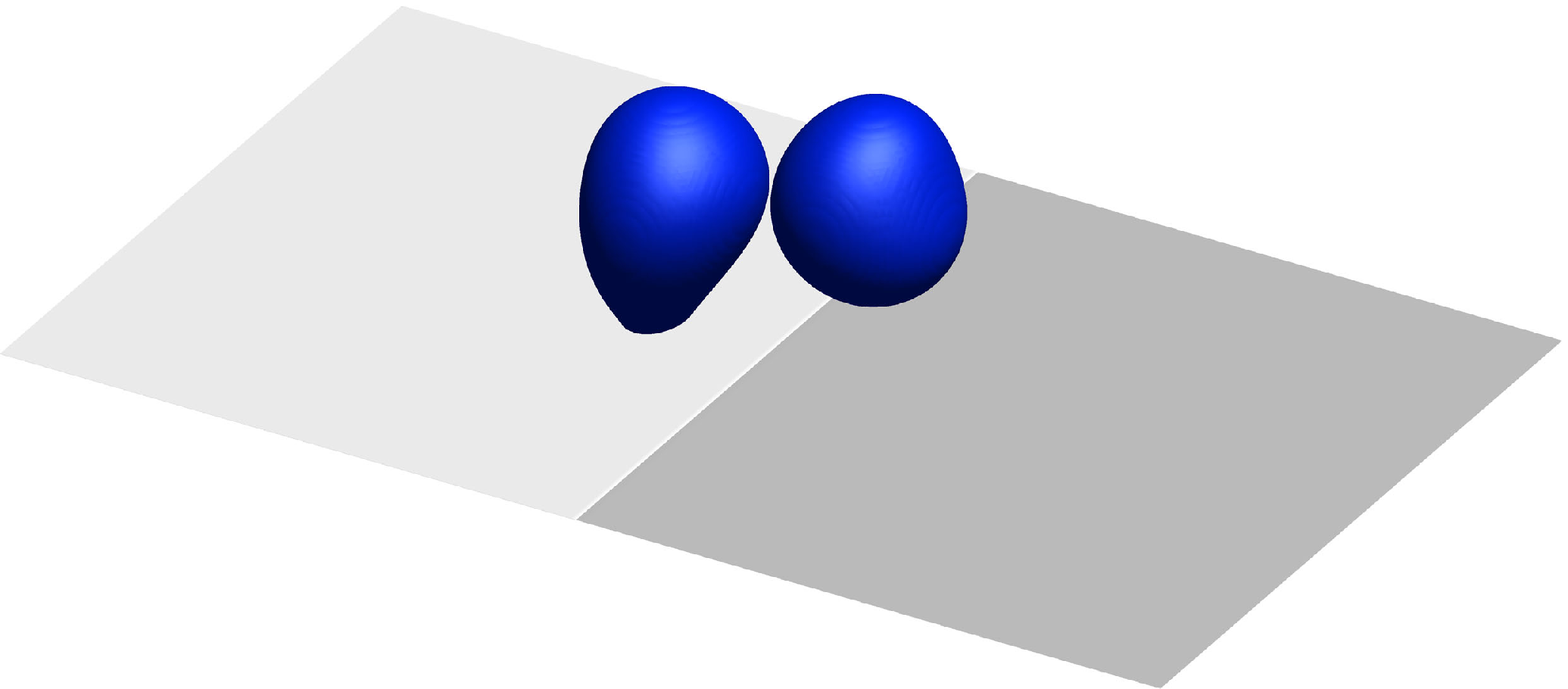}
			\caption*{$ L^{*}=1.0 $}
		\end{minipage}	
		\begin{minipage}[c]{0.2\textwidth}
			\includegraphics[width=\textwidth]{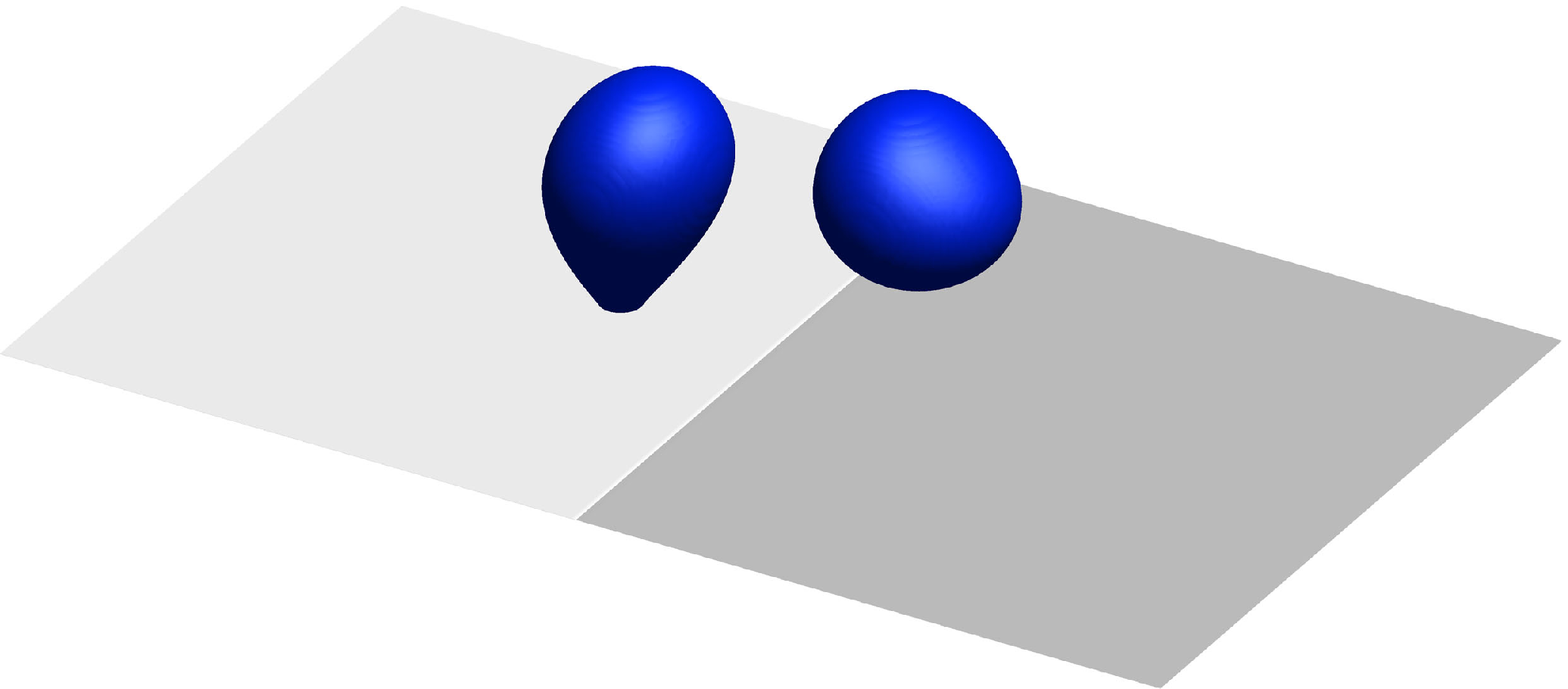}
			\caption*{$ L^{*}=1.0625 $}
		\end{minipage}

		\caption{The morphologies of double droplets impinging upon the heterogeneous surface under different droplet spacing $ L^{*} $  at $\Delta \theta = 20 \degree $ and $We=52.48$.} 
		\label{fig15}
	\end{figure}

Fig. \ref{fig15} shows the morphological evolution of the double impacting droplets with different spacing at $ \Delta \theta = 20 \degree $ and $ We=52.48 $.  Due to the same wettability difference and impact velocity of the double droplets, the initial spreading is exhibited similarly.  With the increase of droplet spacing, the contact and merging of two droplets takes longer, and the velocity of the droplet edge decreases gradually when the double droplets contact (see Fig. \ref{fig15}(b)). Finally, the droplet completely migrates to the right hydrophobic surface, and it can be concluded that there are three kind of  detachment patterns during the detachment process (see Fig. \ref{fig15}(f)).

\begin{figure}[H]
		\centering
		
	 	\begin{minipage}[c]{0.1\textwidth}
	 		\centering
			\caption*{(a) Pattern \textrm{I} }
		\end{minipage}
		\begin{minipage}[c]{0.25\textwidth}
			\includegraphics[width=\textwidth]{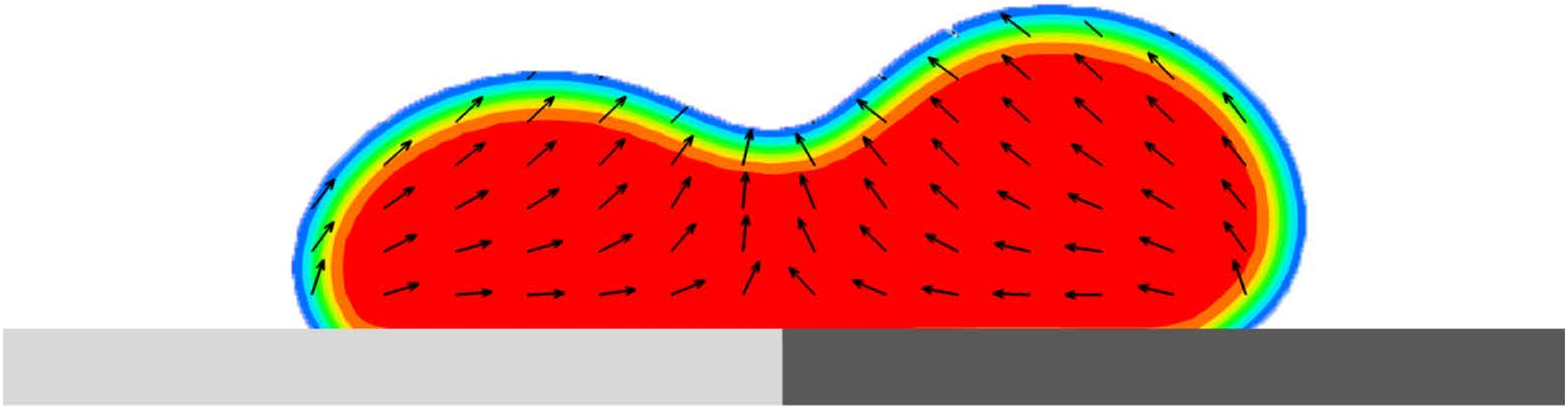}
		\end{minipage}
		\begin{minipage}[c]{0.25\textwidth}
			\includegraphics[width=\textwidth]{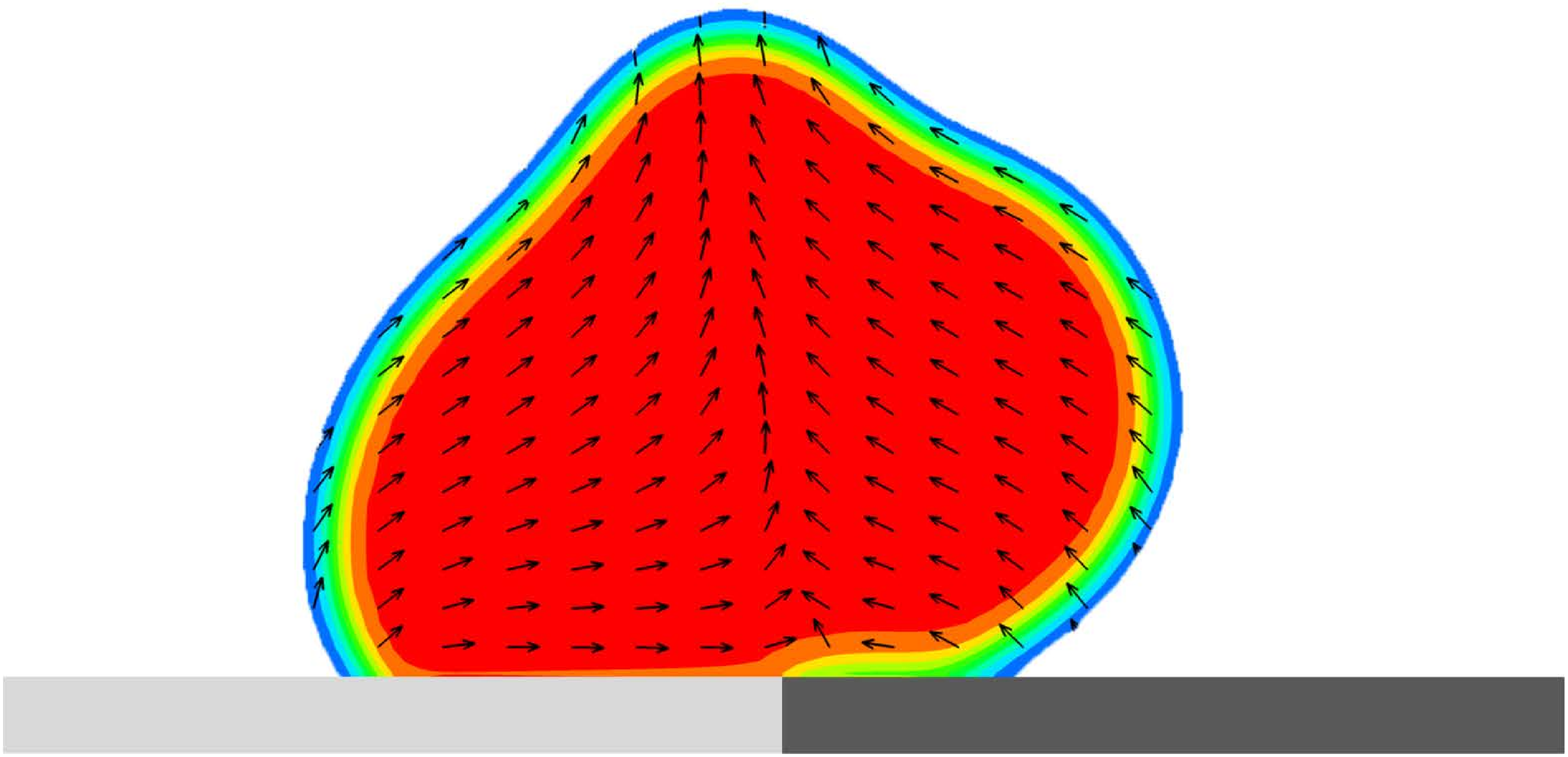}
		\end{minipage}
		\begin{minipage}[c]{0.25\textwidth}
			\includegraphics[width=\textwidth]{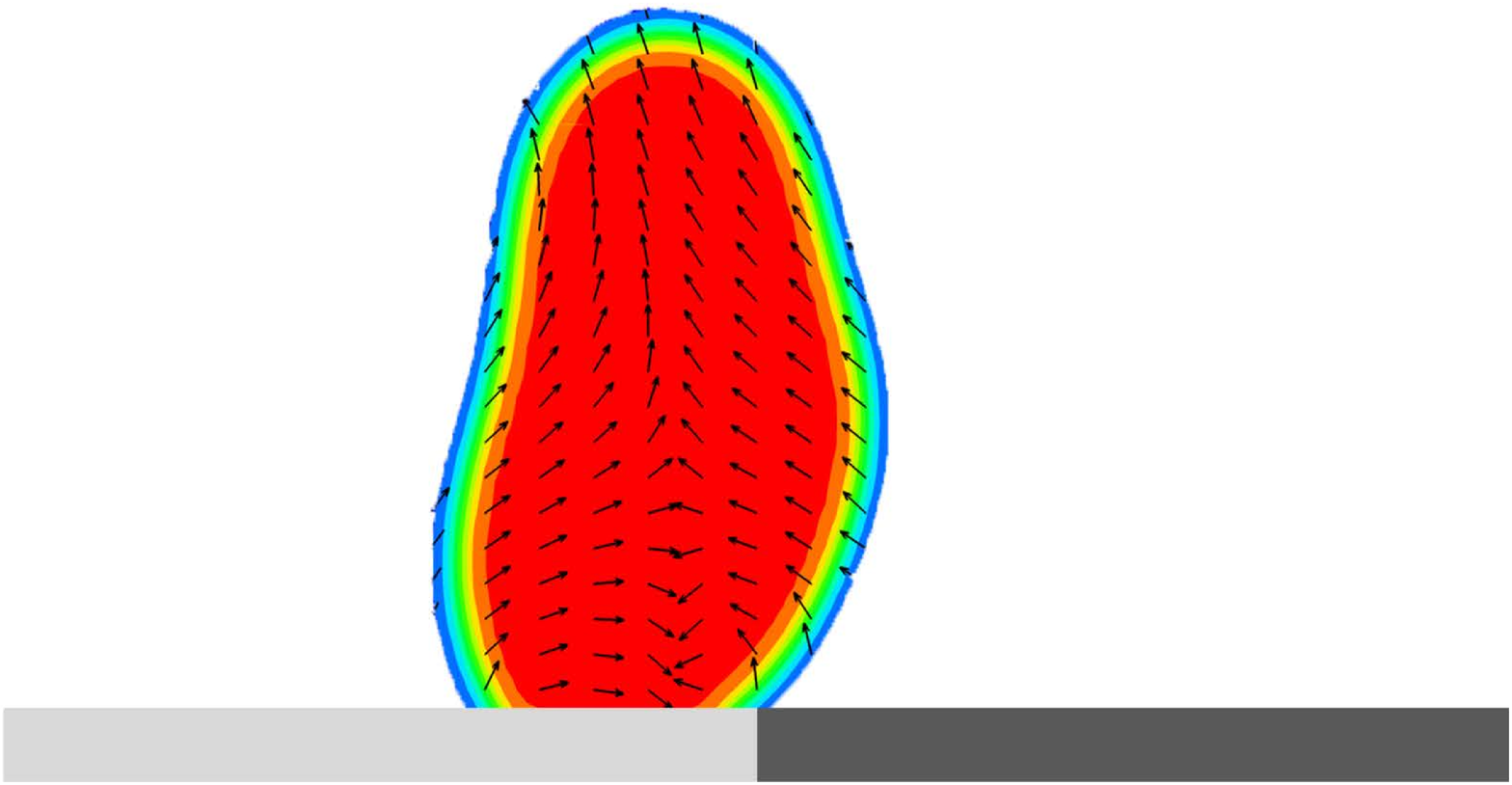}
		\end{minipage}

		\begin{minipage}[c]{0.1\textwidth}
			\centering
			\caption*{(b) Pattern \textrm{II}  }
		\end{minipage}
		\begin{minipage}[c]{0.25\textwidth}
			\includegraphics[width=\textwidth]{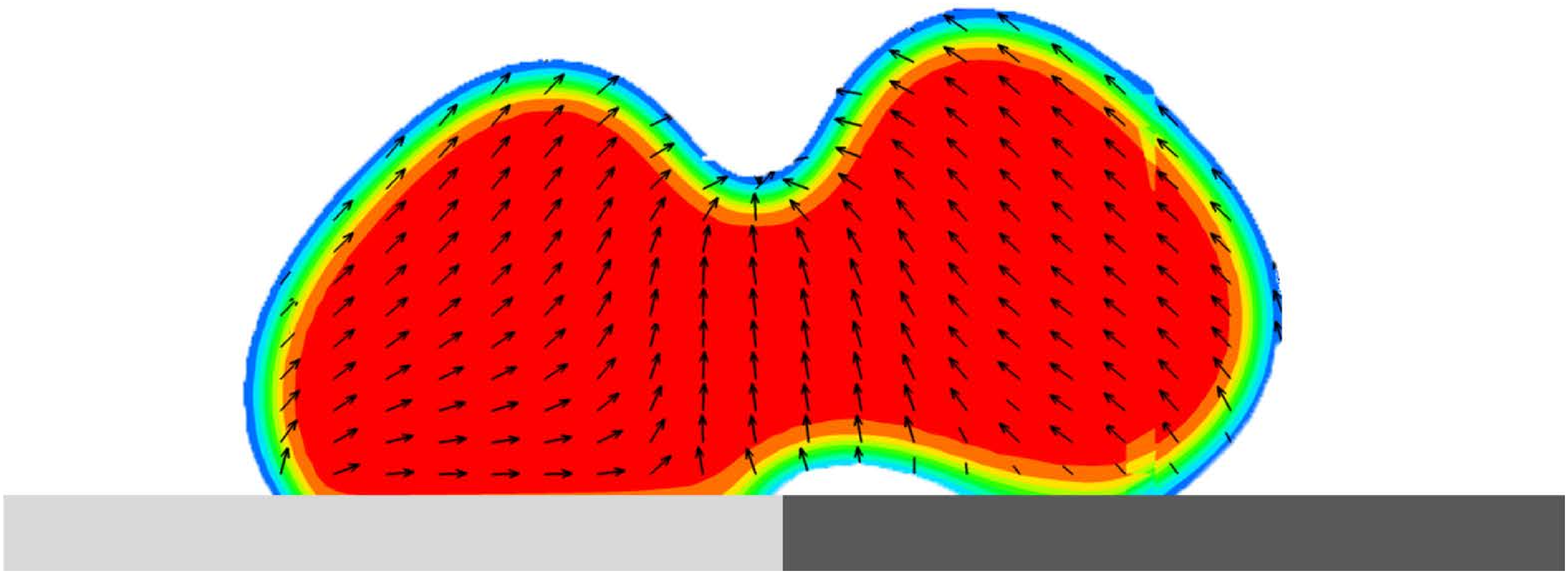}
		\end{minipage}
		\begin{minipage}[c]{0.25\textwidth}
			\includegraphics[width=\textwidth]{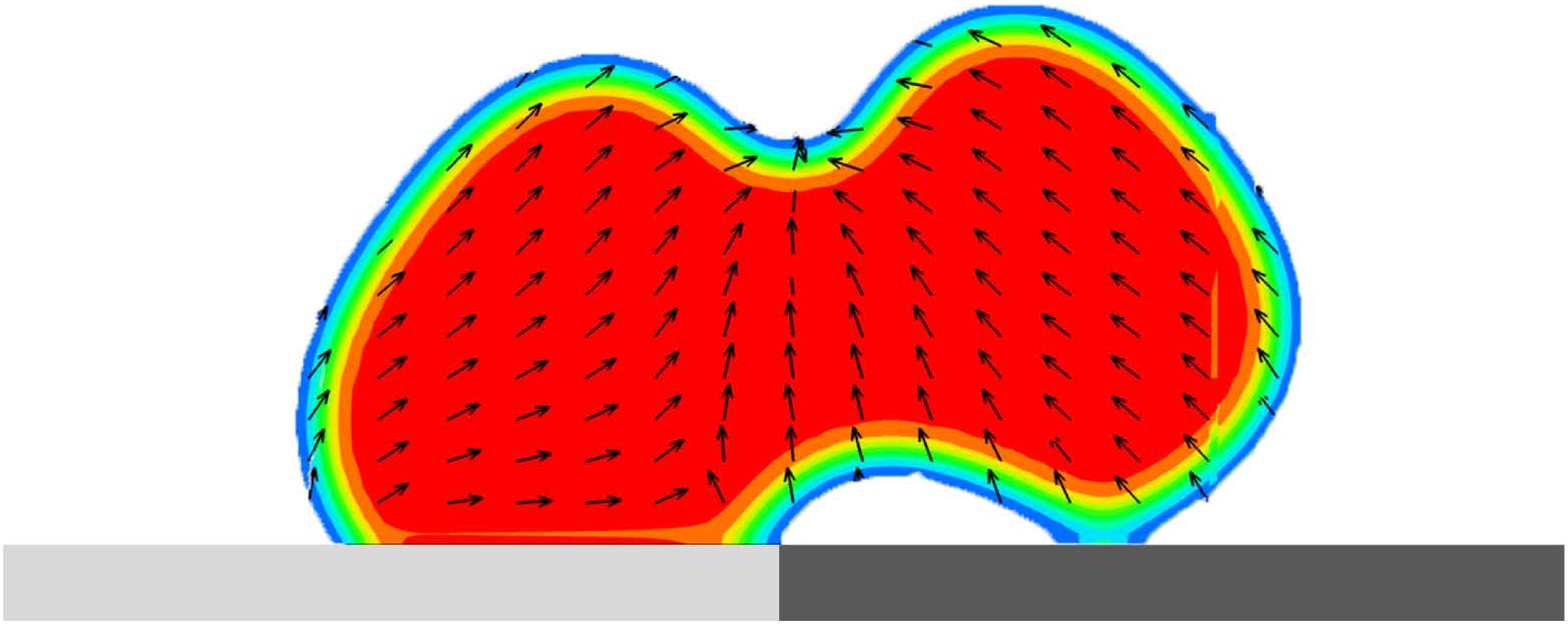}
		\end{minipage}
		\begin{minipage}[c]{0.25\textwidth}
			\includegraphics[width=\textwidth]{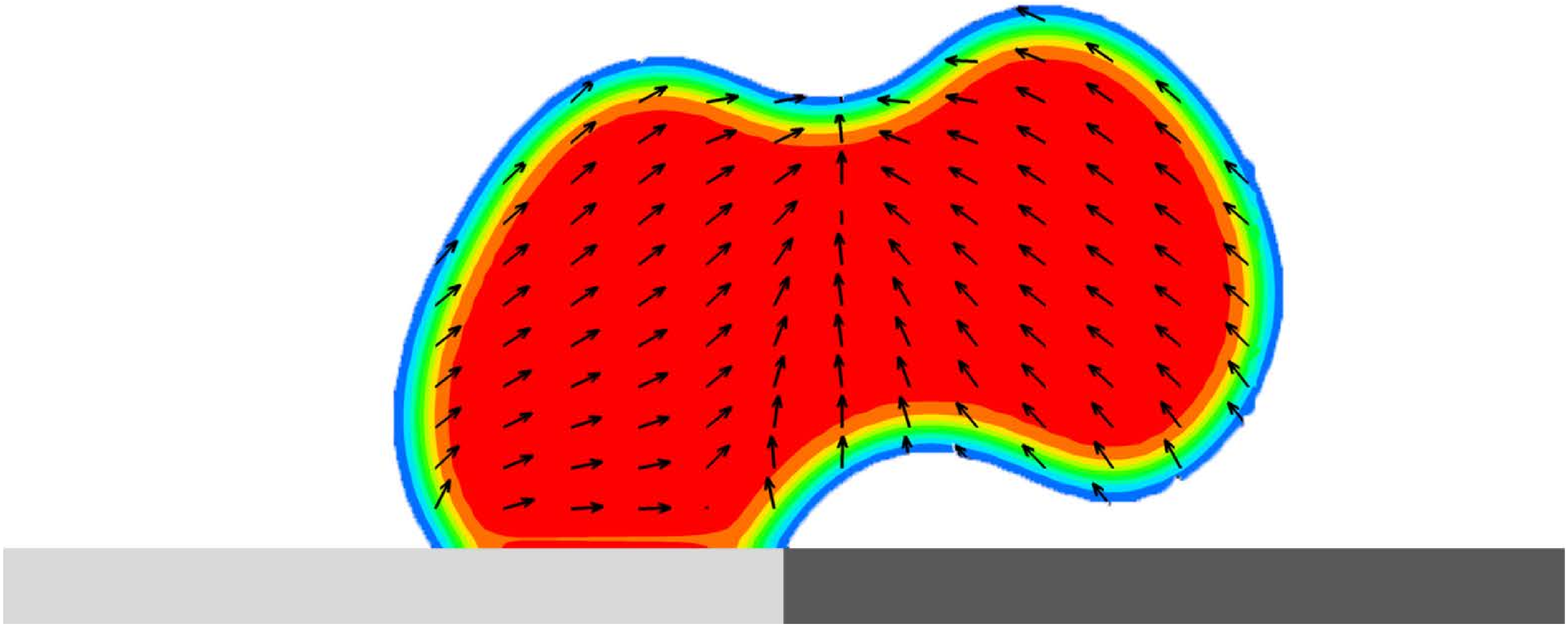}
		\end{minipage}

		\begin{minipage}[c]{0.1\textwidth}
			\centering
			\caption*{(c) Pattern \textrm{III}  }
		\end{minipage}
		\begin{minipage}[c]{0.25\textwidth}
			\includegraphics[width=\textwidth]{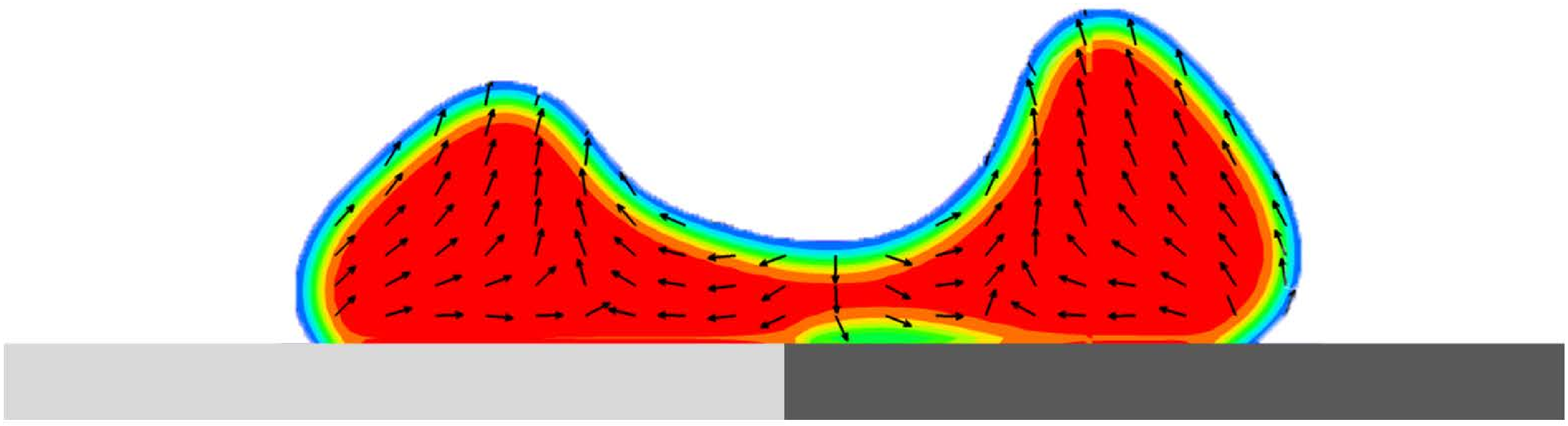}
		\end{minipage}
		\begin{minipage}[c]{0.25\textwidth}
			\includegraphics[width=\textwidth]{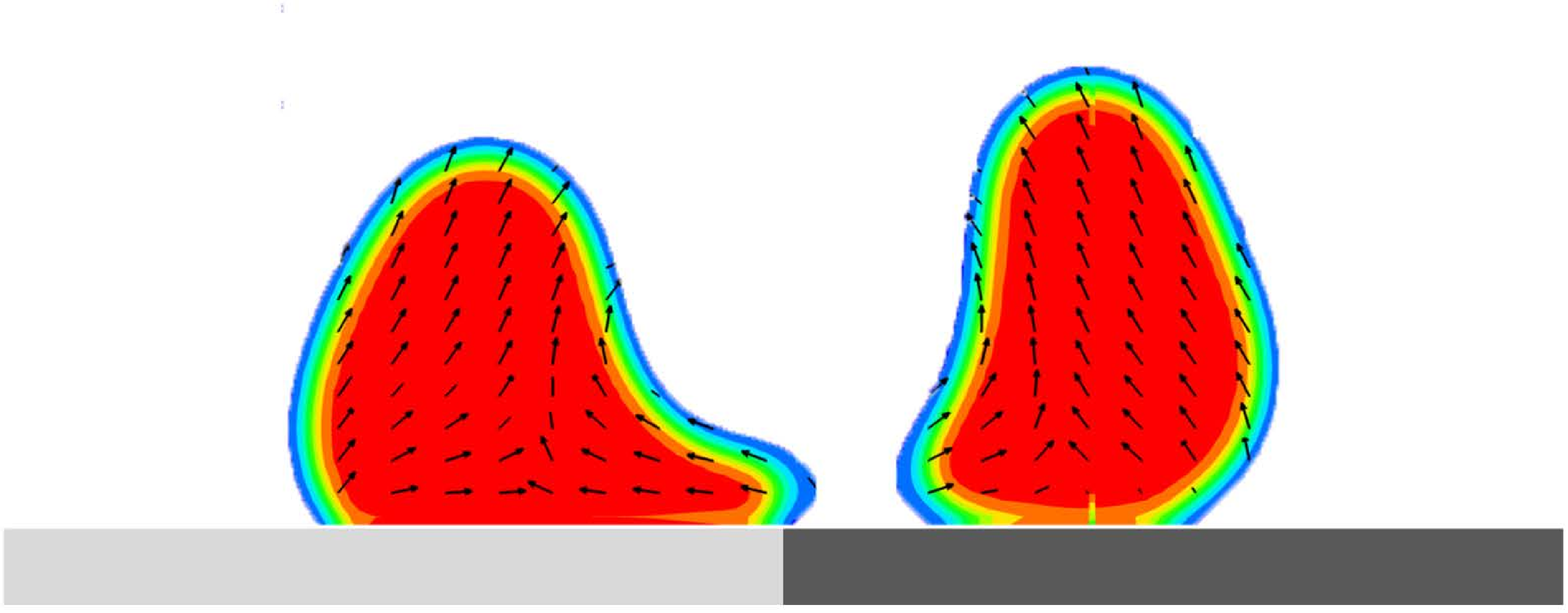}
		\end{minipage}
		\begin{minipage}[c]{0.25\textwidth}
	 		\includegraphics[width=\textwidth]{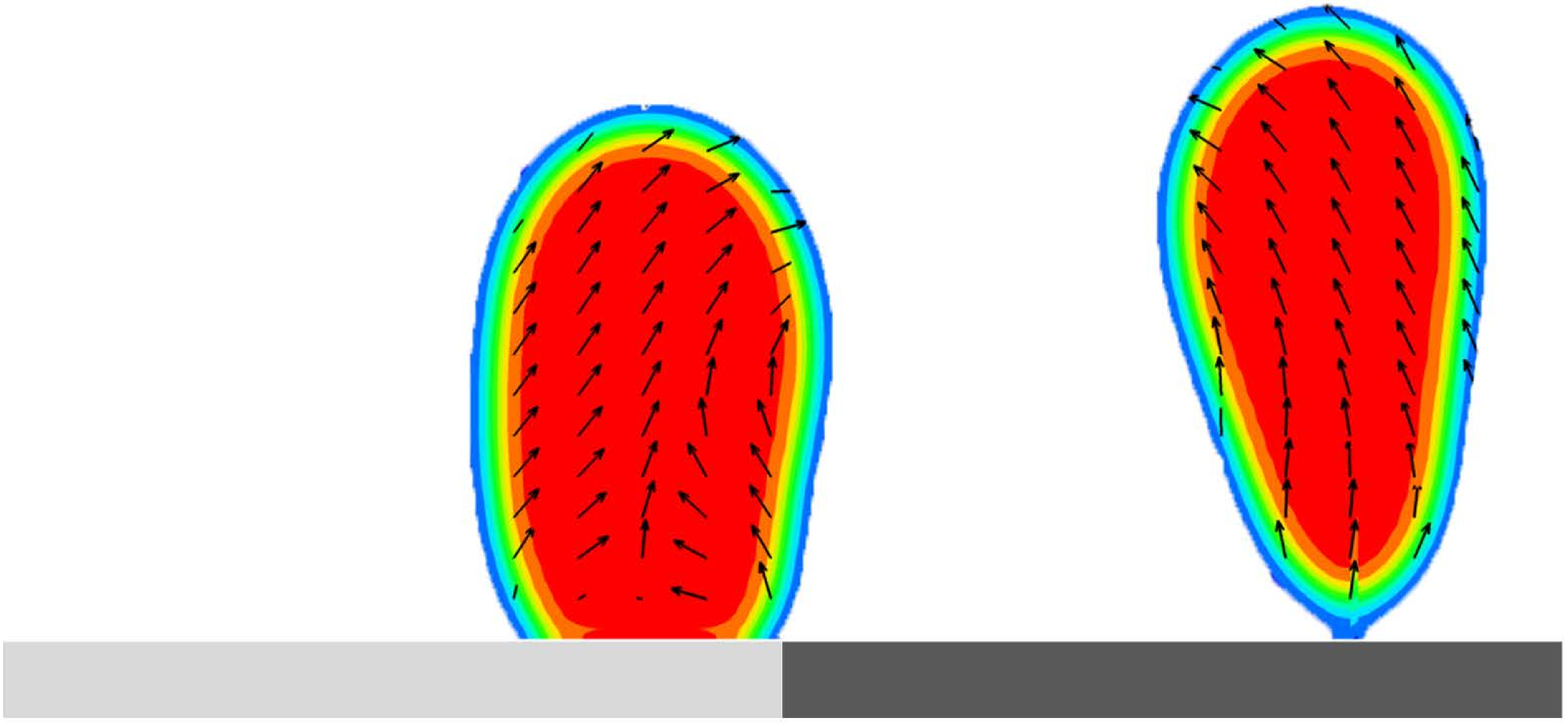}
	 	\end{minipage}

		\caption{Velocity field inside the droplet at different detachment patterns.} 
		\label{fig16}
	\end{figure}

These three detachment patterns snapshots are shown in Fig. \ref{fig16}. For a small horizontal spacing $L^{*}=0.75$, the earlier coalescence leads to the strong interaction between the two droplets. The opposite spreading flow drives the contact part of the two droplets spreading outward and upward to form an growing connecting ridge as seen in Fig. \ref{fig15}. The liquid will flow upward and form two obvious bulges on both sides of the borderline line. Subsequently, The two bulges gradually rise and close to each other as the liquid around the retraction point spreading upward. The droplets then coalesce to the new retraction point and gradually break away from the low-wetting surfaces. During the detachment process, the right three-phase contact line gradually contract to the borderline, which is denoted as pattern \textrm{I} as show in Fig. \ref{fig16}(a). It can be seen from the velocity vector distribution of the inside of the droplet that there is still a large rightward velocity vector near the hydrophobic substrate compared to the other two separation patterns, which is mainly due to the strong interaction force between the droplets. For $L^{*}=0.875$, the morphology evolution of droplet is similar to $L^{*}=0.75$. However, the liquid bridge connecting the bulges is longer, resulting in the two bulges not condensing into the middle to form a new retraction point in the contraction process. With the progress of contraction process, the velocity vector at the connecting ridge is basically vertical upward so that the air layer appears near the contact line of the hydrophobic substrate. As the air layer gradually expands to the right side, the right droplet is separated from the low wettability surface. This detachment pattern is named as pattern \textrm{II} as shown in Fig. \ref{fig16}(b). For a larger horizontal spacing $L^{*}=1.0625$, the edge velocity of the double droplets is smaller and the strength of coalescence becomes weaker, an elongated connecting ridge is formed between the two droplets and the droplets form a dumbbell shape. Subsequently, the liquid around the retraction point gathers and continuously increases as show in Fig. \ref{fig15}, resulting in the liquid bridge gradually stretching until it breaks into two sub-droplets, and the two sub-droplets rebound to the opposite side. For $L^{*}=1.0$, the two sub-droplets secondary consolidation is observed in \ref{fig15}(f). In addition, due to the high hydrophobicity of the right side, the breakpoint of the droplet is located in the right region and the right sub-droplet bounces higher and faster. The liquid bridge splits into two sub-droplets and the sub-droplets are detached from the low wetting substrate by rebounding. This detachment pattern is named pattern \textrm{III} as shown in Fig. \ref{fig16}(c).

	\begin{figure}[ht]
		\centering
		\includegraphics[width=0.5\textwidth]{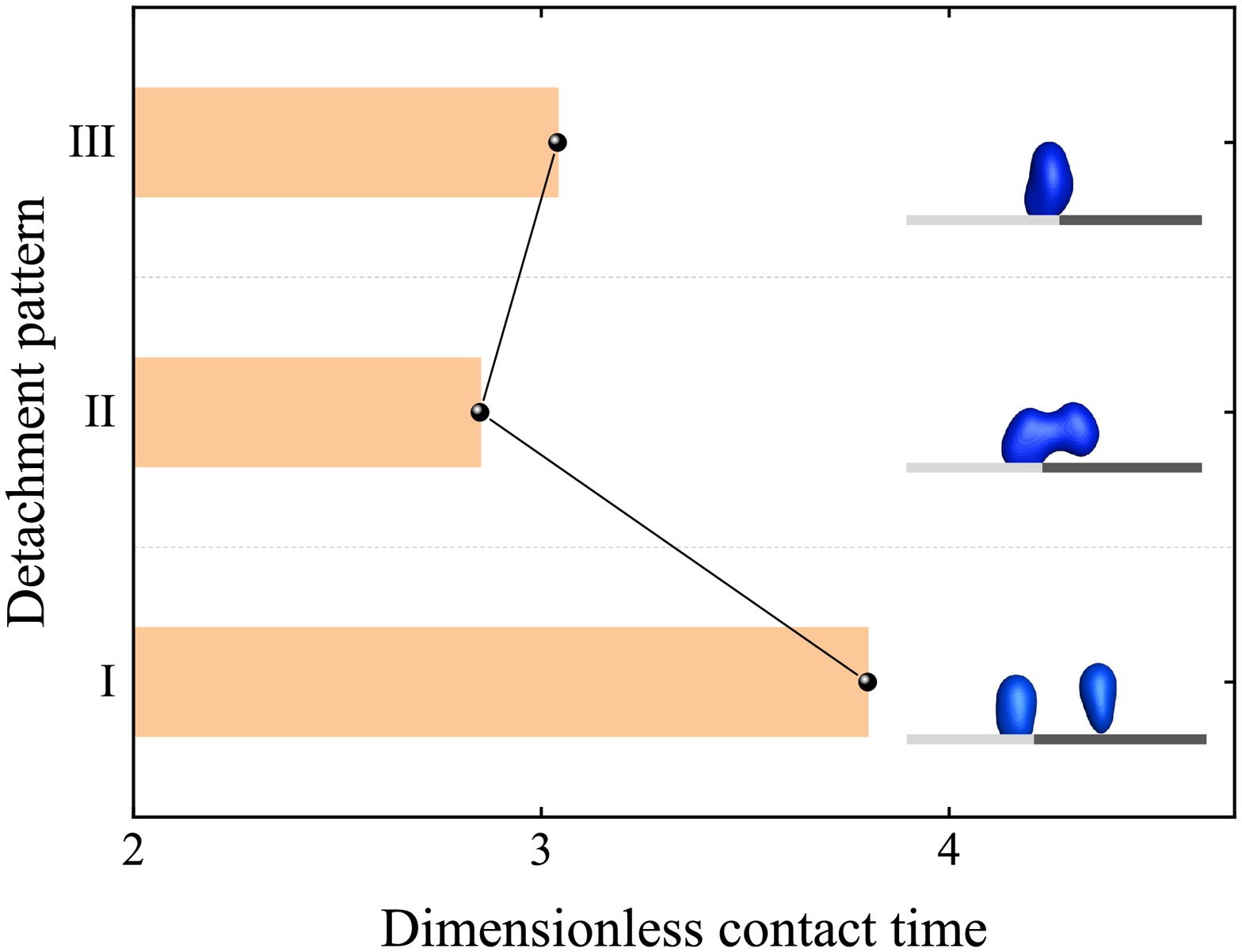}
		\caption{Dimensionless contact time for three different detachment patterns.}
		\label{fig17}
	\end{figure}

Fig. \ref{fig17} illustrates the contact time under the three detachment patterns. It can be seen that the contact time of the detachment pattern \textrm{II} is shorter than that of the other two patterns. It can be seen from Fig. \ref{fig16}(b) that the upward force exerted by the ridge leads to the gradual expansion of the air layer to the right, and only one foot contacts with the surface in the low wetting area. The high-momentum fluid rises continuously under the action of the upward force exerted by the formed ridges, while driving the lower fluid higher and migrating to the left side. These occurrences would significantly decrease the contact time. For the detachment pattern \textrm{I}, due to the small droplet spacing, the two contraction points of the droplet prematurely merge into a new contraction point, which causes the high momentum fluid to continuously stretch vertically upward the liquid. The three-phase contact line of droplet moves considerably slowly in the process of upward pulling. Moreover, due to the lack of a connecting ridge, the recoil effect of the left side. These facts lead to a longer contact time of pattern \textrm{I}.


\section{Conclusion}

In this study, the behaviors of double droplets simultaneously impacting the surface with wettability difference is numerically investigated using a three dimenisonal MRT pseudopotential LB model. After the verification of thermodynamic consistency, Laplace test and single droplet impacting on the homogeneous surface, the motion and deformation of the droplet on the inhomogeneous surface are widely simulated. We obtained different migration, splitting and rebound behavior by changing the wettability difference, Weber number and droplet spacing. The main conclusions are as follows.

Based on the present numerical results, we can conclude that the unbalanced external Young’s
force caused by the wettability difference of the surface will lead to the directional migration of the droplets towards the high wettability region and eventually stay in the high wettability region completely. In the process of contraction, the droplet is in an out-of-phase motion state. With the increase of the wettability difference, the recoil velocity of the droplet in the high-wettability side gradually decreases, and the hindering effect on the right part of the droplet is smaller, so that the contact time decreases with the increase of the wettability difference. In particular, the droplets migrate to the high-wettability surface and rebound directionally for $ \Delta \theta = 20 \degree $. Secondly, the initial kinetic energy of the droplet and the maximum spreading factor of the droplet gradually increased with the increase of the Weber number. The directional rebound of the droplet occurs for large Weber number, while the secondary spreading of the droplet occurs in the wetting stage for small Weber number. Since the occurrence of secondary spreading increases the contact time of droplets, the contact time of droplets decreases with the increase of Weber number. Furthermore, the phase diagram of droplet bounce reveals that the bounce tends to occur in the case of high Weber number and high hydrophobicity,  while the directional rebound occurs on the heterogeneous surface and the rebound direction on the homogeneous surface is vertical upward.  By changing the droplet spacing, we found that increasing the droplet spacing forms an elongated liquid bridge, which breaks in the asymmetric contraction process, resulting in the droplet splitting into two sub-droplets. However, the air layer appears during the droplet contraction for a small droplet spacing, which appears in the low wettability area near the boundary.  As the air layer expands, the right part of the droplet gradually breaks away from the high hydrophobicity area. In addition, there are three separation modes, and the contact time of the detachment pattern \textrm{II} is shorter than that of the other two patterns.

\section*{Conflict of interest}
	We declare that we have no financial and personal relationships with other people or organizations that can inappropriately influence our work, there is no professional or other personal interest of any nature or kind in any product, service and/or company that could be construed as influencing the position presented in, or the review of, the manuscript entitled.

\section*{Acknowledgements}
	This work is financially supported by the National Natural Science Foundation of China (Grant No. 12002320), and the Fundamental Research Funds for the Central Universities (Grant No. CUGGC05).


\end{document}